\documentclass[11pt]{article}
\usepackage[centertags]{amsmath}
\usepackage{amsfonts}
\usepackage{amssymb}
\usepackage{amsthm}
\usepackage{epsf}
\usepackage[]{cite}

\begin{document}
\begin{center}
$\,$\\[10mm]
{\Large \bf \noindent
Nonequilibrium Critical Phenomena and \\[2mm]
Phase Transitions into Absorbing States}\\[15mm]
Haye Hinrichsen
\footnote{email: {\tt hinrichs@comphys.uni-duisburg.de}}\\[5mm]
{\it \small Theoretische Physik, Fachbereich 10 \\
        Gerhard-Mercator-Universit\"at Duisburg \\
        D-47048 Duisburg, Germany} \\[20mm]
May 2000\\[13mm]
\end{center}
{\bf Abstract:} \\ This review addresses recent developments in
nonequilibrium statistical physics. Focusing on phase transitions
from fluctuating phases into absorbing states, the universality
class of directed percolation is investigated in detail. The
survey gives a general introduction to various lattice models of
directed percolation and studies their scaling properties,
field-theoretic aspects, numerical techniques, as well as possible
experimental realizations. In addition, several examples of
absorbing-state transitions which do not belong to the directed
percolation universality class will be discussed. As a closely
related technique, we investigate the concept of damage spreading.
It is shown that this technique is ambiguous to some extent,
making it impossible to define chaotic and regular phases in
stochastic nonequilibrium systems. Finally, we discuss various
classes of depinning transitions in models for interface growth
which are related to phase transitions into absorbing states.
\\[13mm]
{ \footnotesize  submitted to {\it Advances in Physics}} \\[5mm] 
% 05.70.Ln: nonequilibrium thermodynamics, irreversible processes
% 64.60.Ht: dynamic critical phenomena
% 64.60.Ak: percolation

{\newpage $\ $\\ \parskip 1.5mm \tableofcontents}
\newpage \normalsize \pagestyle{plain} \parskip 2mm

%==============================================================================
%               MACROS
%==============================================================================

% DEFINITIONS OF SYMBOLS:

\def\annh{\alpha}
\def\coupling{\mu}
\def\crit{\kappa}
\def\cutoff{{\Omega}}
\def\diff{D}
\def\noise{\zeta}
\def\disorder{\eta}
\def\namp{\Gamma}
\def\nupar{\nu_\parallel}
\def\nuperp{\nu_\perp}
\def\renorm{{\prime\prime}}
\def\scalarprod{\hspace{-1mm}\cdot\hspace{-1mm}}
\def\scalefac{\Lambda}
\def\smalldev{{l}}
\def\deviation{{\Delta}}
\def\timederivative{\partial_t}
\def\velocity{{v}}
\def\vacancy{\text{\O}}
\def\xvec{\text{{\bf x}}}
\def\sumstate{{\langle 1|}}

% ------ MACROS FOR SIMPLE HEADLINES AND SMALL FIGURE CAPTIONS-------

\def\headline#1{\vspace{3mm}\noindent{\bf #1}\\*[0.3mm] \noindent}
\def\smallcaption#1{\caption{\footnotesize #1}}
\renewcommand{\figurename}{\footnotesize Fig.}

% --------------------- FORMAT FOR APPENDICES -----------------------

\renewcommand{\appendix}{%
\renewcommand{\theequation}{\thesection.\arabic{equation}}%
\renewcommand{\section}{%
\setcounter{equation}{0}%
\secdef\Appendix\sAppendix}%
\setcounter{section}{0}%
\addcontentsline{toc}{section}{Appendices}{}%
\renewcommand{\thesection}{\Alph{section}}}

\newcommand{\Appendix}[2][?]{%
\refstepcounter{section}%
\addcontentsline{toc}{subsection}%
{\protect\numberline{\thesection} #1}%
{\flushleft\Large\bfseries\appendixname\ \thesection:\ #2\par}%
\sectionmark{#1}\vspace{\baselineskip}}

\newcommand{\sAppendix}[1]{%
{\flushleft\large\bfseries\appendixname\par\flushleft#1\par}%
\vspace{\baselineskip}}

%==============================================================================
\section{Introduction}
%============================================================= =================

Random behavior is a common feature of complex physical systems.
Although systems in nature generally evolve according to
well-known physical laws, it is in most cases impossible to
describe them by means of {\it ab initio} methods since details of
the microscopic dynamics are not fully known. Instead, it is often
a good approximation to assume that the individual degrees of
freedom behave randomly according to certain probabilistic rules.
For this reason methods of statistical mechanics become essential
in order to study the physical properties of complex systems. In
this approach a physical system is described by a reduced set of
dynamical variables while the remaining degrees of freedom are
considered as an effective noise with a certain postulated
distribution. The actual origin of the noise, which may be related
to chaotic motion, thermal interactions or even quantum-mechanical
fluctuations, is usually ignored. Thus, statistical mechanics
deals with stochastic models of systems that are much more
complicated in reality.

A complete description of a stochastic model is provided by the
probability distribution $P_t(s)$ to find the system at time $t$
in a certain configuration $s$. For systems at thermal equilibrium
this probability distribution is given by the stationary Gibbs
ensemble $P_{eq}(s) \sim e^{-{\cal H}(s)/k_BT}$, where ${\cal
H}(s)$ denotes the microscopic Hamiltonian~\cite{Gibbs02}. In
principle, the Gibbs ensemble allows us to compute the expectation
value of any time-independent observable by summing over all
accessible configurations of the system. However, in most cases it
is very difficult to perform the configurational sum. In fact,
although numerous exact solutions have been found~\cite{Baxter82},
the vast majority of stochastic models cannot yet be solved
exactly. In order to investigate such nonintegrable systems,
powerful approximation techniques such as series
expansions~\cite{Guttmann89} and renormalization group
methods~\cite{Amit84} have been developed. Thus, in equilibrium
statistical mechanics, we have a well-established theoretical
framework at our disposal.

From the physical point of view it is particularly interesting to
investigate stochastic systems in which the microscopic degrees of
freedom behave collectively over large
scales~\cite{Liggett85,Spohn91}. Collective behavior of this kind
is usually observed when the system undergoes a continuous phase
transition. The best known example is the order-disorder
transition in the two-dimensional Ising model, where the typical
size of ordered domains diverges when the critical temperature
$T_c$ is approached~\cite{McCoyWu73}. In most cases the emerging
long-range correlations are fully specified by the symmetry
properties of the model under consideration and do not depend on
details of the microscopic interactions. This allows phase
transitions to be categorized into different {\em universality
classes}. The notion of universality was originally introduced by
experimentalists in order to describe the observation that several
apparently unrelated physical systems may be characterized by the
same type of singular behavior near the transition. Since then
universality became a paradigm of the theory of equilibrium
critical phenomena. As the number of possible universality classes
seems to be limited, it would be an important theoretical task to
provide a complete classification scheme, similar to the periodic
table of elements. The most remarkable breakthrough in this
respect was the application of conformal field theory to
equilibrium critical phenomena~\cite{Polyakov70,Cardy87,Henkel99},
leading to a classification scheme of continuous phase transitions
in two dimensions.

\newpage
In nature, however, thermal equilibrium is rather an exception
than a rule. In most cases the temporal evolution starts out from
an initial state which is far away from equilibrium. The
relaxation of such a system towards its stationary state depends
on the specific dynamical properties and cannot be described
within the framework of equilibrium statistical mechanics. Instead
it is necessary to deal with a probabilistic model for the
microscopic dynamics of the system. Assuming certain transition
probabilities, the time-dependent probability distribution
$P_t(s)$ has to be derived from a differential equation, the
so-called Fokker-Planck or master equation. Nonequilibrium
phenomena are also encountered if an external current runs through
the system, keeping it away from thermal
equilibrium~\cite{SchmittmannZia95}. A simple example of such a
driven system is a resistor in an electric circuit. Although the
resistor eventually reaches a stationary state, its probability
distribution will no longer be given by the Gibbs ensemble. As a
physical consequence the thermal noise produced by the resistor is
no longer characterized by a Gaussian distribution. Similar
nonequilibrium phenomena are observed in catalytic reactions,
surface growth, and many other phenomena with a flow of energy or
particles through the system. Since nonequilibrium systems do not
require detailed balance, they exhibit a potentially richer
behavior than equilibrium systems. However, as their probability
distribution cannot be expressed solely in terms of an energy
functional ${\cal H}(s)$, the master equation has to be solved,
being usually a much more difficult task. Therefore, compared to
equilibrium statistical mechanics, the theoretical understanding
of nonequilibrium processes is still at its beginning.

The simplest nonequilibrium situation is encountered if a single
or several particles in a potential are subjected to a random
force. Important examples are the Kramers and the Smoluchowski
equations describing the evolution of the probability distribution
for Brownian motion of classical particles in an external field
(for a review see~\cite{Risken88}). But even more complicated
systems, for example one-dimensional tight-binding fermion systems
as well as electrical lines of random conductances or
capacitances, can be described in terms of discrete
single-particle equations~\cite{ABSO81}. A more complex situation,
on which we will focus in the present work, emerges in stochastic
lattice models with many interacting degrees of freedom. A
well-known example is the Glauber model~\cite{Glauber63} which
describes the spin relaxation of an Ising system towards the
stationary state. The corresponding master equation in one
dimension was solved exactly by Felderhof~\cite{Felderhof71}, who
mapped the time evolution operator onto a quantum spin chain
Hamiltonian that can be treated by similar methods as in
Ref.~\cite{LSM61}.

One motivation for today's interest in particle hopping models
originates in the study of superionic conductors in the
70's~\cite{Salamon79,DFP80}. In the superionic conductor AgI, for
example, the Ag$^+$ ions may be viewed as particles moving
stochastically through a lattice of I$^-$ ions. Each lattice site
can be occupied by at most one Ag$^+$ ion, i.e., the particles
obey an exclusion principle. It was observed experimentally that
the conductivity of AgI changes abruptly when the temperature is
increased, indicating an underlying order-disorder phase
transition of Ag$^+$ ions. Assuming short-range interactions, this
phase transition was explained in terms of a model for diffusing
particles on a lattice~\cite{KHKR98}. Subsequently, particle
hopping models have been generalized to so-called
reaction-diffusion models by including various types of particle
reactions or external driving forces~\cite{Kampen81}.  It should
be noted that particles in a reaction-diffusion model do not
always represent physical particles. Moreover, the reactions are
not always of chemical nature. For example, in models for traffic
flow individual cars are considered as interacting
particles~\cite{SSNI95}. Similarly, electronic excitations of
certain polymer chains may be viewed as particles subjected to a
stochastic temporal evolution~\cite{KFS93}.

The dynamic properties of a reaction-diffusion model on a lattice
are fully specified by its master
equation~\cite{KadanoffSwift68,Doi76,GrassbergerScheunert80}. In a
few cases it is possible to solve the master equation exactly.
During the last decade there has been an enormous progress in the
field of exactly solvable nonequilibrium processes. This
development was mainly triggered by the observation that the
Liouville operator of certain (1+1)-dimensional reaction-diffusion
models is related to Hamiltonians of previously known quantum spin
systems. For example, as first realized by Alexander and
Holstein~\cite{AlexanderHolstein78}, the symmetric exclusion
process can be mapped exactly onto the Schr\"odinger equation of a
Heisenberg ferromagnet. This type of mapping was extended to
various other one-dimensional reaction-diffusion processes by
Alcaraz {\it et~al.}~\cite{ADHR94}, allowing exact methods of
many-body quantum mechanics such as the Bethe ansatz and
free-fermion techniques to be applied in nonequilibrium
physics~\cite{Schuetz98b}. Moreover, novel algebraic techniques
have been developed in which the stationary state of certain
reaction-diffusion models is expressed in terms of products of
non-commuting algebraic objects~\cite{DerridaEvans97}.

In spite of this remarkable progress, the majority of
reaction-diffusion models cannot be solved exactly. It is
therefore necessary to use approximation techniques in order to
describe their essential properties. The oldest approximation
method is the law of mass action, where the reaction rate of two
reactants is assumed to be proportional to the product of their
concentrations. This mean field approach is justified if diffusive
mixing of particles is much stronger than the influence of
correlations produced by the reactions. Mean-field techniques have
been applied successfully to a large variety of reaction-diffusion
systems. The study of pattern formation in nonlinear
reaction-diffusion models, for example, is essentially based on a
mean-field approach~\cite{Mikhailov94}. However, as has already
been realized by Smoluchowski~\cite{Smoluchowski17}, fluctuations
may be extremely important in low-dimensional systems where the
diffusive mixing is not strong enough~\cite{Privman97}. For
example, if particles of one species diffuse and annihilate by the
reaction $A+A\rightarrow \vacancy$, the standard mean-field
approximation predicts an asymptotic decay of the particle
concentration as $\rho(t) \sim t^{-1}$. In one dimension, however,
the density is found to decay as $\rho(t) \sim t^{-1/2}$. This
slow decay is due to fluctuations produced by the dynamics,
leading to spatial anticorrelations of the particles. The
existence of such fluctuation effects has been confirmed
experimentally by measuring the luminescence of annihilating
excitons on polymer chains~\cite{KFS93}.

As in equilibrium statistical mechanics, nonequilibrium phenomena
are particularly interesting if the system undergoes a phase
transition, leading to a collective behavior of the particles over
long distances. There is a large variety of phenomenological
nonequilibrium phase transitions in nature, ranging from
morphological transitions of growing
surfaces~\cite{BarabasiStanley95} to traffic jams~\cite{WSB96}. It
turns out that the concept of {\em universality}, which has been
very successful in the field of equilibrium critical phenomena,
can be applied to nonequilibrium phase transitions as well.
However, the universality classes of nonequilibrium critical
phenomena are expected to be even more diverse as they are
governed by various symmetry properties of the evolution dynamics.
On the other hand, the experimental evidence for universality of
nonequilibrium phase transitions is still very poor, calling for
intensified experimental efforts.

In the present work we will focus on nonequilibrium phase
transitions in models with so-called absorbing states, i.e.,
configurations that can be reached by the dynamics but
cannot be left. The most important universality class of
absorbing-state transitions is {\em directed percolation}
(DP)~\cite{Kinzel83}. This type of transition occurs, for example,
in models for the spreading of an infectious disease. In these models
the lattice sites are considered as individuals which can be healthy
or infected. Infected individuals may either recover by themselves
or infect their nearest neighbors. Depending on the infection rate,
the spreading process may either survive or evolve into a passive
state where the infection is completely eliminated. In the limit
of large system sizes the two regimes of survival and extinction
are separated by a continuous phase transition. As in equilibrium
statistical mechanics, the critical behavior close to the
transition is characterized by diverging correlation lengths
associated with certain critical exponents. Similar spreading
processes with the same exponents can be observed in models for
catalytic reactions, percolation in porous media, and even in
certain hadronic interactions. It turns out that all these phase
transitions belong generically to a single universality class,
irrespective of microscopic details of their dynamic rules.
In view of its robustness, the DP class may therefore be as
important as the Ising universality class in equilibrium
statistical mechanics. Amazingly, directed percolation is one of
very few critical phenomena which cannot be solved exactly in one
spatial dimension. Although DP is easy to define, its critical
behavior is highly nontrivial. This is probably one of the reasons
why DP continues to fascinate theoretical physicists.

The present review addresses several aspects of nonequilibrium
phase transition\footnote{The review is based on a Habilitation
thesis submitted by the author to the Free University of Berlin in
June 1999.}. In the following Section we introduce elementary
concepts of nonequilibrium statistical mechanics such as the
master equation, reaction diffusion processes, Monte Carlo
simulations, as well as the most important analytical methods and
approximation techniques. The third Section discusses the problem
of directed percolation, including a comprehensive introduction to
DP lattice models, basic scaling concepts, approximation
techniques, as well as field-theoretic methods.  In view of the
robustness of DP, it is particularly interesting to search for
non-DP phase transitions which usually emerge in presence of
additional symmetries. These exceptional universality classes,
which have attracted considerable attention during the last few
years, will be reviewed in Sec.~4. Sec.~5 discusses a simulation
technique, called damage spreading, which has been used in the
past to search for chaotic behavior in random processes. It is
shown that this technique suffers of severe conceptual problems,
making it impossible to define chaotic phases. We also discuss the
critical behavior of damage spreading transitions which are
closely related to phase transitions into absorbing states.
Finally, we turn to depinning transitions in models of growing
interfaces which are related to nonequilibrium phase transitions
into absorbing states as well. As it is not intended to cover the
whole field of nonequilibrium critical phenomena, we will not
address various related topics such as self-organized critical
phenomena~\cite{BTW87}, modified reaction-diffusion
processes~\cite{Privman97}, the dynamics of reacting
fronts~\cite{BCH96}, driven diffusive
systems~\cite{SchmittmannZia95}, and recent results on spontaneous
symmetry breaking and phase separation in one-dimensional
systems~\cite{EKKM98}. For further reading we will give references
to related fields. Supplementary information concerning the
definition of tensor products, the derivation of the effective
action of Reggeon field theory, Wilson's shell integration, and
the one-loop integrals for DP are given in the
appendices~\ref{APPVECSEC}-\ref{INTEGRALS}. For easy reference we
also append a list of frequently used symbols and abbreviations.

\newpage
%##############################################################################
        \section{Stochastic many-particle systems}
%##############################################################################
%
In this Section we discuss elementary concepts of nonequilibrium
statistical mechanics. In order to introduce basic notions, we
first consider the example of a simple random walk. Turning to
many-particle systems we introduce the asymmetric exclusion
process which is a model for biased diffusion of many particles on
a one-dimensional line. Moreover, we explain the standard mean
field approach to reaction-diffusion processes. In order to
demonstrate the importance of fluctuations, two simple lattice
models with particle reactions will be discussed, namely
coagulation $2A\rightarrow A$ and pair annihilation $2A
\rightarrow \vacancy$. It turns out that in one dimension the
temporal evolution of these systems differs significantly from the
mean field prediction, proving that fluctuations may play an
important role. Furthermore, we review basic concepts of numerical
simulation techniques comparing different update schemes. Finally
we turn to certain analytical methods by which reaction diffusion
models can be solved exactly. In particular we discuss a recently
introduced algebraic technique which allows the stationary state
of certain nonequilibrium models to be expressed in terms of
products of noncommutative operators.

%==============================================================================
\subsection{The one-dimensional random walk}
%==============================================================================

In order to introduce basic concepts of nonequilibrium statistical
physics, let us first consider a simple symmetric random walk on a
one-dimensional line. The `configuration' of this dynamical system
at time $t$ is characterized by the position of the walker $s(t)$.
A random walk may be defined either on a continuous manifold or on
a lattice. If both position $s$ and time $t$ are discrete
variables, an unbiased random walk may be realized by the random
process
\begin{equation}
\label{SEM}
s(t+1) = s(t) + X(t) \ .
\end{equation}
In this expression $X(t)=\pm 1$ is a fluctuating random variable
with correlations
\begin{equation}
\label{SimpleRandomWalk}
\langle X(t) \rangle = 0 \ , \qquad
\langle X(t) X(t^\prime) \rangle = \delta_{t,t^{\prime}}
\ ,
\end{equation}
where $\langle \ldots \rangle$ denotes the average over many
realizations of randomness.
While the individual space-time trajectory of a random walker
is not predictable, the probability distribution $P_t(s)$ to
find the walker after $t$ time steps
at position $s$ evolves deterministically
according to the so-called {\em master equation}
\begin{equation}
\label{ME}
P_{t+1}(s) = \frac12 \Bigl(P_t(s-1) + P_t(s+1)\Bigr) \ .
\end{equation}
Assuming the particle to be initially located at the origin
$P_0(s)=\delta_{s,0}$, this difference equation is solved by
\begin{equation}
P_t(s) =
\begin{cases}
\frac{1}{2^t}\binom{t}{(t+s)/2} & \text{ if $t+s$ even, } \\
0          & \text{ if $t+s$ odd. }
\end{cases}
\end{equation}
If space and time are continuous, the motion of a random walker
may be described by a stochastic {\em Langevin equation}
\begin{equation}
\timederivative s(t) = \noise(t) \ ,
\end{equation}
where, according to the central limit theorem, $\noise(t)$ is
a Gaussian white noise with zero mean and correlations
$\langle \noise(t) \noise(t^\prime) \rangle = \namp \delta(t-t^\prime)$.
The Langevin equation may be regarded as a continuum
version of Eq.~(\ref{SimpleRandomWalk}).
Starting from the origin $s(0)=0$ the mean square
displacement of the random walker grows as
$\langle s^2(t) \rangle = \int_0^tdt_1\int_0^tdt_2
\langle \noise(t_1)\noise(t_2) \rangle = \namp t$. The resulting
probability distribution
\begin{equation}
P_t(s)=\frac{1}{2 \pi \namp t} e^{-s^2/(2\namp t)}
\end{equation}
is a solution of the {\em Fokker-Planck} equation~\cite{Risken88}
\begin{equation}
\timederivative P_t(s) = \frac{\namp}{2} \partial_s^2 \, P_t(s)
\end{equation}
which can be seen as a variant of the master equation
in a continuum~(\ref{ME}).

The example of a random walk is particularly simple as it involves
only one degree of freedom. In order to describe systems with many
particles, it would seem natural to introduce several degrees of
freedom $s_1,s_2,\ldots$, where $s_n$ denotes the position of the
$n$-th particle. However, this approach is restricted to systems
with a conserved number of particles. For systems with
non-conserved particle number it is more convenient to introduce
local degrees of freedom for the number of particles located at 
certain positions in space.

%==============================================================================
\subsection{The master equation}
%==============================================================================
%
%
\label{MASTER} Stochastic systems with many particles are usually
defined on a $d$-dimensio\-nal Euclidean manifold representing the
physical `space'. Attached to this manifold are local degrees of
freedom characterizing the configuration of the system.
Depending on whether the spatial manifold is continuous or
discrete, the local degrees of freedom are introduced as
continuous fields or local variables residing at the lattice
sites. Furthermore, a time coordinate $t$ is introduced which may
be interpreted as an additional dimension of the system.
Therefore, stochastic models are said to be defined in $d+1$
dimensions. Since $t$ may be continuous or discontinuous, we have
to distinguish between models with asynchronous and synchronous
dynamics.

%--------------------------------------------------------
\headline{Asynchronous dynamics}
%--------------------------------------------------------
%
Stochastic models with continuous time evolve by {\em asynchronous
dynamics}, i.e., transitions from a state $s$ into another state
$s^\prime$ occur spontaneously at a given rate $w_{s \rightarrow
s^\prime}\geq 0$ per unit time. It can be shown that in the limit
of very large systems sizes the temporal evolution of the
probability distribution $P_t(s)$ evolves deterministically
according to a master equation with appropriate initial
conditions~\cite{KadanoffSwift68,Doi76,GrassbergerScheunert80}.
The master equation is a linear partial differential equation
describing the flow of probability into and away from a
configuration~$s$:
\begin{equation}
\label{MasterEquation}
\frac{\partial}{\partial t} P_t(s) =
\underbrace
{\sum_{s^\prime} w_{s^\prime \rightarrow s} P_t(s^\prime)}_{\text{gain}} -
\underbrace
{\sum_{s^\prime} w_{s \rightarrow s^\prime} P_t(s)}_{\text{loss}} \ .
\end{equation}
The gain and loss terms balance one another so that the
normalization $\sum_s P_t(s)=1$ is conserved. Since the temporal
change of $P_t(s)$ is fully determined by the actual probability
distribution at time $t$, the master equation describes a Markov
process, i.e., it has no intrinsic memory. Moreover, it is
important to note that the coefficients $w_{s \rightarrow
s^\prime}$ are {\em rates} rather than probabilities. Thus, they
may be larger than $1$ and can be rescaled by changing the time
scale.

Using a vector notation (see Appendix~\ref{APPVECSEC})
the master equation (\ref{MasterEquation}) may be written as
\begin{equation}
\label{CompactNotation}
\timederivative |P_t\rangle = - {\cal L} |P_t\rangle \ ,
\end{equation}
where $|P_t\rangle$ denotes a vector whose components are the
probabilities $P_t(s)$. The Liouville operator ${\cal L}$
generates the temporal evolution and is defined through the matrix
elements
\begin{equation}
\langle s^\prime|{\cal L}|s\rangle =
-w_{s \rightarrow s^\prime} + \delta_{s,s^\prime}
\sum_{s^{\prime\prime}} w_{s \rightarrow s^{\prime\prime}}
 \ .
\end{equation}
A formal solution of the master equation is given by
$|P_t\rangle=\exp(-{\cal L}t)|P_0\rangle$,
where $|P_0\rangle$ denotes the initial probability distribution.
Therefore, in order to determine $|P_t\rangle$, the Liouville operator
has to be diagonalized which is usually a nontrivial task.

Apart from very few exceptions, stochastic processes are
irreversible and therefore not invariant under time reversal.
Hence the Liouville operator ${\cal L}$ is generally
non-hermitean. Moreover, it may have complex conjugate
eigenvalues, indicating oscillatory behavior. Oscillating modes
are not only a mathematical artifact, but can be observed
experimentally in certain chemical reactions such as the
Belousov-Zhabotinski reaction~\cite{Scott94}. Due to the
positivity of rates, the real part of all eigenvalues is
nonnegative, i.e., the amplitude of excited eigenmodes decays
exponentially in time. The spectrum of the Liouville operator
includes at least one zero mode ${\cal L}|P_s\rangle=0$,
representing the stationary state of the system. Moreover,
probability conservation can be expressed as $\sumstate
P_t\rangle=1$, where $\sumstate$ denotes the sum vector over all
configurations (cf. Appendix~\ref{APPVECSEC}). Consequently the
Liouville operator obeys the equation $\sumstate {\cal L}=0$,
i.e., the sum over each column of ${\cal L}$ vanishes.

%--------------------------------------------------------
\headline{Synchronous dynamics}
%--------------------------------------------------------
%
If the time variable $t$ is a {\em discrete} quantity, the
model evolves by synchronous dynamics, i.e.,
all lattice sites are simultaneously updated
according to certain transition probabilities
$p_{s \rightarrow s^\prime} \in [0,1]$. The corresponding
master equation is a linear recurrence relation
\begin{equation}
\label{ParallelEquation}
P_{t+1}(s) = P_{t}(s) +
\underbrace
{\sum_{s^\prime} p_{s^\prime \rightarrow s} P_t(s^\prime)}_{\text{gain}} -
\underbrace
{\sum_{s^\prime} p_{s \rightarrow s^\prime} P_t(s)}_{\text{loss}} \ ,
\end{equation}
which can be written in a compact form as a linear map
\begin{equation}
\label{TransferMatrix}
|P_{t+1}\rangle = {\cal T} |P_t\rangle \ ,
\end{equation}
where  ${\cal T}$ is the so-called transfer matrix. A formal
solution is given by $|P_{t}\rangle = {\cal T}^{\, t}
|P_0\rangle$. As can be verified easily, the conservation of
probability $\sumstate P_t\rangle=1$ implies that $\sumstate {\cal
T}=\langle 1|$, i.e., the sum over each column of the transfer
matrix is equal to $1$.

There has been a long debate which of the two update schemes is
the more `realistic' one. For many researchers models with
uncorrelated spontaneous updates appear to be more `natural' than
models with synchronous dynamics where all particles move
simultaneously according to an artificial clock cycle. On the
other hand, many computational physicists prefer stochastic
cellular automata with synchronous dynamics since they can be
implemented efficiently on parallel computers. However, as a
matter of fact, in both cases the dynamic rules are simplified
descriptions of a much more complex physical process. Therefore,
it would be misleading to consider one of the two variants as
being more `natural' than the other. Instead, the choice of the
dynamic procedure should depend on the specific physical system
under consideration. Very often both variants display essentially
the same physical properties. In some cases, however, they lead to
different results. For example, models for traffic flow with
synchronous updates turn out to be more realistic than random
sequential ones. Another example is polynuclear growth (see
Sec.~\ref{PNGSECTION}), where a roughening transition occurs only
when synchronous updates are used.

%==============================================================================
\subsection{Diffusion of many particles: The asymmetric exclusion process}
%==============================================================================

%
%
\begin{figure}
\epsfxsize=125mm
\centerline{\epsffile{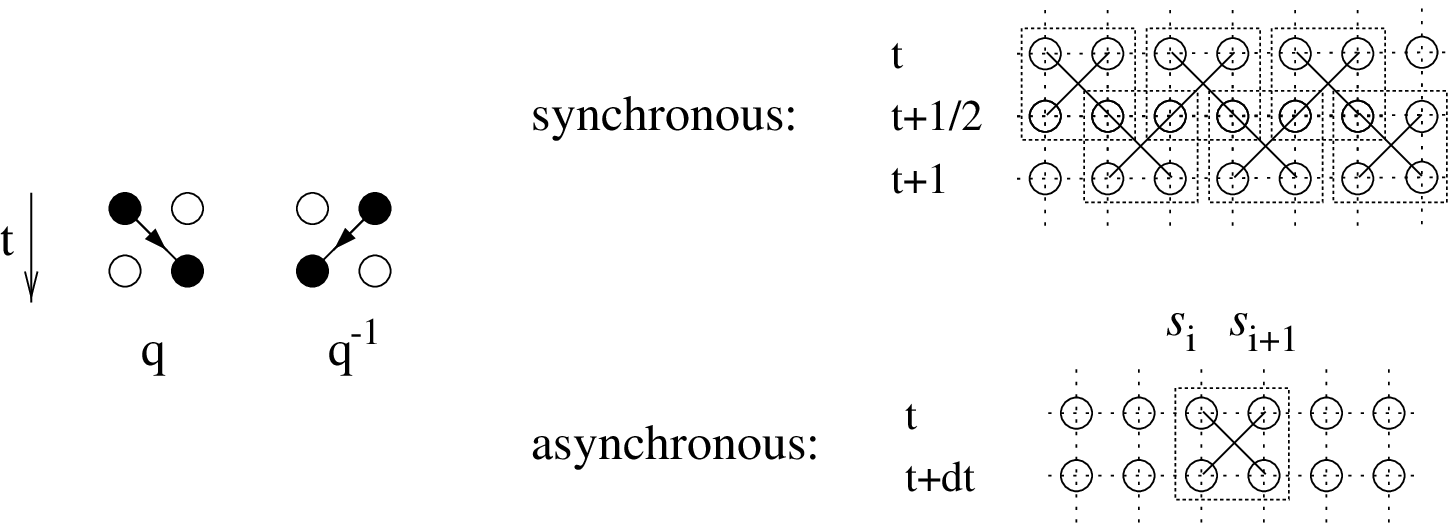}}
\smallcaption{
\label{FIGEXCLUSION}
The partially asymmetric exclusion process in one dimension
with synchronous (sublattice-parallel) and
asynchronous (random-sequential) dynamics.
}
\end{figure}

One of the simplest stochastic many-particle models is the partially
asymmetric exclusion process on a one-dimensional chain with $N$
sites~\cite{Liggett85}. In this model hard-core particles move
randomly to the right (left) at rate $q$ ($q^{-1}$).
An exclusion principle is imposed, i.e., each lattice site may
be occupied by at most one particle. Therefore, attempted moves
are rejected if the target site is already occupied.
In the following we assume closed
boundary conditions, i.e., particles cannot leave or enter
the system at the ends of the chain. The configuration
$s=\{s_1,s_2,\ldots,s_N\}$ of the system
is given in terms of local
variables $s_i$, indicating presence ($s_i=1$) or
absence ($s_i=0$) of a particle at site $i$.

%--------------------------------------------------------
\headline{Exclusion process with asynchronous dynamics}
%--------------------------------------------------------
%
Let us first consider the exclusion process with asynchronous
dynamics. In this case a pair of sites $i$ and $i+1$ is randomly
selected. If only one of the two sites is occupied, the particle
moves with probability $q/(q+q^{-1})$ to the right and with
probability $q^{-1}/(q+q^{-1})$ to the left, as shown in
Fig.~\ref{FIGEXCLUSION}. Each update attempt corresponds to a time
increment of $1/(N(q+q^{-1}))$. Thus, the transition rates are
defined by
\begin{equation}
\begin{split}
w_{s \rightarrow s^\prime}=
\sum_{i=1}^{N-1}
\Bigl( \prod_{j=1}^{i-1} \delta_{s_j,s_j^\prime}\Bigr)
\Bigl( \prod_{j=i+2}^{N} \delta_{s_j,s_j^\prime}\Bigr)
\Bigl(& \
q \,\delta_{s_i,1}\,\delta_{s_{i+1},0}\,
    \delta_{s_i^\prime,0}\,\delta_{s_{i+1}^\prime,1}
+\\& \, \,
q^{-1}\,\delta_{s_i,0}\,\delta_{s_{i+1},1}\,
   \delta_{s_i^\prime,1}\,\delta_{s_{i+1}^\prime,0}
\Bigr)
\ .
\end{split}
\end{equation}
The corresponding Liouville operator can be written as
\begin{equation}
{\cal L} = \sum_{i=1}^{N-1} {\bf 1} \otimes {\bf 1} \otimes \ldots
\underbrace{\otimes {\cal L}_i \otimes}_{i\text{-th position}}
\ldots \otimes {\bf 1} =: \sum_{i=1}^{N-1} {\cal L}_{i}\,,
\end{equation}
where ${\bf 1}$ denotes a $2\times 2$ unit matrix and
${\cal L}_{i}$ is a $4\times 4$ matrix generating
particle hopping between sites $i$ and $i+1$.
In the standard basis (see Appendix~\ref{APPVECSEC})
this matrix is given by
\begin{equation}
\label{ExclusionProcessGenerator}
{\cal L}_{i}=
\begin{pmatrix}
0 & 0 & 0 & 0 \\
0 & q^{-1} & -q & 0 \\
0 & -q^{-1} & q & 0 \\
0 & 0 & 0 & 0
\end{pmatrix} \ .
\end{equation}
As can be verified easily, a stationary state of the system
is given (up to normalization) by the tensor product
\begin{equation}
\label{stat}
|P_s\rangle =
\begin{pmatrix} 1 \\ 1 \end{pmatrix} \otimes
\begin{pmatrix} q^{-1} \\ q \end{pmatrix} \otimes
\ldots \otimes
\begin{pmatrix} q^{-N} \\ q^N \end{pmatrix}
= \bigotimes_{j=1}^N \begin{pmatrix} q^{-j} \\ q^j \end{pmatrix} \ .
\end{equation}
Since the vector $|P_s\rangle$ can be written as a tensor product,
the local variables $s_i$ are completely uncorrelated. Such a
state is said to have a {\em product measure}.

As the total number of particles is conserved in the asymmetric
exclusion process, the dynamics decomposes into independent {\em
sectors}. In fact, the Liouville operator commutes with the
particle number operator
\begin{equation}
M=\sum_{i=1}^N m_i\,, \qquad
m_i=\begin{pmatrix}
1 & 0 \\
0 & 0\end{pmatrix} \, .
\end{equation}
Obviously the vector $|P_s\rangle$ is a superposition of solutions
belonging to different sectors, i.e., it represents a whole {\it
ensemble} of stationary states. To obtain a physically meaningful
solution for a given number of particles, the vector $|P_s\rangle$
has to be projected onto the corresponding sector. In this sector
the stationary system evolves through certain configurations
with specific weights given by the normalized components 
of the projected vector.

%--------------------------------------------------------
\headline{Exclusion process with synchronous dynamics}
%--------------------------------------------------------
%
The asymmetric exclusion process with synchronous updates may be
realized by introducing two half time steps. In the first half
time step the odd sublattice is updated whereas the
even sublattice is updated in the second half time step (see
Fig.~\ref{FIGEXCLUSION}). Note that the use of sublattice-parallel 
updates admits local dynamic rules\footnote{In the exclusion
process with fully parallel updates local moves may overlap,
leading to subtle long-range correlations (see
Ref.~\cite{ERS99}).}. Assuming the number of sites $N$ to be
odd, the corresponding transfer matrix reads
\begin{equation}
{\cal T}=({\cal T}_2 \otimes {\cal T}_4 \otimes {\cal T}_6
\otimes \ldots \otimes {\cal T}_{N-1}) \,
({\cal T}_1 \otimes {\cal T}_3 \otimes {\cal T}_5 \otimes \ldots
\otimes {\cal T}_{N-2}) \ ,
\end{equation}
where
\begin{equation}
{\cal T}_{i}= \frac{1}{q+q^{-1}}
\begin{pmatrix}
q+q^{-1} & 0 & 0 & 0 \\
0 & q & q & 0 \\
0 & q^{-1} & q^{-1} & 0 \\
0 & 0 & 0 & q+q^{-1}
\end{pmatrix}
\end{equation}
is the local hopping matrix.
Again the product state~(\ref{stat}) is
a stationary eigenvector of the transfer matrix.
Thus, both the asynchronous and the synchronous exclusion
process have exactly the same stationary properties.

%--------------------------------------------------------
\headline{Asymmetric diffusion in a continuum}
%--------------------------------------------------------
%
Let us finally turn to asymmetric diffusion on a continuous
manifold. In principle it would be possible to trace 
trajectories of individual particles. However, it is much more
convenient to characterize the state of the system by a
density field $\rho(\xvec,t)$, rendering the coarse-grained density
of particles at position $\xvec$ at time $t$. The Langevin
equation of such a system may be written as
\begin{equation}
\timederivative \rho(\xvec,t) = D[\rho(\xvec,t)] + U[\rho (\xvec,t)] +
\noise(\xvec,t)\,,
\end{equation}
where $D$ is a sum of linear differential operators describing
spatial couplings, $U$ a potential for on-site particle
interactions, and $\noise(\xvec,t)$ a noise term taking the
stochastic nature of Brownian motion into account. Since particles
do not interact in the present case, the potential $U$ vanishes.
Moreover, as will be shown below, the noise $\noise(\xvec,t)$ is
irrelevant on large scales and can be neglected. Thus, the
resulting Langevin equation reads
\begin{equation}
\label{DiffusionEquation} \timederivative \rho(\xvec,t) =
\frac{q+q^{-1}}{2} \partial_x^2 \rho(\xvec,t) + \frac{q-q^{-1}}{2}
\partial_x \rho(\xvec,t) \ .
\end{equation}
The second term describes the bias of the diffusive motion which may
be eliminated in a co-moving frame. Notice that this equation is
linear and does not incorporate the exclusion principle. 
In a co-moving frame it reduces to the ordinary diffusion equation.

The diffusion equation provides a simple example of dynamic
scaling invariance. As can be verified easily, the equation
$\timederivative\rho(\xvec,t) = D \nabla_x^2 \rho(\xvec,t)$ is
invariant under rescaling of space {\em and} time
\begin{equation}
\label{SimpleRescaling}
\xvec \rightarrow \scalefac \xvec\,, \qquad
t \rightarrow \scalefac^z t \,,
\end{equation}
where $z$ is the so-called {\em dynamic exponent}. Since space
and time are different in nature, the exponent $z$ is
usually larger than $1$. The value $z=2$ indicates diffusive
behavior.

%==============================================================================
\subsection{Reaction-diffusion processes}
%==============================================================================

\begin{figure}
\epsfxsize=120mm \centerline{\epsffile{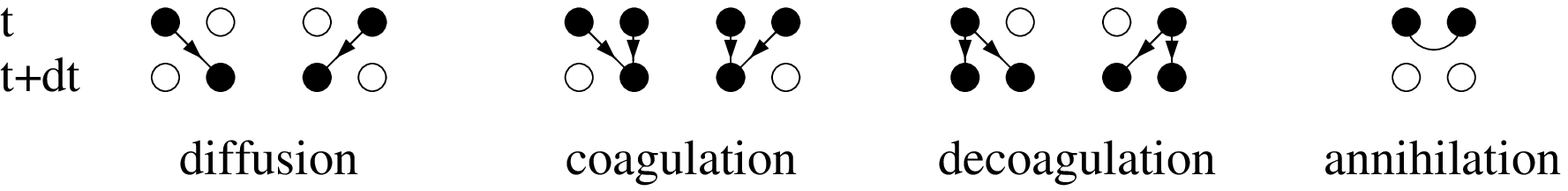}} \smallcaption{
\label{FIGDECOAG} Nearest-neighbor processes in reaction-diffusion
models with only one particle species. }
\end{figure}

\label{REAC} 
Reaction-diffusion processes are stochastic models
for chemical reactions in which particles are predominantly
transported by thermal diffusion. Usually a chemical reaction in a
solvent or on a catalytic surface consists of a complex sequence
of intermediate steps. In reaction-diffusion models these
intermediate steps are ignored and the reaction chain is replaced
by simplified probabilistic transition rules. The involved atoms
and molecules are interpreted as {\em particles} of several
species, represented by capital letters $A,B,C,\ldots$ . These
particles neither carry a mass nor an internal momentum, instead
the configuration of a reaction-diffusion model is completely
specified by the position of the particles. On a lattice with
exclusion principle such a configuration can be expressed in terms
of local variables $s_i=0,1,2,3,\ldots,$ representing a vacancy
$\vacancy$ and particles $A,B,C,\ldots$, respectively. Sometimes
it is even not necessary to keep track of all substances involved
in a chemical reaction. For example, if a molecule of a gas phase
is adsorbed at a catalytic surface, this process may be
effectively described by spontaneous particle creation $\vacancy
\rightarrow A$ without modeling the explicit dynamics in the gas
phase. Therefore, the number of particles in reaction-diffusion
models is generally not conserved.

Apart from spontaneous particle creation $\vacancy \rightarrow A$
many other reactions are possible.
{\em Unary} reactions are spontaneous transitions of individual
particles, the most important examples being
\begin{equation*}
\hspace{20mm}
\begin{array}{ll}
A \rightarrow \vacancy \hspace{10mm} & \text{self-destruction,} \\
A \rightarrow 2A    & \text{offspring production or
decoagulation,} \\ A \rightarrow B     & \text{transmutation,} \\
A \rightarrow A+B   & \text{induced creation of particles.}
\end{array}
\end{equation*}
On the other hand, {\em binary} reactions require two particles
to meet at the same place (or at neighboring sites).
Here the most important examples are
\begin{equation*}
\begin{array}{ll}
2A \rightarrow \vacancy \hspace{10mm} & \text{pair annihilation,}
\\ 2A \rightarrow A    & \text{coagulation (coalescence),} \\ A+B
\rightarrow \vacancy& \text{two-species annihilation,} \\ A+B
\rightarrow B   & \text{induced adsorption of particles.}
\end{array}
\end{equation*}
In addition, the particles may diffuse with certain rates in the
same way as in the previously discussed exclusion process. A
process is called {\em diffusion-limited} if diffusion becomes
dominant in the long-time limit, i.e., the diffusive moves become
much more frequent than reactions. This happens, for example, in
reaction-diffusion models with binary reactions when the particle
density is very low. On the other hand, if particle reactions
become dominant after very long time, the process is called {\em
reaction-limited}.

In the following we will focus on simple reaction-diffusion
models with only one type of particles
(see Fig.~\ref{FIGDECOAG}). They can be considered
as two-state models since each site can either be occupied
by a particle ($A$) or be empty ($\vacancy$). Examples include
the so-called {\em coagulation model} in which
particles diffuse at rate $\diff$, coagulate
at rate $\lambda$ and decoagulate at rate $\crit$:
\begin{equation}
\label{Decoag}
A\vacancy \underset{\diff}{\leftrightarrow} \vacancy A
\ , \qquad
AA \underset{\lambda}{\rightarrow} A\vacancy,\vacancy A
\ , \qquad
A\vacancy,\vacancy A \underset{\crit}{\rightarrow} AA
\ .
\end{equation}
Using the same notation as in Eq.~(\ref{ExclusionProcessGenerator}), the
corresponding nearest-neighbor transition matrix is given by
\begin{equation}
\label{DecoagGenerator}
{\cal L}_{i}^{\text{coag}}=
\begin{pmatrix}
0 & 0 & 0 & 0 \\
0 & \diff+\crit & -\diff & -\lambda \\
0 & -\diff & \diff+\crit & -\lambda \\
0 & -\crit & -\crit & 2\lambda
\end{pmatrix} \ .
\end{equation}
The special property of this model lies in the fact that the
empty state cannot be reached by the dynamic processes.
An exact solution of the coagulation model will be
discussed in Sec.~\ref{SECEXACT}.

Another important example is the {\em annihilation model}
where particles diffuse and annihilate:
\begin{equation}
A\vacancy \underset{\diff}{\leftrightarrow} \vacancy A \ , \qquad
AA \underset{\annh}{\rightarrow} \vacancy\vacancy \ .
\end{equation}
The corresponding interaction matrix is given by
\begin{equation}
\label{AnnhGenerator}
{\cal L}_{i}^{\text{annh}}=
\begin{pmatrix}
0 & 0 & 0 & -\annh \\
0 & \diff & -\diff & 0 \\
0 & -\diff & \diff & 0 \\
0 & 0 & 0 & \annh
\end{pmatrix} \  .
\end{equation}
In the annihilation model the number of particles is conserved
modulo $2$. As will be shown in Sec.~\ref{SECEXACT}, both models
are equivalent and can be related by a similarity transformation.

%
%==============================================================================
\subsection{Mean field approximation}
%==============================================================================
%
%
In many cases the macroscopic properties of a reaction-diffusion
process can be predicted by solving the corresponding mean field
theory. In chemistry the simplest mean field approximation is
known as the `law of mass action': For a given temperature the
rate of a reaction is assumed to be proportional to the product of
concentrations of the reacting substances. This approach assumes
that the particles are homogeneously distributed. It therefore
ignores any spatial correlations as well as instabilities with
respect to inhomogeneous perturbations. Thus, the homogeneous mean
field approximation is expected to hold on scales where diffusive
mixing is strong enough to wipe out spatial structures. Especially
in higher dimensions, where diffusive mixing is more efficient,
the mean field approximation provides a good description. It
becomes exact in infinitely many dimensions, where all particles
can be considered as being neighbored.

The mean field equations can be constructed directly by
translating the reaction scheme into a differential equation for
gain and loss of the particle density $\rho(t)$. For example, in
the mean field approximation of the coagulation model
(\ref{Decoag}) the process $A \stackrel{\kappa}{\rightarrow} 2A$
takes place with a frequency proportional to $\crit\rho(t)$,
leading to an increase of the particle density. Similarly, the
coagulation process $2A \stackrel{\lambda}{\rightarrow} A$
decreases the number of particles with a frequency proportional to
$\lambda\rho^2(t)$. Ignoring diffusion, the resulting mean field
equation reads
\begin{equation}
\label{CoagMF}
\timederivative \rho(t) = \crit \rho(t) - \lambda \rho^2(t) \ ,
\end{equation}
where $\crit$ and $\lambda$ are the rates for decoagulation and
coagulation, respectively. In contrast to the master equation this
differential equation is {\em nonlinear}. For $\crit>0$ it has
two fixed points, namely an unstable fixed point at $\rho=0$ and a
stable fixed point at $\rho=\crit/\lambda$. The physical meaning of
the two fixed points is easy to understand. The empty system
remains empty, but as soon as we perturb the system by adding a
few particles, it quickly evolves towards a stationary active
state with a certain average concentration $\rho>0$. This active
state is then stable against perturbations (such as adding or
removing particles).

The mean field equation~(\ref{CoagMF}) can also be used to predict
{\it dynamic}  properties of the system.
Starting from a fully occupied lattice $\rho(0)=1$ the
time-dependent solution is given by
\begin{equation}
\label{MFDecay}
\rho(t) = \frac{\crit}{\lambda-(\lambda-\crit) e^{-\crit t}} \ .
\end{equation}
In the limit of a vanishing decoagulation rate $\crit \rightarrow
0$, the two fixed points merge into a marginal one. As in many
physical systems this leads to a much slower dynamics. In fact,
for $\crit=0$ Eq.~(\ref{MFDecay}) turns into
\begin{equation}
\label{CoagMasslessDecay}
\lim_{\crit \rightarrow 0} \rho(t) = \frac{1}{1+\lambda t} \ ,
\end{equation}
i.e., the particle density decays asymptotically according to a
power law as
\begin{equation}
\rho(t) \sim t^{-1}\,.
\end{equation}
The stability of the mean field solution with respect to
inhomogeneous perturbations may be studied by adding a term for
diffusion
\begin{equation}
\label{CoagInhMF}
\timederivative \rho(\xvec,t) = \crit \rho(\xvec,t) - \lambda \rho^2(\xvec,t) +
\diff \nabla^2 \rho(\xvec,t)\ .
\end{equation}
In the present case the diffusive term suppresses perturbations
with short wavelength and therefore stabilizes the homogeneous
solutions. However, in certain chemical reactions with several
particle species such a diffusive term may have a destabilizing
influence. The study of mean-field instabilities is the starting
point for the theory of {\em pattern formation} which has become
an important field of statistical physics~\cite{Mikhailov94}. A
very interesting application is the Belousov-Zhabotinski
reaction~\cite{Scott94} that produces rotating spirals in a Petri
dish.

It may be surprising that even simple reaction-diffusion processes
are described by {\em nonlinear} mean field rate equations,
whereas the corresponding master equation is always linear.
However, mean field and Langevin  equations are always defined in
terms of  coarse-grained particle densities involving many
local degrees of freedom. These coarse-grained densities, which
can be thought of as observables in configuration space, may
evolve according to a nonlinear laws. A similar paradox occurs in
quantum physics: Although the Schr\"odinger equation is strictly
linear, most observables evolve in a highly nonlinear way.

%
%
%==============================================================================
\subsection{The influence of fluctuations}
%==============================================================================
%
\label{FLUCSEC}
Although the mean field equation (\ref{CoagInhMF}) includes a term
for particle diffusion, it still ignores fluctuation effects
and spatial correlations. However,
especially in low-dimensional systems,
fluctuations may play an important role and
are able to entirely change the physical properties of
a reaction-diffusion process.

In order to demonstrate the influence
of fluctuations, let us consider the coagulation process
$2A \rightarrow A$. The full Langevin equation for this
process reads
\begin{equation}
\label{CoagLangevin}
\timederivative \rho(\xvec,t) = - \lambda \rho^2(\xvec,t) +
\diff \nabla^2 \rho(\xvec,t) + \noise(\xvec,t)\ ,
\end{equation}
where $\noise(\xvec,t)$ is a noise term which accounts for the
fluctuations of the particle density at position $\xvec$ at time $t$.
Clearly, the noise amplitude depends on the magnitude of
the density field $\rho(\xvec,t)$.
In particular, without any particles present,
there will be no fluctuations. According to the central
limit theorem, the noise is expected to be Gaussian with a
squared  amplitude proportional to the frequency of
events leading to a change of the particle number. Since
the particle number only fluctuates when two particles coagulate,
this frequency should be proportional to  $\rho^2(\xvec,t)$.
Following these naive arguments, the noise correlations should
be given by
\begin{eqnarray}
\label{CoagNoise}
\langle \noise(\xvec,t) \rangle &=& 0 \,, \\
\langle \noise(\xvec,t) \noise(\xvec',t')  \rangle &=&
\namp \, \rho^2(\xvec,t) \, \delta^d(\xvec-\xvec') \, \delta(t-t')
\nonumber
\,,
\end{eqnarray}
where $\namp$ denotes the noise amplitude and $d$ the spatial
dimension. The next question is to what extent the macroscopic
behavior of the system will be affected by the noise. Typically
there are three possible answers:
\begin{enumerate}
\item The noise is {\em irrelevant} on large scales so that
the macroscopic behavior is correctly described by the mean field solution.
\item The noise is {\em relevant} on large scales, leading to
a macroscopic behavior that is different from the mean-field prediction.
\item The noise is {\em marginal}, producing (typically logarithmic)
deviations from the mean-field solution.
\end{enumerate}
In order to find out whether the noise is relevant on large scale
we need to introduce the concept of {\em renormalization}~\cite{Cardy96}.
The term `renormalization' refers to  various theoretical
methods investigating the scaling behavior of
physical systems under coarse-graining of
space and time. Roughly speaking, it describes how the parameters
of a system have to be adjusted under coarse-graining of lengths scales
without changing its physical properties. A fixed point
of the renormalization flow is then associated with
certain universal scaling laws of the system. The simplest
renormalization group (RG) scheme ignores the influence of fluctuations.
This approach is referred to as `mean field renormalization'. Approaching
the fixed point, the noise amplitude may diverge, vanish or stay finite,
corresponding to the classification given above. Hence, by studying
mean field renormalization, we can predict whether fluctuations
are relevant or not.

\begin{figure}
\epsfxsize=65mm \centerline{\epsffile{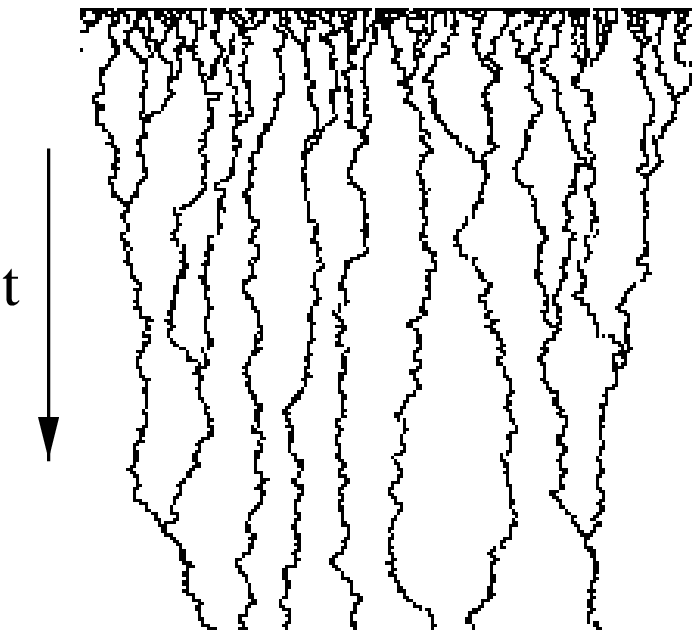}} \smallcaption{
\label{FIGCOAG1} Monte Carlo simulation of the (1+1)-dimensional
coagulation-diffusion model $A+A \rightarrow A$ with
random-sequential updates. The figure shows an individual run
starting with a fully occupied lattice of 200 sites. As can be
seen, the coagulation process is diffusion-limited. }
\end{figure}

In the mean field approximation the Langevin equation~(\ref{CoagLangevin})
may be renormalized by a scaling transformation
\begin{equation}
\label{Scaling}
 \xvec \rightarrow \scalefac \, \xvec \ , \qquad
 t \rightarrow \scalefac^z \, t \ , \qquad
 \rho(\xvec,t) \rightarrow \scalefac^\chi \,
 \rho( \scalefac \xvec,\scalefac^z t)\,,
\end{equation}
where $z$ denotes the dynamic exponent. The exponent $\chi$
describes the scaling properties of the density field itself. If
the particles were distributed homogeneously, the field would
scale as an ordinary density, that is, with the exponent
$\chi=-d$. However, in the coagulation process nontrivial
correlations between particles lead to a different scaling
dimension of the particle distribution. In fact, invariance of
Eqs.~(\ref{CoagLangevin})-(\ref{CoagNoise}) under rescaling
implies that $z=2$ and $\chi=-2$. Therefore, the noise amplitude
scales as
\begin{equation}
\namp \rightarrow \scalefac^{1-d/2} \namp \ ,
\end{equation}
where $d$ is the spatial dimension. Hence in one spatial dimension
fluctuations are relevant whereas they are marginal in two and
irrelevant in $d>2$ dimensions. The value of $d$ where the noise
becomes marginal is denoted as the {\em upper critical dimension}
$d_c$. For the coagulation model the upper critical dimension is
$d_c=2$. Above the critical dimension the mean field approximation
provides a correct description, whereas for $d<d_c$ fluctuation
effects have to be taken into account. This can be done by using
improved mean field approaches, exact solutions, as well as
field-theoretic renormalization group
techniques~\cite{MattisGlasser98}.

A systematic field-theoretic analysis of the coagulation process
$2A\rightarrow A$ leads to an unexpected result: The noise
amplitude $\namp$ in Eq.~(\ref{CoagNoise}) turns out to be
negative~\cite{HowardTauber97}. Consequently, the noise
$\noise(\xvec,t)$ is {\em imaginary}. This result is rather
counterintuitive as we expect the noise to describe density
fluctuations which, by definition, are real. However, since the
noise amplitude is a measure of annihilation events, it is
subjected to correlations that are produced by the annihilation
process itself. In one dimension these correlations are negative,
i.e., particles avoid each other. This simple example demonstrates
that it can be dangerous to set up a Langevin equation by
considering the mean field equation and adding a physically
reasonable noise field. Instead it is necessary to derive the
Langevin equation directly from the microscopic dynamics, as
explained in Ref.~\cite{Lee94}.
%
%
%==============================================================================
\subsection{Numerical simulations}
%==============================================================================
%
\label{NUMERIC} To verify analytical results, it is often helpful
to perform Monte Carlo simulations. In order to demonstrate this
numerical technique, let us again consider the coagulation process
$2A \rightarrow A$ on a one-dimensional chain. For simplicity we
assume the rates for diffusion and coagulation to be equal. This
ensures that particles can move at constant rate irrespective of
the state of the target site. If the target site is empty, it will
be occupied by the moving particle.  On the other hand, if the
target site is already occupied, the two particles will coagulate
into a single one. Such a move from site $i$ to site $j$ may be
realized by the pseudo code instruction
\begin{quote}
\ {\tt Move(i,j) $\{$ if (s[i]==1) $\{$ s[i]=0; s[j]=1; $\}\}$; }
\end{quote}
\noindent
where {\tt s[i]} denotes the occupation variable $s_i=0,1$ at site $i$.
In one dimension particles move randomly to the left and to the right.
Thus a local update at sites $(i,i+1)$ may be realized by the instruction
\begin{quote}
\begin{tabular}{ll}
{\tt Update(i) $\{$}
& {\tt if (rnd(0,1)<0.5) Move(i,i+1); } \\
& {\tt else Move(i+1,i); $\}$; }
\end{tabular}
\end{quote}
\noindent
where {\tt rnd(0,1)} returns a real random number from a flat
distribution between $0$ and~$1$. Since the coagulation
model evolves by asynchronous dynamics it uses so-called
{\em random-sequential} updates, i.e., the update
attempts take place at randomly selected pairs of sites.
A Monte Carlo {\em sweep} consists of $N$ such update attempts:
\begin{quote}
\ {\tt\indent \indent for (i=1; i<=N; i++) Udpate(rndint(1,N-1)); }
\end{quote}
\noindent
where $N$ denotes the lattice size and {\tt rndint(1,N-1)}
returns an integer random number between $1$ and $N-1$.
Since on average each site is updated once, such a
sweep corresponds to a unit time step.
It can be proven that the statistical ensemble of space-time
trajectories generated by random-sequential updates
converges to  the solution of the master
equation~(\ref{MasterEquation}) in the
limit $N \rightarrow \infty$. The above update algorithm
can easily be generalized to more complicated reaction schemes
and higher dimensional lattices.

The coagulation process with synchronous updates may be simulated
by using {\em parallel} updates on alternating sublattices:
\begin{quote}
{\tt \indent\indent for (i=1; i<=N-1; i+=2) Update(i); \\
     \indent\indent for (i=2; i<=N-2; i+=2) Update(i); }
\end{quote}
\noindent
In Monte Carlo simulations most of the CPU time is consumed for
generating random numbers. Therefore, models with parallel updates
are usually more efficient since it is not necessary to determine
random positions for the updates. In addition, models with parallel
updates can be implemented easily on computers with
parallel architecture~\cite{OKVR99}.

\begin{figure}
\epsfxsize=135mm
\centerline{\epsffile{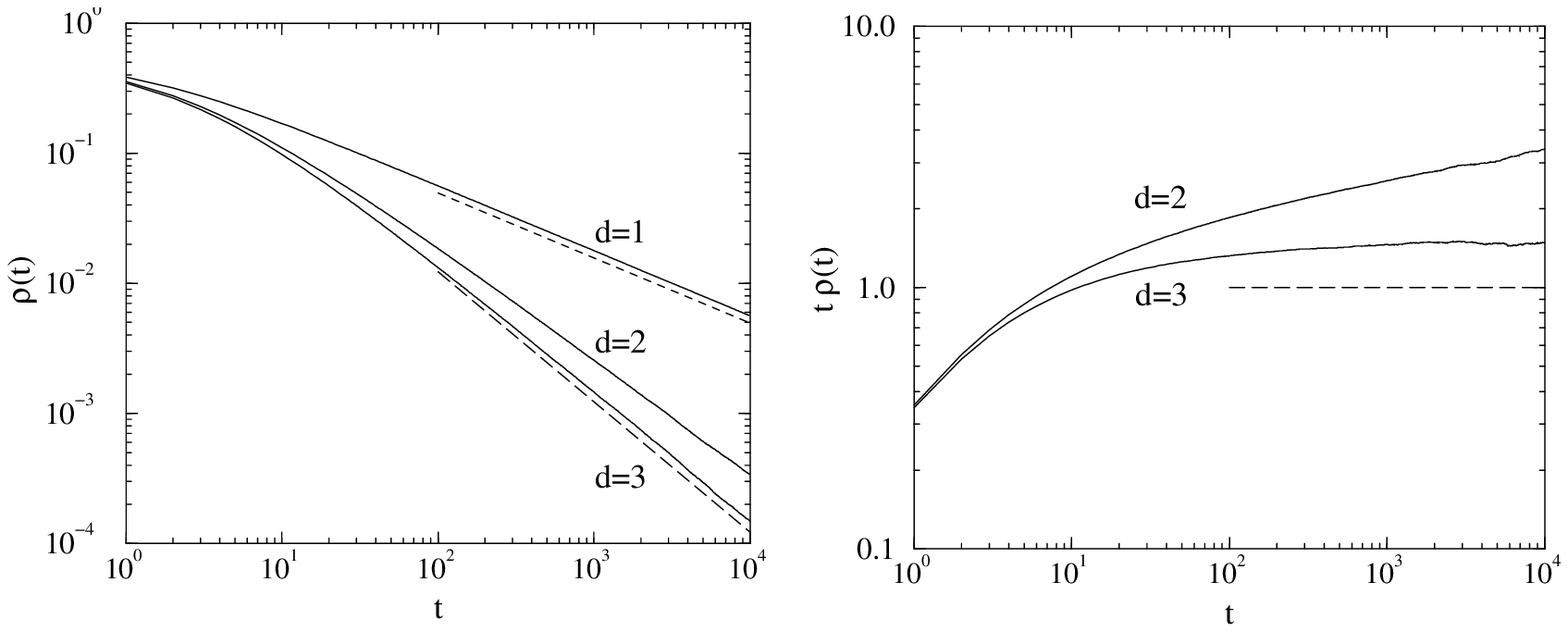}}
\smallcaption{
\label{FIGCOAG2}
Monte Carlo simulation of the coagulation-diffusion model.
Left: Decay of the particle density $\rho(t)$ measured in a system with
$10^5$ sites in one, two, and three spatial dimensions,
averaged over $10^4$ independent runs. The dashed lines indicate the
slopes $-1/2$ and $-1$, respectively. Right: $t \rho(t)$ versus $t$,
illustrating logarithmic corrections in $d=2$ dimensions.
}
\end{figure}

Fig.~\ref{FIGCOAG2} shows the particle density as a function of time
for the coagulation model with random-sequential updates
and closed boundary conditions in various dimensions. The particle
concentration is averaged over $10^4$ independent runs and
plotted in a double-logarithmic representation, where straight
lines indicate power-law behavior. As expected,
the mean-field prediction $\rho(t)\sim 1/t$ is reproduced
in $d>2$ dimensions. In one dimension, however, the graph
suggests the density to decay as
\begin{equation}
\rho(t) \sim t^{-1/2} \qquad (d=1) \ .
\end{equation}
Thus, the simulation result demonstrates that fluctuation
effects can change the asymptotic behavior
(an exact solution will be discussed in Sec. \ref{SECEXACT}).
At the critical dimension $d=d_c=2$ the density $\rho(t)$
deviates slightly from the mean-field prediction, indicating
logarithmic corrections. In fact, as can be shown by
a field-theoretic analysis~\cite{Lee94}, the density
decays asymptotically as
\begin{equation}
\label{AnnhDens}
\rho(t) \sim \left\{ \begin{array}{ll}
    t^{-d/2} &{\rm for \ } d < 2 \ , \\
    t^{-1} \ln t &{\rm for \ } d = d_c = 2 \ , \\
    t^{-1} &{\rm for \ } d > 2 \ .
    \end{array} \right.
\end{equation}
%
%
%
%==============================================================================
\subsection{Exact results}
%==============================================================================
%
\vspace{-3mm}
\label{SECEXACT}
%
%---------------------------------------------------------------------------
\headline{Equivalence of annihilation and coagulation processes}
%---------------------------------------------------------------------------
%
Sometimes it is possible to relate different stochastic processes
by an exact similarity transformation~\cite{Simon95,Santos96,HOS96}.
For example,  the coagulation process $2A \rightarrow A$ and
the annihilation process $2A \rightarrow \vacancy$
defined in Sec.~\ref{REAC} are fully equivalent if
their rates are tuned appropriately.
More precisely, for a particular choice of the rates
it is possible to find a similarity transformation ${\cal U}$ such that
\begin{equation}
{\cal L}^{\text{coag}}={\cal U}{\cal L}^{\text{annh}}{\cal U}^{-1} \ ,
\end{equation}
where we assume the chains to have closed ends, i.e.,
\begin{equation}
{\cal L}^{\text{coag}} =\sum_{i=1}^{N-1} {\cal L}_{i}^{\text{coag}}
\ , \qquad
{\cal L}^{\text{annh}} =\sum_{i=1}^{N-1} {\cal L}_{i}^{\text{annh}}
\ .
\end{equation}
Since ${\cal L}^{\text{coag}}$ and ${\cal L}^{\text{annh}}$ are
non-hermitean operators, the similarity transformation ${\cal U}$
is not orthogonal. However, if ${\cal U}$ exists, the two
operators will have the same spectrum of eigenvalues. As can be
verified easily, the local operators ${\cal L}_{i}^{\text{coag}}$
and ${\cal L}_{i}^{\text{annh}}$ have the eigenvalues
$\{0,0,2\diff+\crit,2\lambda+\crit \}$ and $\{0,0,2\diff,a\}$,
respectively. Therefore, choosing the rates
\begin{equation}
\label{Choice}
\diff=\lambda=1, \qquad \annh=2, \qquad \crit=0 \ ,
\end{equation}
both operators obtain the same spectrum $\{0,0,1,1\}$. Moreover, it
can be shown that they both obey the same commutation relations, namely
the so-called Hecke algebra~\cite{Martin89,MartinRittenberg92}
\begin{align}
\label{HeckeAlgebra}
{\cal L}_{i}^2&={\cal L}_{i} \ , \nonumber \\
{\cal L}_{i} {\cal L}_{i+1} {\cal L}_{i} -
{\cal L}_{i+1} {\cal L}_{i} {\cal L}_{i+1}
&= 2 ({\cal L}_{i}-{\cal L}_{i+1}) \ , \\
[{\cal L}_{i},{\cal L}_{j}] &= 0 \ \ \text{for} \ \ (|i-j|\geq 2) \nonumber
\ .
\end{align}
Since this algebra generates the spectrum of ${\cal L}$, we can
conclude that the spectra of ${\cal L}^{\text{coag}}$ and ${\cal
L}^{\text{annh}}$ coincide for an arbitrary number of sites.
Obviously, the equivalence of the spectra is a necessary condition
for the existence of a similarity transformation between the two
systems. In the present case it is even possible to compute the
similarity transformation explicitly. It turns out that ${\cal U}$
can be expressed in terms of local tensor products (see
Appendix~\ref{APPVECSEC})
\begin{equation}
{\cal U} = u \otimes u \otimes \ldots \otimes u = \bigotimes
_{i=1}^{N} u = u^{\otimes N}\ ,
\end{equation}
where
\begin{equation}
u = \begin{pmatrix} 1 & -1 \\ 0 & 2 \end{pmatrix} \ , \qquad
u^{-1} = \begin{pmatrix} 1 & 1/2 \\ 0 & 1/2 \end{pmatrix} \ .
\end{equation}
Consequently the $n$-point density correlation functions of both
models are related by
\begin{equation}
\langle s_{j_1}s_{j_2}\cdots s_{j_n}\rangle^{\text{coag}} =
2^n \langle s_{j_1}s_{j_2}\cdots s_{j_n}\rangle^{\text{annh}} \ .
\end{equation}
In particular, the particle densities in both models differ by a
factor of $2$:
$$\rho^{\text{coag}}(t)=2\rho^{\text{annh}}(t).$$
It should be noted that only a subset of initial conditions in the
coagulation model can be mapped onto physically meaningful initial conditions
in the annihilation model (see Ref.~\cite{HOS95}).

%---------------------------------------------------------------------------
\headline{Exact mapping between equilibrium and nonequilibrium systems}
%---------------------------------------------------------------------------
%
A remarkable progress has been achieved by realizing that certain
nonequilibrium models can be mapped onto well-studied integrable
equilibrium models. More specifically, it has been shown that the
Liouville operator ${\cal L}$ of a nonequilibrium models may be
related to the Hamiltonian ${\cal H}$ of an integrable quantum
spin systems by similarity transformation~\cite{Felderhof71}. This
allows the nonequilibrium model to be solved by exact techniques
of equilibrium statistical mechanics such as free-fermion
diagonalization, the Bethe ansatz, or other algebraic
methods~\cite{Baxter82}. For example, the exclusion process can be
mapped onto the {\it XXZ} quantum
chain~\cite{AlexanderHolstein78,DFP80,GwaSpohn92}, whereas the
coagulation-decoagulation process is related to the {\it XY} chain
in a magnetic field~\cite{Siggia77,KPWH95,HKP96}. Exact mappings
were also found for higher spin
analogs~\cite{AlcarazRittenberg93}. A complete summary of the
known results can be found in Ref.~\cite{Schuetz98b}.

\begin{figure}
\epsfxsize=120mm
\centerline{\epsffile{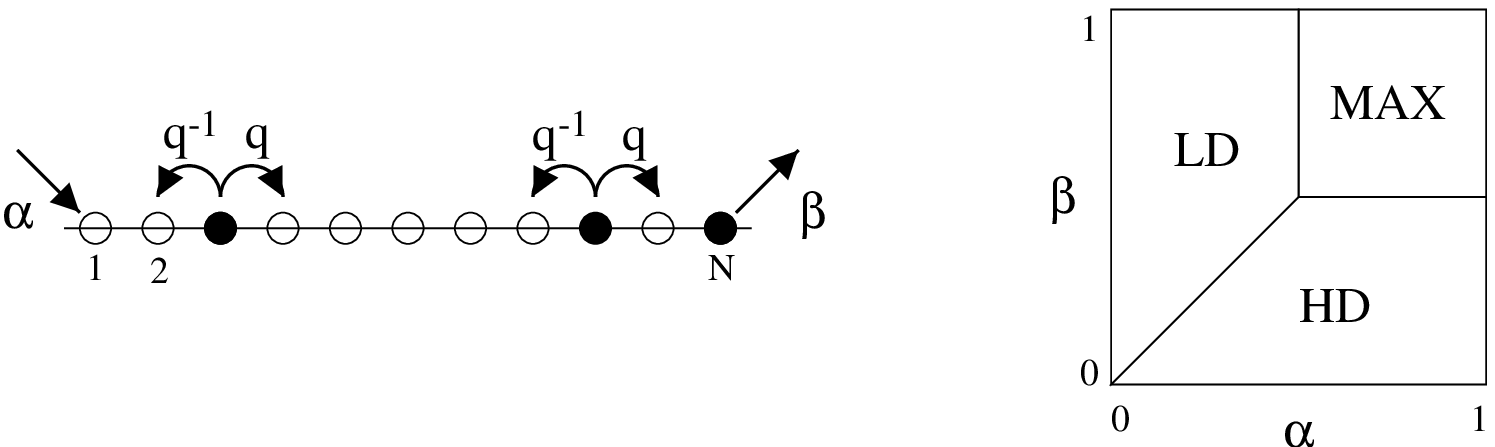}}
\smallcaption{
\label{FIGMPS}
The partially asymmetric exclusion process with  particle
adsorption and desorption at the boundaries. Left: Particles
adsorb at the left boundary at rate $\alpha$,
perform a biased random walk in the bulk of the chain,
and finally desorb from site $N$ at rate $\beta$.
Right: Phase diagram for the totally asymmetric
exclusion process, denoting the low (LD), high (HD)
and maximal (MAX) density phases. Analogous phase diagrams
for partial asymmetry were obtained in Ref.~\cite{Sandow94}.
}
\end{figure}
In order to demonstrate this technique, let us consider the
partially asymmetric exclusion process in one spatial dimension
(see Eq.~\ref{ExclusionProcessGenerator}). As will be shown in the
following, this model is related to the quantum spin Hamiltonian
of the ferromagnetic {\it XXZ} Heisenberg quantum chain with open
boundary conditions ${\cal H}=\sum_{i=1}^{N-1} {\cal H}_i$, where
\begin{equation}
\label{XXZ} {\cal H}_i = -\frac12\,\Bigl( \sigma_i^x\sigma_{i+1}^x
\;+\; \sigma^y_i\sigma_{i+1}^y \;+\;
\frac{q+q^{-1}}2\,(\sigma^z_i\sigma^z_{i+1}-1) \;+\;
\frac{q-q^{-1}}2\,(\sigma^z_i-\sigma^z_{i+1}) \Bigr)\,.
\end{equation}
This quantum chain Hamiltonian generates translations in the
corresponding two-dimen\-sio\-nal {\it XXZ} model in a strongly
anisotropic scaling limit. As can be verified easily, the
Hamiltonian is non-hermitean for $q\neq 1$. Using the standard
basis of Pauli matrices
$\sigma^x=\left(\begin{smallmatrix}
0&1\\1&0 \end{smallmatrix} \right)$,
$\sigma^y=\left(\begin{smallmatrix} 0&-i\\i&0 \end{smallmatrix}
\right)$, and
$\sigma^z=\left(\begin{smallmatrix} 1&0\\0&-1
\end{smallmatrix} \right)$ the interaction matrix is given by
\begin{equation}
{\cal H}_{i}=
\begin{pmatrix}
0 & 0 & 0 & 0 \\
0 & q^{-1} & -1 & 0 \\
0 & -1 & q & 0 \\
0 & 0 & 0 & 0
\end{pmatrix}
\ .
\end{equation}
The {\it XXZ} Heisenberg chain is integrable by means of Bethe
ansatz methods~\cite{Schuetz98b}. The integrability is closely
related to two different algebraic structures. On the one hand,
the Hamiltonian~(\ref{XXZ}) commutes with the generators $K,S^\pm$
of the quantum algebra $U_q[SU(2)]$ (see
Ref.~\cite{PasquierSaleur90})
\begin{equation}
KS^\pm K^{-1}=qS^\pm \ , \qquad
[S^+,S^-]=\frac{K^2-K^{-2}}{q-q^{-1}} \,,
\end{equation}
where
\begin{equation}
K=(q^{\sigma^z/2})^{\otimes N} \ , \qquad
S^\pm = \sum_{k=1}^N (q^{\sigma^z/2})^{\otimes (k-1)} \otimes
\sigma^\pm \otimes (q^{-\sigma^z/2})^{\otimes (N-k)}
\ .
\end{equation}
Roughly speaking, the quantum group symmetry determines the
degeneracies of the eigenvalues of $\cal H$.
On the other hand, the generators ${\cal H}_i$ are a representation
of the Temperley-Lieb algebra~\cite{TemperleyLieb71}
\begin{align}
{\cal H}_i^2&=(q+q^{-1}) {\cal H}_i \ , \nonumber \\
{\cal H}_i&={\cal H}_i{\cal H}_{i\pm 1}{\cal H}_i \ , \\
[{\cal H}_i,{\cal H}_j]&=0 \text{ for } |i-j|\geq 2 \ . \nonumber
\end{align}
This algebra determines the actual numerical value of the energy
levels. As realized by Alcaraz and
Rittenberg~\cite{AlcarazRittenberg93}, the same commutation
relations are satisfied by the transition matrix ${\cal L}_i$ of
the asymmetric exclusion process in
Eq.~(\ref{ExclusionProcessGenerator}). In fact, it can be shown
that the {\it XXZ} chain and the exclusion process with
$\alpha=\beta=0$ are related by
a similarity transformation ${\cal L=UHU}^{-1}$, where ${\cal U}$
can be written as a tensor product of local transformations
\begin{equation}
{\cal U} = \begin{pmatrix} 1&0 \\ 0&q \end{pmatrix} \otimes
\begin{pmatrix} 1&0 \\ 0&q^2 \end{pmatrix} \otimes \ldots \otimes
\begin{pmatrix} 1&0 \\ 0&q^N \end{pmatrix} =
\bigotimes_{i=1}^N \begin{pmatrix} 1&0 \\ 0&q^i \end{pmatrix}
\ .
\end{equation}
In order to illustrate how symmetries of the equilibrium model
translate into physical properties of the stochastic process, let
us consider the quantum group symmetry of the {\it XXZ} model.
Regarding the exclusion process this symmetry emerges as a
conservation of the total number of particles $n$. The generators
$S^\pm$ act as ladder operators between different sectors with a
fixed number of particles. The diagonal operator $K$ is
proportional to $q^{-n}$, weighting the sectors as in a
grand-canonical ensemble, where $q$ plays the role of a fugacity.
The partially asymmetric exclusion process is therefore a physical
realization of a quantum group symmetry with a real-valued
deformation parameter $q$.

A similar mapping relates the coagulation model $2A \rightarrow A$
and the ferromagnetic {\it XY} quantum chain in a magnetic field,
which is defined by the interaction matrix
\begin{equation}
\label{XY}
{\cal H}_i \;=\; -\frac12\,\Bigl(
\sigma_i^x\sigma_{i+1}^x \;+\;
\sigma^y_i\sigma_{i+1}^y \;+\;
\sigma^z_i+\sigma^z_{i+1} \;-\; 2 \Bigr)\,.
\end{equation}
In fact, it can be easily verified that the operators ${\cal H}_i$
satisfy the Hecke algebra~(\ref{HeckeAlgebra}). The {\it XY} chain
is exactly solvable in terms of free fermions~\cite{LSM61}. The
integrability of the model is closely related to a $SU_q(1|1)$
quantum group symmetry. In the {\it XY} chain this symmetry shows
up as a fermionic zero mode, leading to two-fold degenerate energy
levels. In the coagulation model this symmetry emerges as a state
without particles that can neither be reached nor left by the
dynamics. In the (equivalent) annihilation model the symmetry
appears as a parity conservation law.

It should be noted that reaction-diffusion models are usually
related to {\em ferromagnetic} quantum chains, the reason
being that the diffusion process always corresponds to
a ferromagnetic interaction in the quantum spin model.
\begin{figure}
\epsfxsize=130mm \centerline{\epsffile{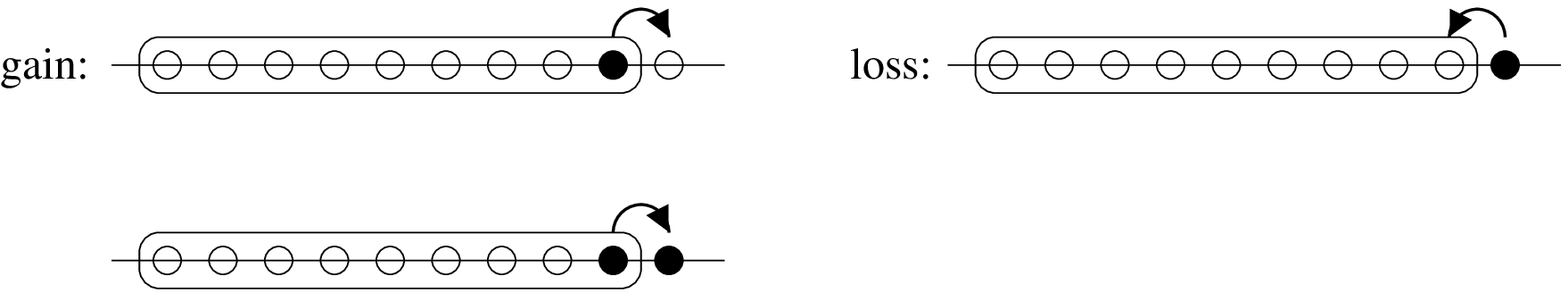}} \smallcaption{
\label{FIGIPDF} Interparticle distribution functions. The figure
illustrates gain and loss processes for the empty-interval
probability $I_\ell(t)$ by diffusion (left) and coagulation
(right). The same processes take place at the left boundary of the
interval.}
\end{figure}

%---------------------------------------------------------------------------
\headline{Interparticle distribution functions}
%---------------------------------------------------------------------------
%
Even if a stochastic model can be mapped onto a known equilibrium
system by a similarity transformation, it is often technically
difficult to derive physical quantities such as density profiles
and correlation functions~\cite{Alcaraz94}. For models with an
underlying fermionic symmetry an alternative approach has been
developed which does not explicitly use a similarity
transformation. Instead it expresses the state of a model in terms
of so-called {\em interparticle distribution functions}
(IPDF)~\cite{Avraham95,AlemanyAvraham95,Avraham97}. Consider, for
example, the coagulation-diffusion process with asynchronous
dynamics on an infinite chain where particles coagulate
$(A+A\rightarrow A)$ and diffuse at unit rates. Let $I_\ell$ be
the probability that an arbitrarily chosen interval of $\ell$
sites contains no particles. In terms of these empty-interval
probabilities the master equation can be written in a particularly
simple form. Since $I_{\ell}-I_{\ell+1}$ is the probability to
find a particle at a neighboring site next to the interval of
length $\ell$, it is possible to rewrite diffusion and coagulation
in terms of gain and loss processes (see Fig.~\ref{FIGIPDF})
\begin{equation}
\label{IPDFEq}
\timederivative I_\ell(t) =
\underbrace{I_{\ell-1}(t)-I_{\ell}(t)}_{\text{gain}} \, - \,
\underbrace{I_{\ell}(t)+I_{\ell+1}(t)}_{\text{loss}} \
\end{equation}
with $I_0(t) \equiv 1$. It is important to note that this
particularly simple form requires the rates for diffusion and
coagulation to be identical. This ensures that the gain processes
do not depend on whether the target site is already occupied by a
particle. If the two rates are different, higher-order
probabilities for several adjacent intervals have to be included,
resulting in a coupled hierarchy of equations. The IPDF method
exploits the fact that this complicated hierarchy of equations
decouples for a particular choice of the rates.

By solving the above equation we can compute the particle density
$\rho(t)$ which is given by the probability for an empty interval
of length $1$ to be absent, i.e.,
\begin{equation}
\rho(t)=1-I_1(t) \, .
\end{equation}
In order to determine the asymptotic behavior of $\rho(t)$
let us consider the continuum limit of Eq.~(\ref{IPDFEq})
\begin{equation}
(\timederivative -\partial_\ell^2) \; I(\ell,t) =  0\ , \qquad
I(0,t)=1 \ ,
\end{equation}
where $\rho(t) =\left. \partial_\ell  I(\ell,t) \right|_{\ell=0}$.
This equation has the solution
$I(\ell,t)=1-\text{erf}(x/\sqrt{t})$. In the long time limit, the
particle density therefore decays algebraically as 
\begin{equation}
\rho(t)\sim t^{-1/2}\,,
\end{equation}
confirming the numerical result of Sec.~\ref{NUMERIC}.
It is interesting to compare this result with the mean field
approximation (\ref{CoagMasslessDecay}) which lead to the
incorrect result $\rho(t)\sim t^{-1}$. Therefore, the above exact
solution demonstrates that fluctuations may influence the entire
temporal evolution of a stochastic process.

The IPDF technique~\cite{Avraham95,Avraham97} was extended to the
coagulation-decoagulation model by including the inverse
reaction $A\rightarrow A+A$. Other exact solutions revealed
phenomena such as anomalous kinetics, critical ordering,
nonequilibrium dynamic phase transitions, as well as the existence
of Fisher
waves~\cite{DoeringAvraham88,DoeringAvraham89,BDA89,ABD90,DBH91}.
The IPDF technique was also used to study the finite-size scaling
behavior of coagulation processes~\cite{DoeringBurschka90,KPWH95}.
Even anisotropic systems~\cite{HKP96} and models with homogeneous
or localized particle input~\cite{DoeringAvraham89,HRS97} have
been solved. Nevertheless the IPDF method is a rather special
technique which seems to be restricted to models with an
underlying fermionic symmetry.

%==============================================================================
\subsection{Experimental verification of fluctuation effects}
%==============================================================================
%
%
\begin{figure}
\epsfxsize=75mm \centerline{\epsffile{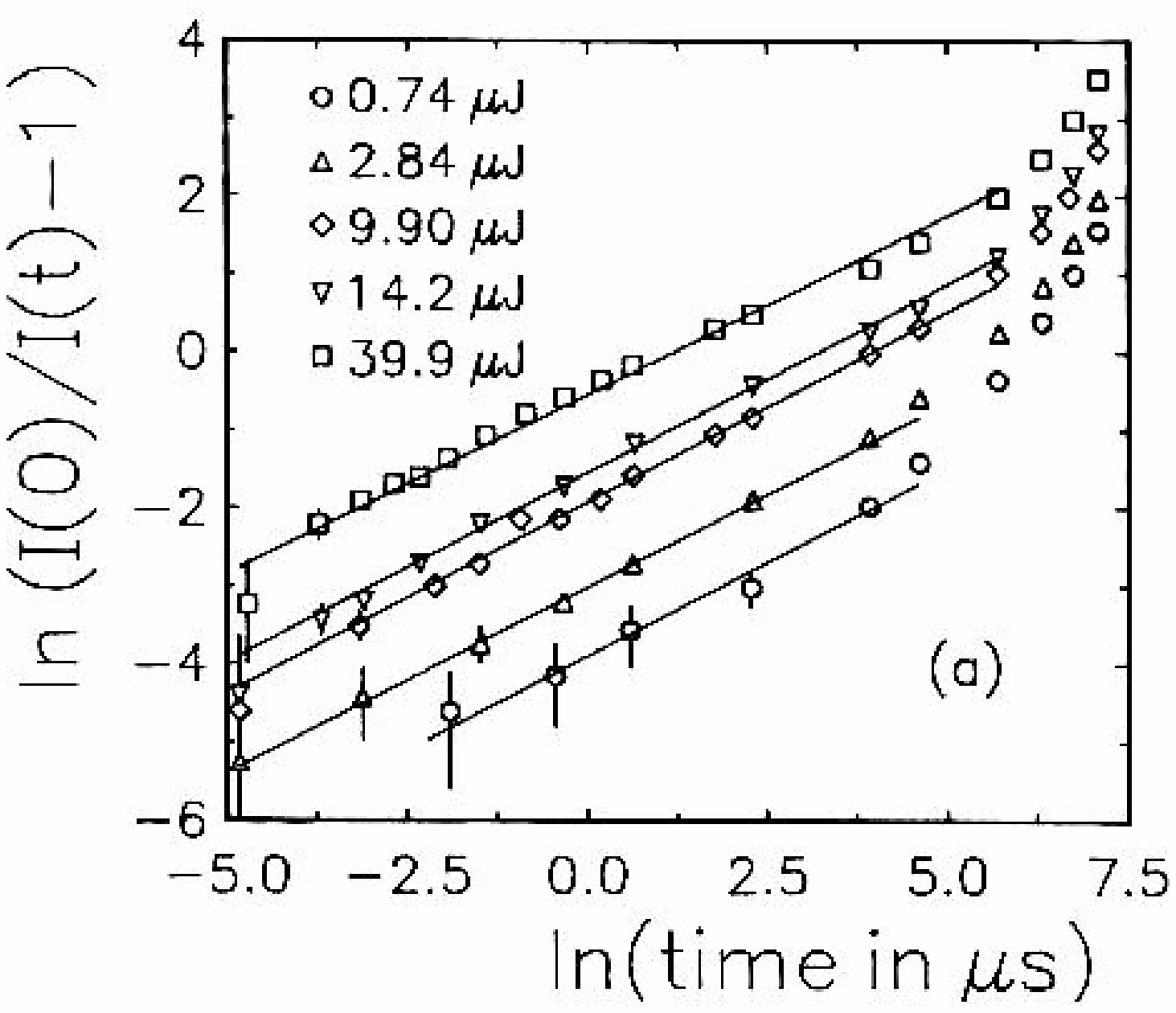}} \smallcaption{
\label{FIGTMMC} Decay of the luminescence of TMMC after excitation
by a laser pulse for various energies (figure reprinted from
Ref.~\cite{KroonSprik97}). The straight lines are best fits to
Eq.~(\ref{FitTMMC}). }
\end{figure}
The preceding exact calculation proves that the particle
concentration in a one-dimensio\-nal coagulation process decays as
$\rho(t) \sim t^{-1/2}$. This result differs significantly from
the mean-field prediction $\rho(t)\sim1/t$. Therefore, the
coagulation model provides one of the simplest examples where
fluctuation effects change the entire temporal behavior of a
reaction-diffusion process.

It is quite remarkable that this result could be verified
experimentally by analyzing the kinetics of laser-induced excitons
on tetramethylammonium manganese trichloride
(TMMC)~\cite{KFS93,KroonSprik97}. TMMC is a crystal consisting
of parallel manganese chloride chains. Laser-induced electronic
excitations of the Mn$^{2+}$ ions, so-called excitons, migrate
along the chain and may be interpreted as quasi-particles. The
chains are separated by large tetramethylammonium ions so that the
exchange of excitons between different chains is suppressed by a
factor of $10^4$. Therefore, the polymer chains can be considered
as one-dimensional systems. Because of exciton-phonon induced
lattice distortions the motion of excitons is diffusive. Moreover,
when two excitons meet at the same lattice site, the Mn$^{2+}$ ion
is excited to twice the excitation energy. Subsequently, the ion
relaxes back to a simply exited state by the emission of phonons.
Thus, the fusion of excitons can be viewed as a coagulation
process $2A \rightarrow A+${\it heat} on a one-dimensional
lattice.

The concentration of quasiparticles can be measured indirectly by
detecting the luminescence intensity $I(t)$ which is proportional
to the number of excitons. Eq.~(\ref{IPDFEq}) predicts that $I(t)
\simeq I(0)/(1+\alpha t^\delta)$, where $\alpha$ fixes the time
scale and $\delta=1/2$. This equation can be rewritten as
\begin{equation}
\label{FitTMMC} \log \left[ \frac{I(0)}{I(t)}-1\right] = \delta
\log t + \log \alpha \ .
\end{equation}
The experimental results are shown in Fig.~\ref{FIGTMMC}. The best
fits according to Eq.~(\ref{FitTMMC}) yield estimates of about
$\delta=0.48(3)$, being in perfect agreement with the
theoretical prediction $\delta=1/2$. In other experiments the
polymers chains are confined to small pores. Here the excitons
perform both annihilation and coagulation processes. The estimates
$\delta=0.55(4)$~\cite{PrasadKopelman89} and
$\delta=0.47(3)$~\cite{KLS90} are again in agreement with the
theoretical result.

To summarize, the experimental investigation of excitons on
polymer chains confirms that the concept of stochastic
reaction-diffusion processes is well justified in order to
quantitatively predict the behavior of certain complex systems. In
addition, these experiments prove that fluctuations effects do
exist in nature and may change the physical properties of the
system in agreement with the theoretical prediction.
%
%
%==============================================================================
\subsection{Dynamic processes approaching thermal equilibrium}
%==============================================================================

\label{THEQU}
Stochastic dynamic processes also play an important role in the
context of equilibrium models. As outlined in the Introduction,
equilibrium statistical mechanics deals with many-particle
systems in contact with a thermal reservoir
(heat bath) of temperature $T$. In the long-time limit
such a system approaches a statistically stationary state where
it evolves through certain configurations according to a well-defined
probability distribution $P_{eq}(s)$. The key property of equilibrium
models is the existence of an energy functional ${\cal H}$ associating
each configuration $s$ with a certain energy ${\cal H}(s)$.
The equilibrium distribution $P_{eq}(s)$ is then given by the canonical
ensemble~\cite{Gibbs02}
\begin{equation}
P_{eq}(s)=\frac{1}{Z} \exp(-{\cal H}(s)/k_BT) \ ,
\end{equation}
where $T$ is the temperature, $k_B$ the Boltzmann constant,
and $Z$ the partition sum. This probability distribution
can be used to determine averages of certain
macroscopic observables by summing over all accessible states.
It is important to note that equilibrium statistical
mechanics does not involve any dynamical aspect.
In other words, it is irrelevant {\em how} the
system evolves through different configurations, one is only interested
in the relative frequency of certain configurations to be visited
in the stationary state.

Although there is no `time' in equilibrium statistical physics,
one may use dynamic random processes as a tool to {\em generate} the
equilibrium ensemble $P_{eq}(s)$ of a particular equilibrium model. More
precisely, such a dynamic process evolves into a stationary state
$P_\infty(s) := \lim_{t \rightarrow \infty} P_t(s) $ that coincides
with the equilibrium ensemble $P_{eq}(s)$. Generally there is a large
variety of dynamic random processes that can be used to generate the
stationary ensemble of a particular equilibrium model. Let us, for example,
consider the ferromagnetic Ising model on a $d$-dimensional
square lattice~\cite{McCoyWu73}. Its energy functional is given by
\begin{equation}
\label{IsingModel}
{\cal H}(\sigma)=-J \sum_{<i,j>} \sigma_i \sigma_j \ ,
\end{equation}
where the sum runs over pairs of adjacent sites, $J$ is a coupling
constant, and $\sigma_i=\pm 1$ denotes the local spin at site $i$.
The equilibrium ensemble of the Ising model may be generated by a
dynamic process with synchronous dynamics mimicking the contact of
the system with a thermal reservoir. These dynamic rules --
usually referred to as {\em heat bath} dynamics -- are defined
through the transition probabilities
\begin{equation}
\label{IsingTransitionProb}
p_{\sigma \rightarrow \sigma^\prime} = \prod_i p_i(\sigma)
\ , \qquad
p_i(\sigma) = \frac{e^{h_i(\sigma)}}{e^{h_i(\sigma)}+e^{-h_i(\sigma)}}
\ , \qquad
h_i(\sigma)=\frac{1}{k_B T} \sum_{j} \sigma_j \ ,
\end{equation}
where $\sigma$ denotes the actual state of the model and $j$ runs
over the nearest neighbors of $i$. In order to verify the
coincidence of the stationary distribution $P_\infty(\sigma)$ and
the equilibrium ensemble of the Ising model, it is sufficient to
prove that the dynamic processes are ergodic and obey {\em
detailed balance}
\begin{equation}
\label{DetailedBalance}
P_{eq}(\sigma) p_{\sigma \rightarrow \sigma^\prime} =
P_{eq}(\sigma^\prime) p_{\sigma^\prime \rightarrow \sigma} \ .
\end{equation}
Detailed balance means that the probability currents between two
states are exactly equal in both directions, i.e., the currents
cancel each other in the stationary state. Heat bath dynamics is
only one out of infinitely many dynamic processes generating
the Ising equilibrium ensemble. Examples include Glauber,
Metropolis, and Kawasaki dynamics, as well as the Swendsen-Wang
and Wolf cluster algorithms. Although these stochastic models have
very different dynamic properties, they all evolve towards the
same stationary state which is just the equilibrium state of the
Ising model.

%==============================================================================
\subsection{Matrix product states}
%==============================================================================

\label{MPS} For the majority of reaction-diffusion models it is
quite difficult or even impossible to solve the master equation
analytically. In some cases, however, it is still possible to
compute the {\em stationary state} of the system. In recent years
a powerful algebraic approach has been developed by which
$n$-point correlation functions of certain nonequilibrium systems
can be computed exactly (see Ref.~\cite{DerridaEvans97} for a
general review). This approach generalizes states with product
measure to so-called matrix product states (MPS) by replacing
real-valued probabilities with non-commutative operators.
Representing these operators in terms of matrices it is possible
to compute correlation functions by evaluating certain matrix
products.

%---------------------------------------------------------------------------
\headline{MPS for the exclusion process}
%---------------------------------------------------------------------------
%
In order to introduce the matrix product technique, let us
consider the partially asymmetric exclusion process in one spatial
dimension with asynchronous dynamics and particle adsorption
(desorption) at the left (right) boundary. The Liouville operator
of this system is given by
\begin{equation}
{\cal L}={\cal S}_1 + {\cal S}_N + \sum_{i=1}^{N-1} {\cal L}_i \,,
\end{equation}
where ${\cal L}_i$ describes the hopping in the bulk while ${\cal
S}_1$ and ${\cal S}_N$ are the surface contributions for
adsorption and desorption, respectively. In the standard basis
(see Appendix~\ref{APPVECSEC}), these operators read
\begin{equation}
\label{ASYMEXCL} {\cal L}_{i}=
\begin{pmatrix}
0 & 0 & 0 & 0 \\ 0 & q^{-1} & -q & 0 \\ 0 & -q^{-1} & q & 0 \\ 0 &
0 & 0 & 0
\end{pmatrix}
\ , \qquad {\cal S}_{1}=
\begin{pmatrix}
\alpha & 0 \\ -\alpha & 0
\end{pmatrix}
\ , \qquad {\cal S}_{N}=
\begin{pmatrix}
0 & -\beta \\ 0 & \beta
\end{pmatrix}
\ .
\end{equation}
The partially asymmetic exclusion process with particle input and
output is particularly interesting in presence of a current,
that is, for $q \neq 1$. Depending on $\alpha$ and $\beta$, the
system is in a phase of low, high, or maximal density (see
Fig.~\ref{FIGMPS}). It seems to be quite surprising that this
simple model could be solved only a few years ago by recursion
techniques~\cite{DDM92,SchuetzDomany93} and, at the same time, by
using the matrix product method~\cite{DEHP93,Sandow94}.

The solution is particularly simple along
the coexistence line between the high and
low density phases $\alpha+\beta=q-q^{-1}$.
Here the stationary state $|P_s\rangle$
may be written as
\begin{equation}
\label{ProductState}
|P_s\rangle = \frac{1}{Z}\,
\begin{pmatrix} e \\ d \end{pmatrix}
\otimes \ldots \otimes \begin{pmatrix} e \\ d \end{pmatrix}
= \begin{pmatrix} e \\ d \end{pmatrix} ^{\otimes N} \ ,
\end{equation}
where $e=1/\alpha$ and $d=1/\beta$. This state has a product
measure, i.e., all spatial correlations vanish. The constant
$Z=(e+d)^N$ normalizes the probability distribution. The
stationarity ${\cal L}|P_s\rangle =0$ can be verified by proving
the relations
\begin{equation}
\label{ProductCancellation}
\begin{split}
{\cal S}_1 \begin{pmatrix} e \\ d \end{pmatrix} =
\begin{pmatrix} 1 \\ -1 \end{pmatrix} \ , \qquad
{\cal S}_N \begin{pmatrix} e \\ d \end{pmatrix} =-
\begin{pmatrix} 1 \\ -1 \end{pmatrix}\ , \\
{\cal L}_i \left[ \begin{pmatrix} e \\ d \end{pmatrix}
\otimes \begin{pmatrix} e \\ d \end{pmatrix} \right]=-
\begin{pmatrix} 1 \\ -1 \end{pmatrix} \otimes
\begin{pmatrix} e \\ d \end{pmatrix} +
\begin{pmatrix} e \\ d \end{pmatrix} \otimes
\begin{pmatrix} 1 \\ -1 \end{pmatrix} \ ,
\end{split}
\end{equation}
which provide the following cancellation mechanism:
First the action of ${\cal S}_1$ generates a factor
$\left(\begin{smallmatrix} 1\\-1 \end{smallmatrix} \right)$
at the leftmost position
in the product~(\ref{ProductState}). This factor is
then commuted to the right by successive action of
${\cal L}_1, \ldots, {\cal L}_{N-1}$
until it is adsorbed at the right boundary by the action
of ${\cal S}_N$. This cancellation mechanism proves that
the homogeneous product state~(\ref{ProductState}) is
stationary along the coexistence line $\alpha+\beta=q-q^{-1}$.

For $\alpha+\beta \neq q-q^{-1}$ the stationary state has no
longer the form of a simple product state. However, as will be
shown below, it can be expressed as a {\em matrix product state}.
To this end we replace the probabilities $e$ and $d$ in
Eq.~(\ref{ProductState}) by noncommutative operators $E$ and $D$.
These operators act in an {\em auxiliary space} which is different
from the configurational vector space of the lattice model.
Introducing boundary vectors $\langle W|$ and $| V \rangle$ with
$\langle W | V \rangle \neq 0$ a matrix product state may be
written as
\begin{equation}
\label{MatrixProductState}
|P_s\rangle = \frac{1}{Z} \, \langle W|
\begin{pmatrix} E\\ D \end{pmatrix} ^{\otimes N}
| V \rangle\ .
\end{equation}
Notice that $\langle W|$ and $| V \rangle$ are vectors in the
auxiliary space while $|P_s\rangle$ denotes the stationary
probability distribution in configuration space. By selecting the
matrix element $\langle W| \ldots | V \rangle$, products of the
operators can be mapped to real-valued probabilities\footnote{The
matrix product technique can also be applied to models with
periodic boundary conditions. In this case the matrix elements
$\langle W|\ldots|V\rangle$ have to be replaced by a trace
operation $\text{Tr}[\ldots ]$.}. The normalization constant is
given by $Z=\langle W|C^N|V \rangle$, where $C=D+E$. The
cancellation mechanism
\begin{equation}
\label{MatrixProductCancellation}
\begin{split}
\langle W|
{\cal S}_1 \begin{pmatrix} E \\ D \end{pmatrix} =
\langle W|
\begin{pmatrix} 1 \\ -1 \end{pmatrix} \ , \qquad
{\cal S}_N \begin{pmatrix} E \\ D \end{pmatrix}
| V \rangle =-
\begin{pmatrix} 1 \\ -1 \end{pmatrix}
| V \rangle
\ , \\
{\cal L}_i \left[ \begin{pmatrix} E \\ D \end{pmatrix}
\otimes \begin{pmatrix} E \\ D \end{pmatrix} \right]=-
\begin{pmatrix} 1 \\ -1 \end{pmatrix} \otimes
\begin{pmatrix} E\\ D \end{pmatrix} +
\begin{pmatrix} E\\ D \end{pmatrix} \otimes
\begin{pmatrix} 1 \\ -1 \end{pmatrix}
\end{split}
\end{equation}
is the same as in Eq.~(\ref{ProductCancellation}), leading to the
algebra
\begin{equation}
\label{BulkAlgebraExcl}
DE=D+E
\end{equation}
and the boundary conditions
\begin{equation}
\langle W| E=\alpha^{-1} \langle W|  \ , \qquad
D | V \rangle = \beta^{-1} | V \rangle \,.
\end{equation}
This ansatz is not restricted to the exclusion process but can be
applied to any reaction-diffusion process with random-sequential
updates. It converts the dynamic rules of the model into a set of
algebraic relations and boundary conditions. The problem of
calculating the stationary state is then shifted to the problem of
finding a matrix representation of the algebra.  In the present
case the quadratic algebra~(\ref{BulkAlgebraExcl}) can be mapped
onto a bosonic algebra for which an infinite-dimensional matrix
representation exists~\cite{DEHP93}. For particular values of
$\alpha$ and $\beta$, however, there are also finite-dimensional
representations~\cite{EsslerRittenberg96,MallickSandow97}. For
example, along the coexistence line $\alpha+\beta=q-q^{-1}$ the
product state~(\ref{ProductState}) is nothing but a
one-dimensional representation of the algebra. Moreover, it can be
shown that for $\alpha+\beta+\alpha\beta q=q+q^{-1}$ a
two-dimensional matrix representation is given by
\begin{figure}
\epsfxsize=90mm \centerline{\epsffile{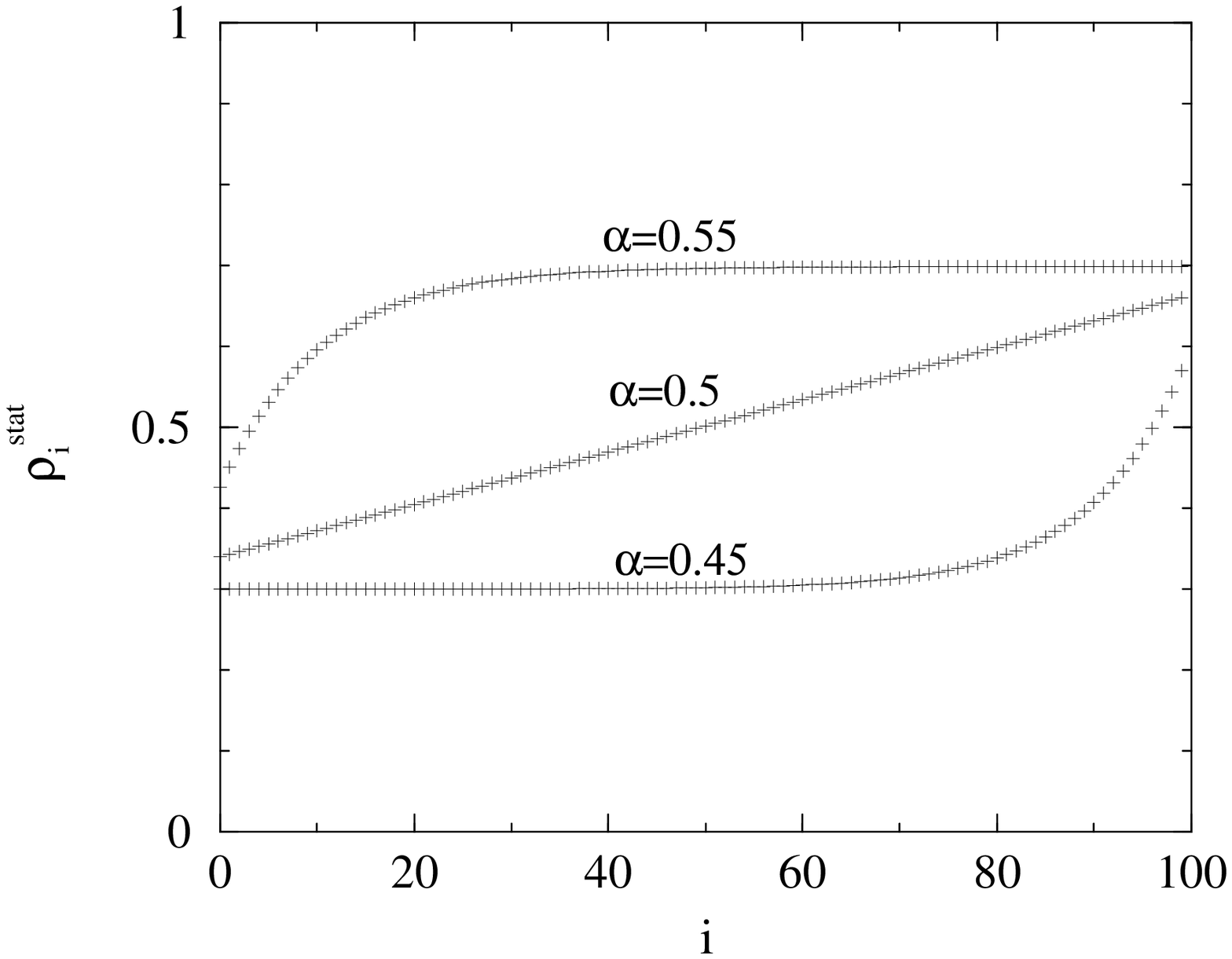}} \vspace{-3mm}
\smallcaption{ \label{FIGPROFILE} Exact stationary density profile
for the partially asymmetric exclusion process with  particle
adsorption and desorption at the boundaries for $q=2$ and $N=100$
in the low density phase $\alpha=0.45$, along the coexistence line
$\alpha=0.5$, and in the high density phase $\alpha=0.55$. }
\end{figure}
\begin{equation}
\begin{split}
E = \begin{pmatrix} \frac{1}{\alpha} & 0 \\
1 & \frac{1+\alpha q}{\alpha q^2} \end{pmatrix} \ , \qquad
D = \begin{pmatrix} \frac{1}{\beta} & -1 \\
0 & \frac{1+\beta q}{\beta q^2} \end{pmatrix} \ , \qquad
\langle W| = (1\,\, 0) \ , \qquad
| V \rangle = \begin{pmatrix} 1\\0 \end{pmatrix} \ .
\end{split}
\end{equation}
For a given matrix representation the stationary particle concentration
profile $\rho_i^{stat}$ can be computed by evaluating the matrix product
\begin{equation}
\rho_i^{stat} = \frac
{\langle W| C^{i-1} D C^{N-i} | V \rangle }
{\langle W| C^N | V \rangle }
\ .
\end{equation}
In the case of the above $2\times2$ representation
we obtain the exact result
\begin{equation}
\rho_i^{stat}=
\frac
{
\frac{1}{\alpha q} \lambda_1^{N-1} -
\frac{\alpha q}{\beta^2 (1+\alpha q) } \lambda_2^{N-1} +
\lambda_1^{i-1} \lambda_2^{N-i}
}
{
\frac{\beta(1+\alpha q)}{\alpha q} \lambda_1^N -
\frac{\alpha q}{\beta(1+\alpha q)} \lambda_2^N
}\ ,
\end{equation}
where
\begin{equation}
\lambda_1=\frac{q-q^{-1}}{\alpha\beta(1+\alpha q)} \ , \qquad
\lambda_2=\frac{q-q^{-1}}{\alpha\beta(1+\beta q)}
\end{equation}
are the eigenvalues of $C$. The density profile is shown in
Fig.~\ref{FIGPROFILE} for various values of~$\alpha$,
visualizing the transition between the low and the
high density phase. Similarly, one can compute
two-point correlation functions
\begin{equation}
\langle s_i s_j \rangle = \frac
{\langle W| C^{i-1} D C^{j-i-1} D C^{N-j} | V \rangle }
{\langle W| C^N | V \rangle } \ .
\end{equation}
In this expression the matrix $C$ plays the role of a transfer
matrix between the sites $i$ and~$j$. Therefore, the long-distance
behavior of correlation functions in the bulk will be governed by
the largest eigenvalue of $C$. In particular, for any
finite-dimensional representation of the algebra, the correlation
functions in the stationary state will decay exponentially. Only
infinite-dimensional representations can lead to long-range
correlations with power-law decay in the stationary state.

It is also possible to apply the matrix product technique to the
totally asymmetric exclusion process with sublattice-parallel
updates~\cite{SchadschneiderSchreckenberg93,Schuetz93a,YKT94},
where a different cancellation mechanism is
needed~\cite{Hinrichsen96,RSS96,HoneckerPeschel97,RSSS98}.
Recently, the matrix product method could even be extended to the
case of fully parallel updates~\cite{ERS99}.

%---------------------------------------------------------------------------
\headline{MPS for models with particle reactions}
%---------------------------------------------------------------------------
%
Most models which have been solved so far by using the matrix
product method are diffusive systems, i.e., they describe
stochastic transport of particles. Up to now only one exception is
known where particles {\em react} with each other, namely the
anisotropic decoagulation model with closed boundary conditions
and random sequential updates. As will be explained in the
following, this model exhibits a boundary-induced phase transition
and can be solved by using a generalized matrix product
ansatz~\cite{HSP96}.

\noindent
The anisotropic decoagulation model is defined by the following
dynamic rules:
$$
\begin{array}{lccl}
\text{diffusion:} &&
\vacancy A \underset{q}{\rightarrow} A \vacancy \, , &
A \vacancy \underset{q^{-1}}{\rightarrow} \vacancy A \, , \\
\text{coagulation:} &&
AA \underset{q}{\rightarrow} A \vacancy  \, , &
AA \underset{q^{-1}}{\rightarrow} \vacancy A \, , \\
\text{decoagulation:} &&
\vacancy A \underset{\crit q}{\rightarrow} A A \, ,  &
A \vacancy \underset{\crit q^{-1}}{\rightarrow} A A \, . \\
\end{array}
$$
The corresponding Liouville operator reads
\begin{equation}
\label{AnisoCoag}
{\cal L}_i \;=\; \left(
\begin{array}{cccc}
0 & 0 & 0 & 0 \\
0 & (\crit+1) q & -q^{-1} & -q^{-1} \\
0 & -q & (\crit+1) q^{-1} & -q \\
0 & -\crit q & -\crit q^{-1} & q+q^{-1}
\end{array} \right)\,.
\end{equation}
Thus, the model is controlled by two parameters, namely the anisotropy
$q$ and the decoagulation rate~$\crit$. The phase diagram displays
two phases, a low-density phase for $\crit < q^2-1$ and a
high-density phase for $\crit > q^2-1$. From the physical point of
view these phases are different from those observed in the
asymmetric exclusion process since here the number of particles is
not conserved. Notice that the rates for diffusion and coagulation
coincide. Moreover, all reactions have the same bias. This special
choice ensures that the model is integrable. Various exact results
have been obtained by using IPDF and free-fermion
techniques~\cite{DoeringAvraham88,ADHR94,PRS94,KPWH95}. At the
critical point $\crit_c=q^2-1$, the relaxational spectrum becomes
massless and algebraic long-range correlations can be
observed~\cite{HKP96}.

In order to express the stationary state of the model as a matrix
product state of the form~(\ref{MatrixProductState}), the
cancellation mechanism of Eq.~(\ref{MatrixProductCancellation})
has to be generalized. This can be achieved by replacing the
vector $\left(\begin{smallmatrix} 1\\-1 \end{smallmatrix} \right)$
by $\left(\begin{smallmatrix} \bar{E}\\\bar{D} \end{smallmatrix}
\right)$, where $\bar{E}$ and $\bar{D}$ are two additional
operators acting in the auxiliary space. The generalized
cancellation mechanism reads
\begin{equation}
\label{GeneralizedCancellation}
\begin{split}
\langle W|
{\cal S}_1 \begin{pmatrix} E \\ D \end{pmatrix} =
\langle W|
\begin{pmatrix} \bar{E} \\ \bar{D} \end{pmatrix} \ , \qquad
{\cal S}_N \begin{pmatrix} E \\ D \end{pmatrix}
| V \rangle =-
\begin{pmatrix} \bar{E} \\ \bar{D} \end{pmatrix}
| V \rangle
\ , \\
{\cal L}_i \left[ \begin{pmatrix} E \\ D \end{pmatrix}
\otimes \begin{pmatrix} E \\ D \end{pmatrix} \right]=-
\begin{pmatrix} \bar{E} \\ \bar{D} \end{pmatrix} \otimes
\begin{pmatrix} E\\ D \end{pmatrix} +
\begin{pmatrix} E\\ D \end{pmatrix} \otimes
\begin{pmatrix} \bar{E} \\ \bar{D} \end{pmatrix} \ .
\end{split}
\end{equation}
This ansatz leads to the bulk algebra
\begin{equation}
\label{BulkAlgebra}
\begin{split}
0 = E\bar{E}-\bar{E}E \ , \\
(\crit+1) q \,ED \,-\, q^{-1} DE \,-\, q^{-1} DD =
E\bar{D}-\bar{E}D \ , \\
-q\,ED \,+\, (\crit+1) q^{-1} DE \,-\, q\,DD =
D\bar{E}-\bar{D}E \ , \\
-\crit q \, ED \,-\, \crit q^{-1} DE \,+\, (q+q^{-1})\,DD =
D\bar{D}-\bar{D}D \ ,
\end{split}
\end{equation}
and the boundary conditions
\begin{equation}
\label{BoundaryRelations}
\langle W | \bar{E} = \langle W | \bar{D} =
\bar{E} | V \rangle = \bar{D} | V \rangle \;=\; 0\,.
\end{equation}
A trivial one-dimensional representation of this algebra is given by
$E=C=1,\;\bar{E}=\bar{C}=0$, describing a system without particles.
In the symmetric case $q=1$ there is another one-dimensional
representation $E=1,\;C=\gamma^2,\;\bar{E}=\bar{C}=0$, corresponding to
a homogeneous product state with particle density $\rho=\crit/(1+\crit)$.
For $q \neq 1$ and $\crit \neq q^2-1$ we find the four-dimensional
representation
\begin{align}
&E =
\begin{pmatrix}
q^{-2} & q^2-\gamma^{-2} & q^2-1 & q^2(1-\gamma^2) \\
0 & \gamma^{-2} & 0 & \gamma^2-q^2 \\
0 & 0 & 1 & \gamma^2(q^2-1) \\
0 & 0 & 0 & q^2
\end{pmatrix}
, \qquad
C =
\begin{pmatrix}
q^{-2} & 0 & 0 & 0 \\
0 & 1 & 0 & 0 \\
0 & 0 & \gamma^2 & 0 \\
0 & 0 & 0 & q^2
\end{pmatrix}
,
\nonumber \\[2mm]
&\langle W | =
\Bigl(
\frac{1}{1-q^2 \gamma^2}\,,\;\;
0\,,\;\;
\frac{q^2}{q^2 \gamma^2-1}\,,\;\;
\frac{a\,(q^2-q^{-2})(\gamma^2-q^2)\gamma^2 \,-\, q^2\gamma^2}
{(\gamma^2-1)(q^2+1)}
\Bigr)
\ , \\[1mm]
&| V \rangle =
\Bigl(
\frac{b\,(q^4-1)(q^2\gamma^2-1)\,+\,q^4}{q^2+1}
 \,,\;\;
0\,,\;\;
\frac{q^2(\gamma^2-1)}{\gamma^2-q^2}\,,\;\;
\frac{(\gamma^2-1)q^2}{\gamma^4-\gamma^2q^2}
\Bigr)
\ . \nonumber
\end{align}
Using this representation, it is easy to compute
the stationary particle density
\begin{equation}
\textstyle
\rho_j^{stat} = \frac{
\gamma^{2N} \Bigl( (\gamma^2-1)+(q^2-1)\gamma^2(q\gamma)^{-2j} \Bigr) -
q^{2N}\Bigl((\gamma^2-1)q^{2-4j}+(q^2-1)(q/\gamma)^{-2j}\Bigr)}
{\gamma^2\,(\gamma^{2N} + \gamma^{-2N}-q^{2N}-q^{-2N}) }
\end{equation}
which is in agreement with the result obtained by using the
IPDF method~\cite{HKP96}. For $\crit \neq q^2-1$
a similar four-dimensional matrix representation can be constructed.

\begin{figure}
\epsfxsize=135mm
\centerline{\epsffile{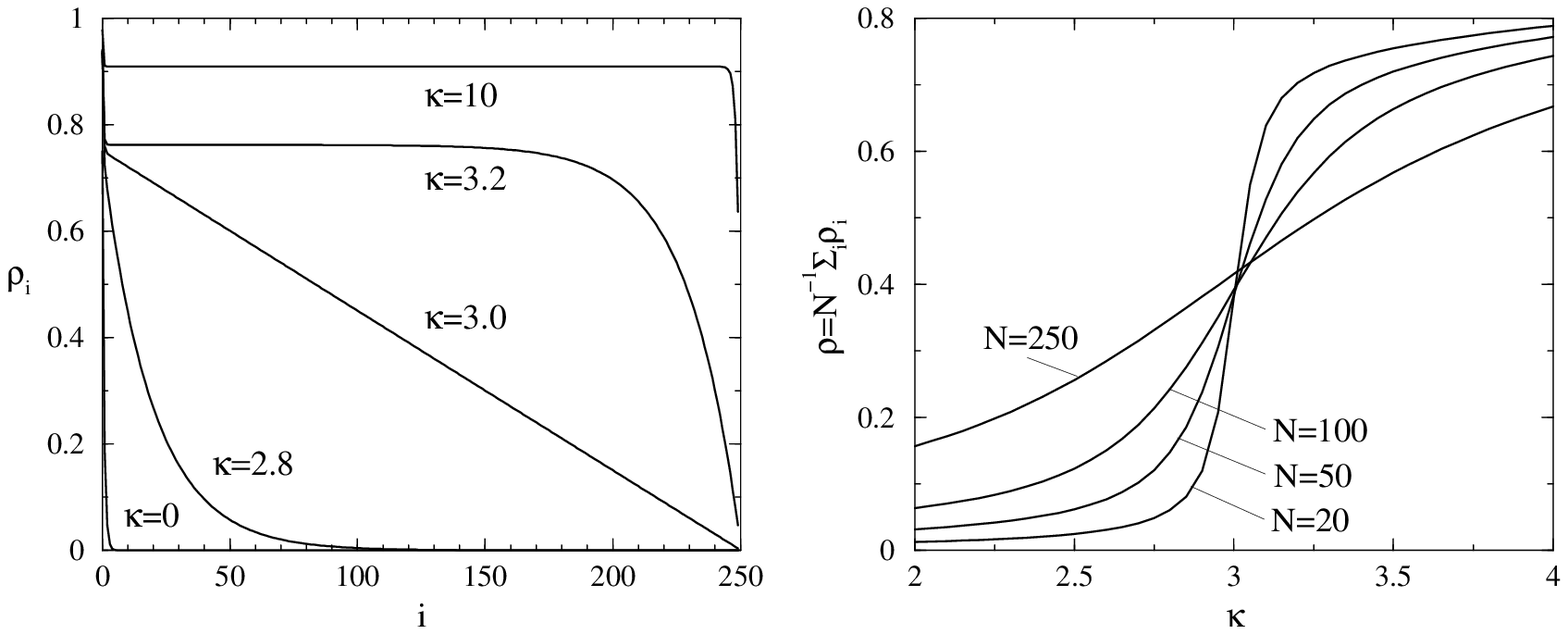}}
\smallcaption{
\label{FIGMPSCOAG}
The anisotropic coagulation-decoagulation model with bias $q=2$.
Left: Density profiles in a system with $N=250$ sites for various
values of the decoagulation rate $\crit$. Right: Density of particles
averaged over the entire system as a function of $\crit$ for various
system sizes, illustrating the first-order phase transition.
}
\end{figure}

In the thermodynamic limit the anisotropic coagulation model
exhibits a first-order phase transition (see
Fig.~\ref{FIGMPSCOAG}). If the decoagulation rate $\crit$ is small
enough, the particles are swept towards one of the boundaries
where they coagulate. The stationary particle density is therefore
zero in the thermodynamic limit. Increasing $\crit$ this region
grows until its size diverges at a critical value $\crit_c=q^2-1$.
Above $\crit_c$ the decoagulation process is strong enough to
maintain a non-vanishing density of particles in the bulk. It
should be emphasized that this type of phase transition is induced
by the boundaries. In particular, there is no such transition if
periodic boundary conditions are used.

The matrix product technique has also been applied to various
other systems such as valence-bond-state models~\cite{KSZ92},
spin-one quantum antiferromagnets~\cite{FNW89,KSZ93}, hard-core
diffusion of oppositely charged particles~\cite{EFGM95}, systems
with fixed~\cite{HinrichsenSandow97} or moving
impurities~\cite{Mallick96,DerridaEvans99}, as well as $n$-state
diffusion processes~\cite{ADR98,Karimipour99}. Furthermore, a dynamic matrix
product ansatz has been introduced by which time-dependent
properties of the exclusion process can be
described~\cite{StinchcombeSchuetz95,Schuetz98a}.
Although the full range of possible applications is not yet known,
the matrix product technique seems to be limited to a few classes
of models. By definition the method is restricted to
one-dimensional systems. Moreover, there seems to be a subtle
connection between matrix product states and integrability. This
connection is not yet fully understood. In
Ref.~\cite{SandowKrebs97} it was shown that the stationary state
of any reaction-diffusion model can be expressed in terms of a
MPS. However, in this generic case the corresponding matrix
representation depends on the system size and is therefore useless
from the practical point of view. A systematic classification
scheme for matrix product states is not yet known. In this context
it is interesting to note that the method of dynamic density
matrix renormalization allows finite-dimensional matrix
representation to be detected. As shown in
Ref.~\cite{KaulkePeschel98}, the existence of a finite-dimensional
MPS is indicated by the fact that the density matrix has only a
finite number of nonvanishing eigenvalues. Thus, by scanning the
spectrum over a certain range of the systems parameter space, it
is possible to search systematically for finite-dimensional matrix
product representations.

\newpage % for condmat-version

%##############################################################################
\section{Directed percolation}
%##############################################################################
%
\label{DPSection}
Spreading processes are encountered in many different situations
in nature as diverse as epidemics~\cite{Mollison77},
forest fires~\cite{Albano94a}, and transport in random
media~\cite{HavlinAvraham87,BouchaudGeorges90}. Spreading
phenomena are usually characterized by two competing processes.
For example, in an infectious disease the spreading agent (bacteria)
may multiply and infect neighboring individuals. On the
other hand, infected individuals may recover, decreasing
the total amount of the spreading agent. Depending on the
relative rates for infection and recovery, two different
situations may emerge. If the infection process dominates,
the epidemic disease will spread  over the entire population,
approaching a stationary state in which infection and recovery
balance one another. However, if recovery dominates,
the total amount of the spreading agent continues to decrease and
eventually vanishes.

Theoretical interest in models for spreading stems mainly from the
emerging phase transition between survival and extinction. The
simplest model exhibiting such a transition is directed
percolation (DP)\footnote{ For further reading in this field we
recommend the review on directed percolation by
Kinzel~\cite{Kinzel83,Kinzel85}, a summary of open problems by
Grassberger~\cite{Grassberger97}, and the relevant chapters in a
recent book by Marro and Dickman~\cite{MarroDickman98}.}. In DP
sites of a lattice can either be active (infected) or inactive
(healthy). Depending on a parameter controlling the balance 
between infection and recovery, activity may either spread
over the entire system or die out after some time. In the latter
case the system becomes trapped in a completely inactive state,
the so-called {\em absorbing} state of the model. Since the
absorbing state can only be reached but not be left, it is
impossible to obey detailed balance, i.e., DP is a
nonequilibrium process. The transition between the active and the
inactive phase is continuous and characterized by universal
critical behavior.

In many respects, the nonequilibrium critical behavior of DP is
similar to that of equilibrium models. In particular, it is
possible to use the concept of scale invariance and to identify
various critical exponents. As in equilibrium statistical
mechanics, these exponents allow phase transitions of different
lattice models to be categorized into universality classes. As we
will see below, DP is a fundamental class of nonequilibrium phase
transitions, playing a similar role as the Ising
universality class in equilibrium statistical mechanics. Although
DP is very robust and easy to simulate, its critical behavior
turns out to be highly nontrivial. This is what makes DP so
fascinating.

In this Section we give a general introduction to DP,
discussing the most important lattice models,
basic scaling concepts, finite-size properties,
as well as mean-field approaches. We also summarize
various approximation techniques such
Monte Carlo simulations, series expansions, and numerical
diagonalization. Introducing basic field-theoretic methods
we discuss the critical behavior of DP at surfaces,
the early-time behavior, the influence of fractal initial
conditions, persistence probabilities, and the
influence of quenched disorder. Finally we review
possible experimental realizations of DP
and discuss the question why it is so difficult to verify
the critical exponents in experiments.

\newpage
%==============================================================================
\subsection{Directed percolation as a spreading process}
%==============================================================================
%
\vspace{-3mm}
\label{DPINTROSEC}
%
%---------------------------------------------------------------------------
\headline{From isotropic to directed percolation}
%---------------------------------------------------------------------------
%
Although DP is often regarded as a dynamic process, it was
originally defined as a geometric model for connectivity in
random media which generalizes isotropic (undirected)
percolation~\cite{Essam80,StaufferAharony92}. Such a random
medium could be a porous rock in which neighboring pores are
connected by channels of varying permeability. An
important question in geology would be how deep the water
can penetrate into the rock.

\begin{figure}
\epsfxsize=110mm
\centerline{\epsffile{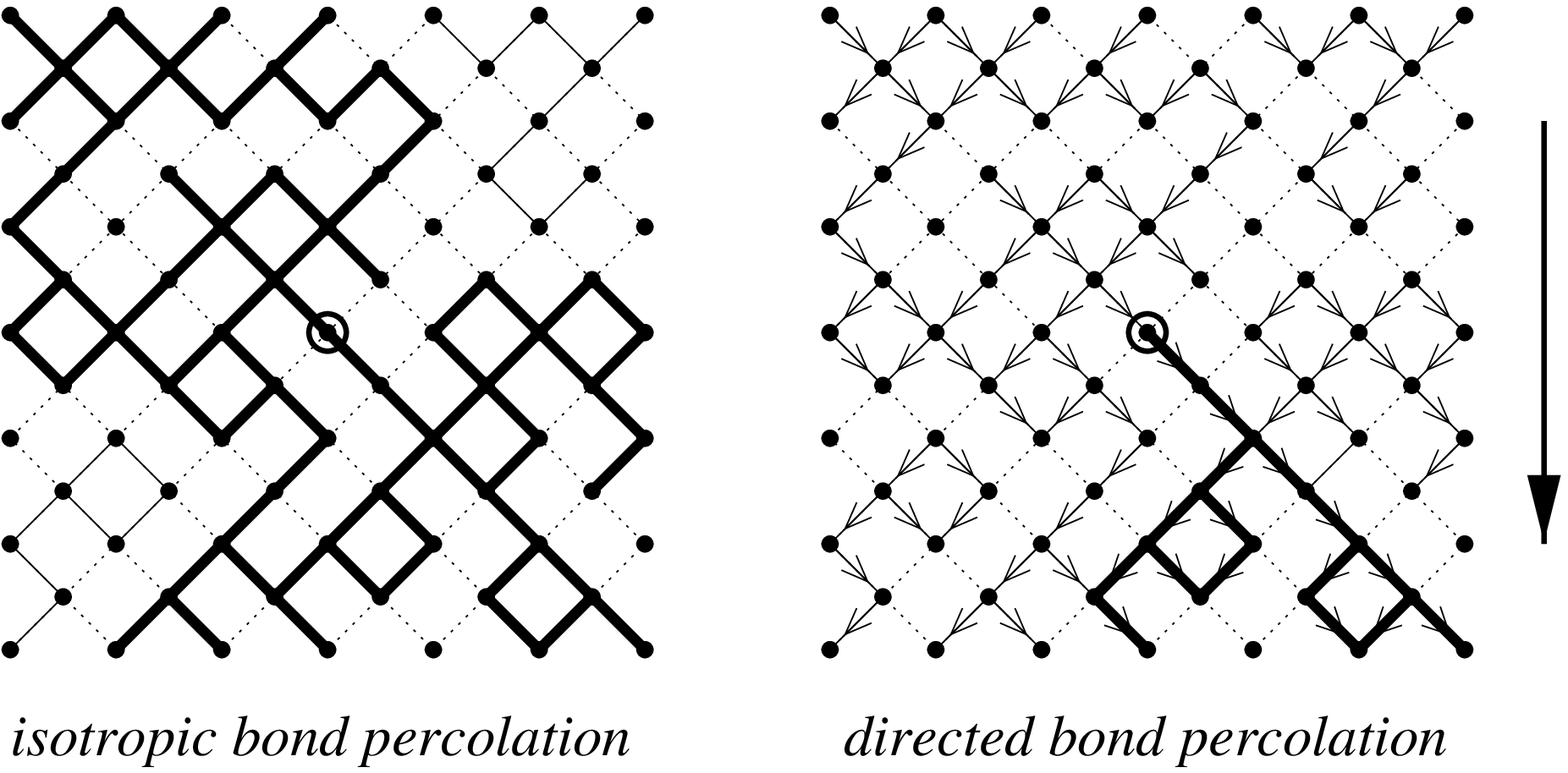}}
\vspace{2mm}
\smallcaption{
\label{FIGISODIR}
Isotropic and directed bond percolation on a diagonal square lattice with
free boundary conditions. Open (closed) bonds are represented by solid
(dotted) lines. In both cases a cluster, indicated by bold lines,
is generated from the lattice site in the center.
In the directed case the spreading agent is
restricted to follow sense of the arrows, leading to a directed
cluster of connected sites.
}
\end{figure}

In ordinary percolation the water propagates isotropically in all
directions of space.  One of the simplest models for isotropic
percolation is {\it bond percolation} on a $d$-dimensional square
lattice, as shown in the left part of Fig.~\ref{FIGISODIR}. 
In this model the channels of the porous medium are
represented by bonds between adjacent sites of a square lattice
which are open with probability~$p$ and otherwise
closed\footnote{Alternatively, we could have blocked sites instead
of bonds with a certain probability. The resulting model, called
{\em site percolation}, exhibits the same type of universal
critical behavior at the transition.}. For simplicity it is
assumed that the states of different bonds are uncorrelated.
Clearly, if $p$  is sufficiently large, the water will percolate
through the medium over arbitrarily long distances. However, if
$p$ is small enough, the penetration depth is expected to be
finite so that large volumes of the material will be impermeable.
Both regimes are separated by a continuous phase transition.

The left part in  Fig.~\ref{FIGISODIR} shows a typical configuration of
open and closed bonds in two dimensions. As can be seen,
each site generates a certain {\it cluster} of connected sites
corresponding to the maximal spreading range if the water
was injected into a single pore. The site from where
the cluster is generated is called the {\it origin} of the cluster.
The size (or mass) of a cluster is defined as the number
of connected sites. Notice that different origins
may generate the same cluster. Consequently the whole lattice
decomposes into a set of disjoint clusters.

\newpage
{\it Directed percolation}, introduced in 1957 by Broadbent and
Hammersley~\cite{BroadbentHammersly57}, is an anisotropic variant
of isotropic percolation. As shown in the right part of
Fig.~\ref{FIGISODIR}, this variant introduces a specific direction
in space. The channels (bonds) function as `valves' in a way that
the spreading agent can only percolate along the given direction,
as indicated by the arrows. For example, we may think of a porous
medium in a gravitational field that forces the water to propagate
downwards\footnote{This assumption is highly idealized since water
is a conserved quantity. 
Moreover, the water can even flow against the gravity
field (see Sec.~\ref{DPEXPERIM}).}. Thus, filling in the spreading
agent at a particular site, the resulting cluster of wet sites is
a subset of the corresponding cluster in the isotropic case (see
Fig.~\ref{FIGISODIR}). Since in DP each site generates an
individual cluster, a decomposition of the lattice
into disjoint clusters is no
longer possible. As in the case of isotropic percolation, DP
exhibits a continuous phase transition.

The phase transitions of isotropic and directed percolation are
similar in many respects. They both can be characterized by an
order parameter $P_\infty$ which is defined as the probability
that a randomly selected site generates an infinite cluster. If
$P_\infty$ is finite, the spreading agent is able to percolate
over arbitrarily long distances wherefore the system is said to be
in the wet phase. If $P_\infty$ vanishes, the system is in the
so-called dry phase where the spreading range is finite. Both
isotropic and directed percolation are trivial in one dimension:
Since an infinite cluster on a line requires {\em all} bonds to be
open, the wet phase consists of a single point $p=1$. Another
trivial case is the limit of infinitely many dimensions.  Since
each site is connected with infinitely many neighbors, an infinite
cluster will be generated for any $p>0$. Consequently the inactive
phase consists of single point in phase space, namely $p=0$. In
finite dimensions $2 \leq d < \infty$ there is a continuous
phase transition separating the wet phase from 
the dry phase at some critical value $0<p_c<1$.
In the supercritical phase $p>p_c$ the medium is permeable
($P_\infty>0$) while in the subcritical phase $p<p_c$
the medium becomes impermeable ($P_\infty=0$).
As expected, the critical threshold $p_c$ for directed
percolation is larger than in the isotropic case.

Although isotropic and directed percolation have several
common features, their critical behavior near the phase
transition turns out to be different. In the isotropic case, the
critical properties are in all directions the same (apart from
lattice effects which are usually irrelevant
on large scales) and hence the emerging long-range
correlations are rotationally invariant. Because of a duality
symmetry, the critical point of isotropic bond percolation is 
$p_c=1/2$~\cite{StaufferAharony92}. 
Moreover, the critical exponents are given by simple rational numbers.
Contrarily, the critical properties of DP reflect
the anisotropy in space, leading to different 
critical exponents. In contrast to the isotropic case,
the numerical values of the critical point and the exponents 
of DP are not yet known analytically and seem to be given
by irrational numbers (see Sec.~\ref{ABSEXP}).

%---------------------------------------------------------------------------
\headline{Interpretation of directed percolation as a dynamic process}
%---------------------------------------------------------------------------
%
Regarding the given direction as `time', directed percolation may
be interpreted as a $d$+1-dimensional dynamic process that
describes the spreading of some non-conserved agent\footnote{ Note
that DP differs from `dynamic percolation' which is used as a
epidemic processes with immunization (see Sec.~\ref{RELSEC}.}. For
example, as already mentioned before, DP may be viewed as a simple
model for epidemic spreading of some infectious disease without
immunization~\cite{Harris74}. In recent years the dynamic
interpretation has become increasingly popular, partly because the
time-dependent formulation is the natural realization of DP on a
computer. In what follows we will adopt the dynamic
interpretation. However, one should keep in mind that the
geometric definition in terms of directed paths is fully
equivalent.

\begin{figure}
\epsfxsize=140mm
\centerline{\epsffile{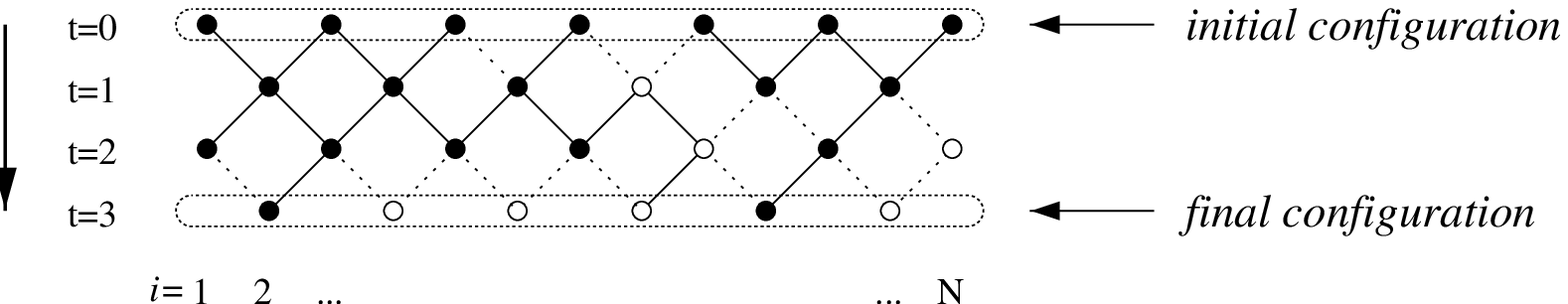}}
\vspace{2mm}
\smallcaption{
\label{FIGBONDPERC}
Directed bond percolation in 1+1 dimensions
interpreted as a time-dependent stochastic process.
Open (closed) bonds are indicated by solid (dashed) lines.
Filled (hollow) circles denote active (inactive) sites.
The configuration of the horizontal row at $t=0$ is
the initial state. Starting from a fully occupied
initial state the model `evolves' through intermediate configurations
according to the dynamic rules of Eq.~(\ref{BondPercCA})
and reaches a final state at $t=3$.
}
\end{figure}

The dynamic interpretation of DP is illustrated in
Fig.~\ref{FIGBONDPERC}, where the lattice sites of a
(1+1)-dimensional directed bond percolation process are enumerated
horizontally by a spatial coordinate $i$ and vertically by a
discrete time variable $t$. Since we use a diagonal square
lattice, odd and even time steps have to be distinguished. A local
binary variable $s_i(t)$ is attached to each site. $s_i=1$ means that
the site is active (occupied, wet) while $s_i=0$ denotes an
inactive (empty, dry) site. The set $s=\{s_i\}$ at a given time $t$
specifies the {\it configuration} of the system.

For a given configuration at time $t$, the next configuration at
time $t+1$ can be determined as follows. For each pair of
bonds between the sites $(i\pm 1,t)$ and $(i,t+1)$ two random number
$z_i^\pm \in (0,1)$ are generated. A bond is considered
to be open (with probability $p$) if $z_i^\pm < p$,
leading to the update rule
\begin{equation}
\label{BondPercCA}
s_{i}(t+1)=
\left\{
\begin{array}{cl}
1 \;&\; \text{if $s_{i-1}(t)=1$ and $z_i^-<p$} \ ,\\
1 \;&\; \text{if $s_{i+1}(t)=1$ and $z_i^+<p$} \ , \\
0 \;&\; \text{otherwise} \ .
\end{array}
\right.
\end{equation}
Thus, directed bond percolation can be considered as a Markov
process with parallel dynamics.
As in any dynamic system, we have to specify the initial state.
Common initial states are the fully occupied lattice,
random initial conditions,
and configurations with a single particle at the origin
(also called `active seed').

\begin{figure}
\epsfxsize=110mm
\centerline{\epsffile{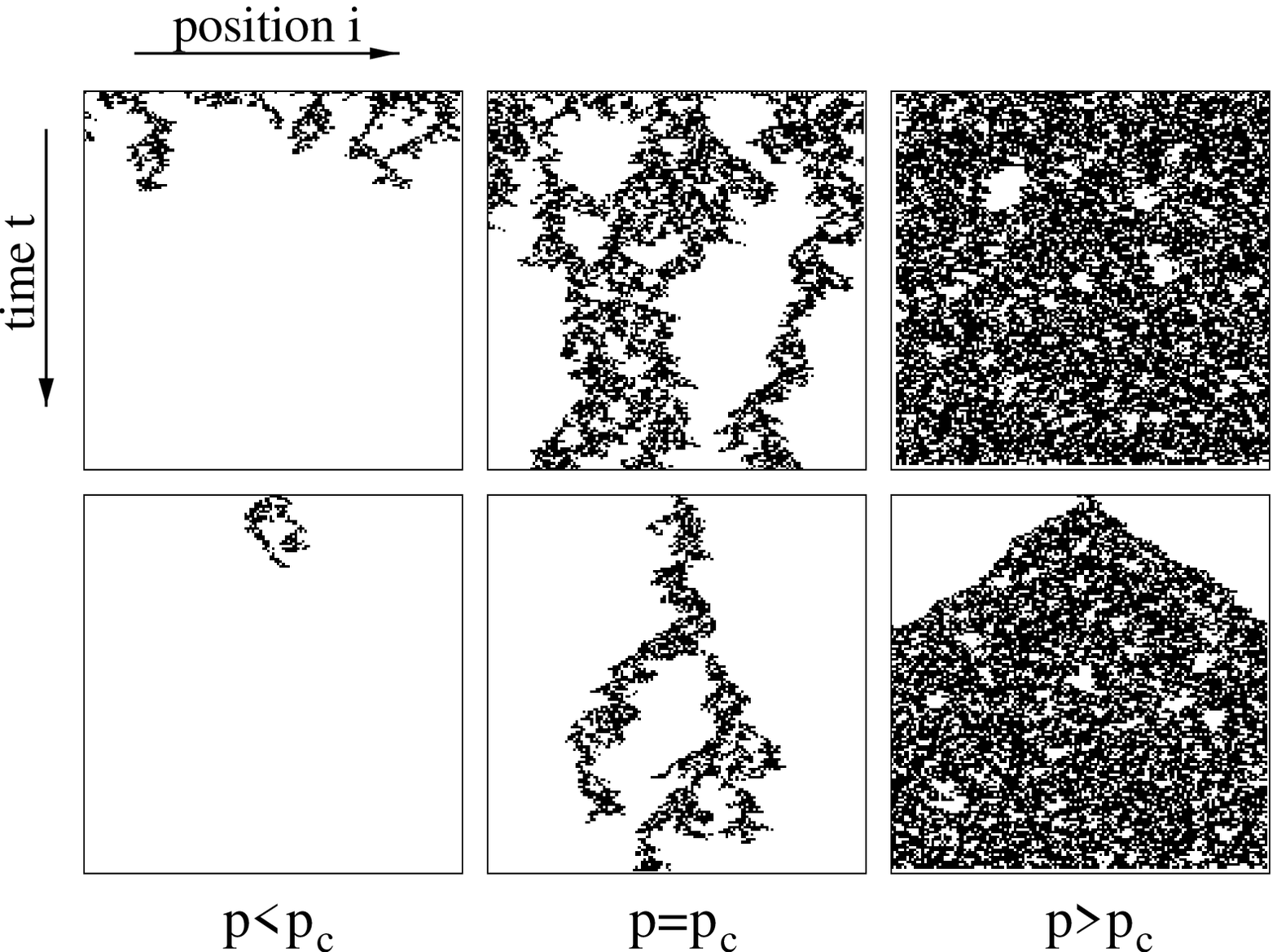}}
\smallcaption{
\label{FIGDPDEMO}
Directed bond percolation in 1+1 dimensions starting from
random initial conditions (top) and from a single active seed (bottom).
Each horizontal row of pixels represents four updates. As can be
seen, critical DP is a reaction-limited process.
}
\end{figure}

Even very simple numerical simulations
demonstrate that the temporal evolution of a DP process changes
significantly at the phase transition. Typical space-time histories
for random initial conditions are shown in the upper part of
Fig.~\ref{FIGDPDEMO}. For $p<p_c$ the number of particles decreases
exponentially until the system reaches the absorbing state,
whereas for $p>p_c$ the average particle number saturates
at some constant value. At the critical point the mean particle
number decays very slowly and the emerging clusters of active
sites remind of fractal structures. A similar behavior can be
observed  if the DP process starts from a single seed
(see lower part of Fig.~\ref{FIGDPDEMO}). For $p<p_c$ the
average number of particles first grows for a short time and then
decays exponentially. For $p>p_c$ there is a finite probability
that the resulting cluster is infinite. 
In this case activity spreads within a
certain triangular region, the so-called spreading cone.
At $p=p_c$ a critical cluster is generated from a single seed, whose scaling
properties will be discussed in Sec.~\ref{ABSSCALESEC}.

It is often helpful to regard DP as a reaction-diffusion process
of interacting particles. Associating active sites with particles $A$
and inactive sites with vacancies $\vacancy$, a DP process corresponds
to the reaction-diffusion scheme
\begin{equation}
\label{DPReactionDiffusion}
\begin{array}{ll}
\text{self-destruction:} & A \rightarrow \vacancy \ , \\
\text{diffusion:} & \vacancy+A \rightarrow A+\vacancy \ , \\
\text{offspring production:} & A \rightarrow 2A \ , \\
\text{coagulation:} & 2A  \rightarrow A \ . \\
\end{array}
\end{equation}
To understand this reaction-diffusion scheme, let us again
consider the example of directed bond percolation. Depending
on the configuration of the bonds, each active site (particle) may activate
two neighboring sites of the subsequent row (next time step).
If both bonds are closed, the particle self-destructs.
If only one bond is open, the particle will diffuse to the left or
to the right with equal probability, whereas an offspring is produced
when both bonds are open. On the other hand,
if two particles reach the same target site,
they coalesce into a single particle, giving rise to the reaction
$2A \rightarrow A$. This process limits the maximal density of
active sites. In fact, as will be shown below, the coagulation
process is the essential nonlinear ingredient of DP.
In `fermionic' models with an exclusion principle
it is automatically included. However, in `bosonic' models allowing
for an infinite number of particles per site one would have to
add this process explicitly.
%
%

%==============================================================================
\subsection{Lattice models for directed percolation}
%==============================================================================
%
\label{LATTICESEC}
In the literature there is a vast variety of DP models
following the spirit of the above reaction-diffusion scheme.
As we will see below, they all exhibit the same type of
critical behavior at the transition.
The common feature of all these models is the existence of
an {\it absorbing state}, i.e.,
a configuration that the model can reach but from where
it cannot escape. In most cases, the absorbing state is
just the empty lattice. The existence of an absorbing state
implies that certain microscopic processes are forbidden
(for example, spontaneous creation of particles $\vacancy \rightarrow A$).
In the sequel we will discuss three examples, namely
the Domany-Kinzel cellular automaton, the contact process, and
the Ziff-Gulari-Barshad model for heterogeneous catalysis.

\begin{figure}
\epsfxsize=110mm \centerline{\epsffile{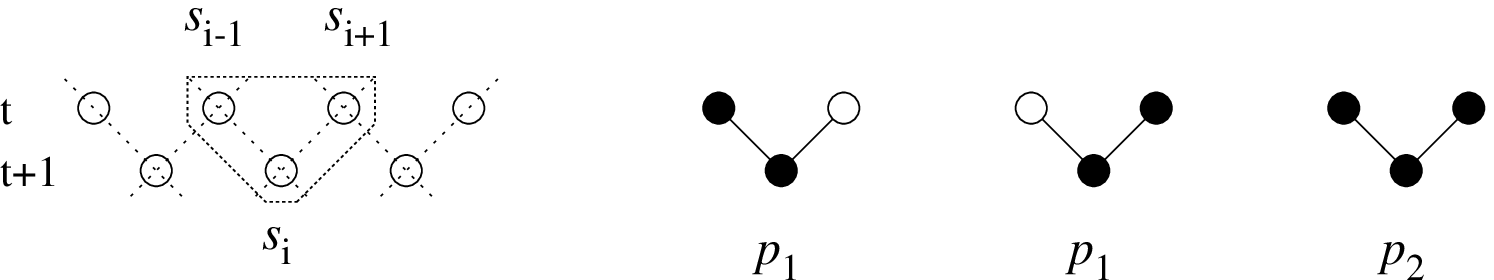}}
\smallcaption{ \label{FIGDKUPDATE} Transition probabilities in the
(1+1)-dimensional Domany-Kinzel cellular automaton. }
\end{figure}
%
%

%---------------------------------------------------------------------------
\headline{The Domany-Kinzel cellular automaton}
%---------------------------------------------------------------------------
%
Cellular automata are algorithms that map a configuration of a
lattice onto a new configuration. The automaton evolves in time
by iteration of the map. Thus, the time variable~$t$ is discrete.
Usually the map can be  decomposed into independent local
updates. Since these updates can be processed simultaneously,
cellular automata can efficiently  be implemented on parallel computers.
Depending on the type of updates, we distinguish between
deterministic and stochastic cellular automata.
A general classification of stochastic cellular automata was
presented by Wolfram~\cite{Wolfram83}.

\begin{figure}
\epsfxsize=100mm
\centerline{\epsffile{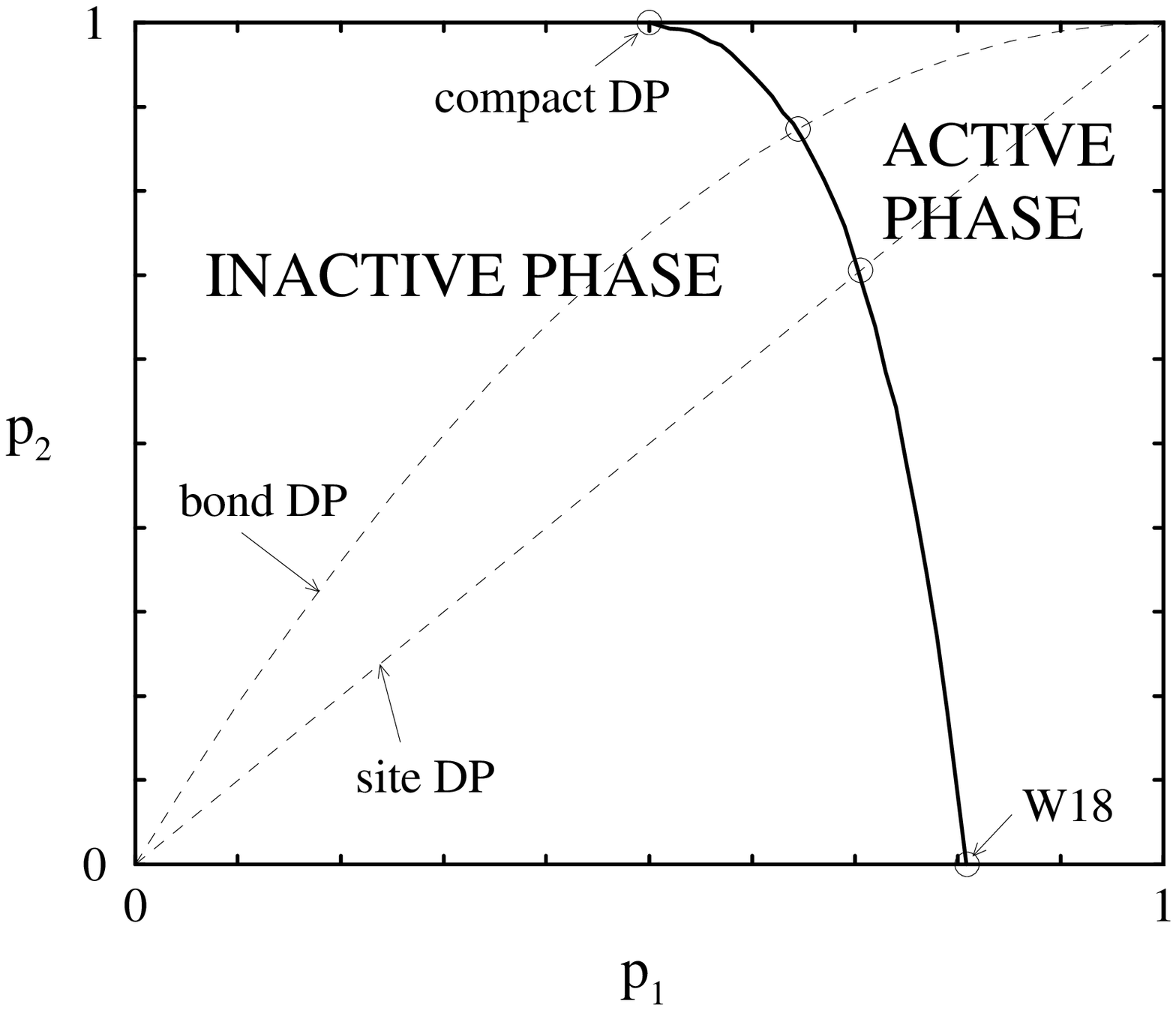}}
\vspace{-4mm}
\smallcaption{
\label{FIGDKPHASE}
Phase diagram of the Domany-Kinzel model.
}
\end{figure}

Various stochastic cellular automata are known to exhibit a DP
transition from a fluctuating phase into an absorbing state. One of
the simplest models in this class is the (1+1)-dimensional
Domany-Kinzel (DK) model~\cite{DomanyKinzel84,Kinzel85}. It is
defined on a diagonal square lattice and evolves by parallel
updates according to certain conditional transition probabilities
$P[s_i(t+1)|s_{i-1}(t),s_{i+1}(t)]$. These probabilities depend on
two parameters and are defined by
\begin{align}
\label{DKTransitionProbabilities}
&P[1|0,0]=0 \nonumber \ , \\
&P[1|0,1]=P[1|1,0]=p_1  \ , \\
&P[1|1,1]=p_2 \ , \nonumber
\end{align}
where $P[0|\cdot,\cdot]=1-P[1|\cdot,\cdot]$. The
corresponding update scheme may be realized by the
following algorithm (see Fig.~\ref{FIGDKUPDATE}): For each
site $i$ we generate a uniformly distributed random
number $z_i(t) \in (0,1)$ and set
\begin{equation}
\label{DKUpdateAlgorithm}
s_i(t+1) =
\begin{cases}
1 & \text{if} \ s_{i-1}(t) \neq s_{i+1}(t)
    \hspace{8mm} \text{and} \ z_i(t)<p_1 \ , \\
1 & \text{if} \ s_{i-1}(t) = s_{i+1}(t) =1 \ \text{and} \ z_i(t)<p_2 \ , \\
0 & \text{otherwise} \ .
\end{cases}
\end{equation}
In contrast to directed bond percolation, the DK model depends on
{\it two} percolation probabilities $p_1$ and $p_2$. The corresponding
phase diagram is shown in Fig.~\ref{FIGDKPHASE}. It comprises
an active and an inactive phase, separated by a phase transition line
(the solid line in the figure).
In the active phase a fluctuating steady state exists on the
infinite lattice, whereas in the inactive phase the model always
reaches the absorbing state.
The DK model includes three special cases.
The previously discussed case of directed bond
percolation corresponds to the choice $p_1=p$ and $p_2=p(2-p)$.
Another special case is directed site percolation~\cite{Kinzel83},
corresponding to the choice $p_1=p_2=p$. The third special case
$p_2=0$ is equivalent to the rule `W18' of Wolfram's classification
scheme~\cite{Wolfram83}.
Numerical estimates for the corresponding critical points
are summarized in Table~\ref{TABDK}.

\begin{table}
\footnotesize
\begin{center}
\begin{tabular}{||c||c|c|c||}
\hline\\[-4mm]
\small
transition point & $p_{1,c}$    & $p_{2,c}$     & Ref. \\
\hline
Wolfram rule 18 & 0.801(2)  & 0             & \cite{ZebendePenna94} \\
site DP     & 0.705489(4)   & 0.705489(4)   & \cite{TretyakovInui95} \\
bond DP     & 0.6447001(1)  & 0.8737620(2)  & \cite{Jensen96a} \\
compact DP  & 1/2       & 1     & \cite{Kinzel83}\\
\hline
\end{tabular}
\end{center}
\vspace{-3mm} \smallcaption{ Special transition points in the
(1+1)-dimensional Domany-Kinzel model. \label{TABDK} }
\end{table}

\begin{figure}
\epsfxsize=90mm
\centerline{\epsffile{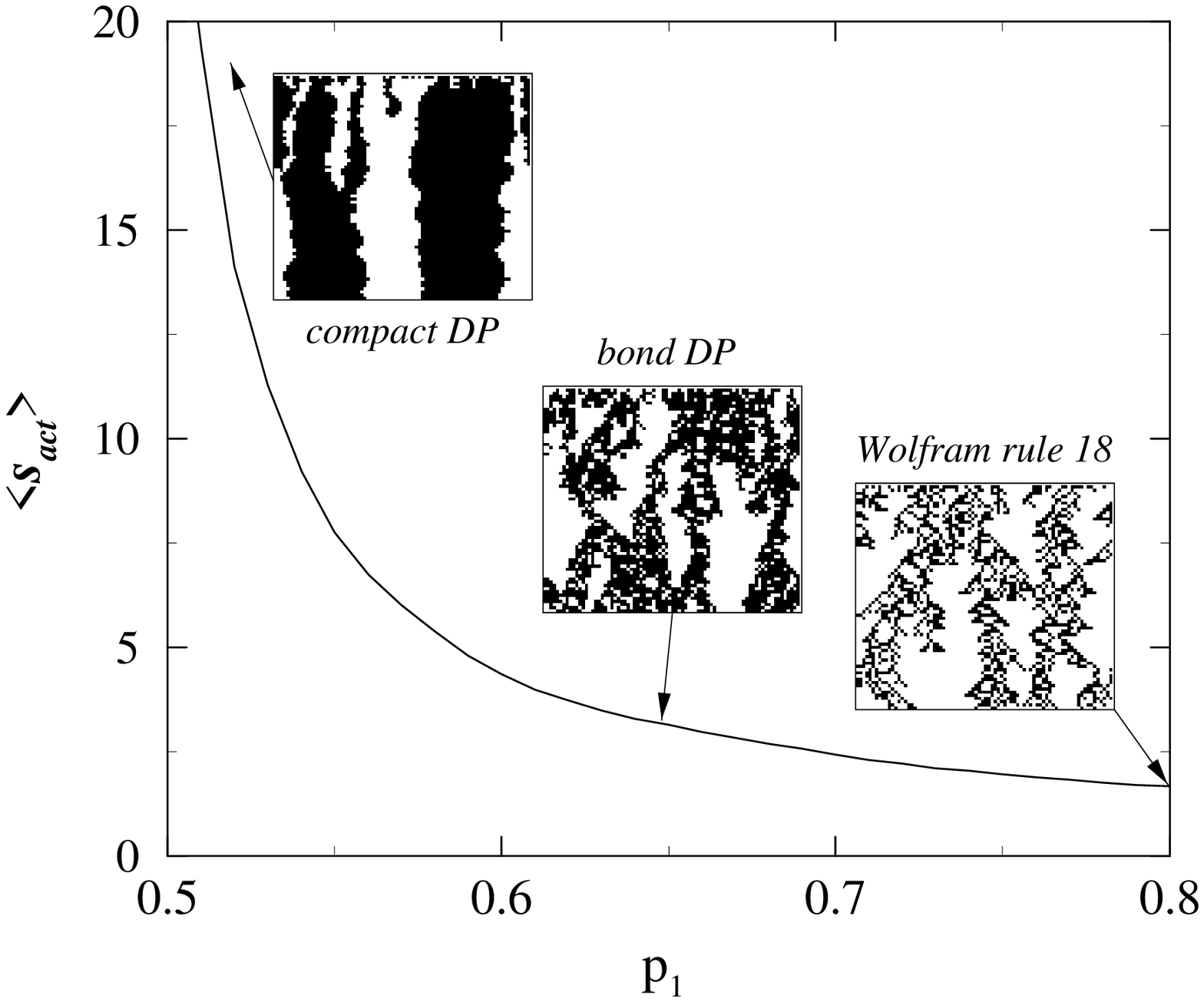}}
\vspace{-4mm}
\smallcaption{
\label{FIGILSIZE}
Numerical estimates for the average size of active spots
$\langle S_{act} \rangle$ in the Domany-Kinzel model
measured along the phase transition line. The
insets show typical clusters for three special cases
discussed in the text.}
\end{figure}

There is strong numerical evidence that the critical behavior
along the whole phase transition line (except for its upper
terminal point) is that of DP. This means that all these
transition points exhibit the same type of {\em long-range}
correlations. The short-range correlations, however, are
non-universal and may change when moving along the phase
transition line. In order to understand the significance of
short-range correlations from the physical point of view, let us
consider spatial configurations (snapshots) of a critical DP
cluster. Such configurations are typically characterized by
localized spots of activity separated by large voids in between.
Approaching the phase transition line the average size of inactive
voids diverges, whereas the mean size of active spots remains
finite and converges to a certain value $\langle S_{act} \rangle$.
By moving along the transition line of the DK model
the asymptotic average size
$\langle S_{act} \rangle$ of active spots varies. As shown in
Fig.~\ref{FIGILSIZE}, it is minimal for $p_2=0$, grows
monotonically with $p_2$, and finally diverges at the terminal
point $p_2=1$, where the system crosses over to a different type
of critical behavior. Thus, by moving along the phase transition
line the nonuniversal short-range properties change while the
long-range properties remain unaffected.

The exceptional behavior at the upper terminal point of the phase
transition line is due to an additional symmetry between active
and inactive sites along the line $p_2=1$~\cite{Kinzel83}. Here
the DK model has {\it two} symmetric absorbing states, namely the
empty and the fully occupied lattice. As the corresponding
symmetry transformation maps $p_1$ to $1-p_1$, the phase
transition line must end in the terminal point
$(p_1,p_2)=(1/2,1)$. As shown in the corresponding inset of
Fig.~\ref{FIGILSIZE}, the resulting clusters of active sites are
compact. Therefore, this special case is referred to as {\it
compact directed percolation} (CDP)~\cite{Essam89}. Unfortunately
this expression is misleading since CDP stands for a universality
class which is completely different from DP. As can be verified
easily, the DK model at the terminal point is
equivalent to the (1+1)-dimensional voter
model~\cite{Liggett85} or the Glauber-Ising model at zero
temperature. Alternatively, one may describe CDP as a
pair-annihilation process of diffusing kinks separating inactive
and active domains (cf. Sec.~\ref{REAC}). 
We may therefore identify the critical behavior 
of CDP with the exactly solvable universality class of
diffusing-annihilating random walks~\cite{Lee94}. We will come
back to CDP in Sec.~\ref{RELSEC}.

In more than one spatial dimension the DK model may be defined
by local updates with conditional probabilities
$P(s_i(t+1)|n_i(t))$ depending 
on the number $n_i(t)=\sum_{j \in <i>}s_j(t)$
of active neighboring sites.
Thus, the model is controlled by $2d$ parameters
$p_1,\ldots,p_{2d}$:
\begin{eqnarray}
&& P[1|0]=0 \ ,\nonumber \\
&& P[1|n]=p_n \ .
\qquad (1\leq n \leq 2d)
\end{eqnarray}
Notice that for $d=1$ this definition is compatible with
the usual definition of the DK model in 
Eq.~(\ref{DKTransitionProbabilities}).
The special case of directed bond percolation corresponds
to the choice $p_n=1-(1-p)^n$ while for equal parameters
$p_n=p$ one obtains directed site percolation
in $d$+1 dimensions.
\begin{figure}
\epsfxsize=130mm \centerline{\epsffile{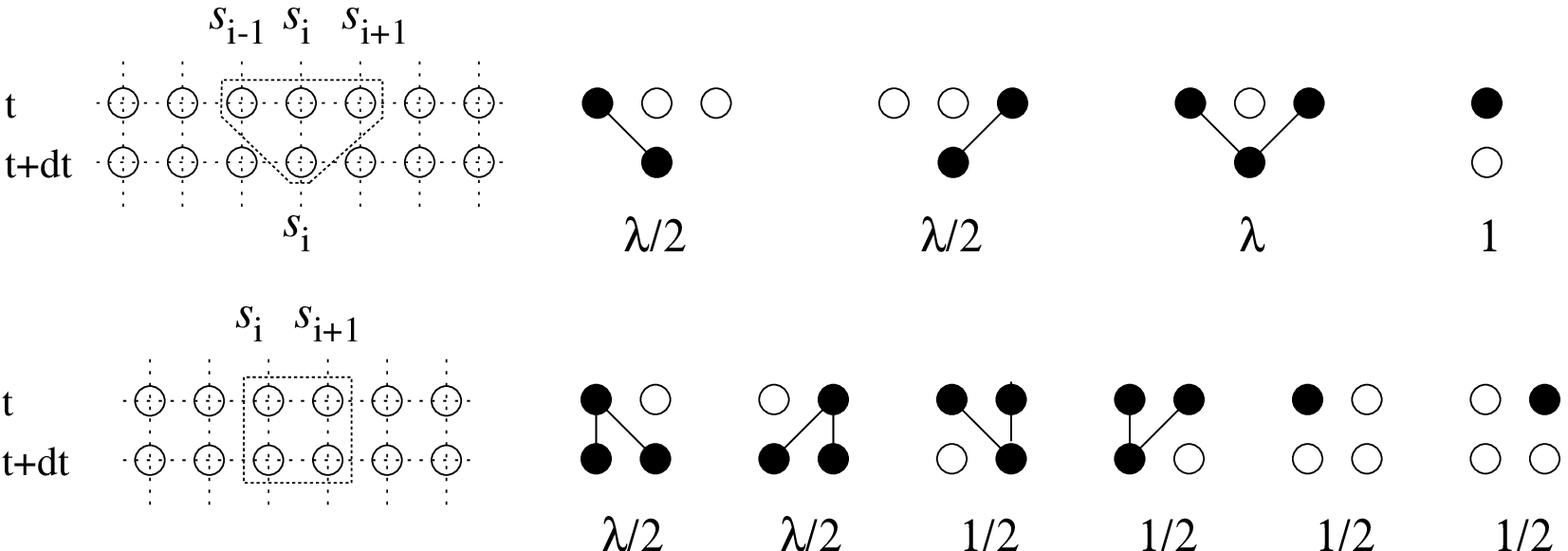}} 
\vspace{2mm}
\smallcaption{ \label{FigureCP} 
Stochastic processes in the
(1+1)-dimensional contact process. Infected sites (bold dots)
infect their neighbors at rate $\lambda/2$ and recover at
rate $1$. The upper part shows the definition of the rules
according to Eq.~(\ref{CPRules}). The lower part shows an
equivalent definition as a two-site process (see text). }
\end{figure}
%
%
%---------------------------------------------------------------------------
\headline{The contact process}
%---------------------------------------------------------------------------
%
Another important lattice model for DP is the contact process.
In contrast to cellular automata this model uses
{\it asynchronous} updates. The contact process was first
introduced by Harris~\cite{Harris74} as a model for epidemic
spreading without immunization (for a review
see Ref.~\cite{MarroDickman98}). Here the lattice sites
represent infected and healthy individuals.
Infected individuals can either heal themselves
or infect their nearest neighbors. Depending on the relative rates
of infection and recovery, the epidemic disease may
either spread over the whole population or vanish after some
time. In contrast to the DK model, infection and healing processes are
assumed to occur spontaneously without correlation
in space and time, i.e. spatially
separated processes are not synchronized. To mimic
this kind of asynchronous dynamics, the contact
process uses {\it random sequential} instead of parallel
updates (cf. Sec.~\ref{MASTER}).

The contact process is defined on a $d$-dimensional square
lattice whose sites can be either active
($s_i(t)=1$) or inactive ($s_i(t)=0$). For each attempted update a
site $i$ is selected at random. Depending on its state $s_i(t)$
and the number of active neighbors $n_i(t)=\sum_{j \in <i>}s_j(t)$
a new value $s_i(t+dt)=0,1$ is assigned according to certain transition rates
$w[s_i(t) \rightarrow s_i(t+dt),n_i(t)]$. In the standard
contact process these rates are defined by
\begin{equation}
\label{CPRules}
\begin{split}
w[0 \rightarrow 1,n] &= \lambda n/2d
\ , \\
w[1 \rightarrow 0,n] &=  1
\ .
\end{split}
\end{equation}
Here the parameter $\lambda$ controls the infection rate and plays
the role of the percolation probability. For the (1+1)-dimensional
case the dynamic processes are sketched in the upper row of
Fig.~\ref{FigureCP}. Monte Carlo simulations and series expansions
suggest that the phase transition in 1+1 dimensions takes place at
the critical point $\lambda_c \simeq 3.29785(8)$
\cite{DickmanJensen91,JensenDickman93a,DickmanSilva98}.

Whereas computational physicist often prefer the DK model for
simulations, the contact process is more popular in the mathematical
community because it is easy to write down the
corresponding master equation.
Following the notation of Sec.~\ref{MASTER}, the master
equation of the 1+1 dimensional contact process
with periodic boundary conditions is given by
\begin{equation}
\begin{split}
\timederivative  P_t(s_1,\ldots,s_N) =
\sum_{i=1}^N \ (2s_i-1) \ \bigl\{
&
\ \ \lambda s_{i-1} \, P_t(s_1,\ldots,s_{i-2},1,0,s_{i+1},\ldots,s_N) \\[-4mm]
&
+ \lambda s_{i+1} \, P_t(s_1,\ldots,s_{i-1},0,1,s_{i+2},\ldots,s_N) \\
&
- P_t(s_1,\ldots,s_{i-1},1,s_{i+1},\ldots,s_N)
\bigr\}
\,,
\end{split}
\end{equation}
where $P_t(s_1,\ldots,s_N)$ denotes the probability to find the system
at time $t$ in the configuration $\{s_1,\ldots,s_N\}$.
As can be seen, the master equation involves only two-site
interactions, as illustrated in the lower row of
Fig.~\ref{FigureCP}.
Using the vector notation of Eq.~(\ref{CompactNotation})
the corresponding Liouville operator
${\cal L}_{CP}=\sum_i {\cal L}_i$ is given by
\begin{equation}
\label{CPHamiltonian}
{\cal L}_i(\lambda) = \frac{1}{2}
\left(
\begin{array}{cccc}
0 & -1 & -1 & 0 \\
0 & 1+\lambda & 0 & -1 \\
0 & 0 & 1+\lambda & -1 \\
0 & -\lambda & -\lambda & 2
\end{array}
\right)
\ .
\end{equation}
Notice that this operator does not satisfy simple algebraic relations
as in Eq.~(\ref{HeckeAlgebra}), indicating that DP is a highly
nonintegrable process. Finite-size spectra of ${\cal L}_{CP}$
will be analyzed in Sec.~\ref{ABSEXP}.

%---------------------------------------------------------------------------
\headline{The Ziff-Gulari-Barshad model for heterogeneous catalysis}
%---------------------------------------------------------------------------
%
Many catalytic reactions such as the oxidation of carbon monoxide on a
platinum surface mimic the reaction scheme of directed percolation.
The key property of these reactions is the existence of
catalytically poisoned states where the system becomes trapped
in a frozen state. Thus, poisoned states play the role of absorbing
configurations. A simple model for surface catalysis
of the chemical reaction CO + O $\rightarrow$ CO$_2$
was introduced in 1986 by Ziff, Gulari,
and Barshad~(ZGB)~\cite{ZGB86}. The model describes a gas
composed of CO and O$_2$ molecules with fixed concentrations
$y$ and $1-y$, respectively, which
is brought into contact with a catalytic material. The
catalytic surface is represented by a square lattice
whose sites can be either vacant~($\vacancy$),
occupied by a CO molecule, or occupied by an O atom. The
ZGB model evolves by random sequential updates according
to the following probabilistic rules:
\begin{enumerate}
\item   CO molecules fill any vacant site at rate $y$.
\item   O$_2$ molecules dissociate on the surface into two
    O atoms and fill pairs of adjacent vacant sites at rate $1-y$.
\item   Neighboring CO molecules and O atoms recombine
    instantaneously to CO$_2$ and desorb from the surface, leaving
    two vacancies behind.
\end{enumerate}
On the lattice the three processes correspond to the reaction scheme
\begin{alignat}{2}
\vacancy & \rightarrow  \text{CO} &&\qquad
\text{at rate $y$ \ ,} \nonumber \\
\vacancy+\vacancy & \rightarrow  \text{O}+\text{O} &&\qquad
\text{at rate $1-y$ \ ,}  \\
\text{O} + \text{CO} & \rightarrow \vacancy + \vacancy &&\qquad
\text{at rate $\infty$ \ .} \nonumber
\end{alignat}
\begin{figure}
\epsfxsize=75mm
\centerline{\epsffile{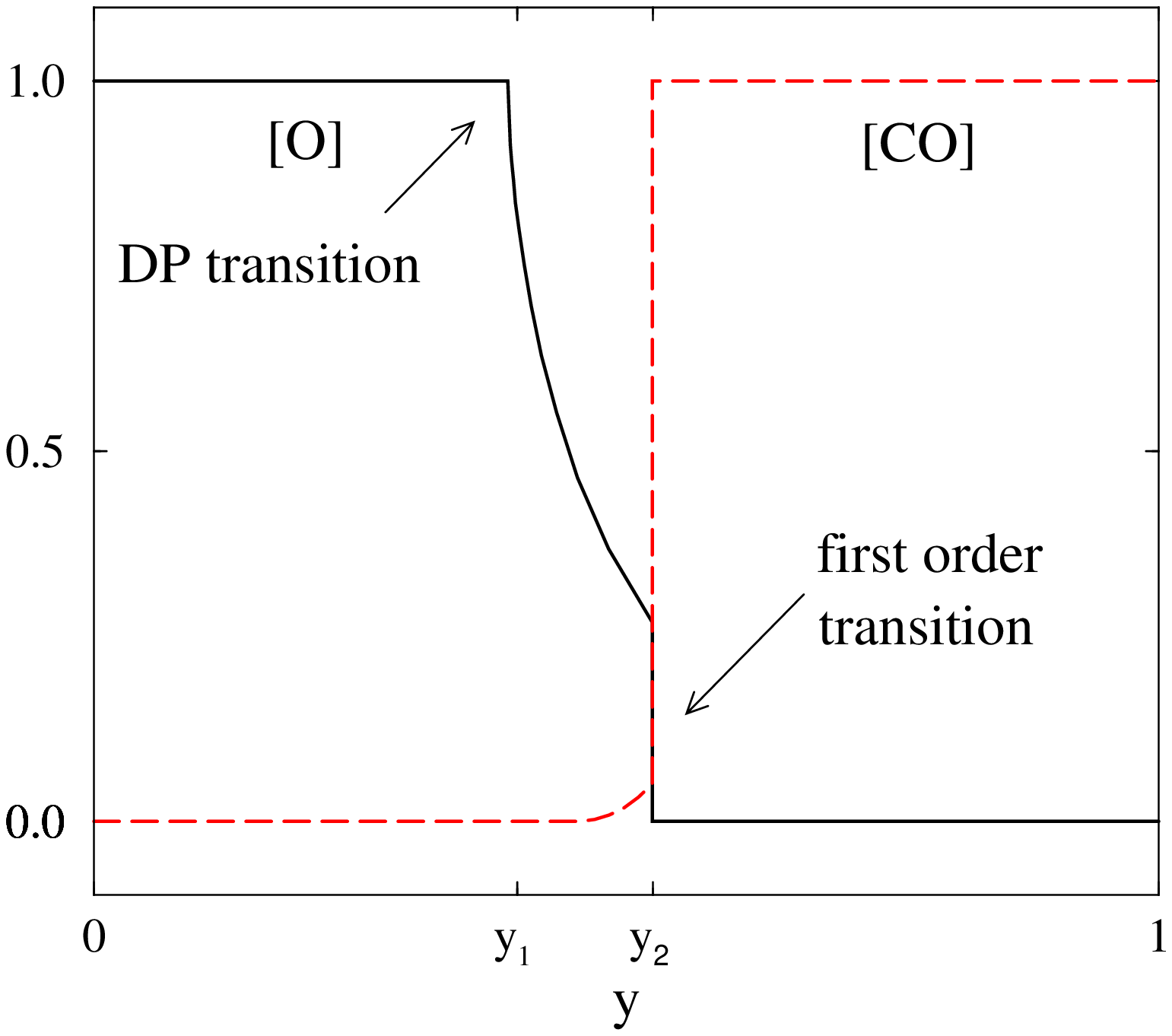}}
\smallcaption{
\label{FIGZGB}
Schematic phase diagram of the Ziff-Gulari-Barshad model.
The concentrations of oxygen (solid line) and
carbonmonoxide (dashed line) are plotted versus the CO adsorption rate~$y$.
}
\end{figure}
Clearly, this reaction is irreversible and thus the dynamic
processes do not obey detailed balance.
Moreover, if the whole lattice  is covered
either with pure CO or O, the system is
trapped in a poisoned absorbing state. As shown in Fig.~\ref{FIGZGB},
the ZGB model can be in three different phases.
For $y<y_1 \simeq 0.389$ the system evolves
into the O-poisoned state whereas for $y>y_2 \simeq 0.525$
it always reaches the CO-poisoned state.
In between the model is catalytically active.
The model exhibits two different phase transitions,
a continuous one at $y=y_1$ and a discontinuous one
at $y=y_2$. Grinstein {\it et~al.}~\cite{GLB89} expected
the continuous transition to belong to the DP universality class.
In order to verify this hypothesis, extensive
numerical simulations were performed.
Initially it was believed that the critical exponents
were different from those of DP~\cite{Meakin90}, while later the
transition at $y=y_1$ was found to belong to DP~\cite{JFD90}.
Very precise estimates of the critical exponents were recently
obtained in Ref.~\cite{VoigtZiff97}, confirming the
existence of a DP transition in the ZGB model.
DP exponents were also obtained in a simplified version of the
ZGB model~\cite{ABW90}. However, so far it has been impossible
to observe DP exponents in experiments. We will come back to
this problem in Sec.~\ref{DPEXPERIM}.

%==============================================================================
\subsection{Phenomenological scaling theory}
%==============================================================================

\label{ABSSCALESEC}
In equilibrium statistical physics continuous phase transitions
are usually characterized by universal scaling laws. For example,
the magnetization order parameter in the ordered phase of the
two-dimensional Ising models vanishes close to the critical point as
$|T-T_c|^\beta$, where $\beta$ is a universal exponent.
Similarly the correlation length $\xi$, which is the
characteristic macroscopic length scale of the model, diverges
as $\xi \sim |T-T_c|^{-\nu}$.
At the critical point the correlation length is infinite, i.e.,
there is no macroscopic length scale in the system.
As a consequence, the system is invariant under suitable scaling
transformations. It turns out that a 
very similar picture emerges in the nonequilibrium
case. In the following we introduce a phenomenological scaling
theory that can be applied to DP and other types of phase
transitions into absorbing states.

\begin{figure}
\epsfxsize=120mm
\centerline{\epsffile{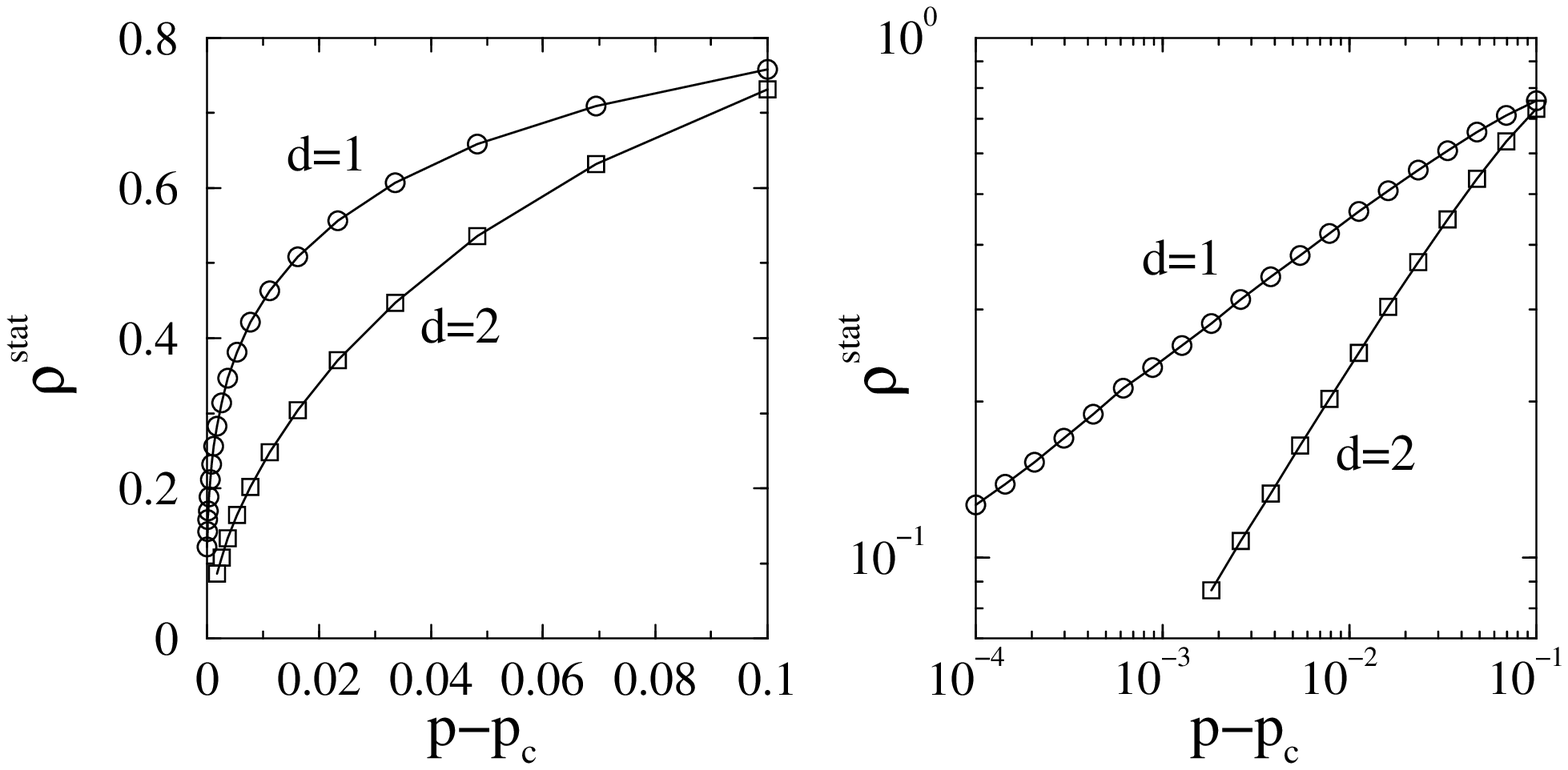}}
\vspace{2mm}
\smallcaption{
\label{FIGDPTRANS}
Stationary density $\rho^{stat}$ in the active phase of directed
bond percolation in 1+1 and 2+1 dimensions, plotted in a
linear and a double-logarithmic representation.
}
\end{figure}
%
%
%---------------------------------------------------------------------------
\headline{The critical exponents $\beta$, $\nuperp$, and $\nupar$}
%---------------------------------------------------------------------------
%
The order parameter of a spreading process is the density of active sites
\begin{equation}
\rho(t)=\langle \frac{1}{N} \sum_i s_i(t) \rangle  \,,
\end{equation}
where $\langle\ldots\rangle$ denotes the ensemble average.
Let us first consider the case of an infinite system.
In the active phase $\rho(t)$
decays and eventually saturates at some stationary
value $\rho^{stat}$. The stationary density varies continuously
with $p-p_c$ and vanishes at the critical point (see Fig.~\ref{FIGDPTRANS}).
Close to the transition the order parameter varies according to
a power law
\begin{equation}
\label{StatDensity}
\rho^{stat} \sim (p-p_c)^\beta
\, ,
\end{equation}
where $\beta$ is the critical exponent associated with
the particle density. In a double-logarith\-mic representation
the power law behavior manifests itself as a straight line
with slope~$\beta$. As can be seen in Fig.~\ref{FIGDPTRANS},
the value of $\beta$ depends on the dimensionality of the system.
The numerical value $\beta \simeq 0.277$ in 1+1 dimensions is comparatively
small, indicating a significant change of $\rho^{stat}$ near the transition.
In 2+1 dimensions a larger value $\beta \simeq 0.58$ is observed.

In addition, spreading processes are characterized by certain correlation
lengths. In contrast to equilibrium models without any dynamical aspect,
nonequilibrium critical phenomena involve `time' as an additional dimension.
Since `time' and `space' are different in character, we have
to distinguish spatial and temporal properties,
denoting them by the indices $\perp$ and $\parallel$, respectively.
In fact, nonequilibrium phase transitions are usually characterized
by {\it two independent} correlation lengths, namely a
spatial length scale $\xi_\perp$ and a temporal length scale $\xi_\parallel$.
Close to the transition, these length scales are expected to diverge as
\begin{equation}
\label{ScalingLengths}
\xi_\perp \sim |p-p_c|^{-\nuperp}\,,
\qquad
\xi_\parallel \sim |p-p_c|^{-\nupar}
\end{equation}
with generally different critical
exponents $\nuperp$ and $\nupar$. In the scaling regime
the two correlation lengths are related by $\xi_\parallel \sim \xi_\perp^z$,
where $z=\nupar/\nuperp$ is the so-called dynamic exponent.
In many models the triplet 
($\beta$, $\nuperp$, $\nupar$) is the fundamental
set of bulk exponents that labels the universality class.
Other critical exponents are usually related to these three
exponents by simple scaling relations (see below).
Nonequilibrium phase transitions in different physical systems
are believed to belong to the same universality class if their
critical exponents coincide\footnote{In addition, it should be proven
that universal scaling functions coincide as well.}. In fact, the DK model,
the contact process, the ZGB model, and a vast variety
of other DP models are characterized by the same triplet of exponents.

\begin{figure}
\epsfxsize=145mm \centerline{\epsffile{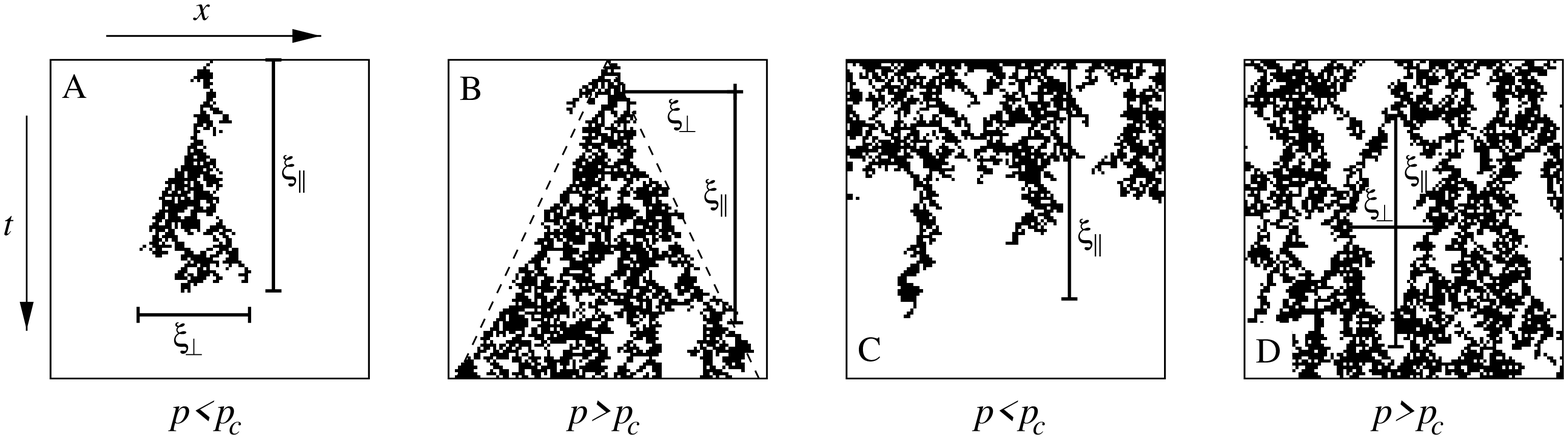}} \vspace{2mm}
\smallcaption{ \label{FIGSCALE} Interpretation of the correlation
lengths $\xi_\perp$ and $\xi_\parallel$ in an almost critical
(1+1)-dimensional DP process below (left) and above criticality
(right). In panels A and B a cluster is grown from a single active
seed while in panel C a fully occupied lattice is used as initial
state. Panel D shows a stationary DP process in the active
phase. The indicated length scales $\xi_\perp$ and $\xi_\parallel$
must be interpreted as averages over many independent
realizations. }
\end{figure}

Fig.~\ref{FIGSCALE} illustrates the physical meaning of the
correlation lengths $\xi_\perp$ and $\xi_\parallel$.
As in equilibrium statistical mechanics, they are
present below and above the critical point.
In the inactive phase clusters originating from a single seed
have the typical form of a droplet (panel A). Averaging
over many independent realizations the lateral size and the
lifetime of such droplets are proportional to $\xi_\perp$ and $\xi_\parallel$,
respectively. Above criticality the surviving clusters grow within
a spreading cone (panel B) whose opening angle is determined by
the ratio $\xi_\perp/\xi_\parallel$. The correlation lengths can
also be seen if homogeneous initial conditions are used. In the
inactive phase the scaling length $\xi_\parallel$ plays the role of a
typical decay time (panel C),
while in the stationary state of the active phase the correlation
lengths appear as the average sizes of {\it inactive} islands (panel D).
This interpretation can be easily generalized to higher dimensions.

As suggested by the scaling properties of the
density~(\ref{StatDensity}) and the correlation lengths
(\ref{ScalingLengths}), a spreading process should be invariant
under dilatation $\xvec \rightarrow \Lambda \xvec$ accompanied by
an appropriate rescaling of time and the deviation from
criticality $\deviation=p-p_c$:
\begin{equation}
\label{Rescaling}
\xvec \rightarrow \Lambda \xvec
\, , \qquad
t \rightarrow \Lambda^z t
\, , \qquad
\deviation \rightarrow \Lambda^{-1/\nuperp} \deviation
\, , \qquad
\rho \rightarrow \Lambda^{-\beta/\nuperp} \rho
\, .
\end{equation}
This allows scale-invariant combinations to be constructed
such as $t/x^z$, $\deviation t^{1/\nupar}$
and $\deviation x^{1/\nuperp}$. As we will see below,
universal scaling functions can only depend on such scale-invariant
ratios.

%---------------------------------------------------------------------------
\headline{Scaling theory for phase transitions into absorbing states}
%---------------------------------------------------------------------------
%
So far we have seen that the stationary density in the active phase
scales as $\rho^{stat}\sim \deviation^\beta$, where $\deviation=p-p_c$
denotes the distance from the critical point. A very similar
quantity is the ultimate survival probability
$P_\infty$ that a randomly chosen
site belongs to an infinite cluster (cf. Sec.~\ref{DPINTROSEC}).
In the active phase this probability is finite and scales as
\begin{equation}
\label{PScaling}
P_\infty \sim \deviation^{\beta^\prime} \
\end{equation}
with some critical exponent $\beta'$. Although
$\beta$ and $\beta^\prime$ coincide in the case of DP,
they may be different in more general contexts, for example,
in models with many absorbing states. Therefore, phase transitions
into absorbing states are generally described by {\em four} exponents
$\beta$, $\beta^\prime$, $\nuperp$, and $\nupar$. The different
roles of $\beta$ and $\beta^\prime$ become apparent in a field-theoretic
formulation (see Sec.~\ref{FTHSEC}). It turns out that $\beta$
is associated with the particle annihilation operator. Therefore,
this exponent emerges whenever a particle density
is `measured' in some final state. The exponent
$\beta^\prime$, on the other hand, is associated with the creation
of particles and thus plays a role whenever particles are `introduced'.
This happens, for example, if an initial configuration is specified.
In correlation functions, which involve creation as well as
annihilation operators,
both exponents are expected to appear.

Turning to time-dependent scaling properties
in an infinitely large system, there are two
important complementary quantities, namely the particle density
$\rho(t)$ starting from a fully occupied lattice,
and the survival probability $P(t)$ that a cluster
grown from a single seed is still active after $t$ time steps.
Following the usual scaling concept of equilibrium
statistical mechanics, both quantities
are expected to scale as
\begin{equation}
\label{DensityScaling}
\rho(t) \simeq t^{-\alpha}  f( \deviation \, t^{1/\nupar})
\ , \qquad
P(t) \simeq t^{-\delta}  g( \deviation \, t^{1/\nupar}) \ ,
\end{equation}
where $\alpha$ and $\delta$ are certain
critical exponents for decay and survival,
respectively~\cite{GrassbergerTorre79,MGT97}.
$f$ and $g$ are {\em universal scaling functions},
i.e., they have the same functional form in all
DP models. For small arguments they both tend to a
constant, whereas for large arguments they scale
in a way that the time dependence drops out:
\begin{equation}
f(\zeta) \sim \zeta^{\alpha \nupar} \, , \qquad
g(\zeta) \sim \zeta^{\delta \nupar} \, . \qquad\qquad
(\zeta \rightarrow \infty) \
\end{equation}
In the active phase the two quantities therefore saturate at
$\rho^{stat}=\rho(\infty)\sim\deviation^{\alpha\nupar}$ and
$P_\infty=P(\infty)\sim\deviation^{\delta\nupar}$.
Comparison with Eqs.~(\ref{StatDensity}) and~(\ref{PScaling}) yields
\begin{equation}
\alpha = \beta/\nupar \ , \qquad
\delta = \beta^\prime/\nupar \ .
\end{equation}
An important quantity that combines both creation and annihilation
of particles is the {\it pair connectedness function}
$c(\xvec^\prime,t^\prime,\xvec,t)$, which is defined as the probability that
the sites $(\xvec^\prime,t^\prime)$ and $(\xvec,t)$ are connected
by a directed path of open bonds.
Since the pair connectedness function is translationally
invariant in space and time, we may also write
$c(\xvec^\prime,t^\prime,\xvec,t) \equiv c(\xvec-\xvec^\prime,t-t^\prime)$.
Starting from an initial condition with a single active
site at the origin $\xvec^\prime=0$, the pair connectedness function
$c(\xvec,t)$  is just the density of active sites in the
resulting clusters averaged over many realizations of randomness.
Because of scaling invariance, the pair connectedness function
$c(\xvec,t)$ obeys the scaling form~\cite{GrassbergerTorre79}
\begin{equation}
\label{ScalingPairconn}
c(\xvec,t) \sim t^{\theta-d/z}\,
F(x/t^{1/z}, \deviation \, t^{1/\nupar}) \ ,
\end{equation}
where $d$ denotes the spatial dimension and $z=\nupar/\nuperp$.
The so-called {\em critical initial slip exponent}
$\theta$ describes the growth of the average number of
particles as a function of time (see Sec.~\ref{BCSEC}).
In order to determine $\theta$ we note that
in the active phase $\deviation>0$ surviving clusters will
create an average density $\rho^{stat}\sim \deviation^\beta$
in the interior of the spreading cone (cf. panel B of Fig.~\ref{FIGSCALE}).
Thus the autocorrelation function $c(0,t)$ should saturate at the value
\begin{equation}
\label{AutoCorr2}
c(0,\infty) = \lim_{t \rightarrow \infty} c(0,t) \sim
\deviation^{\beta+\beta^\prime} \,.
\end{equation}
On the other hand, the scaling form~(\ref{ScalingPairconn}) implies
that $c(0,t)$ saturates in the active phase at a constant
with the scaling behavior\footnote{To prove this relation, notice that
$F(0,\zeta) \sim \zeta^{-(\theta-d/z)\nupar}$ for
large values of $\zeta$.}
\begin{equation}
\label{AutoCorr1}
c(0,\infty) \sim \deviation^{\nupar(d/z-\theta)}
\,.
\end{equation}
Comparing the two expressions we obtain the
{\it generalized hyperscaling relation}~\cite{MDHM94}
for phase transitions into absorbing states
\begin{equation}
\label{GeneralizedHyperscaling}
\theta - \frac{d}{z} = -\frac{\beta+\beta^\prime}{\nupar}
\,.
\end{equation}
It should be noted that the scaling argument~(\ref{AutoCorr2})
relies on the assumption that the cluster spreads around the
origin, i.e., the spreading cone {\em surrounds} the origin.
In sufficiently high spatial dimensions, however, the cone
becomes sparse and diffuses away from the origin so that
the autocorrelation function $c(0,\infty)$ vanishes. For example,
in a DP process this happens above the upper critical dimension $d_c=4$.
In fact, the generalized hyperscaling
relation~(\ref{GeneralizedHyperscaling}) turns out to be valid only
{\em below} the upper critical dimension of the spreading process
under consideration.

The scaling theory outlined above assumes the system size to be infinite.
For finite systems sizes the scaling functions also depend on
the invariant ratio $\xi_\perp^d/N=t^{d/z}/N$,
where $N=L^d$ is the total number of sites. The generalized
scaling forms read
\begin{eqnarray}
\label{FSDensityScaling}
\rho(t) &\sim& t^{-\beta/\nupar} \,
f( \deviation \, t^{1/\nupar}, \, t^{d/z}/N) \ ,
\\
\label{FSScalingSurvival}
P(t) &\sim& t^{-\beta^\prime/\nupar} \,
g(\deviation \, t^{1/\nupar}, \, t^{d/z}/N) \ ,
\\
\label{FSScalingPairconn}
c(\xvec,t) &\sim& t^{-(\beta+\beta^\prime)/\nupar}\,
F(x/t^{1/z}, \deviation \, t^{1/\nupar}, \, t^{d/z}/N) \ .
\end{eqnarray}
%
%

%---------------------------------------------------------------------------
\headline{Derived scaling properties}
%---------------------------------------------------------------------------
%
The scaling behavior of various other quantities
can be derived directly from the scaling
relations~(\ref{FSDensityScaling})-(\ref{FSScalingPairconn}).
For example, the mean {\em cluster mass} $M$ is given by
the total integral of the pair connectedness function
\begin{equation}
\label{MeanClusterMass}
M = \int d^d x \int_0^{\infty} dt \, c(\xvec,t) \ .
\end{equation}
Inserting the scaling relation (\ref{ScalingPairconn})
and substituting the scaling variables we obtain a scaling
law for the average cluster mass measured in an infinite system
below criticality:
\begin{equation}
M  \sim \int d^d x \int_0^{\infty} dt \, t^{\theta-d/z}\,
F(x/t^{1/z}, \deviation \, t^{1/\nupar}) \sim
|\deviation|^{-\nupar(1+\theta)} \ .
\end{equation}
Similarly, the {\em mean survival time} $T$,
the {\em mean spatial volume} $V$, and the {\em mean size} $S$
of a cluster in the inactive phase are given by
\begin{align}
T &= \int dt P(t)  = \int dt \,
t^{-\delta} \, G(\deviation \, t^{1/\nupar}) \sim
|\deviation|^{-\nupar(1-\delta)} \,, \\
V &= \int dt \, P(t) t^{d/z-1} =
\int dt \, t^{d/z-\delta-1}\,G(\deviation \, t^{1/\nupar}) \sim
|\deviation|^{-\nupar(d/z-\delta)} \,, \\
S &= \int dt \, P(t) t^{d/z} =
\int dt \, t^{d/z-\delta}\,G(\deviation \, t^{1/\nupar}) \sim
|\deviation|^{-\nupar(d/z+1-\delta)} \,.
\end{align}
For these quantities, we obtain the following scaling relations:
\begin{alignat}{2}
M &\sim |\deviation|^{-\gamma} \ , \qquad &
\gamma & = \nupar(1+\theta) =
\nupar+d\nuperp-\beta-\beta^\prime \ , \\
\label{TScaling}
T &\sim |\deviation|^{-\tau} \ , \qquad &
\tau &= \nupar(1-\delta) = \nupar-\beta^\prime \ , \\
V &\sim |\deviation|^{-v} \ , \qquad &
v &= \nupar(d/z-\delta) = d\nuperp-\beta^\prime \ , \\
S &\sim |\deviation|^{-\sigma} \ , \qquad &
\sigma &= \nupar(d/z+1-\delta) =
\nupar+d\nuperp-\beta^\prime \ .
\end{alignat}

\newpage

%---------------------------------------------------------------------------
\headline{Spreading processes in an external field}
%---------------------------------------------------------------------------
%
Let us finally consider a spreading process in an external field
$h$. Using the particle interpretation, such a field may be realized
by spontaneous creation of particles $\vacancy \rightarrow A$ at
rate $h$ during the temporal evolution. Clearly, spontaneous
particle creation destroys the absorbing state and therefore the
transition itself. That is, the external field drives the system
away from criticality. For small $h$ the resulting distance from
criticality obeys certain scaling laws.

In principle the presence of an external field requires to
introduce another independent critical exponent 
for the coupling constant. In the case
of DP, however, this exponent is not independent, it is rather identical
with the mean cluster size exponent $\gamma$. To understand this
relation, let us consider the stationary state of a subcritical
DP process in presence of a weak field. Obviously, a site can 
only become active if it is connected with at least one other site 
backwards in time where a particle was spontaneously created by
the external field. Since the number of such sites is equal to the
cluster size, the probability to become active
is given by $\rho^{stat} \sim 1-(1-h)^{M(\Delta)}$.
For weak fields, the stationary density is therefore
linear in $h$:
\begin{equation}
\rho^{stat} \sim h\,M(\deviation) \,. \qquad (\deviation<0)
\end{equation}
Consequently, the susceptibility of a supercritical DP
process scales as
\begin{equation}
\chi = \frac{\partial}{\partial h} \rho^{stat}
\sim |\deviation|^{-\gamma} \ .
\end{equation}
Invariance under rescaling~(\ref{Rescaling}) requires
the external field to change as
\begin{equation}
h \rightarrow \Lambda^{\beta'/\nuperp-z-d}h =
\Lambda^{-\sigma/\nuperp} h \,.
\end{equation}
Thus, at criticality the stationary response of a DP process is given by
\begin{equation}
\label{FieldResponse}
\rho^{stat} \sim h^{\beta/\sigma} \ .
\end{equation}
More generally, we may extend the scaling forms
(\ref{FSDensityScaling})-(\ref{FSScalingPairconn}) by including
the scale-invariant argument
$h t^{\sigma/\nupar}$. For example,
the density $\rho(t)$ evolves as
\begin{equation}
\rho(t) \sim t^{-\beta/\nupar} \,
f( \deviation \, t^{1/\nupar}, \,
t^{d/z}/N, \, h t^{\sigma/\nupar}) \ .
\end{equation}
%
%

%---------------------------------------------------------------------------
\headline{Time reversal symmetry of directed percolation}
%---------------------------------------------------------------------------
%
As shown in Ref.~\cite{GrassbergerTorre79},
a special symmetry of DP under time reversal
implies the additional scaling relation
\begin{equation}
\label{ScalingRelationDelta}
\beta = \beta^\prime \ .
\end{equation}
This is the reason why DP is characterized by only three 
instead of four critical exponents.
In order to understand this duality symmetry from the physical
point of view, let us consider the special
case of directed bond percolation. By reversing
the arrows shown in the right part of Fig.~\ref{FIGISODIR}
one obtains a directed bond percolation process
that evolves `backwards' in time. Obviously, the
reversed process follows {\em exactly} the same probabilistic
rules as the original one. Moreover, if two sites were connected by
a directed path in the original process, they will
also be connected in the reversed process. Hence, if the
reversed process was started from a fully occupied lattice
at $t>0$, the resulting active sites at $t=0$ would be precisely
those sites which --  in the original process -- would generate
clusters that are still alive at time $t$.
Therefore, in the case of directed bond percolation we obtain
\begin{equation}
\label{DualitySymmetry}
P(t) = \rho(t) \, ,
\end{equation}
i.e., the survival probability of a single seed $P(t)$
is {\em exactly} equal to the
density of active sites $\rho(t)$ in a DP process
starting with a fully occupied lattice. Thus, in the active phase
the two quantities  saturate at the
same value $P_\infty = \rho^{stat}$ wherefore the
corresponding critical exponents $\beta$ and $\beta^\prime$
have to be identical. It should be emphasized that this
time reversal symmetry of DP is nontrivial
and does not hold for other systems such as models
with several absorbing states~\cite{MDHM94} or spreading processes with a
fluctuating background~\cite{WOH98}. Together with
Eq.~(\ref{ScalingRelationDelta}) the generalized hyperscaling
relation~(\ref{GeneralizedHyperscaling}) reduces to the
DP hyperscaling relation
\begin{equation}
\label{HyperscalingRelation}
\theta \ = \ d/z - 2\delta \ = \ \frac{d \nuperp - 2\beta}{\nupar}
\ .
\end{equation}
Consequently, the autocorrelation function $c(0,t)$ for DP
saturates in the active phase at the value
$c(0,\infty) \sim \deviation^{2\beta}$.
%
%

%---------------------------------------------------------------------------
\headline{The DP conjecture}
%---------------------------------------------------------------------------
%
One of the most fascinating properties of DP models is their robustness
with respect to the microscopic dynamic rules. In fact, the DP class
covers a wide range of models. It includes, for example,
the vast majority of spreading models such as the
contact process~\cite{Liggett85,DickmanBurschka88},
epidemic spreading without immunization~\cite{Grassberger82b}, and
forest fire models~\cite{NTG89,Albano94a,Albano95a}.
Moreover, the DP class includes
models for catalytic reactions~\cite{Schloegl72,ZGB86,BrowneKleban90},
interacting particles~\cite{ParkPark95}, as well as
branching-annihilating random walks with odd number of
offspring~\cite{TakayasuTretyakov92,Jensen93b,AHM97}.
Furthermore, certain growth
processes~\cite{KerteszWolf89,Nagatani92}
and coupled map lattices with asynchronous updates~\cite{RBJ97}
display DP behavior. In fact, this list is far from being complete.

The variety and robustness of DP models led Janssen and Grassberger to
the conjecture that a model should belong to the DP universality class
if the following conditions hold~\cite{Janssen81,Grassberger82}:
\begin{enumerate}
\item The model displays a continuous phase transition from a {\em fluctuating}
      active phase into a {\em unique} absorbing state.
\item The transition is characterized by a {\em positive
      one-component} order parameter.
\item The dynamic rules involve only {\em short-range} processes.
\item The system has no special attributes such as additional
      symmetries or quenched randomness.
\end{enumerate}
Although this conjecture has not yet been proven rigorously,
it is highly supported by numerical evidence. In fact, DP seems
to be even more general and may be identified in
systems that violate some of the four conditions,
for example in certain models with
non-unique~\cite{Albano95b,MGDL96,MGDL97,MGD98}
or fluctuating passive states~\cite{MDHM94}. Even complicated
spreading processes with several spreading agents and
multicomponent order parameters were shown to exhibit DP behavior
~\cite{MeakinScalapino87,GLB89,JFD90,PKKAR93,Albano95b,HAM97,ZRP93}.
Some models with infinitely many absorbing states,
that were initially thought to belong to different
universality classes, were later found to be in the DP class as
well (see Sec.~\ref{RELSEC}).
Not only the bulk exponents $\beta,\nuperp,\nupar$ are
universal but also other quantities as, for example, scaling
functions and higher moments of the order
parameter~\cite{DickmanSilva98}.

It is remarkable that DP is one of very few critical phenomena in
1+1 dimensions which has not yet been solved exactly.
Despite of its simplicity and robustness it seems to be impossible
to compute the critical exponents exactly. In fact,
the numerical estimates suggest that the critical exponents
are given by {\it irrational} numbers rather than simple rational values.
The lack of analytical results may be related to the fact that DP --
in contrast to ordinary (isotropic) percolation -- is not conformally
invariant since there is no symmetry between `space' and `time'.
Attempts to replace conformal invariance by an
anisotropic scaling theory
have not yet been successfully applied to DP~\cite{Henkel97}.

Only few exceptions of DP are known so far. They all violate at least
one of the four conditions listed above. For example, a different
universality class emerges when the system has two or more
{\it symmetric} absorbing states (see Sec.~\ref{SEVABSSEC}).
Another example is the activated random walk model with a
conserved order parameter (see Sec.~\ref{ARWSEC})
Different universal properties are also encountered in 
models where activity spreads over long distances by L\'evy 
flights (see Sec.~\ref{ANOMALSEC}).

%==============================================================================
\subsection{Estimation of the critical exponents}
%==============================================================================
%
\label{ABSEXP}
Directed percolation is one of very few critical phenomena
whose critical exponents in 1+1 dimensions are not known
exactly. However, thanks to extensive numerical simulations,
transfer matrix techniques, series expansions, and field-theoretical
calculations the critical exponents have been estimated in various
dimensions to an extremely high accuracy. This subsection briefly
summarizes the available methods and the most precise estimates.

%---------------------------------------------------------------------------
\headline{Mean-field approximation}
%---------------------------------------------------------------------------
%
In order to estimate the critical exponents
$\beta$, $\nuperp$, and $\nupar$ of directed percolation,
let us first consider a simple mean-field (MF) approximation.
Denoting by $\rho(t)$ the density of active sites at time $t$
averaged over the entire system, the MF rate equation
for the contact process~(\ref{CPRules}) reads
\begin{equation}
\label{SimpleDPMF}
\timederivative  \rho(t) =
(\lambda-1) \rho(t) - \lambda \rho^2(t)\,.
\end{equation}
This equation has two stationary solutions, namely $\rho^{stat}=0$ and
$\rho^{stat}=(\lambda-1)/\lambda$. Hence the mean-field
critical point is $\lambda_c=1$.
The solution $\rho^{stat}=0$ represents the absorbing state
from where the system cannot escape. In the inactive phase
$\lambda<\lambda_c$ the absorbing state is stable while the other
solution with negative density is unphysical. In the active phase
$\lambda>\lambda_c$ the absorbing state becomes unstable against
small perturbations while the second solution represents a stable
active state. Near criticality the stationary density vanishes linearly
as $\rho^{stat} \sim \lambda-\lambda_c$. Therefore, the mean field
density exponent is given by $\beta^{MF}=1$.
On the other hand, Eq.~(\ref{SimpleDPMF}) implies
the density to decay in the inactive phase asymptotically as
$\rho(t) \sim  \exp(-|\lambda-\lambda_c|t)  \sim \exp(-t/\xi_\parallel) $,
hence $\xi_\parallel \sim |\lambda-\lambda_c|^{-1}$ and $\nupar^{MF}=1$.

In order to determine the spatial scaling exponent $\nu_{\perp}^{MF}$,
the mean-field rate equation~(\ref{SimpleDPMF}) has to be extended
by a term for particle diffusion
\begin{equation}
\label{ExtendedDPMF}
\timederivative  \rho(\xvec,t) =
\deviation \rho(\xvec,t)
- \lambda \rho^2(\xvec,t)
+ \diff \nabla^2 \rho(\xvec,t)\,,
\end{equation}
where $\diff$ is the diffusion constant
and $\deviation=\lambda-\lambda_c$ is the deviation from criticality.
In a lattice model this term corresponds to nearest-neighbor interactions.
According to Eq.~(\ref{DensityScaling}) the density $\rho(\xvec,t)$
changes under rescaling~(\ref{Rescaling}) as
\begin{equation}
\label{FieldRescaling}
\rho(\xvec,t) \rightarrow \Lambda^{-\beta/\nuperp}
\rho(\Lambda \xvec, \Lambda^z t)
\ .
\end{equation}
Simple dimensional analysis shows that Eq.~(\ref{ExtendedDPMF})
is invariant under rescaling if
\begin{equation}
\label{DPMF}
\beta^{MF}=1
\,,\qquad
\nu_{\perp}^{MF}=\frac{1}{2}
\,,\qquad
\nupar^{MF}=1
\,.
\end{equation}
The above mean field approximation becomes exact in the limit of
infinitely many dimensions. In a finite-dimensional contact process
it is not clear whether the mean field approximation still
applies, it is even not obvious that a continuous phase
transition still exists. However, Liggett \cite{Liggett85} was able to
rigorously prove the existence of a phase transition
for a contact process in $d \geq 1$ dimensions. As will be shown
below, the mean field exponents turn out to be exact for
$d \geq 4$, where $d_c=4$ is the upper critical
dimension of DP. Note that the mean field exponents satisfy
the hyperscaling relation~(\ref{HyperscalingRelation}) precisely
in $d=4$ dimensions.

In order to go beyond the standard mean field approximation in
low-dimensional systems spatial correlations have to be taken into
account. An improved mean field approximation for the contact
process in 1+1 dimensions was developed by Ben-Naim and
Krapivsky~\cite{BenNaimKrapivsky94}, who expressed the temporal
evolution of empty intervals on the lattice by an infinite
hierarchy of differential equations. This approach is very similar
to the IPDF technique introduced in Sec.~\ref{SECEXACT}
Approximating the probability to find pairs of neighboring empty
intervals by the product of single-interval probabilities, they
derived a set of equations which can be solved exactly. In this
approximation the critical exponents are given by
$\beta=\frac{1}{2}$, $\nu_{\perp}=1$, and
$\nupar=z=\frac{3}{2}$. \'Odor could improve these
estimates by using a generalized mean field approximation combined
with coherent anomaly techniques ~\cite{Odor95}, reaching an
accuracy of almost $1 \%$.

\begin{figure}
\epsfxsize=150mm \centerline{\epsffile{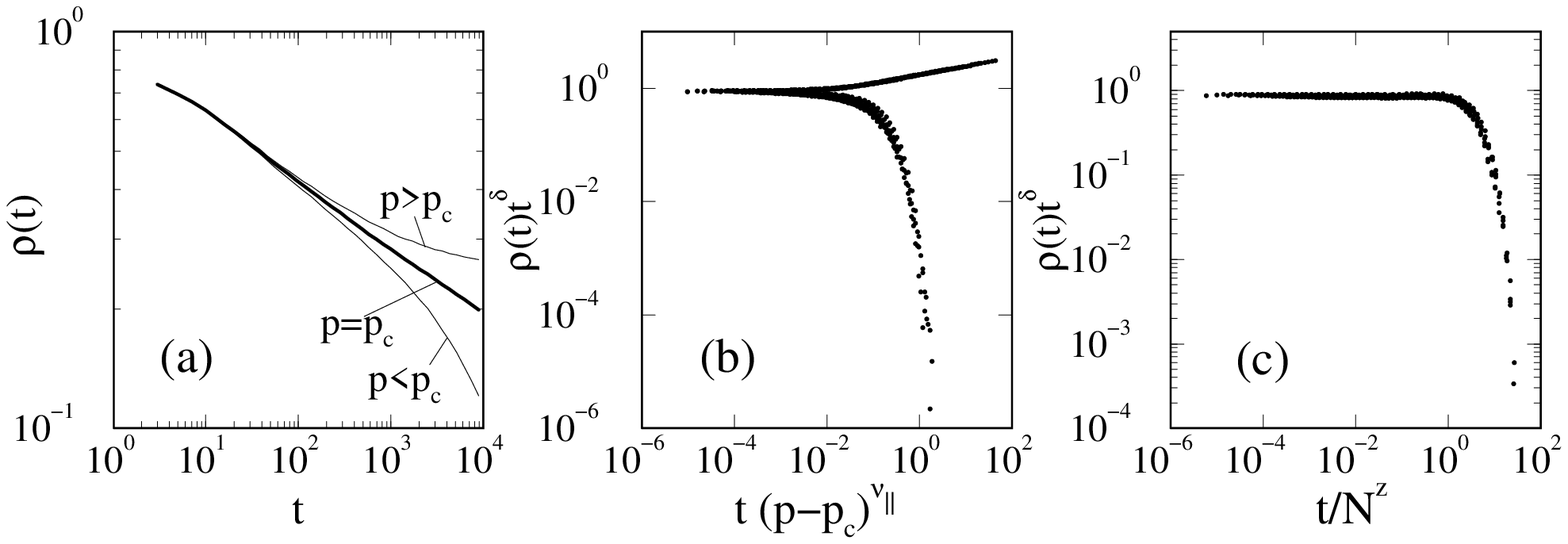}} \smallcaption{
\label{FIGORDMC} Ordinary Monte Carlo simulations of a
(1+1)-dimensional directed bond percolation process starting from
a fully occupied lattice. Part (a) shows the particle density as a
function of time and demonstrates the determination of the
critical point. Parts (b) and (c) show data collapses for
off-critical and finite-size simulations, respectively (see text).
}
\end{figure}
%
%

%--------------------------------------------------------
\headline{Monte Carlo simulations with homogeneous initial conditions}
%--------------------------------------------------------
%
In order to study nonequilibrium phase transitions quantitatively,
numerical techniques such as Monte Carlo simulations
have become an important tool. Because of the steadily
growing computer capacity the critical exponents can nowadays be
estimated within a few per cent, in some cases even up to
four digits. Further progress is expected as reaction-diffusion
models can easily be simulated on parallel computers
with a large number of simple processors~\cite{OKVR99}.

The simplest numerical method that allows the critical exponents
to be estimated is a Monte Carlo (MC) simulation starting
with a fully occupied lattice. This technique is based on the
scaling properties of Eq.~(\ref{FSDensityScaling}).
In a first series of simulations the critical percolation
threshold has to be determined by measuring deviations from
the asymptotic power-law decay $\rho(t)\sim t^{-\delta}$ in a sufficiently
large system. To this end $\rho(t)$ is plotted versus~$t$
in a double-logarithmic graph (see Fig.~\ref{FIGORDMC}a).
Positive (negative) curvature for large~$t$ indicates that the
system is still in the active (inactive) phase. It should be
carefully analyzed to what extent the estimate depends on the
system size used in the simulation. If finite-size effects play
a role, extrapolation techniques should be used in order to
improve the estimate~\cite{HenkelSchutz88}.

Having determined $p_c$ and $\delta$ the
exponent $\beta$ may be estimated by
measuring the stationary density of active sites
$\rho^{stat}\sim\deviation^\beta$ in the active phase.
However, this type of estimate is known to be quite
inaccurate since the equilibration time to reach the
stationary state grows rapidly as the critical point is approached.
This {\it critical slowing down} can be controlled by plotting
$\rho(t) \, t^\delta$ versus $t \, \deviation^{\nupar}$ for different
values of $\deviation$ and tuning $\nupar$ in a way that
all curves collapse (see Fig.~\ref{FIGORDMC}b).
The exponent $\beta$ is then given by
$\beta=\delta \nupar$. Finally, the exponent $\nuperp$ can
be determined by finite-size simulations. According to
Eq.~(\ref{FSDensityScaling}), $\rho(t) \, t^\delta$
has to be plotted against $t/N^{z/d}$ for various system sizes
(see Fig.~\ref{FIGORDMC}c).
By tuning~$z$ the data points collapse onto a single curve
which gives an estimate for $\nuperp=\nupar/z$.

%--------------------------------------------------------
\headline{Monte Carlo simulations with localized initial conditions}
%--------------------------------------------------------
%
More accurate estimates for the critical exponents can be obtained
by {\it dynamic} simulations starting from a single particle
(active seed)~\cite{GrassbergerTorre79}. This technique
exploits the scaling properties of the pair-connectedness
function $c(\xvec,t)$. Starting from a single particle,
one measures the survival probability $P(t)$,
the number of active sites $N(t)$,
and the mean square of spreading  from the
origin $R^2(t)$ averaged over surviving runs.
According to Eqs.~(\ref{FSScalingSurvival}) and~(\ref{FSScalingPairconn})
these quantities obey the scaling forms
\begin{align}
P(t) &\simeq t^{-\delta} \,
g(\deviation \, t^{1/\nupar}, \, t^{d/z}/N) \ ,
\\
\label{ClusterParticleNumber}
N(t) &= \int d^dx \, c(\xvec,t) \simeq
t^\theta \, \bar{F}(\deviation t^{1/\nupar}, \, t^{d/z}/N) \ ,
\\
\label{MeanSquareSpreading}
R^2(t) &= \langle \xvec^2(t) \rangle =
\frac{1}{N(t)} \, \int d^dx \, \xvec^2 \, c(\xvec,t) \simeq
t^{2/z} \, \tilde{F}(\deviation t^{1/\nupar} , \, t^{d/z}/N) \ .
\end{align}
\begin{figure}
\epsfxsize=150mm
\centerline{\epsffile{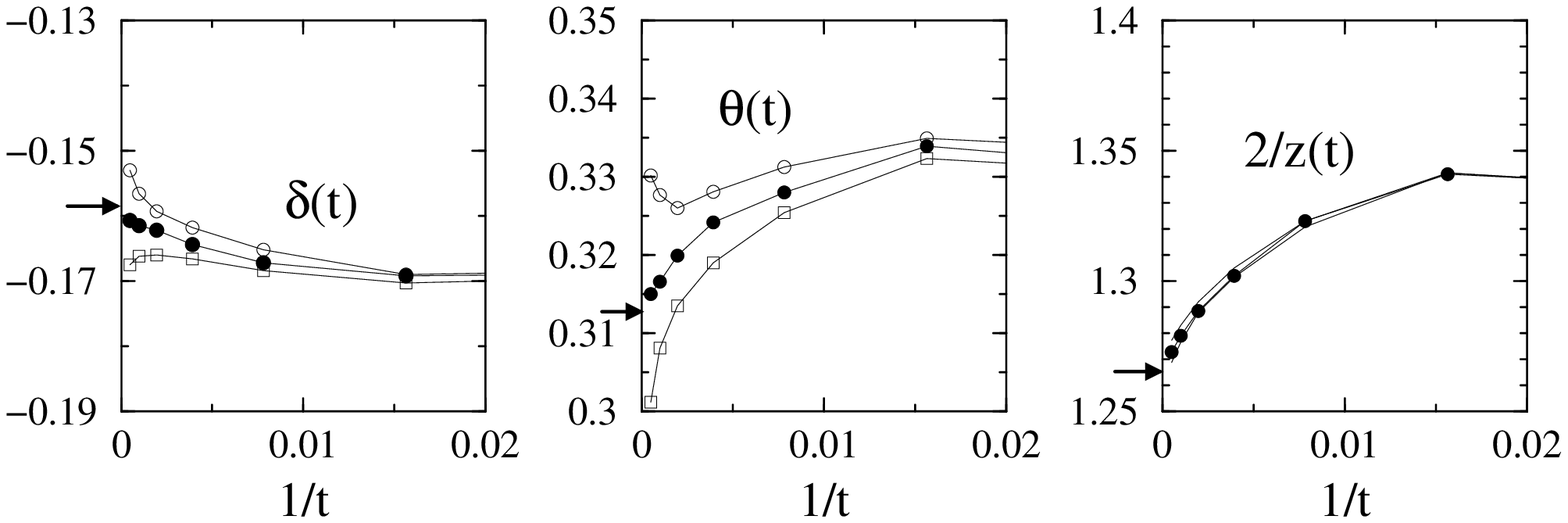}}
\smallcaption{
\label{FIGDYNMC}
Dynamic Monte Carlo simulations of a directed bond percolation process.
The effective exponents $\delta(t)$, $\theta(t)$, and $2/z(t)$ are
shown as a function of time for $p=0.6446, 0.6447, 0.6448$.
The arrows mark the actual values of the critical exponents.
}
\end{figure}
At criticality, they are expected to display asymptotic power laws
\begin{equation}
\label{quantities}
P(t) \sim t^{-\delta} \,, \qquad
N(t) \sim t^\theta \,, \qquad
R^2(t) \sim t^{2/z} \,,
\end{equation}
i.e., they show straight lines in double logarithmic plots.
Off criticality, the lines are curved, allowing a precise
determination of the percolation threshold $p_c$.
Technically it is often useful to consider local slopes of these
curves by introducing {\it effective exponents}
\begin{equation}
-\delta(t) \;=\; \frac{\log_{10}\bigl(P(t)/P(t/b)\bigr)}{\log_{10} b}
\end{equation}
and similarly $\theta(t)$ and $2/z(t)$, where $\log_{10} b$ is the
distance used for estimating the slope. Plotting the local slopes
as functions of $1/t$, the curves may be extrapolated to $t
\rightarrow \infty$, as illustrated in Fig.~\ref{FIGDYNMC}. The
same method works also in higher dimensional
systems~\cite{Grassberger89a}. In order to improve the estimates,
it is useful to eliminate the curvature of the data points at
criticality by plotting the quantities~(\ref{quantities}) against
$1/t^\delta$ instead of $1/t$.

Since the spatial size of the growing
cluster at a given time is finite, the simulation
can be accelerated considerably by storing the coordinates
of active particles in a dynamically generated list. Especially at
criticality, where the density of particles is low,
such algorithms are much more efficient.
Moreover, finite-size effects are eliminated completely.

%--------------------------------------------------------
\headline{Numerical diagonalization}
%--------------------------------------------------------
%
%
\begin{figure}
\epsfxsize=120mm \centerline{\epsffile{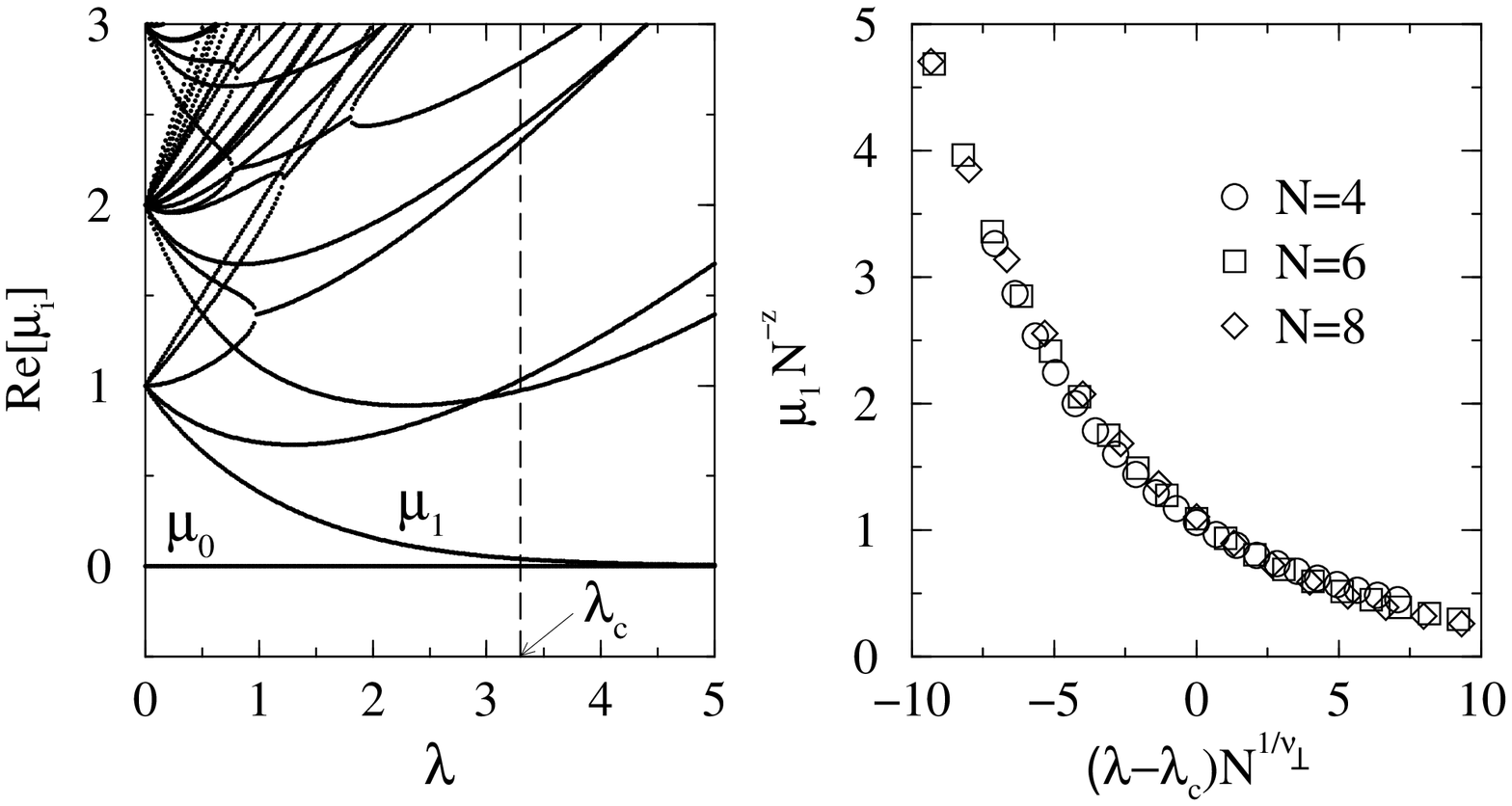}}
\smallcaption{ \label{FIGTRANSFERMAT} Numerical diagonalization of
the Liouville operator of the contact process. 
Left panel: Low-lying part of the
relaxational spectrum of the (1+1)-dimensional contact process on
a lattice with $8$ sites and periodic boundary conditions. Right
panel: Data collapse according to the scaling
form~(\ref{FirstEVScaling}). }
\end{figure}
The critical exponents may also be approximated by numerical
diagonalization of the evolution operator. Although the resulting
estimates are usually less accurate than those obtained by other
methods, this technique is of conceptual interest. Let us, for
example, consider the (1+1)-dimensional contact process on a
finite lattice with $N$ sites and periodic boundary conditions
which is defined by the Liouville operator~(\ref{CPHamiltonian}).
Solving the eigenvalue problem
\begin{equation}
{\cal L}_{CP}(\lambda) \ |\psi_i\rangle = \mu_i \ |\psi_i\rangle
\end{equation}
we obtain a spectrum of eigenvalues $\{\mu_i\}$, as shown in
the left panel of Fig.~\ref{FIGTRANSFERMAT}. As in all reaction-diffusion
models, the lowest eigenvalue $\mu_0$ vanishes. The 
corresponding stationary state $|\psi_0\rangle$ is
the absorbing state of the contact process.
The other eigenvectors represent the relaxational modes of the system.
As can be seen in Fig.~\ref{FIGTRANSFERMAT}, all of them have a
short lifetime except for the first excited state
$|\psi_1\rangle$ whose eigenvalue $\mu_1$
tends to zero as $\lambda$ increases. This eigenvector represents
the active state of the system. In finite systems there is always
a finite probability to reach the absorbing state, hence $\mu_1>0$.
In infinite systems, however, this eigenvalue decreases with
$\lambda$ and vanishes at the critical point.
Since the amplitude of $|\psi_1\rangle$ decays in
the inactive phase as $e^{-\mu_1 t}$, we may identify
$\mu_1^{-1}$ with the temporal scaling length $\xi_\parallel$. In an
infinite system we therefore expect $\mu_1$ to decrease as
$\mu_1 \sim |\lambda-\lambda_c|^{\nupar}$
for $\lambda<\lambda_c$ and to vanish for $\lambda>\lambda_c$.
The corresponding scaling form reads
\begin{equation}
\label{FirstEVScaling}
\mu_1 \sim N^{-z/d} \, h( \deviation \, N^{1/d\nuperp} ) \ ,
\end{equation}
where $\deviation=\lambda-\lambda_c$.
Thus, by plotting $\mu_1 N^{z/d}$ against
$\deviation \, N^{1/d\nuperp}$,
the exponents $z$ and $\nuperp$
can be determined by data collapse,
as demonstrated in the right panel of Fig.~\ref{FIGTRANSFERMAT}.
In order to determine the exponent $\beta$, it would be
necessary to analyze the components of
the eigenvector $|\psi_1\rangle$ with respect to the
particle density in the active phase.
A similar analysis of DP models with parallel updates, which are defined
by transfer matrices instead of Liouville operators,
can be found in Ref.~\cite{ABS91}.

%--------------------------------------------------------
\headline{Density matrix renormalization group methods}
%--------------------------------------------------------
%
%
The method of numerical diagonalization can be improved considerably
by using density matrix renormalization group (DMRG) techniques.
The concept of DMRG was introduced in 1992 by White~\cite{White92} in
the context of equilibrium statistical physics as
a tool for the diagonalization of quantum spin chains.
The main idea is to prolongate a given spin chain by
inserting additional spins and to reduce the resulting configuration
space by a suitable projection mechanism, keeping only the most
relevant eigenstates. This renormalization procedure is then repeated 
many times and the spectrum of the iterated Hamiltonian is analyzed.
Recently DMRG techniques have also been applied to various (1+1)-dimensional 
nonequilibrium systems~\cite{Hieida98,CHS99,CHS00} (see
\cite{KaulkePeschel98} for a general overview). 
The method yields surprisingly accurate results. 
For example, Carlon {\it et al.}~\cite{CHS99} were able to estimate 
the critical exponents of DP by 
$\beta/\nuperp=0.249(3)$, $\nuperp=1.08(2)$, and $z=1.580(1)$,
deviating from the currently accepted values by less than $1.5 \%$.

%--------------------------------------------------------
\headline{Series expansions}
%--------------------------------------------------------
%
The most precise estimates of the DP exponents in 1+1
dimensions have been obtained by series
expansions~\cite{JensenDickman93a}. This technique is very similar
to low- or high-temperature expansions in equilibrium statistical
physics. As an example let us consider the (1+1)-dimensional
contact process. Its Liouville operator~(\ref{CPHamiltonian}) may
be separated into two parts ${\cal L}(\lambda)={\cal L}_0+\lambda
{\cal L}_1$, where ${\cal L}_0$ and ${\cal L}_1$ describe
spontaneous self-destruction $A\rightarrow \vacancy$ and offspring
production $A \rightarrow 2A$, respectively. The basic idea is to
regard $\lambda$ as a small perturbation and to express physical
quantities as power series in $\lambda$. To this end it is useful
to introduce the Laplace transform of the probability distribution
$|P_t\rangle$ and to expand it in powers of $\lambda$:
\begin{equation}
\label{SubcriticalExpansion}
|\tilde{P}(s)\rangle =
\int_0^\infty\,dt\,e^{-st} \, |P_t \rangle =
\sum_{n=0}^{\infty} \, \lambda^n \, |\tilde{P}_n(s)\rangle
\ .
\end{equation}
By applying ${\cal L}(\lambda)$ from the left one can easily derive
the recursion relation
\begin{equation}
\label{SeriesRecursion}
(s-{\cal L}_0) \, |\tilde{P}_n(s)\rangle =
\left\{
\begin{array}{cc}
|P_0\rangle             & \text{ if } n=0 \\
{\cal L}_1|\tilde{P}_{n-1}(s)\rangle    & \text{ if } n\geq 1
\end{array}
\right.
\ ,
\end{equation}
where $|P_0\rangle$ denotes the initial particle configuration.
Hence, if the process started from a configuration with a single particle,
the vector $|\tilde{P}_n(s)\rangle$ describes an ensemble of configurations
with at most $n$ particles. It is therefore possible to explicitly
construct the vectors $|\tilde{P}_n(s)\rangle$, as described in
detail in Ref.~\cite{JensenDickman93a}.

The above expansion allows the temporal integral of
any observable $X(t)=\sumstate X|P_t\rangle$
to be expressed as a power series in $\lambda$
(for notations see Appendix~\ref{APPVECSEC}):
\begin{equation}
\int_0^\infty \, dt\, X(t) =
\lim_{s \rightarrow 0} \sumstate X|\tilde{P}(s)\rangle =
\sum_{n=0}^\infty \, \lambda^n \, \lim_{s \rightarrow 0}
\sumstate X|\tilde{P}_n(s)\rangle
\ .
\end{equation}
Let us, for example, consider the survival probability $P(t)$
that the system has not yet reached the absorbing state at time~$t$
[cf. Eq.~(\ref{DensityScaling})].
Using the vector formalism this quantity may be written as
\begin{equation}
P(t) = 1-\langle 0|P_t\rangle =
\sumstate P_t\rangle-\langle 0|P_t\rangle
\ ,
\end{equation}
where $\langle0|=(1,0,0,\ldots,0)$ denotes the absorbing state.
The critical exponents can be estimated as follows.
On the one hand, the mean survival time $T$
of clusters in the inactive phase
can be expanded in powers of $\lambda$ by
\begin{equation}
T = \int_0^\infty dt\, P(t) =
\sum_{n=0}^\infty \, \lambda^n \, \lim_{s \rightarrow 0}
\biggl( \sumstate \tilde{P}_n(s)\rangle -
\langle 0|\tilde{P}_n(s)\rangle \biggr)
\ .
\end{equation}
On the other hand, according to Eq.~(\ref{TScaling}) we have
$T \sim (-\deviation)^{\beta^\prime-\nupar}$ so that
\begin{equation}
\label{SeriesLogDer}
\frac{d}{d\lambda}\ln T \simeq
\frac{\nupar-\beta^\prime}{\lambda_c-\lambda}
+ const
\ .
\end{equation}
Therefore, in order to estimate $\lambda_c$ and $\beta-\nupar$,
three steps have to be taken. At first, the vectors
$|\tilde{P}_n(s)\rangle$ have to be determined by iterating
Eq.~(\ref{SeriesRecursion}) up to order $n_{max}$. Although this recursion
relation is quite complicated, it is still simple enough to be implemented
on a computer (for example, in Ref.~\cite{JensenDickman93a} the
iteration was carried out up to order $n_{max}=24$). Next, one has to
express $T$ as a power series in $\lambda$.
Finally, $\lambda_c$ and $\beta-\nupar$ can be estimated
by determining the location and the amplitude of the singularity
in Eq.~(\ref{SeriesLogDer}) and by using a
Pad\'e approximation~\cite{Guttmann89}.
Since the singularity is approached from the inactive phase,
we are dealing with a {\it subcritical} expansion.
Similarly one may also consider the {\it supercritical} case
by expanding ${\cal L}(\mu)=\mu {\cal L}_0+{\cal L}_1$ in powers of $\mu$.

A general review on series expansion can be found in
Ref.~\cite{Guttmann89}. Series expansions were applied to
(1+1)-dimensional DP firstly in Ref.~\cite{EBAB86}, where the
critical exponents could be determined with a relative accuracy of
about $10^{-3}$. Refined simulations~\cite{EBG88} led the authors
to the conjecture that the DP exponents should be given by the
{\it rational} values $\beta=199/720$, $\nuperp=26/15$, and
$\nupar=79/72$. In a sequence of
papers~\cite{JensenDickman93a,JensenDickman93c,JensenGuttmann95,Jensen96a}
the error margins could be further reduced down to
$10^{-4}\ldots10^{-5}$. These improved estimates showed that the
conjectured rational values were incorrect, indicating that the
critical exponents of DP could be given by {\it irrational} numbers.
This should be taken as a warning that critical exponents of
non-integrable systems are usually not given by simple rational values.
Currently, the most precise estimates are
given in Ref.~\cite{Jensen99a}. Series
expansions for DP were also performed in two spatial
dimensions~\cite{JensenGuttmann96}. In addition, the exponents
were found to be independent of the type of lattice under
consideration. For easy reference we listed the most precise
estimates in Table~\ref{TABEXP}.
%
%
%%%%%%%%%%%%%%%%%%%% TABLE %%%%%%%%%%%%%%%%%%%%%%%%
%
\begin{table}
\footnotesize
\begin{center}
\begin{tabular}{||c||c|c|c|c|c|c||}
\hline\\[-4mm]
critical
& MF    & IMF   &
      $d=1$ &
      $d=2$ &
      $d=3$ &
      $d=4-\epsilon$ \\
exponent
&       & \cite{BenNaimKrapivsky94} &
      \cite{Jensen99a} &
      \cite{VoigtZiff97} &
      \cite{Jensen92} &
      \cite{BronzanDash74} \\ \hline

$\beta$
& $1$   & $1/2$ & $0.276486(8)$ & 0.584(4) & 0.81(1) &
$1-\epsilon/6-0.01128\,\epsilon^2$ \\

$\nuperp$
& $1/2$ & $1$   & $1.096854(4)$ & 0.734(4) & 0.581(5) &
$1/2+\epsilon/16+0.02110\,\epsilon^2$ \\

$\nupar$
& $1$   & $3/2$ & $1.733847(6)$ & 1.295(6) & 1.105(5) &
$1+\epsilon/12+0.02238 \,\epsilon^2$ \\

$z$
& $2$   & $3/2$ & $1.580745(10)$ & 1.76(3) & 1.90(1) &
$2-\epsilon/12-0.02921 \,\epsilon^2$ \\ \hline

$\delta$
& $1$ & $1/2$ & $0.159464(6)$ & $0.451$ & $0.73$ &
$1-\epsilon/4-0.01283 \,\epsilon^2$ \\

$\theta$
& $0$ & $1/2$ & $0.313686(8)$ & $0.230$ & $0.12$ &
$\epsilon/12+0.03751 \,\epsilon^2$ \\

$\gamma$
& $1$ & $3/2$ & $2.277730(5)$ & $1.60$ & $1.25$ &
$1+\epsilon/6+0.06683 \,\epsilon^2$ \\

$v$
& $1$ & $1/2$ & $0.82037(1)$  & $0.88$ & $0.94$ &
$1-\epsilon/12+0.03317 \,\epsilon^2$ \\

$\sigma$
& $2$ & $2$ & $2.554216(13)$ & $2.18$ & $2.04$ &
$2+\epsilon^2/18$  \\
\hline

\end{tabular}
\end{center}
\smallcaption{
\label{TABEXP}
Estimates for the critical exponents of directed percolation
obtained by mean field (MF), improved mean field (IMF), numerical,
as well as field-theoretical methods.
}
\end{table}

%--------------------------------------------------------
\headline{Field-theoretical approximations}
%--------------------------------------------------------
%
By a field-theoretical renormalization group calculation
(see Sec.~\ref{FTHSEC}) it is possible to compute fluctuation
corrections of the critical exponents close to the
upper critical dimension $d_c=4$ in powers of
$\epsilon = d_c-d$. In a two-loop
approximation~\cite{BronzanDash74,Janssen81}
these corrections are given by
\begin{eqnarray}
\label{TwoLoopResults}
\beta&=&1-\epsilon/6+\left(\frac{11}{12}-\frac{53}{6}\ln\frac{4}{3}\right)
(\epsilon/12)^2  + \text{0}(\epsilon^3) \,,\nonumber \\
\nuperp&=&\frac12+\epsilon/16+\left(\frac{107}{32}-\frac{17}{16}\ln\frac{4}{3}\right)
(\epsilon/12)^2  + \text{0}(\epsilon^3) \,,\\
\nupar&=&1+\epsilon/12+\left(\frac{109}{24}-\frac{55}{12}\ln\frac{4}{3}\right)
(\epsilon/12)^2  + \text{0}(\epsilon^3)\,. \nonumber
\end{eqnarray}
For $d \leq 2$ these approximations are quite inaccurate.
However, in three spatial dimensions, where numerical simulations
and series expansions are difficult to perform, the two-loop
approximations are regarded as the most precise estimates available.

%==============================================================================
\subsection{Field-theoretic formulation of directed percolation}
%==============================================================================

\label{FTHSEC} The robustness of the DP universality class can be
partly understood by studying the corresponding field theory. It
is interesting to note that the field theory of DP was first
discovered in a quite different field of physics, namely in the
context of hadronic interactions at ultra-relativistic energies.
In order to predict the cross sections of such particles at high
energies quantitatively, a field-theoretic approach, called {\it
Reggeon field theory}, was developed in the 70's (see Refs.
\cite{AbarbanelBronzan74,ABBS75,ABSW75,ABSS76,FHFS76,BFM78}, a
general review is given in~\cite{Moshe78}). Surprisingly it took
almost another ten years to realize that Reggeon field theory was
nothing but a field-theoretic realization of the contact
process~\cite{GrassbergerSundermeyer78,Obukhov80,CardySugar80},
sometimes also called Gribov process
\cite{Gribov67,GribovMigdal68}. In the following we sketch 
the main ideas of a field-theoretic approach to DP.

%---------------------------------------------------------------------------
\headline{The DP Langevin equation}
%---------------------------------------------------------------------------
%
The Langevin equation of motion for directed percolation can
be derived directly from the master equation for the
contact process~\cite{Janssen81} and reads
\begin{equation}
\label{DPLangevinEquation}
\timederivative\rho(\xvec,t) \;=\;
\crit \rho(\xvec,t) -
\lambda \rho^2(\xvec,t)  +
\diff \nabla^2 \rho(\xvec,t) +
\noise(\xvec,t)
\ .
\end{equation}
It differs from the mean field equation~(\ref{ExtendedDPMF})
by a density-dependent Gaussian noise field $\noise(\xvec,t)$,
which is defined by its correlations
\begin{equation}
\label{DPNoise}
\begin{split}
\langle \noise(\xvec,t) \rangle &= 0 \,, \\
\langle \noise(\xvec,t) \noise(\xvec',t')  \rangle &=
\namp \, \rho(\xvec,t) \, \delta^d(\xvec-\xvec') \, \delta(t-t')
\ .
\end{split}
\end{equation}
Since the amplitude of $\noise(\xvec,t)$ is proportional to
$\sqrt{\rho(\xvec,t)}$, the noise is said to be {\it
multiplicative}. This ensures that the absorbing state
$\rho(\xvec,t)=0$ does not fluctuate. The square-root behavior
stems from the definition of $\rho(\xvec,t)$ as a coarse-grained
density of active sites averaged over some mesoscopic box size.
Only active sites in this box give rise to fluctuations of the
density, generating a bounded uncorrelated noise. The noise field
$\noise(\xvec,t)$ can be viewed as the sum of all these noise
contributions in the box. According to the central limit theorem,
if the number of particles in the box is sufficiently high,
$\noise(\xvec,t)$ tends to a Gaussian distribution with an
amplitude proportional to the square root of the number of active
sites in the box. This type of noise has to be distinguished from
other nonequilibrium systems with multiplicative noise where the
noise amplitude is proportional to the field $\rho(\xvec,t)$
itself without square root~\cite{GMT96,Munoz98}. These systems do
not belong to the DP class, rather they are related to the KPZ
universality class~\cite{KPZ86}. The DP Langevin equation was also
tested numerically in Ref.~\cite{Dickman94}, confirming that the
critical exponents are in agreement with those of ordinary DP
lattice models. Note that in contrast to the annihilation process
discussed in Sec.~\ref{FLUCSEC}, the noise~(\ref{DPNoise})  is
real due to positive density correlations in the bulk.

The Langevin equation~(\ref{DPLangevinEquation}) can be seen as a
minimal equation needed to describe DP. It may also include
higher order terms such as $\rho^3(\xvec,t)$
or $\nabla^4\rho(\xvec,t)$, but these contributions turn out to be
irrelevant under renormalization group transformations. The same
applies to higher-order contributions to the noise. These additional
terms account for (nonuniversal) short-range correlations while
they are irrelevant on large scales. In fact,
the robustness of DP originates in the irrelevance of
higher-order terms in the Langevin equation.

%---------------------------------------------------------------------------
\headline{Relation to Reggeon field theory}
%---------------------------------------------------------------------------
%
In field-theoretic calculations it is often more convenient to
characterize the dynamic system by a partition sum $Z$.
The sum is carried out
over all realizations of the field $\phi(\xvec,t)=\rho(\xvec,t)$
and the noise $\noise(\xvec,t)$, weighted by
an appropriate effective action. More precisely,
the partition sum is defined
as the integral over all realizations of the field
$\phi(x,t)$ and the noise $\noise(\xvec,t)$ which satisfy the
Langevin equation. Therefore, we may write the integrand as
a $\delta$-function with Eq.~(\ref{DPLangevinEquation})
as its argument:
\begin{equation}
Z \sim \int D\noise \, P[\noise] \int D\phi \,I[\phi] \,\,\,
\delta\Bigl( \timederivative\phi - \diff \nabla^2 \phi -
\crit \phi + \lambda\phi^2 - \noise \Bigr)
\ .
\end{equation}
Here $D\noise$ and $D\phi$ denote functional integration,
$P[\noise]$ is the probability distribution of the noise field,
and $I[\phi]$ stands for an appropriate Jacobian which turns out to
be irrelevant in the present problem. As shown in
Appendix~\ref{ACTION}, it is possible to integrate the noise by
introducing a Martin-Siggia-Rosen
response field $\tilde{\phi}(\xvec,t)$.
The resulting action $S=S_0+S_{int}$ with
\begin{eqnarray}
\label{DPFreePart}
S_0[\phi,\tilde{\phi}] &=&
\int d^dx\,dt \,\,  \tilde{\phi}(\xvec,t) \,
\Bigl(\tau \timederivative - \diff \nabla^2  - \crit \Bigr)\,\phi(\xvec,t)
\ ,
\\
\label{DPInteraction}
S_{int}[\phi,\tilde{\phi}] &=&
\frac{\namp}{2}\,
\int d^dx\,dt \,\, \tilde{\phi}(\xvec,t)
\Bigl(\phi(\xvec,t) - \tilde{\phi}(\xvec,t) \Bigr)
\phi(\xvec,t)
\ ,
\end{eqnarray}
is the effective action of Reggeon field theory~\cite{Moshe78}.
In momentum space it may also be written as
\begin{eqnarray}
\label{ActionS0}
S_0[\phi,\tilde{\phi}] &=&
\int d_{k\omega}\,\, \tilde{\phi}(-k,-\omega)\,
(-i\tau\omega + \diff k^2-\crit) \phi(k,\omega)
\\
\label{ActionSint}
S_{int}[\phi,\tilde{\phi}] &=&
\frac{\namp}{2}\int d_{k\omega} \int d_{k'\omega'}\,\,
\phi(-k,-\omega)\,\tilde{\phi}(-k',-\omega')\, \\
&& \hspace{3mm} \times
\Bigl[ \phi(k+k',\,\omega+\omega') -
\tilde{\phi}(k+k',\,\omega+\omega') \Bigr]\,, \nonumber
\end{eqnarray}
where $d_{k\omega} = (2\pi)^{-d-1}d^dk\,d\omega$.
Formally the free part of the action can be expressed as
\begin{align}
S_0[\phi,\tilde{\phi}] &=
\frac12 \int  d_{k\omega}\,\,
\vec{\Phi}(-k,-\omega) \, {\cal S}_0(k,\omega) \, \vec{\Phi}(k,\omega) \,,
\\
{\cal S}_0(k,\omega) &= \left( \begin{array}{cc}
0 & \diff k^2 - \crit + i \tau \omega \\
\diff k^2 - \crit - i \tau \omega  & 0
\end{array}
\right)  \nonumber \,,
\end{align}
where $\vec{\Phi}=(\phi,\tilde{\phi})$.
Introducing external currents $J(k,\omega)$ and $\tilde{J}(k,\omega)$
we can rewrite the partition sum as
\begin{align}
Z[J,\tilde{J}] &\sim  \int D\phi\,D\tilde{\phi}\, I'[\phi,\tilde{\phi}]\,
\exp\Bigl( -S_0[\phi,\tilde{\phi}] - S_{int}[\phi,\tilde{\phi}]
 + \int d^dx\,dt \, ( J\phi + \tilde{J}\tilde{\phi}) \Bigr) \nonumber \\
 &\sim \exp \Bigl( - {\cal L}_{int} \Bigl[ \frac{\delta}{\delta J},
 \frac{\delta}{\delta \tilde{J}} \Bigr] \Bigr) \,
 \exp \Bigl( \frac12 \int d_{k\omega} \, \vec{J}(-k,-\omega)  \,
 {\cal G}_0(k,\omega) \, \vec{J}(k,\omega) \Bigr) \ ,
\end{align}
where the free propagator ${\cal G}_0={\cal S}_0^{-1}$ is given by
\begin{equation}
\label{FreePropagator}
{\cal G}_0(k,\omega) \;=\;
\left(
\begin{array}{cc}
0 & G_0(k,\omega) \\
G_0(-k,-\omega) & 0
\end{array}\right) \,
\end{equation}
with $G_0(k,\omega)=(\diff k^2 - \crit - i \tau \omega)^{-1}$.
Because of ${\cal G}_0(q,\omega) \neq {\cal G}_0(q,-\omega)$
DP is an irreversible process.

\begin{figure}
\epsfxsize=140mm
\centerline{\epsffile{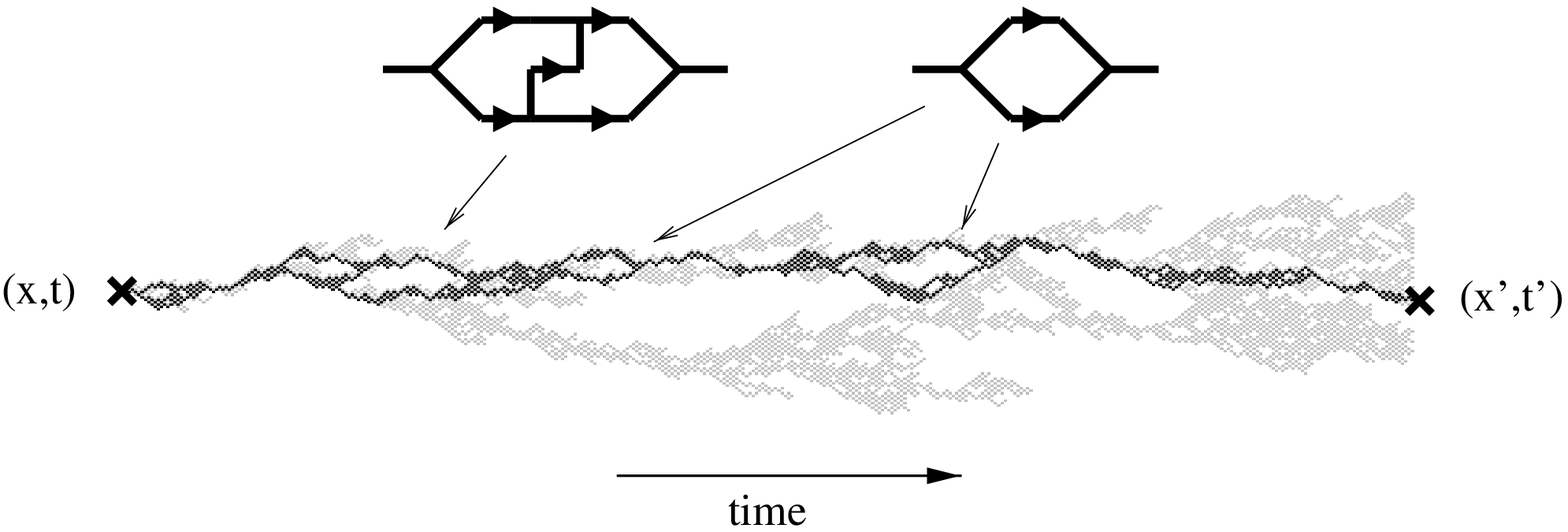}}
\smallcaption{
\label{FIGBACKBONE}
Critical DP cluster (grey) and the backbone of the pair connectedness
function $c(\xvec^\prime,t^\prime,\xvec,t)$, illustrating
the physical meaning of Feynman diagrams in directed percolation.
}
\end{figure}

%---------------------------------------------------------------------------
\headline{Cluster backbone and Feynman diagrams}
%---------------------------------------------------------------------------
%
%
Before turning to field-theoretic renormalization group
techniques let discuss the physical
meaning of Feynman diagrams in directed percolation.
The full propagator of the field theory is the pair connectedness function
$c(\xvec^\prime,t^\prime,\xvec,t)$ which is defined as the probability that
two sites $(\xvec',t')$ and $(\xvec,t)$ are connected
by a directed path of open bonds (see Sec.~\ref{ABSSCALESEC}).
In a given realization of open and closed bonds there may be several
possible directed paths connecting the two sites,
as illustrated in Fig.~\ref{FIGBACKBONE}. The union
of all possible paths constitutes the so-called {\em backbone} of the
pair connectedness function~\cite{HJS95}.
More precisely, the backbone consist of all sites that are
connected with the sites $(\xvec',t')$ {\em and} $(\xvec,t)$
by a directed walk, i.e., we cut off all dangling ends of the
cluster. From the topological point of view
the backbone is a directed graph consisting
of branching and merging lines. Because of the duality
symmetry~(\ref{DualitySymmetry}), it is
statistically invariant under time reversal.

In principle the full propagator is given by a weighted sum 
over all possible backbone configurations. Fluctuation effects
are mainly due to the influence of closed loops.
Above the upper critical dimension $d_c=4$ the degree of spatial 
freedom for propagating lines is so high that the probability to 
merge tends to zero. Thus, the contribution
of closed loops can be neglected and hence the full
propagator is effectively described by the free propagator.
Below the critical dimension, however, loops occur more
frequently and begin 
to play a significant role (see Fig.~\ref{FIGBACKBONE}).

\begin{figure}
\epsfxsize=120mm
\centerline{\epsffile{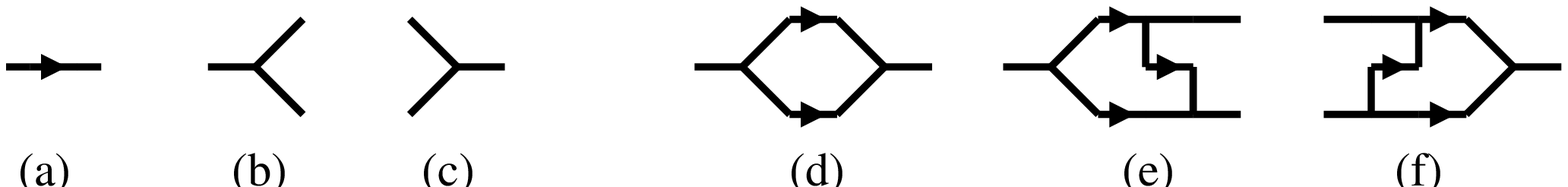}}
\smallcaption{
\label{FIGFEYNMAN}
Feynman diagrams of directed percolation.
Left: Elementary components of the backbone: (a) free propagator,
(b) branching vertex, and (c) merging vertex.
Right: One-loop diagrams for (d) propagator renormalization
and (e,f) vertex renormalization.
}
\end{figure}

The loops of the backbone 
may be associated with the Feynman diagrams 
of Reggeon field theory. As shown in Fig.~\ref{FIGFEYNMAN}(a)-(c), 
the backbone can be decomposed into three elementary components. 
The arrow stands for the free propagator~(\ref{FreePropagator}) while
the diagrams for branching and merging represent the cubic vertices
in Eq.~(\ref{DPInteraction}), associated with the weights $\pm \namp/2$.
Because of self-destruction $A \rightarrow \vacancy$, free paths
have a finite lifetime, as expressed by the bare mass~$\crit$ of
the free propagator. Consequently, the paths in a given
configuration of the backbone have to be weighted by their length.
Moreover, each closed loop carries a weight 
$-\namp^2/4$. 

The negative sign for the weight of
closed loops can be explained as follows.
The pair connectedness function is the sum over all backbone 
configurations $b$ connecting the sites 
$(\xvec',t')$ and $(\xvec,t)$ weighted by their 
probability $P_b$:
\begin{equation}
\label{PCF}
c(\xvec^\prime,t^\prime,\xvec,t) = \sum_b P_b \,.
\end{equation}
In order to find an recurrence relation for 
the probability $P_b$, let us consider 
directed bond percolation on a lattice. Obviously, $P_b$ 
is the weighted sum over all lattice configurations
compatible with the backbone configuration $b$. The
backbone itself contributes with a factor $p^{n_b}$,
where $p$ is the percolation probability and $n_b$ denotes
the number of bonds occupied by the backbone $b$.
Another factor comes from the bonds outside the backbone.
This factor can be expressed as the probability that $b$
is {\em not} contained in a larger backbone~$b'$. Thus, 
$P_b$ satisfies the recurrence relation
\begin{equation}
P_b= p^{n_b}(1 - p^{-n_b}\sum_{b', b \subset b'} P_{b'})\,.
\end{equation}
Since $b'$ contains always more loops than $b$, this relation can
be used to expand the pair connectedness function~(\ref{PCF})
in the number of loops. As can be easily verified, the one-loop
correction carries a negative sign. More generally, it is 
possible to show by an inclusion-exclusion argument that
each closed loop contributes with a negative weight. An analogous proof
for isotropic percolation is explained in detail in
the review article by Essam~\cite{Essam80}.

Thus, apart from the negative weight of closed loops,
the backbone may be interpreted as a graph consisting of
Feynman diagrams. This interpretation is possible 
because in a DP process the field
$\phi(\xvec,t)$ represents the local density of particles.
It should be noted that this is not always true.
Moreover, various authors prefer to shift the field $\phi$
by its mean field expectation value so that the  interpretation
as a density is no longer valid. For this reason we will
continue to use the unshifted fields $\phi$ and $\tilde{\phi}$.

%

%---------------------------------------------------------------------------
\headline{One-loop approximation}
%---------------------------------------------------------------------------
%
Slightly below the upper critical dimension one-loop diagrams start
to contribute to the full propagator while
higher-order diagrams are still strongly suppressed. In this
regime the DP process can be approximated by neglecting higher-order
loop diagrams. The resulting propagator consists of a sum
over $n$ concatenated one-loop diagrams with $n$ running from zero to
infinity. In momentum space the corresponding expression can be
written as a simple geometric series. By
carrying out the integration one obtains
an ultraviolet-divergent expression. 
Hence, in order to regularize the propagator,
an upper cutoff $\cutoff$ has to be introduced in momentum space.
Physically this upper cutoff corresponds to the lattice spacing
of the DP model. In other words, DP needs a lattice; there is no
continuum theory of DP.

\begin{figure}
\epsfxsize=80mm
\centerline{\epsffile{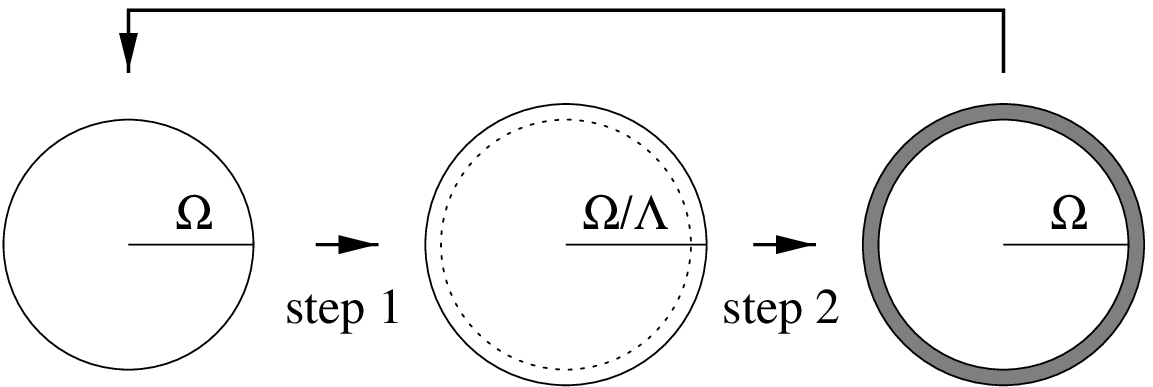}}
\vspace{2mm}
\smallcaption{
\label{FIGWILSON}
Wilson's renormalization group. Step 1: Scaling transformation in momentum
space. Step 2: Integration in a momentum shell (shaded region).
}
\end{figure}

In order to approximate the critical exponents, we use Wilson's
renormalization group scheme~\cite{WilsonKogut74} which consists
of two steps (see Fig.~\ref{FIGWILSON}). At first the theory is
coarse-grained by a scaling transformation $\xvec \rightarrow
\scalefac \xvec$ with $\scalefac<1$, leading to a change of the
coefficients in the effective action and a dilatation of momentum
space (including the cutoff~$\cutoff$). In the second step the
short-range fluctuations are integrated out in a {\em momentum
shell}. This can be done by evaluating the Feynman diagrams in the
range $\cutoff \leq k \leq \cutoff/\scalefac$ and absorbing the
resulting contributions in the coefficients. The total change of
the coefficients determines the RG flow and therefore the critical
exponents.

Let us first consider the scaling transformation.
Because of the duality symmetry of DP (see Sec.~\ref{ABSSCALESEC})
the action $S$ is invariant under the replacement
\begin{equation}
\phi(\xvec,t) \rightarrow -\tilde{\phi}(\xvec,-t)\,,\quad
\tilde{\phi}(\xvec,t) \rightarrow - \phi(\xvec,-t)
\ ,
\end{equation}
implying that $\phi$ and $\tilde{\phi}$ have exactly the same
scaling behavior:
\begin{equation}
\label{DPEquScaling}
x \rightarrow \scalefac \, x\,,
\hspace{5mm}
t \rightarrow \scalefac^z \, t\,,
\hspace{5mm}
\phi(\xvec,t) \rightarrow \scalefac^\chi \, \phi(\scalefac\xvec,\scalefac^z t)\,,
\hspace{5mm}
\tilde{\phi}(\xvec,t) \rightarrow \scalefac^\chi
   \, \tilde{\phi}(\scalefac \xvec,\scalefac^z t) \ .
\end{equation}
Under this scaling transformation the effective action
(\ref{DPFreePart})-(\ref{DPInteraction}) turns into
\begin{eqnarray}
S_0[\phi,\tilde{\phi}] &=& \nonumber
\int d^dx\,dt \,\,  \tilde{\phi}(\xvec,t) \,
\Bigl(\underbrace{\tau \scalefac^{2\chi+d}}_{\tau'}\timederivative -
\underbrace{\diff \scalefac^{2\chi+d+z-2}}_{\diff'} \nabla^2  -
\underbrace{\crit \scalefac^{2 \chi+d+z}}_{\crit'} \Bigr)\,\phi(\xvec,t) \,,
\\
S_{int}[\phi,\tilde{\phi}] &=&
\frac{1}{2}\,\int d^dx\,dt \,\,
\underbrace{\namp\,\scalefac^{3\chi+d+z}}_{\namp'}\,
\phi(\xvec,t) \tilde{\phi}(\xvec,t)
\Bigl(\phi(,t) - \tilde{\phi}(,t) \Bigr)\,.
\end{eqnarray}
Thus, for an infinitesimal dilatation
$\scalefac=1+\smalldev$, the four coefficients rescale as
\begin{eqnarray}
\label{DPLangrangeCoeffRescaling}
\tau' &=& [ 1 + \smalldev \, (2\chi+d) ] \, \tau \nonumber \,,\\
\diff' &=& [ 1 + \smalldev \, (2\chi+d+z-2) ] \, \diff \,, \\
\crit' &=& [ 1 + \smalldev \, (2\chi+d+z) ] \, \crit \nonumber \,,\\
\namp' &=& [ 1 + \smalldev \, (3\chi+d+z) ] \, \namp \,. \nonumber
\end{eqnarray}
\noindent
In the second step of Wilson's RG procedure the one-loop diagrams are
integrated in a momentum shell. The propagator is renormalized by
diagram (d) in Fig.~\ref{FIGFEYNMAN}
\begin{equation}
\label{PropRenorm}
G_0^{-1}(k,\omega)^\renorm =
G_0^{-1}(k,\omega) -
\frac{\namp^2}{2}\,
\int_> d_{k'\omega'}\,\, G_0(\frac{k}{2}+k',\frac{\omega}{2}+\omega')\,
G_0(\frac{k}{2}-k',\frac{\omega}{2}-\omega') \ ,
\end{equation}
where `$>$' denotes integration in the momentum shell
$\cutoff \leq k \leq \cutoff/\scalefac$. This equation
can be rewritten as
\begin{equation}
\label{PropRenorm2}
\crit^\renorm-\diff^\renorm k^2+i \tau^\renorm\omega \;=\;
\crit^\prime-\diff^\prime k^2+i \tau^\prime \omega \,-\,
\frac{{\namp^\prime}^2}{2} J^P
\ ,
\end{equation}
where $J^P$ denotes the integral in Eq.~(\ref{PropRenorm}).
Integrating $J^P$ and expanding the result
to the lowest order in $k$ and $\omega$ yields the series
(see Appendix~\ref{INTEGRALS})
\begin{equation}
J^P \;=\; \frac{\smalldev K_d \cutoff^d}{2 \tau} \,\,
\left( \frac{1}{\cutoff^2 \diff-\crit} \,-\,
\frac{\cutoff^2\diff}{4(\cutoff^2 \diff-\crit)^2} k^2 \,+\,
\frac{i \tau}{2(\cutoff^2 \diff-\crit)^2}\omega \,+\, \ldots \right) \ .
\end{equation}
Therefore, the coefficients in Eq.~(\ref{PropRenorm2}) are renormalized
to one-loop order by
\begin{equation}
\begin{split}
\label{DPRenormMuNuTau}
\tau^\renorm &= \tau^\prime -
\frac{\namp^2 \smalldev K_d}{8(\cutoff^2\diff-\crit)^2}
\,, \qquad\\
\diff^\renorm &= \diff^\prime -
\frac{\namp^2 \smalldev K_d \cutoff^2\diff}
{16 \tau (\cutoff^2\diff-\crit)^2}
\,, \qquad\\
\crit^\renorm &= \crit^\prime -
\frac{\namp^2 \smalldev K_d}{4 \tau
(\cutoff^2\diff-\crit)}  \ .
\end{split}
\end{equation}
Finally, we have to renormalize the coupling constant $\namp$.
Because of the duality symmetry~(\ref{DualitySymmetry}) the cubic vertices
renormalize identically [see diagrams (e) and (f) in Fig.~\ref{FIGFEYNMAN}].
For the cubic vertices it is sufficient to carry out the integration
at $k=\omega=0$:
\begin{equation}
\label{DPRenormGamma}
\namp^\renorm = \namp^\prime - 2 \namp^3
\int_> d_{k\omega} \,\, G_0^2(k,\omega)\,G_0(-k,-\omega)=
\namp^\prime \,-\, \frac{\smalldev \namp^3 K_d}
{2 \tau (\cutoff^2\diff-\crit)^2} \ .
\end{equation}
Adding the changes of the coefficients
under rescaling (\ref{DPLangrangeCoeffRescaling})
and the subsequent shell integration
(\ref{DPRenormMuNuTau})-(\ref{DPRenormGamma})
we obtain the RG flow equations
\begin{eqnarray}
\label{DPLagrangeCoeffTotalChange}
\partial_\smalldev\tau
    &=& \tau \, \Big(2\chi+d-\frac{\namp^2  K_d\cutoff^d}
    {8 \tau (\diff \cutoff^2-\crit)^2} \Big) \nonumber \ , \\
\partial_\smalldev\diff
    &=& \diff \, \Big(2\chi+d+z-2-\frac{\namp^2 K_d\cutoff^d }
    {16 \tau (\diff\cutoff^2-\crit)^2} \Big) \ , \\
\partial_\smalldev\crit
    &=& \crit \, \Big(2\chi+d+z-\frac{\namp^2  K_d\cutoff^d}
    {4 \crit \tau (\diff\cutoff^2-\crit)}\Big) \nonumber \ , \\
\partial_\smalldev\namp
    &=& \namp \, \Big(3\chi+d+z-
    \frac{\namp^2 K_d\cutoff^d} {2 \tau (\diff\cutoff^2-\crit)^2}  \Big)
    \nonumber  \ .
\end{eqnarray}
Two scaling combinations appear in these equations, namely
\begin{equation}
\label{ScalingComb}
S_1=\frac{\namp^2K_d\cutoff^d}
{16\tau(\diff\cutoff^2-\crit)^2}\,, \qquad
S_2=\frac{\namp^2K_d\cutoff^d}
{4\crit\tau(\diff\cutoff^2-\crit)} \,.
\end{equation}
Two of the four parameters $\tau,\diff,\crit,\namp$
can be chosen freely\footnote{On the level of lattice models
such as the DK model or the contact process, this freedom
corresponds to choosing the time scale and a point on the 
phase transition line.}. 
Here we fix the coefficients of the spatial
and temporal derivatives, i.e., we require $\tau$ and $\diff$
to be invariant under RG transformations. Thus the first two equations read
\begin{equation}
\label{ConstantCoefficients}
4-\epsilon+2\chi-2S_1=0
\,, \qquad
2-\epsilon+2\chi+z-S_1=0
\ ,
\end{equation}
where $d=4-\epsilon$. The RG flow is then
described by two differential equations:
\begin{equation}
\label{RGFlowTwo}
\begin{split}
\partial_\smalldev\crit
    &= \crit \, \Big(4-\epsilon+2\chi+z-S_2\Big)
    \;=\; \crit \, \Bigl( 2+S_1-S_2 \Bigr)  \ , \\
\partial_\smalldev\namp
    &= \namp \, \Big(4-\epsilon+3\chi+z-8S_1  \Big)
    \;=\; \namp \, \Bigl(\epsilon/2-6S_1 \Bigr) \ .
\end{split}
\end{equation}
At the fixed point $(S_1^*,S_2^*)$,
$\crit$ and $\namp$ are invariant under RG transformations,
i.e., $2+S_1^*-S_2^*=0$ and $\epsilon/2-6S_1^*=0$.
Therefore, the fixed point is located at
\begin{equation}
S_1^*=\epsilon/12\,, \qquad S_2^*=2+\epsilon/12\,.
\end{equation}
Inserting this solution into Eq.~(\ref{ConstantCoefficients})
we obtain two of the three critical exponents, namely
$\chi = -2+7\epsilon/12$ and $z=2-\epsilon/12$.
The third exponent can be determined by investigating the
RG flow in the vicinity of the fixed point.
Because of Eq. (\ref{ScalingComb}), the fixed point values
for $\crit$ and $\namp$ are given by
\begin{align}
\crit^*&=\frac{4\diff\cutoff^2\epsilon}{24+5\epsilon}
=\frac{\diff\cutoff^2}{6}\epsilon + \text{0}(\epsilon^2)
\,,\\
{\namp^2}^*&=\left(\frac{2\diff(24+\epsilon)}{24+5\epsilon}
\sqrt{\frac{\epsilon\tau}{3K_d}}\right)^2
=\frac{4\diff^2\tau}{3K_d}\epsilon  + \text{0}(\epsilon^2)\,,
\end{align}
\begin{figure}
\epsfxsize=60mm \centerline{\epsffile{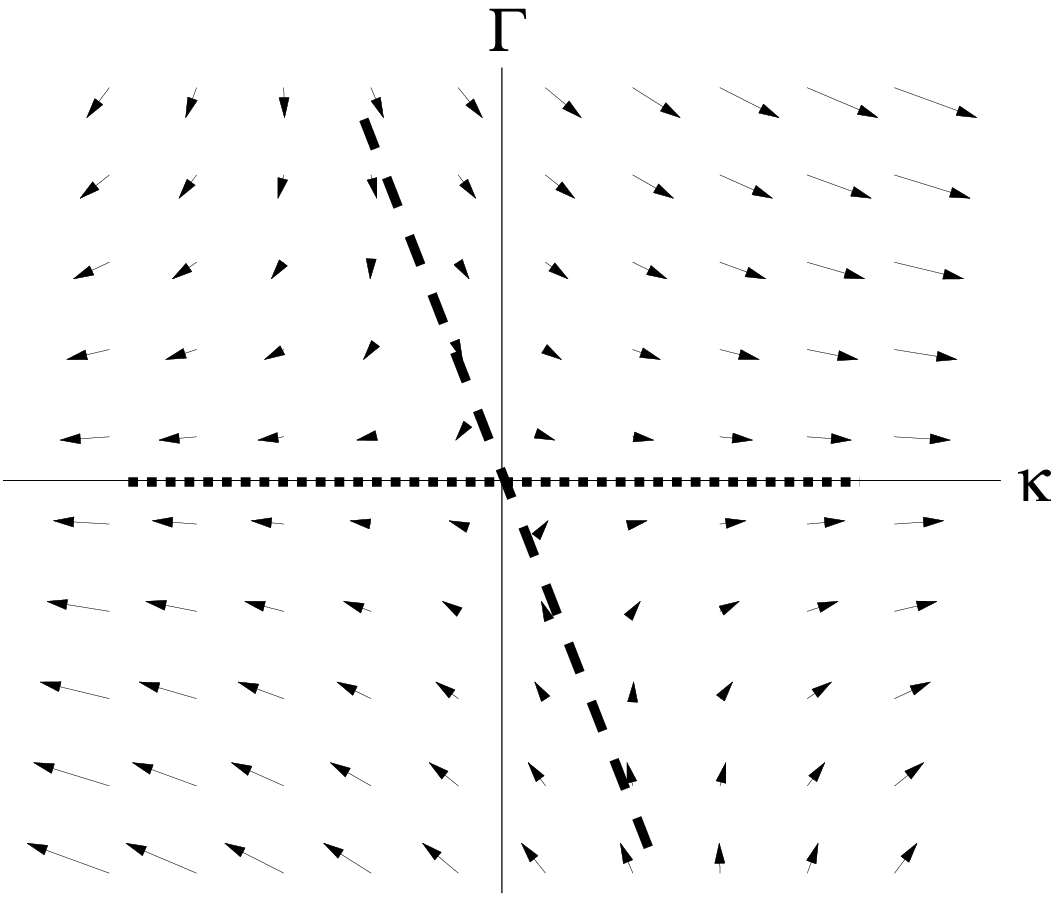}} \vspace{2mm}
\smallcaption{ \label{FIGFLOW} Linearized RG flow near the fixed
point of directed percolation. To reach the fixed point, the
system has to be on the dashed line, i.e., there is one parameter
in the model which has to be tuned to criticality. }
\end{figure}
where we assumed that $\cutoff^d \simeq \cutoff^4$. Close to the
fixed point, the RG flow in Eq.~(\ref{RGFlowTwo}) can be linearized.
As shown in Fig.~\ref{FIGFLOW}, the flow is attractive along
the dashed line and repulsive elsewhere.
To first order in $\epsilon$ the corresponding Jacobian
is triangular. Hence its eigenvalues are given by the
diagonal elements
\begin{equation}
\begin{split}
\left. \partial_\crit\crit  (2+S_1-S_2)
\right|_{\crit=\crit^*,\namp=\namp^*} &=
2-\epsilon/4  + \text{0}(\epsilon^2) \ , \\
\left. \partial_\namp\namp (\epsilon/2-6S_1)
\right|_{\crit=\crit^*,\namp=\namp^*} &=
-\epsilon  + \text{0}(\epsilon^2) \ .
\end{split}
\end{equation}
The positive eigenvalue corresponds to the repulsive
eigenvector (dotted line in Fig.~\ref{FIGFLOW}).
Since the parameter $\crit$ plays the
role of the reduced percolation probability $p-p_c$,
this eigenvalue is equal to $\nuperp^{-1}$,
rendering the third critical exponent.
Because of $\chi=-\beta/\nuperp$ and $z=\nu_{||}/\nuperp$
we thus obtain the critical exponents
\begin{equation}
\begin{split}
\label{OneLoopResults}
\beta &= 1-\epsilon/6  + \text{0}(\epsilon^2)\,,\\
\nuperp &= 1/2+\epsilon/16  + \text{0}(\epsilon^2) \,,\\
\nu_{||}&=1+\epsilon/12  + \text{0}(\epsilon^2)
\ .
\end{split}
\end{equation}
A two-loop approximation of these exponents (see Eq.~(\ref{TwoLoopResults}))
can be found in Ref.~\cite{BronzanDash74}. Although the two-loop result
is quite accurate in 3+1 dimensions, it cannot compete with numerical
methods in lower dimensions. For example,
in 1+1 dimensions the approximation for the
density exponent $\beta$ differs from the known numerical value
by more than $40\%$. Even one-dimensional fermionic field theories,
which have been introduced recently in Ref.~\cite{BOW00},
turn out to be inaccurate.
Therefore, regarding quantitative results,
field-theoretic methods are only of limited interest. However, in
many cases they are extremely useful to understand essential universal
properties of the system. For example, various scaling relations can
only be proven by means of field-theoretic considerations. In fact,
the field-theoretic renormalization group is one of the most powerful
tools of nonequilibrium statistical mechanics.

%==============================================================================
\subsection{Surface critical behavior}
%==============================================================================

\label{BCSEC}

\begin{figure}
\epsfxsize=110mm
\centerline{\epsffile{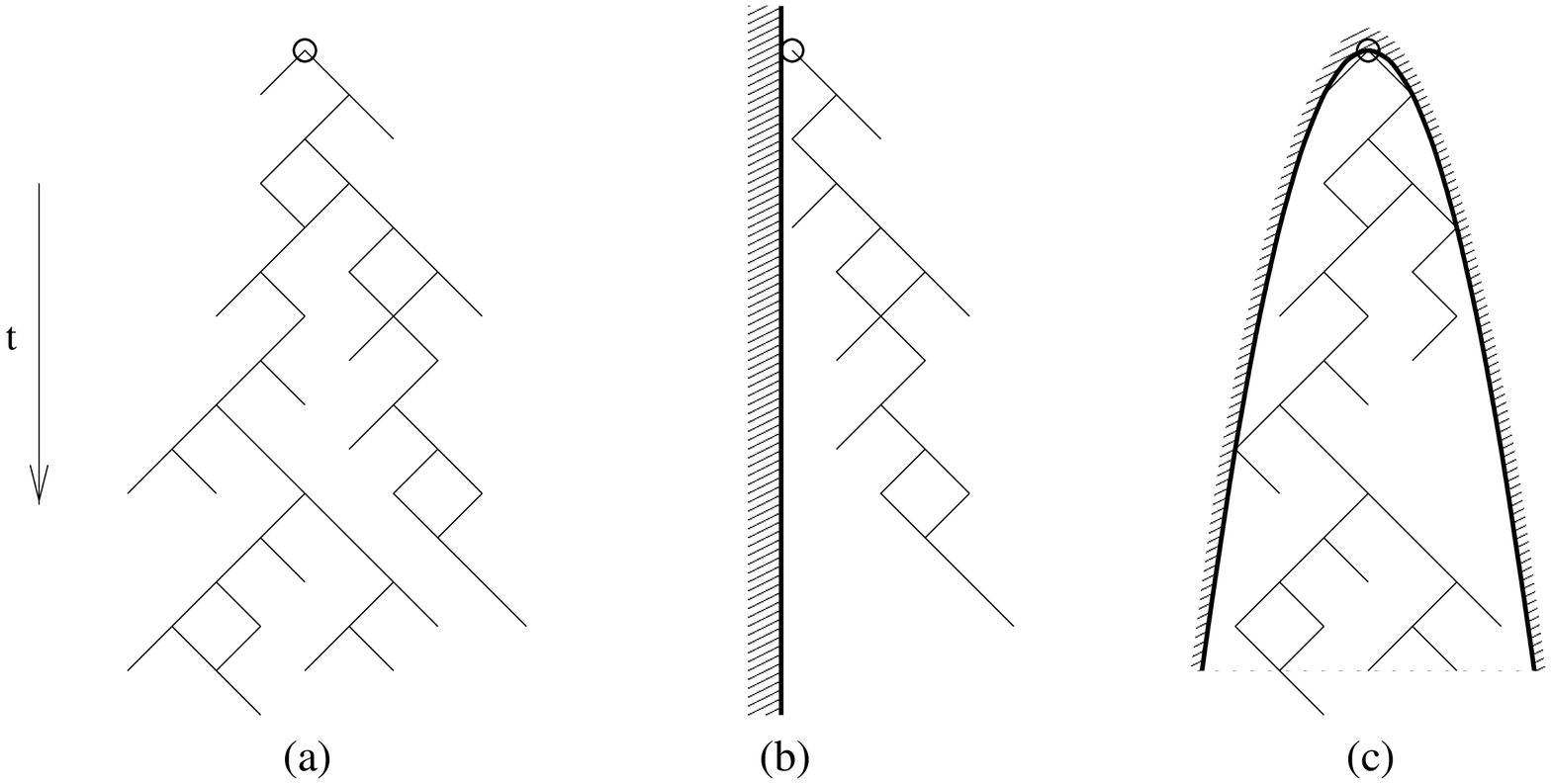}}
\smallcaption{
\label{FIGWALL}
(a) Schematic drawing of a DP cluster.
(b) The same cluster with an absorbing boundary.
(c) The same cluster in a parabola-shaped geometry.
}
\end{figure}
\label{SURFSEC}
As in equilibrium statistical mechanics, nonequilibrium
critical phenomena depend crucially on the boundary conditions
of the system. Because of long-range correlations, the choice
of the boundary conditions may affect the physical properties
of the entire system.

The critical behavior at surfaces of {\em equilibrium} models
has been studied extensively (for a review
see Igl\'oi {\it et~al.}~\cite{IPT93}).
As suggested by Cardy~\cite{Cardy83b}, surface critical phenomena
may be described by introducing an additional
{\em surface exponent} for the order parameter field which
is generally independent of the other bulk exponents.
A similar picture emerges in nonequilibrium
statistical physics. However, since 
in this case there is no symmetry between
space and time, we have to distinguish between spatial,
temporal and mixed surfaces. The simplest
example of a {\em spatial surface} is a semi-infinite system with a wall.
Close to the wall the scaling behavior of the order parameter
is characterized by a surface critical exponent
$\beta_s$ whose value depends on the type of boundary condition.
The most important example of a {\em temporal surface} is
the initial state of a nonequilibrium system. As shown below,
correlations in the initial state may in fact
change the entire evolution of a stochastic process.
Finally, we will consider systems with {\em mixed} boundary
conditions such as DP in a parabola-shaped 
space-time geometry. Mixed boundary conditions
may be viewed as moving boundaries, i.e.,  the system size
varies with time.

%--------------------------------------------------------
\headline{DP with an absorbing wall}
%--------------------------------------------------------
%
In DP an absorbing wall may be introduced by cutting all bonds
crossing a given ($d$-1)-dimensional hyperplane in space
(see Fig.~\ref{FIGWALL}). Hence for $p>p_c$ the stationary density
of active sites close to the wall $\rho^{stat}_s$ is expected to
be smaller than the density in the bulk. In fact, the density
at the wall is found to scale as
\begin{equation}
\rho_s^{stat} \sim (p-p_c)^{\beta_s}\,
\end{equation}
with a surface critical exponent $\beta_s > \beta$. The
problem of an absorbing wall was first studied in the simpler
case of CDP where a surface exponent $\beta_s^{CDP}=2$ was
found~\cite{EssamTanlaKishani94,EssamGuttmann95}.
In a series of papers this scaling theory
was later applied to DP with an absorbing
wall~\cite{JSS88,EGJT96,LSMJ97,FHL98} (for a review see~\cite{HFL00}).
By means of series expansions and numerical
simulations it was observed that the mean survival time $T$
of a cluster in the inactive phase next to the wall scales as
$T \sim \deviation^{-\tau_s}$, where $\deviation$ denotes the distance
from criticality. In 1+1 dimensions the exponent $\tau_s$ was
estimated by $1.0002(3)$, leading to the remarkable conjecture
$\beta_s = \nupar-1$~\cite{EGJT96}. However, very
recent series expansions favor the value
$1.00014(2) \neq 1$, disproving the conjecture~\cite{Jensen99b}.
In fact, in view of dimensional analysis it seems to be unlikely
that $\beta_s$ and $\nupar$ are related by a simple
linear scaling relation. Moreover, in 2+1 dimensions
the numerical value $\tau_s=0.26(2)$ cannot be simply
related to the other exponents. Similarly, the field-theoretic
one-loop result~\cite{FHL98}
\begin{equation}
\tau_s  = -1/2+11\epsilon/48  + \text{0}(\epsilon^2)\,, \qquad
\beta_s = 3/2-7\epsilon/48  + \text{0}(\epsilon^2)
\end{equation}
indicates that the surface exponent is generally independent
of the other exponents~(\ref{OneLoopResults}).

The field-theoretic analysis was also extended to DP
with an absorbing edge~\cite{FHL98}. A closely related application is
the study of spreading processes in narrow channels~\cite{Albano97}.
It is also interesting to study DP with an active wall.
This case is related to the problem of local persistence
and will be discussed below.

%--------------------------------------------------------
\headline{DP clusters in a parabola}
%--------------------------------------------------------
%
Several years before the problem of an absorbing wall was
investigated, Kaiser and Turban considered
the much more complicated problem of DP in a
parabola-shaped geometry~\cite{KaiserTurban94,KaiserTurban95}.
Assuming an absorbing boundary of the form $x=\pm c t^\sigma$
they proposed a general scaling theory. It is based
on the observation that the width $c$ of the parabola $c$
scales as $c \rightarrow \scalefac^{z \sigma -1} c$ under
rescaling~(\ref{Rescaling}). Therefore, the
boundary is relevant for $\sigma>1/z$ and irrelevant otherwise.
To implement this scaling theory,
the scaling forms (\ref{FSDensityScaling})-(\ref{FSScalingSurvival})
have to be extended by an invariant argument of the form
$t^{\sigma-d/z}/c$. The survival probability
of a cluster (\ref{FSScalingSurvival}), for example, has to be
generalized by
\begin{equation}
P(t) \sim t^{-\delta} \, g(\deviation \, t^{1/\nupar}, \, t^{d/z}/N
 , \, t^{\sigma-d/z}/c) \ .
\end{equation}
This scaling form is supported by numerical results and a mean field
approximation~\cite{KaiserTurban95}. The authors also derived a conjecture
for the fractal dimensions
\begin{equation}
d_\parallel(\sigma)=1-z\sigma(d_\parallel-1) \ , \qquad
d_\perp(\sigma)=d_\parallel(\sigma)/\sigma \ ,
\end{equation}
where $d_\parallel=(\beta+\gamma)/\nupar$.

\newpage

%---------------------------------------------------------------------------
\headline{Early-time behavior and critical initial slip}
%---------------------------------------------------------------------------
%
In Sec.~(\ref{ABSEXP}) we reviewed two Monte Carlo techniques for
systems with phase transitions into absorbing states
which differ in their initial state. In simulations starting
with a fully occupied lattice the particle density at criticality
{\em decreases} as $\rho(t)\sim t^{-\beta/\nupar}$.
On the other hand, in dynamic simulations starting from
a single particle (active seed), we observe an {\em increase}
of the average number of particles as $N(t)\sim t^\theta$.
In general the exponent~$\theta$ is independent from
the bulk exponents $\beta,\nuperp,\nupar$.
In the case of DP, however, the duality symmetry under time
reversal (see Eq.~(\ref{DualitySymmetry}))
implies the additional hyperscaling relation
\begin{equation}
\label{HyperscalingRelation2}
\theta = (d\nuperp-2\beta)/\nupar\,.
\end{equation}
\begin{figure}
\epsfxsize=130mm \centerline{\epsffile{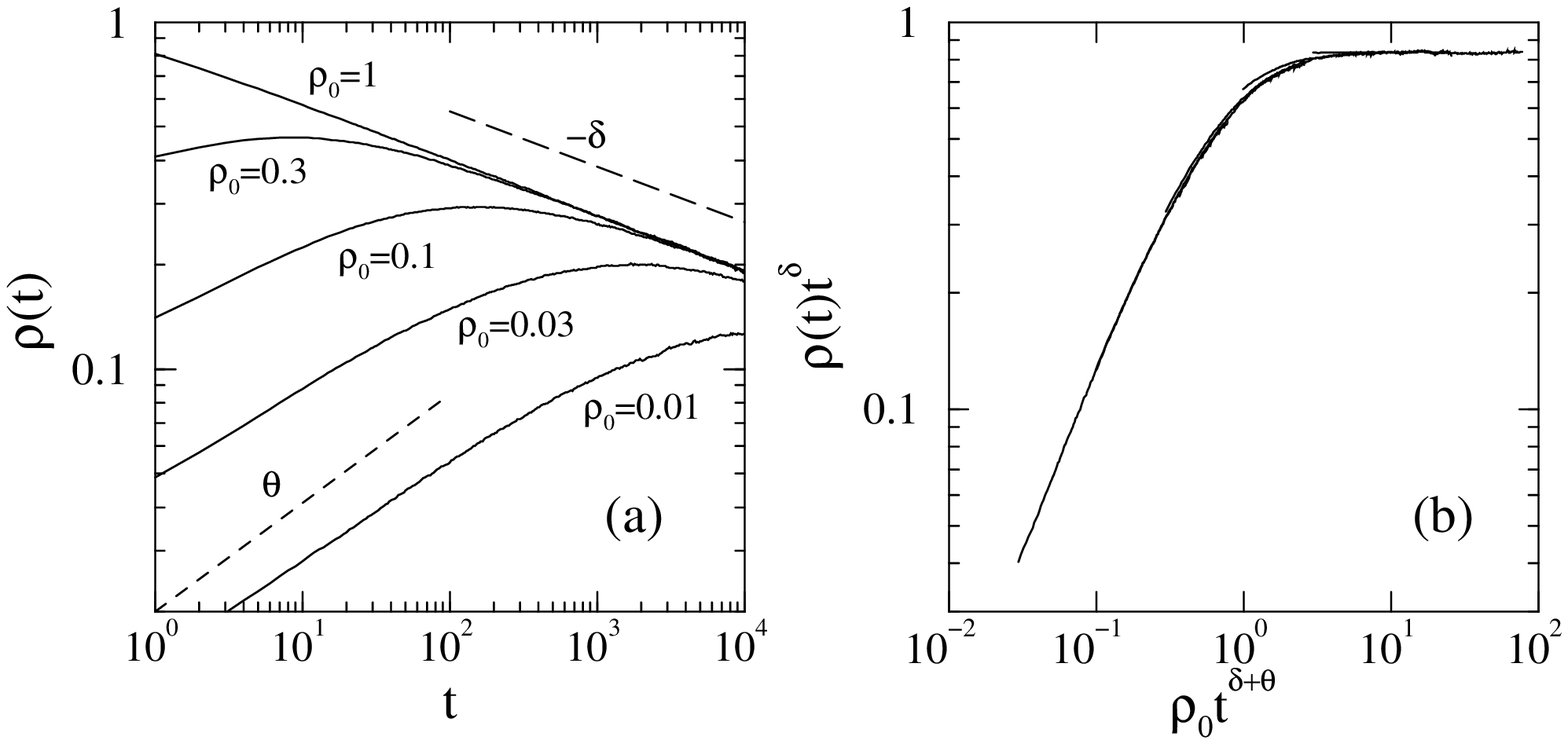}} \smallcaption{
\label{FIGCRSLIP} ``Critical initial slip'' of the particle
density measured in a (1+1)-dimensional directed bond percolation
process at criticality on a lattice with $10^4$ sites. (a):
Particle density $\rho(t)$ for various initial densities $\rho_0$
as a function of time. The dashed lines indicate the slopes
$+\theta$ and $-\delta$. (b): Data collapse of the same data
according to Eq.~(\ref{CrslipScaling}). }
\end{figure}
\noindent
An interesting crossover phenomenon between initial increase and
asymptotic decay of the number of particles emerges when a
critical spreading process starts with a low-density distribution of
active sites. Fig.~\ref{FIGCRSLIP}a shows the temporal behavior of
the density of active sites $\rho(t)$
for various initial densities $\rho_0$.
The density first increases as $\rho(t)\sim t^\theta$
until it reaches a maximum value at time $t_c$ when it
crosses over to the usual
asymptotic decay $\rho(t) \sim t^{-\beta/\nupar}$.
This phenomenon is sometimes referred to as the {\em critical initial slip}
of nonequilibrium systems. As can be seen in Fig.~\ref{FIGCRSLIP}a,
the curves converge to a single one
after sufficiently long time when the memory
of the initial condition is lost. The crossover time $t_c$ depends
on the initial density $\rho_0$ and scales as
\begin{equation}
\label{Crossover}
t_c \sim \rho_0^{-1/(\beta/\nupar+\theta)} \ .
\end{equation}
In finite-size systems near criticality the critical initial
slip may be described by adding the scale-invariant argument
$\rho_0t^{\beta/\nupar+\theta}$ to the
scaling form~(\ref{FSDensityScaling}), i.e.
\begin{equation}
\label{CrslipScaling}
\rho(t) \sim t^{-\beta/\nupar} \,\,
f\Bigl( \deviation \, t^{1/\nupar}, \, t^{d/z}/N,
\, \rho_0 t^{\beta/\nupar+\theta}\Bigr) \,.
\end{equation}
The scaling function $f$ behaves asymptotically as
$f(0,0,u)\sim u$ for $u \rightarrow 0$ and
$f(0,0,u)=const$ for $u \rightarrow \infty$. To verify this
scaling form at criticality, we have plotted
$\rho(t) t^\delta$ versus
$\rho_0 t^{\delta+\theta}$ in Fig.~\ref{FIGCRSLIP}b.
As can be seen, we obtain a convincing data collapse.

The critical initial slip in a DP process can be interpreted
as follows. In a low-density initial state
the active sites are separated by empty intervals
of a certain average size $\xi_0$. As time evolves, they
generate individual clusters of connected sites (see Sec.~\ref{DPINTROSEC}).
Initially these clusters are spatially separated; they do not
interact and the particle number therefore increases as $t^\theta$.
Only a fraction $t^{-\delta}$ of these clusters survives, each of them
spanning a volume of $\xi_\perp^d$. These surviving clusters
start touching each other when  $\xi_\perp^d \sim
\rho_0^{-1} t_c^\delta$. Therefore, we expect the crossover to
take place at $t_c \sim \rho_0^{1/(\delta-d/z)}$. Insertion of
the DP hyperscaling relation~(\ref{HyperscalingRelation2}) leads
to Eq.~(\ref{Crossover}). It is worth being mentioned that
dynamic simulations starting from a single particle represent the
limit $\rho_0 \rightarrow 0$. In this case $t_c$ diverges and the
critical initial slip extends to the entire temporal evolution
of the system.

%
%
%---------------------------------------------------------------------------
\headline{Correlated initial conditions}
%---------------------------------------------------------------------------
%
The previously discussed early-time behavior shows
that initial states with short-range correlations
may affect the temporal evolution of a DP process for a limited
time until the system crosses over to the usual decay of the
particle density. Let us now turn to initial states with {\em long-range}
correlations of the form
\begin{equation}
\langle s_i s_{i+r} \rangle \sim r^{\sigma-d} \ ,
\end{equation}
where $0 \leq \sigma \leq d$ controls the
power-law decay of the correlations
on large scales. In one dimension such states can be generated by
creating uncorrelated empty intervals\footnote{The generation of fractal
distribution is crucial since for $\sigma<d$ the particle density is
zero. Therefore, appropriate cutoffs have to be introduced, as
described in detail in Ref.~\cite{HinrichsenOdor98a}.}
of length~$\ell$ which are algebraically
distributed as $P(\ell) \sim \ell^{-1-\sigma}$.
There are, however, many possibilities to create such states
because higher order correlations can be chosen freely.
Apart from cutoffs, the resulting particle
configurations do not exhibit a specific length scale $\xi_0$,
instead they are characterized by a {\em fractal dimension}
$d_f=\sigma$. It turns out that long-range correlations may
change the {\em entire} temporal evolution of a DP process (similar
phenomena can be observed in other nonequilibrium critical systems
such as in the annihilation model~\cite{AlemanyZanette95}).

\begin{figure}
\epsfxsize=80mm
\centerline{\epsffile{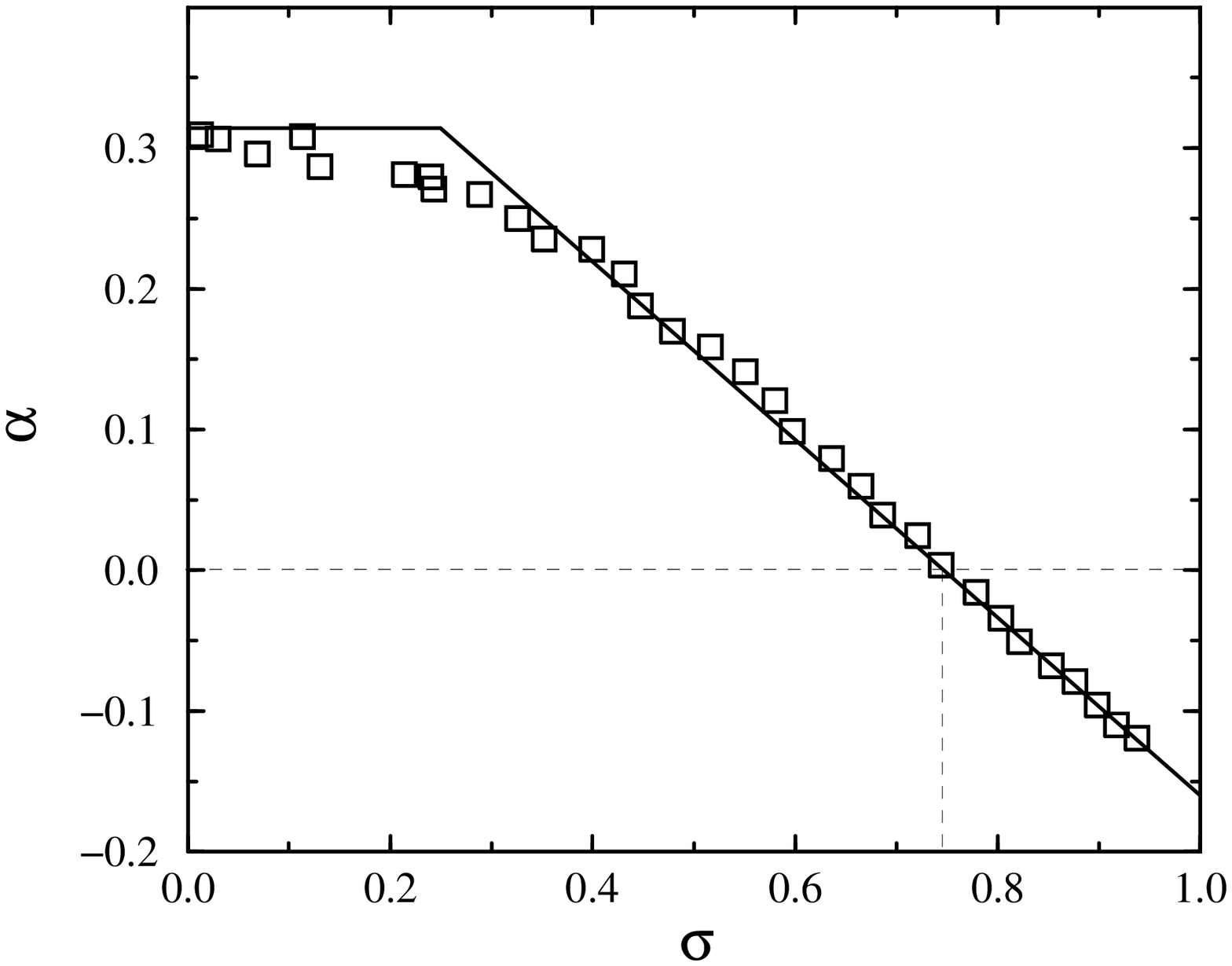}}
\smallcaption{
\label{FIGFRACINIT}
Correlated initial conditions in a DP process.
Numerical estimates for the exponents $\alpha(\sigma)$
in one spatial dimension.
The theoretical prediction (\ref{Prediction})
is shown by the solid line.
The dashed line indicates the `natural' correlations of DP.
}
\end{figure}

For $\sigma=d$ the particles are homogeneously distributed, leading to
the usual long-time behavior $\rho(t) \sim t^{-\beta/\nupar}$.
For $\sigma \rightarrow 0$ the fractal dimension tends to zero,
corresponding to isolated particles where we expect the
density to increase a $\rho(t) \sim t^\theta$. In between
numerical simulations suggest that the decay exponent changes
continuously by
\begin{equation}
\label{Prediction}
\rho(t) \sim \rho_0 t^{\alpha(\sigma)} \ , \qquad
\alpha(\sigma) =
\begin{cases} \theta & \text{ for $\sigma \leq \sigma_c$} \\
\frac{1}{z}(d-\sigma-\beta/\nuperp) & \text{ for $\sigma>\sigma_c$}
\end{cases}
\ ,
\end{equation}
where $\sigma_c=\beta/\nuperp$ plays a role of a critical threshold
above which correlations in the initial state become relevant
(see Fig.~\ref{FIGFRACINIT}).
Below $\sigma_c$ the initial distribution of
particles is so sparse that interactions between
growing clusters turn out to be irrelevant.

The numerical result can be proven by a simple
field-theoretic calculation~\cite{HinrichsenOdor98a}.
In order to take the initial state into account,
the field-theoretic action~(\ref{DPFreePart})-(\ref{DPInteraction}) 
has to be extended
by the term
\begin{equation}
S_{ic} = \coupling \, \int d^dx \, \tilde{\phi}(\xvec,0) \, \phi_0(\xvec) \ .
\end{equation}
Here the initial particle distribution is represented by a field
$\phi_0(\xvec)$ that is coupled to the `creation operator'
$\tilde{\phi}$ at time $t=0$ via a coupling constant $\coupling$.
The scaling behavior of $\phi_0(\xvec)$ depends on the fractal
dimension. Obviously, homogeneous initial conditions
$\phi_0(\xvec)=const$ are invariant under rescaling, whereas a
fully localized initial condition $\phi_0(\xvec)=\delta^d(\xvec)$
has the scaling dimension $-d$. Therefore, the initial density
should scale as
\begin{equation}
\label{InitialCondScaling}
\phi_0(\xvec) \rightarrow \scalefac^{d_f-d} \phi_0(\xvec)
=\scalefac^{\sigma-d} \phi_0(\xvec) \ .
\end{equation}
It is easy to verify that the contribution $S_{ic}$ does not lead
to additional loop corrections in the field theory.
Therefore, the coupling between the system
and the initial condition will not be renormalized. Moreover,
it can be shown that higher-order contributions
of the form $ \tilde{\phi}^k(\xvec,0) \, \phi_0(\xvec)$ with $k>1$
are irrelevant under renormalization~\cite{WOH98,HinrichsenOdor98a}.
Consequently, the coupling constant $\mu$ scales as
\begin{equation}
\coupling \rightarrow \scalefac ^ {\sigma+\chi} \coupling \ ,
\end{equation}
where $\chi=-\beta/\nuperp$. Hence the correlations in the initial
state are relevant if $\sigma>\sigma_c=\beta/\nuperp$.
Scaling invariance of the expression
$\rho(t) \sim \rho_0 t^{\alpha(\sigma)}$  implies
that $\alpha z=d-\sigma+\chi$, completing the proof
of Eq.~(\ref{Prediction}).

Interestingly, the above calculation does not depend
on the specific form of correlations in the initial state
but only on the scaling dimension of the distribution.
That is, no matter how the particles are distributed --
as long as the distribution scales as in
Eq.~(\ref{InitialCondScaling}), the particle density decays
according to the scaling form~(\ref{Prediction}).
Moreover, it is interesting to note that a critical DP
process itself generates two-point correlations
$\langle s_is_{i+r}\rangle\sim r^{-\beta/\nuperp}$, corresponding
to the `natural' fractal dimension $d_{f,\perp}=d-\beta/\nuperp$.
Therefore, choosing `natural' correlations
$\sigma=d-\beta/\nuperp$ the number of
particles remains almost constant (see dashed lines in
Fig.~\ref{FIGFRACINIT}). Similar phenomena have been observed in the
Glauber-Ising model with correlated initial conditions~\cite{BHN91}.

\newpage

%---------------------------------------------------------------------------
\headline{Persistence probability in a DP process}
%---------------------------------------------------------------------------
%
In the past few years it has been realized that certain first passage
quantities of critical nonequilibrium processes
exhibit a power law decay with non-trivial exponents.
One of these quantities is the local persistence probability $P_l(t)$,
defined as the probability that a local variable $s_i(t)$ at a given
site $i$ does not change its state until
time~$t$ during the temporal evolution.
In various systems it was found that
\begin{equation}
P_l(t) \sim t^{-\theta_l} \ ,
\end{equation}
where $\theta_l$ is the so-called {\em local persistence
exponent}~\cite{DBG94,BDG94,Cardy95,DHP96,MSBC96,DHZ96,MenyhardOdor97}.
A similar quantity, the global persistence probability
$P_g(t)$, which is defined as
the probability that the global order parameter
does not change its sign up to time $t$, is also found to decay as a
power law with a {\em global persistence exponent}
$\theta_g$~\cite{MBCS96,SchuelkeZheng97,LeeRutenberg97}. In
general the exponents $\theta_l$ and $\theta_g$ are different and
independent of the usual scaling exponents.
Since the persistence probabilities depend on the history
of  evolution as a whole\footnote{Persistence probabilities may be regarded
as autocorrelation functions involing infinitely many points.}, 
it is generally hard to determine
these exponents analytically. In fact, only a few exact
results have been obtained so far~\cite{BDG94,DHP96,LeeRutenberg97}.
Persistence exponents are known to exhibit certain universal properties.
For example, the local persistence exponent of the two-dimensional Glauber
model in the ordered phase $T<T_c$ does not depend 
on~$T$~\cite{Derrida97,Stauffer97,HinrichsenAntoni98},
whereas it is non-universal with respect to the 
initial magnetization~\cite{DHP96}. Most researchers believe that
persistence exponents are to some extent
`less universal' than ordinary bulk exponents.

In a DP process the local persistence probability
$P_l(t)$ may be defined as the probability that
a particular site never becomes active up to time $t$. Numerical
simulations suggest that the local persistence exponent
is given by
\begin{equation}
\theta_l=1.50(1)
\end{equation}
independent of the initial density of active
sites~\cite{HinrichsenKoduvely98} . Moreover,
the numerical data are in agreement with the scaling form
\begin{equation}
P_l(t,N,\deviation) \sim t^{-\theta_l} \,
f(\deviation t^{1/\nupar}, N^{-z/d} t)\,,
\end{equation}
where $\deviation=p-p_c$ denotes the distance from criticality.
The local persistence probability can also be related to certain return
probabilities in a DP process with an absorbing boundary or an
active source~\cite{HinrichsenKoduvely98}.
Similar measurements of the global persistence probability $P_g(t)$
suggest that $\theta_g>\theta_l$ in agreement with recent
field-theoretic results~\cite{OerdingWijland98}.

%==============================================================================
\subsection{The influence of quenched disorder}
%==============================================================================
%
\label{QUENCHSEC}
In DP models it is usually assumed that the percolation
probability does not vary in space and time. However,
in realistic spreading processes the rate for offspring production is
not homogeneous, rather it fluctuates in space and time.
For example, the local density of open channels in a porous rock
will vary because of inhomogeneities of the material.
Similarly, most spreading processes take place in inhomogeneous
environments.
It is therefore important to investigate how quenched disorder
affects the critical properties of a spreading process.
It turns out that even weak disorder may affect or even
destroy the critical behavior of DP.

In the DP Langevin equation~(\ref{DPLangevinEquation})
the parameter $\crit$ plays the role of the percolation probability.
Quenched disorder may be introduced by random variations
of $\crit$, i.e., by adding another Gaussian noise field $\disorder$:
\begin{equation}
\label{Replacement}
\crit \rightarrow \crit + \disorder(\xvec,t) \,.
\end{equation}
Thus, the resulting Langevin equation reads
\begin{equation}
\label{LangevinEquation}
\timederivative\rho(\xvec,t) \;=\;
\crit \rho(\xvec,t) -
\scalefac \rho^2(\xvec,t)  +
\diff \nabla^2 \rho(\xvec,t) +
\noise(\xvec,t)
+ \rho(\xvec,t)\disorder(\xvec,t) \,.
\end{equation}
The noise $\disorder$ is {\em quenched} in the sense
that quantities like the particle density
are averaged over many independent realizations
of the intrinsic noise $\noise$ while
the disorder field $\disorder$ is kept fixed.
In the following we distinguish three different types of
quenched disorder, namely spatially, temporally
quenched, and spatio-temporally quenched disorder.
The three variants of quenched disorder differ in how far they
affect the critical behavior of DP.

%---------------------------------------------------------------------------
\headline{Spatially quenched disorder}
%---------------------------------------------------------------------------
%
For spatially quenched disorder, the disorder field $\disorder$ is defined
through the correlations
\begin{equation}
\overline{\disorder(\xvec)\disorder(\xvec')}
=\gamma\,\delta^d(\xvec-\xvec') \,,
\end{equation}
where the bar denotes the average over many independent realizations
of the disorder field (in contrast to the ensemble
average $\langle\ldots\rangle$
over realizations of the intrinsic noise $\noise$).
The parameter $\gamma$ is an amplitude controlling the intensity
of disorder. In order to find out whether this type of noise affects the
critical behavior of DP, let us again consider the properties of
the Langevin equation under rescaling. At the critical dimension
$d_c=4$ the additional term $\rho\disorder$ scales as
\begin{equation}
\rho\disorder \rightarrow \scalefac^{-d_c/2-\chi} \rho\disorder\,,
\end{equation}
i.e., spatially quenched disorder is a {\em marginal}
perturbation. Therefore, it may seriously affect the
critical behavior at the transition.
The same result is obtained by considering the
field-theoretic action. Without quench\-ed noise, DP
is described by the action of Reggeon field theory
(see Sec.~\ref{FTHSEC})
\begin{equation}
\label{DPAction}
S = \int d^dx \, dt \, \tilde{\phi}
\Bigl[\timederivative -\crit - \diff\nabla^2
+ \frac{\namp}{2}(\phi-\tilde{\phi})\Bigr] \phi
\end{equation}
where $\phi(\xvec,t)$ represents the local particle density while
$\tilde{\phi}(\xvec,t)$ denotes the Martin-Siggia-Rosen response field.
As shown by Janssen~\cite{Janssen97b}, spatially quenched noise
can be taken into account by adding the term
\begin{equation}
S\rightarrow S + \gamma \int d^dx
\left[ \int dt\,\tilde{\phi}\phi \right]^2 \,.
% dimension 0
\end{equation}
By simple power counting one can prove that this additional term is
indeed a marginal perturbation. Janssen showed by a field-theoretic
analysis that the stable fixed point is shifted to an unphysical
region, leading to runaway solutions of the flow equations
in the physical region of interest. Therefore, spatially quenched
disorder is expected to crucially disturb the critical behavior
of DP. The findings are in agreement with earlier numerical results
by Moreira and Dickman~\cite{MoreiraDickman96} who reported
non-universal logarithmic behavior instead of power laws.
Later Cafiero {\it et~al.}~\cite{CGM98} showed that DP with spatially
quenched randomness can be mapped onto a non-Markovian spreading
process with memory, in agreement with previous results.

\begin{figure}
\epsfxsize=120mm \centerline{\epsffile{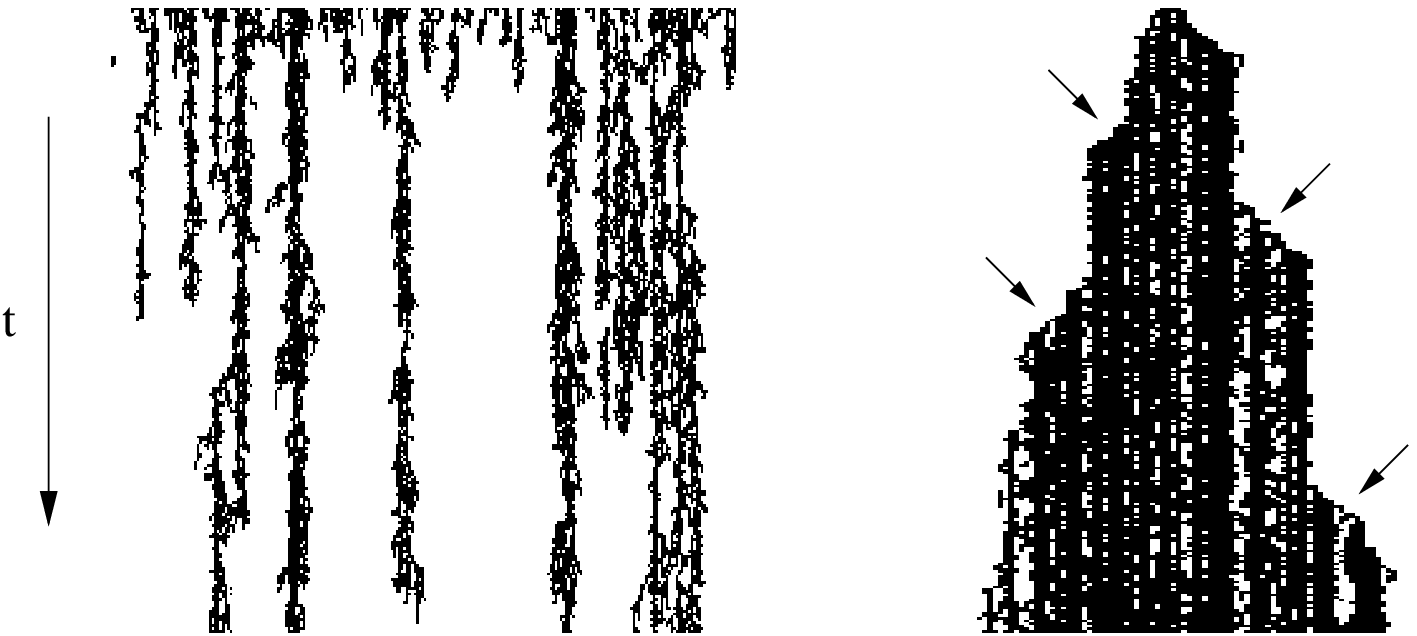}}
\smallcaption{ \label{FIGQUENCHED} (1+1)-dimensional DP with
spatially quenched disorder. Left: In the glassy phase quenched
disorder forces active sites to percolate in narrow `channels'
where the local percolation probability is high. Right:
Supercritical disordered DP process starting from a single seed,
leading to avalanches (marked by the arrows) where the spreading
agent overcomes a local barrier. }
\end{figure}
From a more physical point of view, spatially quenched disorder in
1+1 dimensional DP was studied by Webman {\it
et~al.}~\cite{WACH98}. It turns out that even weak disorder
drastically modifies the phase diagram. Instead of a single
critical point one obtains an intermediate phase of very slow
glassy-like dynamics. The glassy phase is characterized by
non-universal exponents which depend on the percolation
probability and the disorder amplitude. For example, in a
supercritical 1+1 dimensional DP process without quenched disorder
the boundaries of a cluster propagate at constant average velocity
$v$. However, in the glassy phase $v$ decays {\em algebraically}
with time. The corresponding exponent turns out to vary
continuously with the mean percolation probability. The power-law
behavior is due to `blockades' at certain sites where the local
percolation probability is small (see Fig.~\ref{FIGQUENCHED}).
Similarly, in the subcritical edge of the glassy phase, the
spreading agent becomes localized at sites with high percolation
probability. For $d>1$, however, numerical simulations indicate
that a glassy phase does not exist.

%---------------------------------------------------------------------------
\headline{Temporally quenched disorder}
%---------------------------------------------------------------------------
%
Temporally quenched disorder is defined by the correlations
\begin{equation}
\overline{\disorder(t)\disorder(t')}=\gamma\,\delta(t-t') \,.
\end{equation}
In this case the additional term is a {\em relevant} perturbation
which scales as
$\rho\disorder \rightarrow \scalefac^{-z/2-\chi} \rho\disorder$.
Therefore, we expect the critical behavior and the associated
critical exponents to change entirely. In the field-theoretic
formulation this corresponds to adding a term of the form
\begin{equation}
S\rightarrow S + \gamma \int dt \left[ \int d^dx\,\tilde{\phi}\phi \right]^2 \,.
\end{equation}
The influence of temporally quenched disorder was investigated
in detail in Ref.~\cite{Jensen96a}. Employing series expansion
techniques it was demonstrated that the three exponents
$\beta, \nuperp, \nupar$ vary continuously with the
disorder strength. Thus the transition no longer belongs to
the DP universality class.

%---------------------------------------------------------------------------
\headline{Spatio-temporally quenched disorder}
%---------------------------------------------------------------------------
%
For spatio-temporally quenched disorder,
the noise field $\disorder$ is uncorrelated in both space and time:
\begin{equation}
\overline{\disorder(\xvec,t)\disorder(\xvec',t')}=
\gamma\,\delta^d(\xvec-\xvec')\delta(t-t')\,.
\end{equation}
In Reggeon field theory, this would correspond to the addition of the term
\begin{equation}
S \rightarrow S +\gamma \int d^dx \, dt \left[\tilde{\phi}\phi \right]^2 \,,
\end{equation}
being an {\em irrelevant} perturbation. In fact, this noise has
essentially the same properties as the intrinsic noise and can
be considered as being {\em annealed}.
Spatio-temporally quenched disorder is expected in systems where
each time step takes place
in a new spatial environment of the system.
Examples include water in porous media subjected
to a gravitational field as well as systems of flowing sand on an
inclined plane (see Sec.~\ref{DPEXPERIM}).
In these cases the critical behavior of DP
should remain valid on large scales.
\vspace{4mm}

%==============================================================================
\subsection{Related models}
%==============================================================================
%
\label{RELSEC}
Directed percolation plays a role in various other contexts such as in
coupled map lattices, the problem of friendly walkers, real-valued
spreading processes, models with particle conservation, and even in systems
with infinitely absorbing states. In the following we discuss some
of these related models. Moreover, we investigate the special case
of compact directed percolation in more detail.

%---------------------------------------------------------------------------
\headline{Spreading transitions in deterministic systems}
%---------------------------------------------------------------------------
%
%
%
\begin{figure}
\epsfxsize=80mm
\centerline{\epsffile{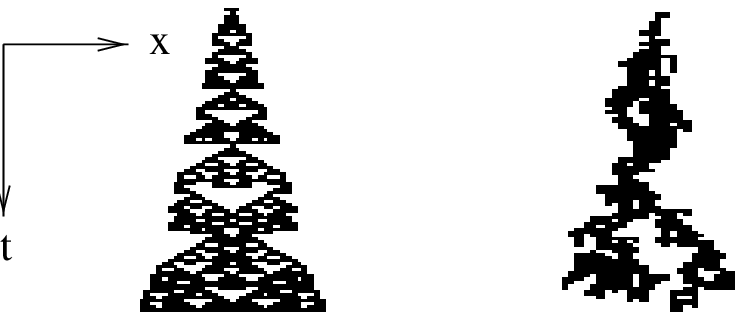}}
\vspace{2mm}
\smallcaption{
\label{FIGMAPS}
Spreading process in a coupled map lattice
for $r=3$ and $\diff=0.57$. Chaotic and
laminar sites are represented by black and
white pixels, respectively. Deterministic
synchronous updates lead to symmetric clusters
(left) with nonuniversal behavior. Asynchronous
updates (right) destroy these correlations,
leading to typical DP clusters.}
\end{figure}
Spreading transitions can also be observed in certain deterministic
lattice models. Instead of using random numbers, these models
employ chaotic maps in order to generate random behavior.
A simple example of such a chaotic map is given by
\begin{equation}
u(t+1)=f\bigl(u(t)\bigr)
\ , \qquad
f(x) = \begin{cases}
rx & \text{if $0\leq x<1/2$} \ , \\
r(1-x) & \text{if $1/2 < x \leq 1$} \ , \\
x & \text{if $1 < x \leq r/2$} \ ,
\end{cases}
\end{equation}
where $r$ is a free parameter. The chaotic motion of $f$ for
$x\leq 1$ is governed by a {\em tent map} of slope~$r$.
However, if $r$ exceeds the value $2$, the map eventually reaches
an absorbing state with $x>1$, the so-called
`laminar' state of the model. In a {\em coupled map lattice}
\cite{ChateManneville88} many of these local maps $u_i(t)$ are coupled
by a diffusive interaction of the form
\begin{equation}
\label{CoupledMaps}
u_i(t+1) = f\bigl(u_i(t)\bigr) \ + \
\frac{\diff}{2}
\biggl[
f\bigl(u_{i-1}(t)\bigr)-2f\bigl(u_{i}(t)\bigr)+f\bigl(u_{i+1}(t)\bigr)
\biggr]
\ ,
\end{equation}
where $\diff$ plays the role of a diffusion constant.
The coupled map lattice evolves deterministically
by synchronous updates. By varying $\diff$ it exhibits
a nonequilibrium phase transition
from a `chaotic' phase into a `laminar' state. The
existence of absorbing states led Pomeau to the
conjecture that the transition should belong to the DP
universality class~\cite{Pomeau86}, hoping that the apparent
randomness of the chaotic maps would effectively lead to stochastic
spreading of activity on large scales. However, subsequent
numerical simulations did not agree with this
conjecture~\cite{HWJ90}, in particular the exponents were found
to depend on~$r$. The non-universal behavior of spreading
transitions in deterministic systems is caused by
subtle correlations emerging as artifacts of the
deterministic update rule. For example, a `cluster' of
chaotic sites starting from a single active seed remains {\em symmetric}
throughout the whole temporal evolution, leading to a
qualitatively different spreading behavior (see Fig.~\ref{FIGMAPS}).
The consequences of these correlations are not yet fully understood.
However, replacing the synchronous dynamics of Eq.~(\ref{CoupledMaps})
by asynchronous updates, the deterministic
correlations are destroyed and the resulting
phase transition is indeed characterized by DP exponents~\cite{RBJ97}.
Similar transitions of two-dimensional coupled map
lattices have been investigated in Ref.~\cite{CLP97}.

\begin{figure}
\epsfxsize=120mm \centerline{\epsffile{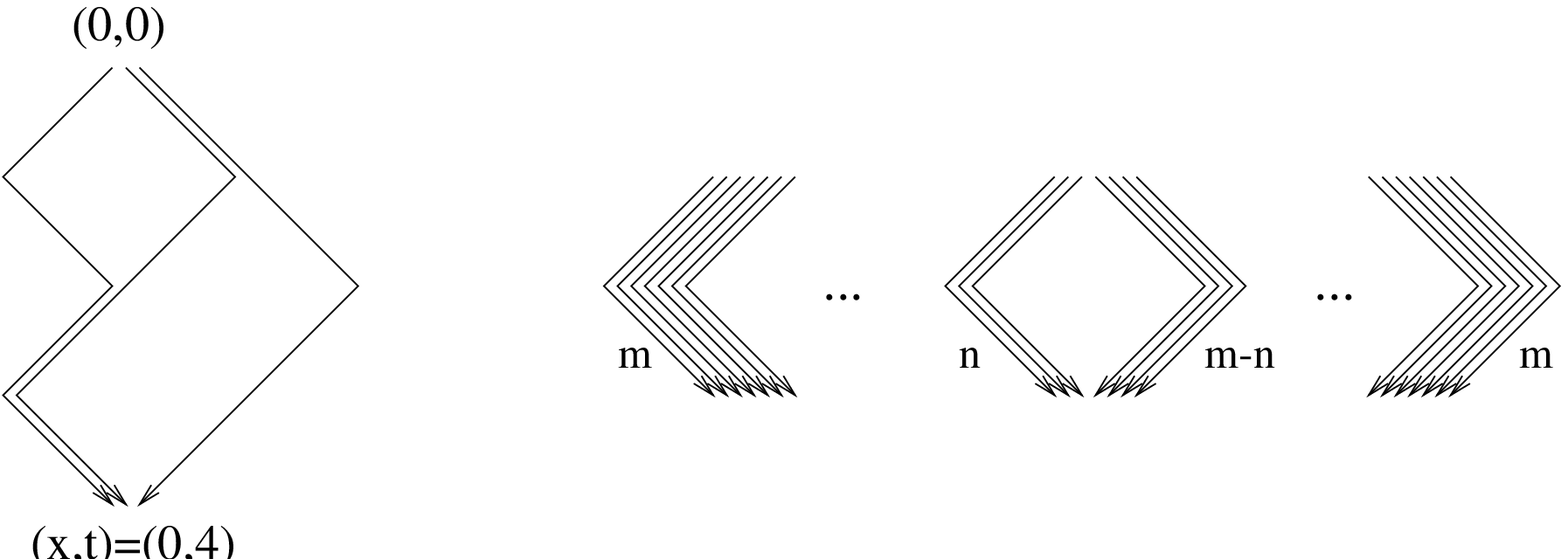}}
\smallcaption{ \label{FIGFRIEND} The problem of friendly walkers
on a diagonal square lattice. Left: A particular realization of
three random walkers traveling from the origin $(0,0)$ to the site
$(0,4)$. Right: All $m+1$ realizations of $m$ random walkers
traveling from $(0,0)$ to $(0,2)$. }
\end{figure}

%---------------------------------------------------------------------------
\headline{DP and the problem of `friendly walkers'}
%---------------------------------------------------------------------------
%
The so-called problem of `friendly walkers' is defined as follows.
Consider the paths of $m$ random walkers on a diagonal square
lattice. All walks originate in site $(0,0)$ and end in
site~$(\xvec,t)$. While traveling the walkers may share the same bonds
but they are not allowed to cross each other (see
Fig.~\ref{FIGFRIEND}). In the partition sum $Z_m(\xvec,t)$ each
possible configuration of random walks is weighted by a factor
$p^k$, where $p>0$ is a parameter and $k$ denotes the number of
bonds used by at least one of the walkers. For $p<1$ it is
therefore advantageous for the walkers to be `friendly' to each
other, i.e., to share the same bonds.

Some time ago, Arrowsmith and Essam~\cite{ArrowsmithEssam90}
suggested a close relationship between DP and the problem
of friendly walkers. More precisely, they showed that the
partition function $Z_m(\xvec,t)$ is related to the
pair-connectedness function $c(\xvec,t)$ of a directed bond
percolation process by~\cite{AME91}
\begin{equation}
c(\xvec,t)=\lim_{m\rightarrow 0} Z_m(\xvec,t) \,,
\end{equation}
where $p$ is the usual percolation probability. Here the limit $m
\rightarrow 0$  has to be performed as a suitable continuation of
polynomial expressions. For example, let us consider $m$ friendly
random walkers traveling from the origin $(0,0)$ to the point
$(\xvec,t)=(0,2)$ (see right part of Fig.~\ref{FIGFRIEND}). There
are $m+1$ possible configurations; two of them use only two bonds
while the others use four bonds. Hence the partition function is
given by $Z_m(0,2)=(m-1)p^4 + 2p^2$. Inserting $m=0$ we obtain
$Z_0(0,2)=2p^2-p^4$. In fact, this expression is exactly equal to
the pair connectedness function $c(0,2)$ in a directed bond
percolation process. This equivalence holds for any $(\xvec,t)$
and also in higher dimensions. Recently this result has been
generalized to friendly walkers with arbitrary
interactions~\cite{CardyColaiori99}.

The problem of friendly walkers may also be interpreted as a flow
of integer numbers on a diagonal square lattice. At the origin
there is a source creating an integer number~$m$. While
traveling on the directed lattice, this integer number may split
up into several parts. Finally there is a sink where all integers
merge into a single one and disappear. Clearly, the integers
represent just the number of friendly walkers sharing the same
bond.

Even more remarkably, it has been shown that DP is related to the
partition sum of a  {\it chiral Potts
model}~\cite{ArrowsmithEssam90,AME91,TsuchiyaKatori98},
generalizing the well-known result of Fortuin and
Kasteleyn for isotropic percolation~\cite{FortuinKasteleyn72}.
However, since the definition of the chiral Potts model is
rather cumbersome, this relation is not
of immediate practical benefit.

%
%---------------------------------------------------------------------------
\headline{DP with real-valued degrees of freedom}
%---------------------------------------------------------------------------
%
DP models are usually defined in terms of discrete local variables
$s_i=0,1$ representing inactive and active sites. An interesting
variant of DP is `self-organized directed percolation', where
real-valued local degrees of freedom are
used~\cite{HansenRoux87,GrassbergerZhang96,MaslovZhang96,DMVZ00}.
To understand the basic mechanism, let us consider directed bond
percolation. Clearly, a given path between two sites is conducting
if all bonds along the path are open, i.e., all random numbers
generated along the path have to be
larger than $p$. Thus, in order to find out whether a
path is conducting, it is only necessary to keep track of the
{\em smallest} random number generated along this path. This number may
be considered as the weight of the path, being a measure of its
weakest link. However, a pair of sites can be connected
by many different paths. For the target site to become active,
at least one of these paths has to be conducting. Therefore, two sites
are connected if the {\em maximum} of all weights is larger than $p$.

\begin{figure}
\epsfxsize=150mm
\centerline{\epsffile{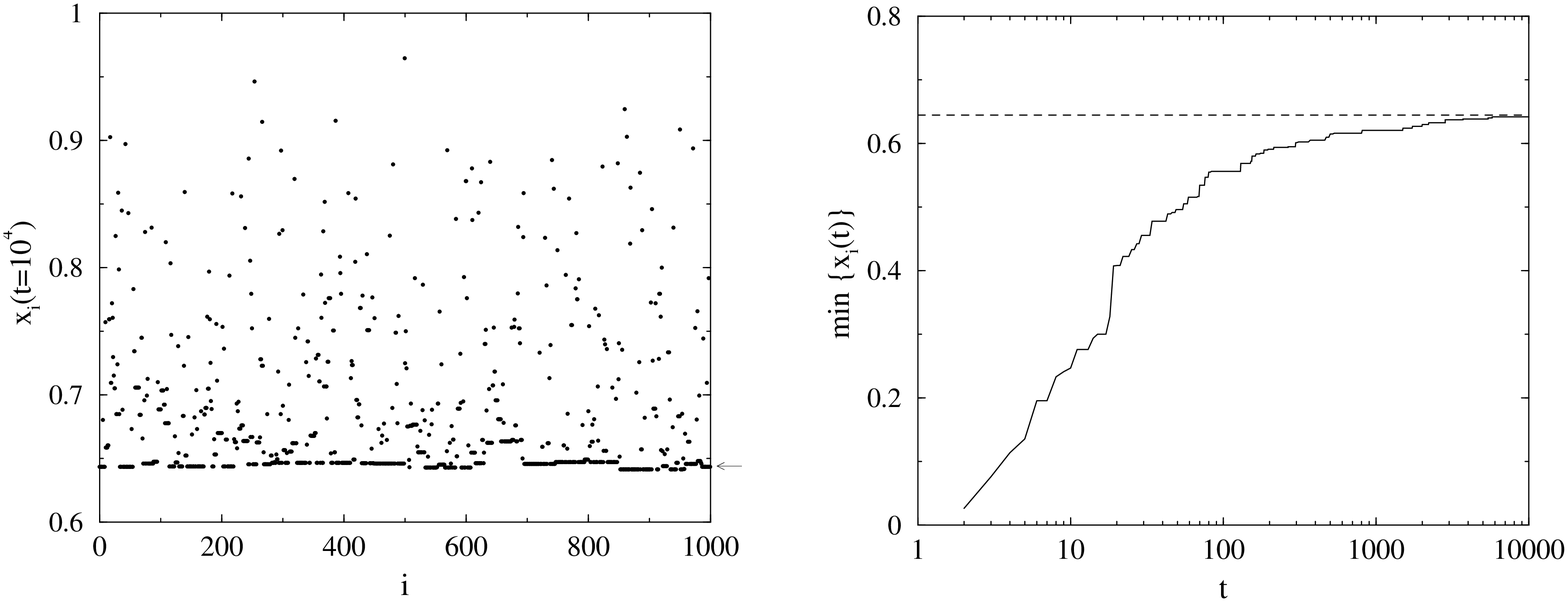}}
\vspace{2mm}
\smallcaption{
\label{FIGREALDP}
Self-organized directed percolation in 1+1 dimensions.
Left: Spatial configuration of the chain after $10^4$ updates. The
arrow indicates the percolation threshold of
directed bond percolation. Right:
Minimum value of all sites as a function of time, approaching
the percolation threshold of bond DP (dashed line).
}
\end{figure}

Interestingly, the maximal weight can be computed by a
local update rule which is defined
in terms of real-valued degrees of freedom $x_i(t) \in [0,1]$.
Starting with the initial condition $x_i(0)=0$
the system evolves according to
\begin{equation}
\label{ContinuousUpdate}
x_i(t+1) = \min\Bigl(\max(z_i^-,x_{i-1}(t)),
                        \max(z_i^+,x_{i-1}(t))\Bigr) \,,
\end{equation}
where we used the notation of Eq.~(\ref{BondPercCA}). A typical spatial
configuration of a (1+1)-dimensional chain after $10^4$ updates is
shown in Fig~\ref{FIGREALDP}. Using this update rule, the binary
state $s_i(t)$ of the corresponding directed bond percolation
process can be retrieved by the projection
\begin{equation}
\label{ContinuousProjection}
s_i(t) = \Theta(p-x_i(t))  \,,
\end{equation}
where $\Theta$ denotes the heaviside step function. Remarkably, the update
rule~(\ref{ContinuousUpdate}) does not involve the percolation
probability~$p$. Instead, it processes all
values of $p$ at once until a particular value of $p$ is
selected by application of the
projection rule~(\ref{ContinuousProjection}).
Thus, `self-organized directed percolation'
can be used as a tool for very efficient
off-critical simulations.
%

%---------------------------------------------------------------------------
\headline{Spreading process with particle conservation}
%---------------------------------------------------------------------------
%
Recently Broeker and
Grassberger~\cite{BroekerGrassberger99} introduced another interesting
`self-organized' variant of DP which is motivated as follows. A
gardener takes care of $N$ plants in a flowerbed. The flowers are
seized with a parasite. Once a plant is struck, it perishes irreversibly.
Moreover, the parasite may spread to neighboring plants. However, if the
number of befallen plants exceeds a certain number, the gardener
replaces one of them, keeping the number of infected plants constant.
In more technical terms, the number of active sites is
{\it conserved} by means of a global update rule. The update consists
of two steps. At first one of the active sites activates a randomly
chosen neighbor, modeling the spreading of the parasite.
If this move was successful, another randomly chosen active site
is deactivated, representing the global control of the gardener.
Clearly, the number of active sites $M$ is conserved, i.e., the
model has no absorbing states.

The number of active sites $M$ is specified by the initial state.
For example, we may start with a compact domain of $M$ active sites
on an infinite lattice. Initially, spreading occurs only at the
edges of the domain. As time proceeds, 
the distribution of active sites becomes more and more
sparse, forming a diffusing cloud. Nevertheless, the cloud
keeps its integrity and reaches a typical size after some time.
Amazingly, the dynamic processes in the
interior of the cloud are those of an almost
critical DP process. In fact, as shown in
Ref.~\cite{BroekerGrassberger99}, most properties
of the cloud can be explained in terms of DP scaling laws.
Considering a small region in the interior of the cloud, the
relation to DP is quite obvious: The two processes
for offspring production $A \rightarrow 2A$ and self-destruction
$A \rightarrow \vacancy$ occur randomly in space, just as in a
contact process with random-sequential updates. However, the global
control adjusts the ratio of the effective rates and drives the
system to criticality.

%---------------------------------------------------------------------------
\headline{Branching Potts interfaces}
%---------------------------------------------------------------------------
%
Recently Cardy~\cite{Cardy98} studied a field-theory for branching
interfaces between ordered domains of a $q$-state Potts model. In
two spatial dimensions these interfaces are one-dimensional
objects. For $q<q_c$ they become fractal with a vanishing
interfacial tension at the critical point, while for $q>q_c$  the
interfacial width diverges at a finite value of the tension,
indicating a first-order transition. In a certain limit, namely $q
\rightarrow -\infty$, the model becomes equivalent to a DP
process. Therefore, the model provides a field theory of directed
percolation that differs from the standard field theory discussed
in Sec.~\ref{FTHSEC}. Although both field theories `intersect' in
one dimension, they are completely different. In particular, the
loop expansion starts out from different critical dimensions,
namely $d_c=4$ for Reggeon field theory and $d_c=2$ for branching
Potts interfaces. Consequently, in the latter case the one-loop
estimates for the critical exponents in $d=1$ are much more
accurate.

\begin{figure}
\epsfxsize=70mm
\centerline{\epsffile{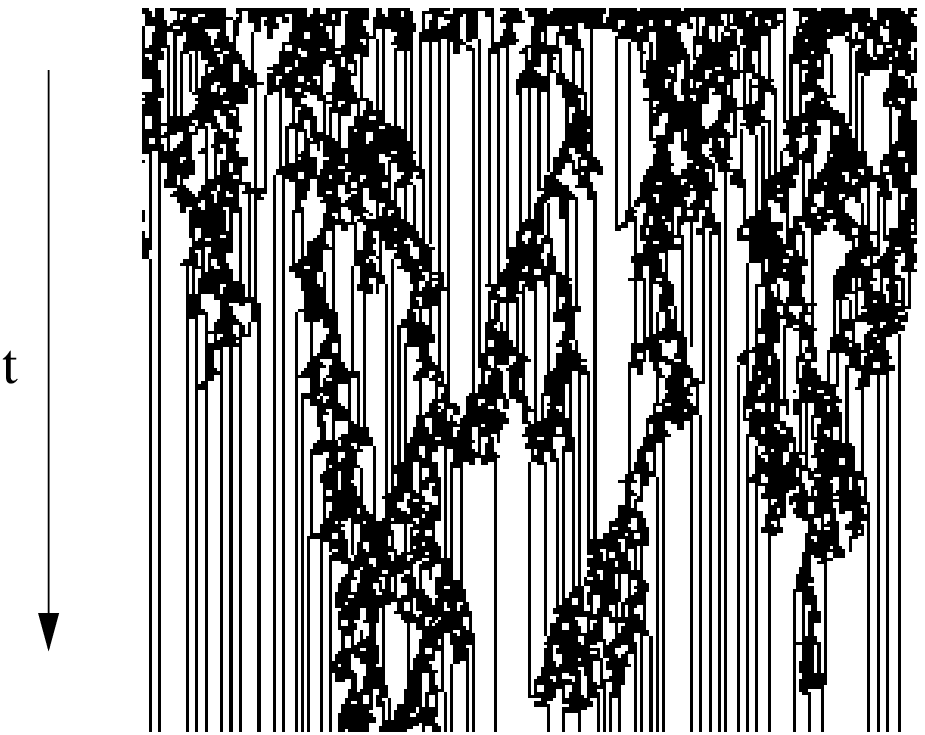}}
\vspace{2mm}
\smallcaption{
\label{FIGPCP}
The pair contact process $2A\rightarrow 3A$, $2A\rightarrow \vacancy$
at criticality.
}
\end{figure}

%---------------------------------------------------------------------------
\headline{DP models with infinitely many absorbing states}
%---------------------------------------------------------------------------
%
According to the DP conjecture, phase transitions into a {\em single}
absorbing state belong generically to the DP universality class.
However, DP behavior may also be observed in models with several
or even infinitely many absorbing states. An interesting
example is the dimer-trimer model for heterogeneous catalysis
introduced by K\"ohler and
ben-Avraham~\cite{KoehlerAvraham91}.
This model generalizes the ZGB model and is defined
by the reaction scheme
\begin{alignat}{2}
\vacancy\vacancy & \rightarrow  AA &&\qquad
\text{at rate $p$ \ ,} \nonumber \\
\vacancy\vacancy\vacancy & \rightarrow  BBB &&\qquad
\text{at rate $1-p$ \ ,}  \\
AB & \rightarrow \vacancy\vacancy &&\qquad
\text{at rate $\infty$ \ .} \nonumber
\end{alignat}
On an infinite lattice this model has infinitely many absorbing states.
For example, configurations of dimers and trimers separated
by single vacant sites are absorbing. The dimer-trimer model displays
a phase transition in 2+1 dimensions. Initially, the values of the
critical exponents were found to be different from those of DP.
Later refined simulations confirmed, however, that the dimer model
still belongs to the DP universality class~\cite{Jensen94c}.
The same result was found in a similar model for catalysis
of dimers and monomers~\cite{YKAK93,Jensen94b}.
Another important example is the
{\it pair contact process} without 
diffusion~\cite{Jensen93a} which is defined
by the reaction scheme
\begin{equation}
\label{PCP}
 2A \rightarrow 3A \,, \qquad 2A \rightarrow \vacancy
\,.
\end{equation}
In this model solitary particles neither react nor diffuse.
Starting from random initial conditions, the critical pair
contact process evolves into certain frozen configurations,
as demonstrated in Fig.~\ref{FIGPCP}. As can be seen, it is
important that single particles are not allowed to diffuse.
In fact, by adding diffusion of individual particles the
critical behavior of the model changes entirely (see Sec.~\ref{PCPD}).

In all models with infinitely many absorbing states and
non-conserved order parameter the critical exponents
$\beta, \nupar, \nuperp$ coincide with those of DP.
This observation suggests an extension of the DP conjecture to
systems with several absorbing states which are characterized
by a non-conserved single-component order
parameter~\cite{JensenDickman93b}.
However, it was realized that the dynamic exponents
$\delta$ and $\theta$ depend on the initial condition and even violate
the usual DP hyperscaling relation~(\ref{HyperscalingRelation}).
Mendes {\it et~al.}~\cite{MDHM94} resolved this problem by introducing the
generalized hyperscaling relation~(\ref{GeneralizedHyperscaling}).
However, the sum $\delta+\theta$ is believed to be independent of the
initial condition.

Recently Mu\~noz~{\it et~al.} proposed a Langevin equation for systems
with infinitely many absorbing states~\cite{MGDL96}.
It differs from the usual DP Langevin
equation~(\ref{DPLangevinEquation}) by an additional term:
\begin{equation}
\label{MultiLangevin}
\begin{split}
\timederivative\phi(\xvec,t) =&
\crit \phi(\xvec,t) -
\lambda \phi^2(\xvec,t)  +
\diff \nabla^2 \phi(\xvec,t) +
\noise(\xvec,t) + \\ &+
\alpha \phi(\xvec,t) \exp \Bigl[-w \int_0^t dt^\prime \,
\phi(\xvec,t^\prime) \Bigr]
\ .
\end{split}
\end{equation}
Here $\alpha$ and $w$ are certain constants (the noise correlations are
assumed to be the same as in Eq.~(\ref{DPNoise})). This Langevin
equation is {\em non-Markovian}, i.e., it has a temporal memory.
The memory is local since the integral correlates fluctuations
at the same position in space. From the physical point of view,
this memory encodes the local realization
of the absorbing state.  As can be seen in Fig.~\ref{FIGPCP},
the emerging inactive domains have a highly inhomogeneous structure which
can be regarded as a fingerprint of the history of the spreading process.
A detailed numerical analysis of the Langevin
equation~(\ref{MultiLangevin}) confirmed that the exponenets $\beta$, 
$\nupar$, and $\nuperp$ do belong to the DP class while $\delta$
and $\theta$ vary with the density of the initial
state~\cite{LopezMunoz97}. In order to explain
the apparent nonuniversality of the spreading exponents,
Grassberger developed a simple toy model that grasps the main properties
of such spreading processes~\cite{GCR97}. In this toy model
the spreading rate at a given site changes irreversibly at the first
encounter with the spreading agent. Although the model does not involve
multiple absorbing states, it displays similar `nonuniversal'
properties.

Another important example for systems with infinitely many
absorbing states is {\em damage spreading} where two copies of a
stochastic system evolve under the same realization
of thermal noise. The concept of damage spreading 
will be discussed in detail in Sec.~\ref{DamageSection}.

%---------------------------------------------------------------------------
\headline{Epidemic processes with immunization}
%---------------------------------------------------------------------------
%
%
%
\begin{figure}
\epsfxsize=60mm
\centerline{\epsffile{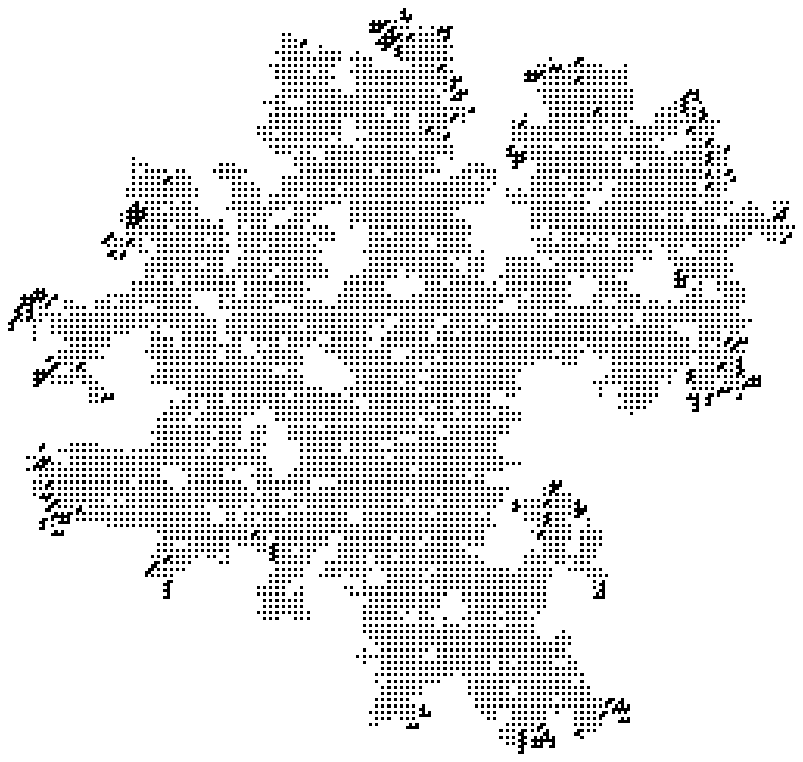}}
\smallcaption{
\label{DYNPERC}
Snapshot of a critical epidemic process with immunisation
grown from a single seed after 200 time steps. Active and immune sites
are represented by bold and thin dots, respectively. 
An zone of activity propagates outwards, leaving a cluster
of immune sites behind. The resulting cluster belongs to the
universality class of isotropic percolation.
}
\end{figure}
As we have seen in Sec.~\ref{DPINTROSEC}, epidemic models without 
immunization belong generically to the DP universality class. 
In most cases, however, an infected individual becomes increasingly
immune after recovery, i.e., the susceptibility for a new infection
decreases. Cardy and Grassberger showed that epidemic models with
immunization are in the same universality class as {\em dynamic
percolation}~\cite{Grassberger83,CardyGrassberger85}. It is 
important to note that dynamic percolation differs significantly
from directed percolation. For example, let us consider a spreading 
process with immunization in 2+1 dimensions starting from an initial 
state where all sites are equally susceptible for infections. 
If a single site in the center is infected, there is a finite 
probability that the disease will spread. However, since
infected sites become increasingly immune, 
a more or less irregular front 
of activity moves away from the origin, leaving behind a certain cluster 
of immune sites (see Fig.~\ref{DYNPERC}). 
The morphology of this cluster depends
on the percolation parameter. In the supercritical case there 
is a finite probability that the front moves to infinity, whereas 
in the subcritical regime the process stops after some time.
At criticality it turns out that the generated cluster has the 
same asymptotic properties as critical clusters of {\em isotropic} 
percolation~\cite{StaufferAharony92} 
(cf. left part of Fig.~\ref{FIGISODIR}). Thus, dynamic percolation
can be used as a tool to generate isotropic percolation clusters
and should not be confused with directed percolation. In particular,
the critical exponents turn out to be different in both cases. 
Interestingly, even a small degree of immunization suffices 
for a (2+1)-dimensional epidemic process to cross over from directed 
to dynamic percolation (together with a shift of the critical point). 
A renormalized field theory of dynamic percolation was studied 
in~\cite{Janssen85}.

%---------------------------------------------------------------------------
\headline{Compact directed percolation}
%---------------------------------------------------------------------------
%
Let us finally come back to
{\it compact directed percolation} (CDP) \cite{Essam89}
which characterizes the critical behavior of the DK model
at the  upper terminal point of the phase transition line
$p_1=1/2$, $p_2=1$ (see Fig.~\ref{FIGDKPHASE}). 
The case $p_2=1$ is special because there are
{\it two symmetric} absorbing states, namely the
dry state $s_1=\ldots=s_N=0$ and the entirely wet state
$s_1=\ldots=s_N=1$. In contrast to DP,  CDP has
a global $Z_2$ symmetry
\begin{equation}
s_i \rightarrow 1-s_i \ , \qquad p_1 \rightarrow 1-p_1 \ .
\end{equation}
Since wet sites cannot spontaneously become dry,
compact islands of active sites are formed.
In 1+1 dimensions CDP is fully equivalent to a zero
temperature Ising model with Glauber dynamics or the
voter model~\cite{Durrett88}. Expressing the dynamic
processes in terms of kinks $X$ between wet and dry domains,
the kinks perform an annihilating random walk
$X+X\rightarrow \vacancy$. Therefore,
(1+1)-dimensional CDP is exactly solvable~\cite{DomanyKinzel84}.
The corresponding critical exponents are given by
\begin{equation}
\begin{split}
\label{CDPExponents}
&\beta=0\,,\qquad
\beta^\prime=1\,,\qquad
\nupar=2\,,\qquad
\nuperp=1\,,\\
&\delta=1/2\,,\qquad
\theta=0\,,\qquad
\tilde{z}=1\,.
\end{split}
\end{equation}
It should be noted that these exponents do not
comply with the usual DP hyperscaling relation
(\ref{HyperscalingRelation}). However, as pointed out in
Ref.~\cite{DickmanTretyakov95}, they satisfy the generalized
hyperscaling relation~(\ref{GeneralizedHyperscaling}).
In fact, as can be verified easily, for CDP the backbone of a
two-point function (see Sec.~\ref{FTHSEC}) is no longer
statistically invariant under time reversal.

Because of the vanishing exponent $\beta$,
the CDP transition is {\em discontinuous}.
In fact, for $p_1<1/2$, $p_2=1$ the empty
lattice is a stable stationary state while for
$p_1>1/2$ the fully occupied lattice is stable.
Various spreading models display a {\it crossover} from CDP to DP.
In these models, the rate for the reaction $A \rightarrow \vacancy$
is very small. Therefore, clusters appear to be compact on
small scales. On larger scales, however, clusters break
up into several active branches, leading to DP
behavior in the asymptotic limit. This type of crossover
has been studied in detail in
Refs.~\cite{ETS94,NguyenRubio95,MDH96,FrojdhNijs97} and may
also play a role in experiments of flowing granular matter
(see Sec.~\ref{DPEXPERIM}).

%==============================================================================
\subsection{Experimental realizations of directed percolation}
%==============================================================================
%
\label{DPEXPERIM}
So far we have seen that directed percolation
is the generic universality class for nonequilibrium phase transitions
into absorbing states. In fact, DP seems to be of
similar importance as the Ising
model in equilibrium statistical mechanics.
Despite this success in theoretical statistical physics,
the critical behavior of DP,  especially the values of
the critical exponents, have not yet been confirmed experimentally.
The lack of experimental evidence is indeed surprising,
especially since a large number of possible experimental
realizations have been suggested in the past.
As Grassberger emphasizes in a summary
on open problems in DP~\cite{Grassberger97}:
\begin{quote}
{\it "...there is still no experiment where the critical
behavior of DP was seen. This is a very strange situation
in view of the vast and successive theoretical efforts made
to understand it. Designing and performing
such an experiment has thus top priority in my
list of open problems."}.
\end{quote}

\noindent
What might be the reason for the apparent lack of experimental
evidence? It seems that the basic features of DP, which can
easily be implemented on a computer, are quite difficult to
realize in nature. One of these idealized
assumptions is the existence of an absorbing state.
In real systems, however, a perfect non-fluctuating state cannot
be realized. For example, a poisoned catalytic
surface is not completely frozen, instead it will always
be affected by small fluctuations.
Although these fluctuations are strongly suppressed, they
could still be strong enough to `soften' the transition,
making it impossible to quantify the critical exponents.

Another reason might be the influence of
{\it quenched disorder} due to spatial or temporal
inhomogeneities. In most experiments frozen randomness
is expected to play a significant role. For example,
a real catalytic surface is  not fully homogeneous
but characterized by certain defects leading to
spatially quenched disorder. As has been shown in Sec.~\ref{QUENCHSEC},
this type of disorder may affect or even destroy the
critical behavior of DP.

In the following we summarize some of the most important experimental
applications which have been proposed so far~\cite{Hinrichsen00a}.
Other experimental applications in systems of growing interfaces will be
discussed in Sec.~\ref{GrowthSection}.

\begin{figure}
\epsfxsize=150mm
%\centerline{\epsffile{pt.eps}}
\centerline{\epsffile{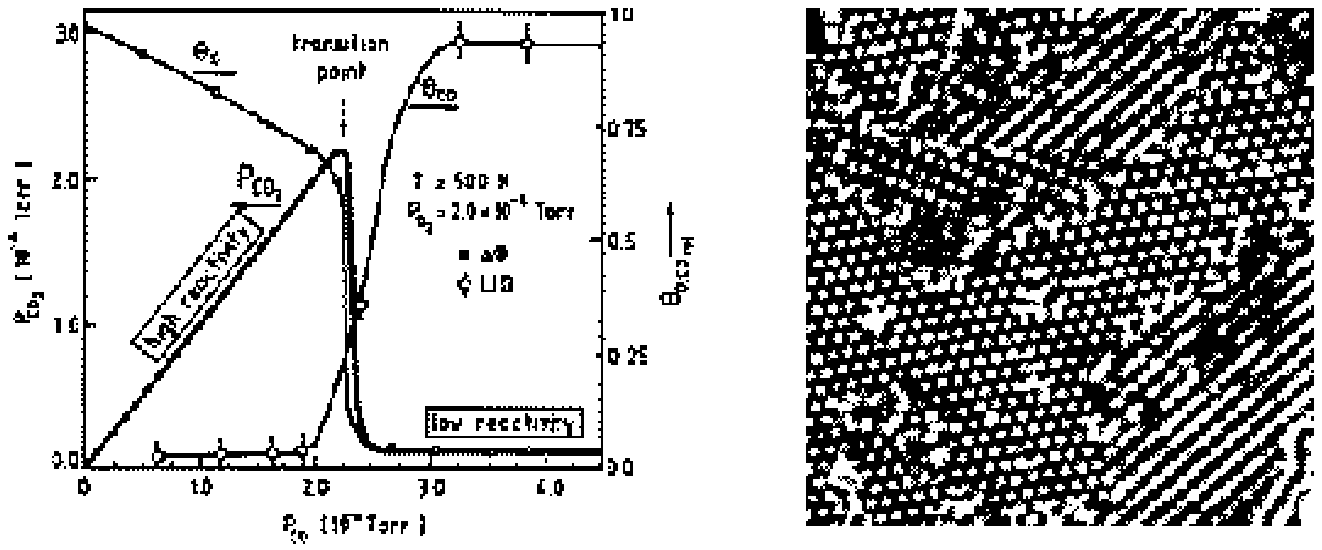}} % reduced size for condmat-version
\vspace{2mm}
\smallcaption{
\label{FIGPT210}
Left: The reaction rate on Pt(210) as a function of $P_{CO}$
(figure reprinted from Ref.~\cite{EMFBCRH89}). Right: STM image
of a catalytic reaction on a Pt(111) crystal (figure reprinted from
Ref.~\cite{WVJZE97}).
}
\end{figure}
%
%
%-------------------------------------------------------------------------
\headline{Catalytic reactions}
%-------------------------------------------------------------------------
%
It is well known that under specific conditions certain
catalytic reactions mimic the microscopic rules of
DP models. For example, as shown in Fig.~\ref{FIGZGB},
the ZGB model for the catalytic
reaction CO + O $\rightarrow$  CO$_2$ on a platinum surface
displays a continuous transition at $y=y_1$ belonging
to DP. In real catalytic reactions, however, only
the discontinuous transition at $y=y_2$ can be observed.
Fig.~\ref{FIGPT210} shows the reaction rates as
functions of the CO pressure measured in a
catalytic reaction on a Pt(210) surface~\cite{EMFBCRH89}.
Although this experiment was designed in order to
investigate the technologically interesting
regime of high activity close to the first-order phase
transition, the results clearly indicate that poisoning with oxygen
does not occur. Instead the reactivity increases almost
{\it linearly} with the CO pressure.
Similar results were obtained for Pt(111) and for other
catalytic materials. Thus, so far there is no
experimental evidence for DP transitions in catalytic reactions.

One may speculate why the DP transition
is obscured or even destroyed under experimental conditions.
On the one hand, the reaction chain in the experiment
is much more complicated than in the ZGB model~\cite{EKE90}.
Moreover, the O-poisoned system might not be a perfect
absorbing state, i.e., the surface can still adsorb CO
molecules although it is already saturated.
Another possibility is thermal (nonreactive) desorption of 
oxygen, acting as an external field which drives
the system away from criticality ~\cite{DickmanPrivate99}
[cf. Eq.~(\ref{FieldResponse})].
Finally, defects and inhomogeneities of the catalytic
material could lead to an effective (spatially quenched) disorder.

For a long time the microscopic dynamics processes were 
difficult to study experimentally. However, in recent years
novel techniques such as scanning tunneling microscopy (STM) led to an
enormous progress in the understanding of catalytic
reactions, pointing at various unexpected
subtleties. For example, it was observed that
reactions preferably take place at the perimeter of
oxygen islands~\cite{WVJZE97}. Furthermore, it was realized
that adsorbed CO molecules on Pt(111) may form
three different rotational patterns representing the
c(4$\times$2) structure of CO on platinum, i.e., there are
several competing absorbing states~\cite{Hinrichsen97}.
Moreover, the STM technique allows one to trace individual molecular
reactions and to determine the corresponding reaction rates.
In addition, the influence of defects such
as terraces on catalytic reactions can be
quantified experimentally~\cite{ZWTE96}. We may therefore expect a
considerable progress in the understanding of catalytic
reactions in near future.

%
%
% FIGURE DEMO DOUADY-DAERR EXPERIMENT
%
\begin{figure}
\epsfxsize=45mm
\centerline{\epsffile{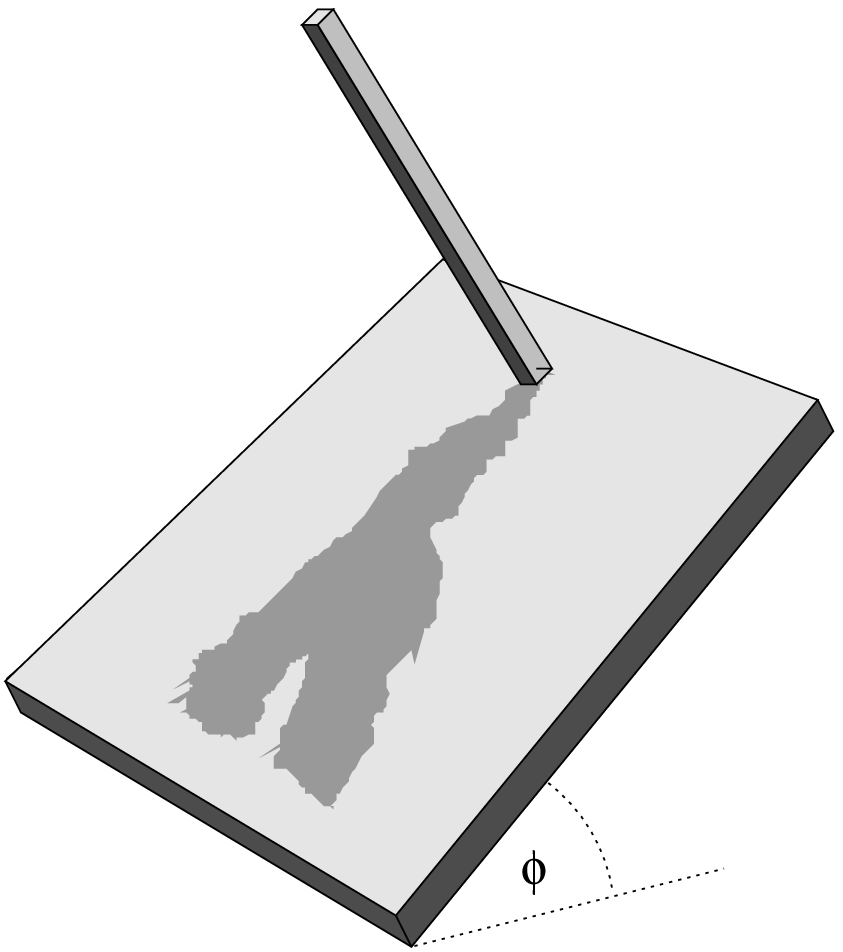}}
\vspace{2mm}
\smallcaption{
Simplified drawing of the Douady-Daerr experiment.
Pouring sand on top of a plane with inclination  $\phi$,
a thin layer settles and remains immobile.
Perturbing the layer locally with a stick
we can release an avalanche of flowing.
\label{FigDouadyDaerr}
}
\end{figure}
%
%
%--------------------------------------------------------------------------
\headline{Flowing granular matter}
%--------------------------------------------------------------------------
%
Recently it has been shown that simple systems of
flowing sand on an inclined plane, such as the experiments
performed by Douady and Daerr~\cite{DouadyDaerr98,DaerrDouady99},
could serve as experimental realizations of DP~\cite{HJRD99}.
In the Douady-Daerr experiment glass beads with a diameter
of $250$-$425 \ \mu$m are poured uniformly
at the top of an inclined plane covered by a rough velvet cloth
(see Fig.~\ref{FigDouadyDaerr}). As the beads flow down,
a thin layer settles and remains immobile. Increasing
the angle of inclination $\phi$ by $\Delta\phi$
the layer becomes dynamically unstable, i.e.,
by locally perturbing the system at the top of the plane
an avalanche of flowing granular matter will be released.
In the experiment these avalanches have the shape of a fairly
regular triangle with an opening angle~$\theta$.
As the increment $\Delta \varphi$ decreases, the value of $\theta$
decreases, vanishing as
\begin{equation}
\tan\theta \sim (\Delta \varphi)^x \,
\label{eq:powerlaw}
\end{equation}
with a certain critical exponent $x$.
The experimental results suggest the value $x=1$~\cite{DaerrDouady99}.

In order to explain the experimentally observed triangular form of the
avalanches, Bouchaud {\it et~al.} proposed a mean-field theory based on
deterministic equations, taking the actual local thickness of the
flowing avalanche into account~\cite{BouchaudCates98}.
This theory predicts the exponent $x=1/2$.
Another explanation assumes that flowing sand may be
interpreted as a nearest-neighbor spreading process~\cite{HJRD99}.
Here the avalanche is considered as a cluster of active sites.
Identifying the vertical coordinate of the plane with time
and the increment of inclination $\Delta \varphi$
with $p - p_c$, the opening angle is expected to scale as
\begin{equation}
\tan \theta \sim \xi_\perp / \xi_\parallel \sim
(\Delta \varphi )^{\nupar - \nuperp } \,,
\label{eq:Dnu}
\end{equation}
where $\nupar$ and $\nuperp$ are the scaling
exponents of the spreading process under consideration.

%
%
% FIGURE DEMO CROSSOVER
%
\begin{figure}
\epsfxsize=80mm
\centerline{\epsffile{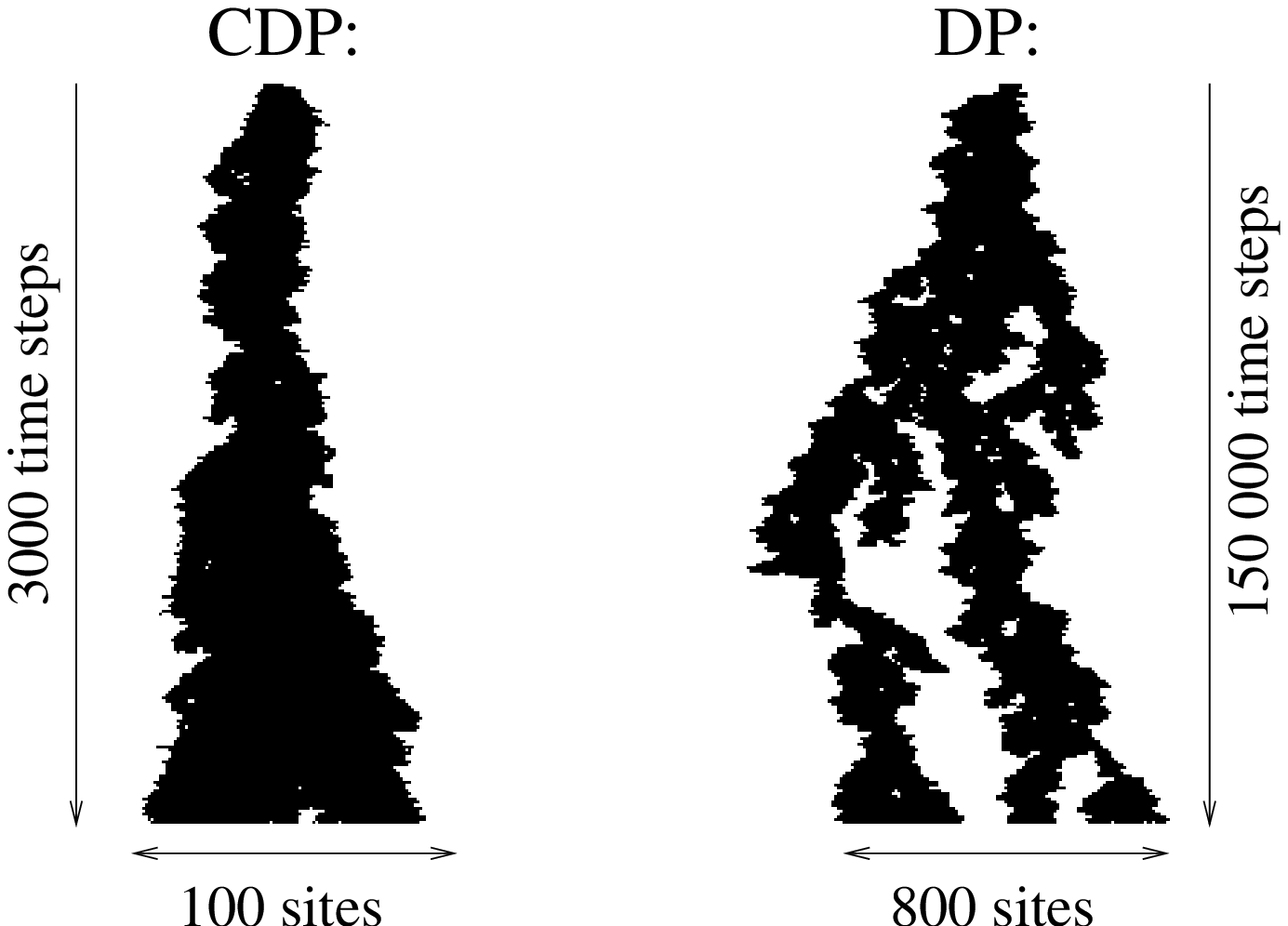}}
\vspace{2mm}
\smallcaption{
Typical clusters generated at
criticality on small and large scales,
illustrating the crossover from CDP to DP.
\label{FigDemoCrossover}
}
\end{figure}

To support this scaling argument, a simple lattice model was
introduced which mimics the physics of flowing sand~\cite{HJRD99}.
The model exhibits a transition from an inactive to an
active phase with avalanches whose compact shapes reproduce
the experimental observations. On laboratory scales the model
predicts a transition belonging to the universality
class of compact directed percolation [see Eq.~(\ref{CDPExponents})],
implying that
\begin{equation}
x=\nupar-\nuperp=1.
\end{equation}
The CDP behavior, however, is only transient and
crosses over to DP after a very long time. Thus
the Douady-Daerr experiment -- performed on sufficiently
large scales -- may serve as a physical realization of DP.
Irregularities of the layers thickness can be considered
as spatio-temporally quenched disorder which is irrelevant on large scales
(see Sec.~\ref{QUENCHSEC}). Thus, in contrast to catalytic reactions,
the problem of quenched disorder does not play a major
role in this type of experiments.

\begin{figure}
\epsfxsize=130mm
%\centerline{\epsffile{daerr.eps}}
\centerline{\epsffile{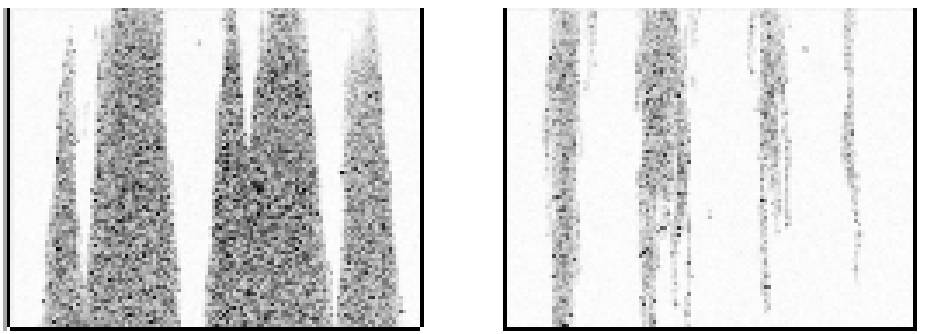}} % reduced size for condmat-version
\vspace{1mm}
\smallcaption{
\label{FIGSPLIT}
Left: Triangular avalanches in the Douady-Daerr experiment.
Right: Avalanches splitting up into several branches for
high angles of inclination (reprinted with kind permission from
A. Daerr).
}
\end{figure}

The crossover from CDP to DP is very slow and presently not
accessible in experiments. To illustrate the crossover,
two avalanches are plotted on different scales
in Fig.~\ref{FigDemoCrossover}. The left one
represents a typical avalanche within the first
few thousand time steps. As can be seen, the cluster appears
to be compact. However, as shown in the right panel
of the figure, the cluster breaks up
into several branches after a very long time.
Recent experimental studies~\cite{Daerr99} confirm that for
high angles of inclination critical avalanches do split
up into several branches (see Fig.~\ref{FIGSPLIT}).
Yet here the avalanches have no well defined front,
the propagation velocity of separate branches rather depends
on their thickness. It is therefore no longer possible to interpret
the vertical axis as a time coordinate.  Another problem
is the kinetic energy of the grains. According to arguments by
Dickman {\it et~al.}~\cite{DMVZ00}, continuous phase transitions
into absorbing states can only be observed if the inertia of
particles can be neglected.

Finally, it is not yet known how the spreading
process depends on correlations in the
initial state. As shown in Sec.~\ref{BCSEC}, such long-range
correlations may change the values of certain dynamic critical
exponents. However, recent studies of a single rolling grain on
an inclined rough plane~\cite{DBW96} support that there are presumably no
long-range correlations due to a `memory' of rolling grains.
By means of molecular dynamics simulations it was shown that
the motion of a rolling grain consists of many small bounces
on each grain of the supporting layer. Therefore, the rolling
grain quickly dissipates almost all of the energy gain from the
previous step and thus forgets its history very fast. For this reason
it seems to be unlikely that quenched disorder of the prepared
layer involves long-range correlations. Therefore, flowing granular
matter seems to  be a promising candidate for an experimental
realization of DP.

%
% FIGURE SPLITTING AVALANCHE
%
%

%--------------------------------------------------------------------------
\headline{Porous media}
%--------------------------------------------------------------------------
%
DP is often motivated as a model for water percolating through a
porous medium in a gravitational field. Due to an external driving
force, the flow in the medium is assumed to be strictly
unidirectional, i.e., the water can only flow downwards (in
contrast to the depinning models of Sec.~\ref{DepinningSubsection}
where the water can flow forth and back). Although this
application seems to be quite natural, it is difficult to realize
experimentally. As shown in Ref.~\cite{HRV97}, porous media in
nature are highly irregular. By cutting sandstone into slices and
digitizing the section images, the porosity distribution and the
local connectivity were measured and averaged over 99 samples. As
expected, the pores have different sizes and are distributed
irregularly. In addition, the percolation probability is found to
depend on the local porosity and the direction in space, i.e.,
sandstone is an anisotropic material. But there are even more
fundamental problems. On the one hand, water is
a conserved quantity, leading to unpredictable long-range
correlations in the bulk. On the other hand, water can always flow against
the gravity field by means of capillary forces. Therefore, it is
quite difficult or even impossible to verify scaling laws in such
experiments and is not yet clear whether the relation to DP is
meaningful or simply a commonly accepted misconcept.

%--------------------------------------------------------------------------
\headline{Epidemics}
%--------------------------------------------------------------------------
%
Another frequently quoted application of DP is the spreading
of epidemics without immunization~\cite{Harris74,Mollison77}.
In an epidemic process infection and recovery resemble
the reaction-diffusion scheme of DP~(\ref{DPReactionDiffusion}).
If the rate of infection is very low, the infectious disease
will disappear after some time. If infections occur more
frequently, the disease may spread and survive for a very
long time. However, spreading processes in nature are usually not
homogeneous enough to reproduce the critical behavior of DP.
Moreover, in many realistic spreading processes short-range
interactions are no longer appropriate. This situation emerges,
for example, when an infectious disease is transported by insects.
Such long-range interactions may be described by L\'evy flights,
leading to  continuously varying critical exponents
(see Sec.~\ref{ANOMALSEC}).

%--------------------------------------------------------------------------
\headline{Forest fires}
%--------------------------------------------------------------------------
%
A closely related problem is the spreading of forest
fires~\cite{Albano94a,Albano95a}. Tephany {\it et~al.} studied the
propagation of flame fronts on a random lattice
both under quiescent conditions and in a wind tunnel~\cite{TND97}.
The experimental estimates of the critical exponents at the
spreading transition are in rough agreement with the
predictions of isotropic and directed percolation, respectively.
However, the accuracy of these experiments remains limited.

%--------------------------------------------------------------------------
\headline{Calcium dynamics}
%--------------------------------------------------------------------------
%
DP transitions may also occur in certain kinetic models for the
dynamics of Calcium ions in living cells. Ca$^{2+}$ ions play an
important physiological role as second messenger for various purposes
ranging from hormonal release to the activation of egg cells by
fertilization~\cite{Berridge93,KRS95}. The cell uses nonlinear
propagation of increasing intracellular Ca$^{2+}$ concentration,
so-called calcium waves, as a tool to transmit
signals over distances that are much longer than the diffusion length.
For example, propagating Ca$^{2+}$ waves can be observed in the
immature {\it Xenopus laevis oocyte}~\cite{LechleiterClapham92}.
So far theoretical work focused mainly on deterministic
reaction-diffusion equations in the continuum, explaining various
phenomena such as solitary and spiral waves~\cite{Jaffe93}.
Recently improved models have been introduced which
take also the stochastic nature of Calcium release into
account~\cite{KeizerSmith98}. As expected,
the transition in one of these models
belongs to the DP universality class~\cite{BFLT00}.
However, from the experimental point of view it seems to be
impossible to confirm or disprove this conjecture. On the one hand,
the size of a living cell is only a few order of magnitude larger
than the diffusion length, leading to strong finite-size effects
in the experiment. On the other hand, inhomogeneities as well as internal
structures of the cell lead to a completely unpredictable form
of quenched noise. Therefore, it seems to 
be impossible to identify the universality
class of the transition in such experiments. It would be rather
an achievement to find clear evidence for the very existence of a phase
transition between survival and extinction of propagating calcium
waves.

%--------------------------------------------------------------------------
\headline{Directed polymers}
%--------------------------------------------------------------------------
%
DP is also related to the problem of directed polymers~\cite{HuseHenley85}.
In contrast to DP, which is defined as a {\em local} process, the directed
polymer problem selects directed paths in a random medium
by {\em global} optimization. Under certain conditions, namely
a bimodal distribution of random numbers, both problems were shown
to be closely related~\cite{PerlsmanHavlin99}. More specifically,
the roughness exponent of the optimal path in a directed polymer
problem is predicted to cross over from the KPZ value 2/3 to
the DP value $\nupar/\nuperp \simeq 0.63$ at
the transition point. Directed polymers were used
to describe the propagation of cracks~\cite{BBL93}.
However, in such experiments it is usually impossible to verify
the tiny crossover from KPZ to DP.

%--------------------------------------------------------------------------
\headline{Turbulence}
%--------------------------------------------------------------------------
%
Finally, DP has also been considered as a toy model for
turbulence. As suggested in Ref.~\cite{Pomeau86}, the front
between turbulent and laminar flow should exhibit the critical
behavior of DP. For example, the velocity of the front should
scale algebraically with a combination of DP exponents.
However, these predictions are based rather on heuristic arguments
than on rigorous results. In fact, in many respects turbulent
phenomena show a much richer behavior than DP. Nevertheless there
are certain similarities between DP and turbulence. Therefore, the
study of DP could be helpful for a better understanding of
turbulent phenomena.

%--------------------------------------------------------------------------
\headline{Summary and outlook}
%--------------------------------------------------------------------------
%
Directed percolation is keeping theoretical physicists
fascinated since more than four decades. Several
reasons make directed percolation so appealing.
First of all, DP is a very simple model in terms of its dynamic
rules. Nevertheless, the DP phase transition turns out to
be highly nontrivial. In fact, DP belongs
to the very few critical phenomena which have not
yet been solved exactly in one spatial dimension. Therefore,
the critical exponents are not yet known analytically.
High-precision estimates indicate that they might be given
rather by irrational than by simple fractional values.
Moreover, DP is extremely robust. It stands for a whole
universality class of phase transitions from a fluctuating phase
into absorbing states. In fact, a large variety of models
displays phase transitions belonging to the DP universality
class. Thus, on the theoretical level, DP plays the role of a
standard universality class similar to the Ising model
in equilibrium statistical mechanics.

In spite of its simplicity, no experiment is known  which
confirms the values of the critical exponents quantitatively.
An exception may be the wetting experiment
performed by Buldyrev {\it et~al.}  (see below in
Sec.~\ref{DepinningSubsection}),
where the value of the roughness exponent $\alpha$ coincides with
$\nuperp/\nupar$ within less than 10\%.
However, since the results of similar experiments are scattered
over a wide range,  further experimental effort in this
direction would be needed in order to confirm the
existence of DP in this type of systems.

Apart from difficulties to realize a non-fluctuating absorbing
state, a fundamental problem of DP experiments is the emergence
of quenched disorder caused by inhomogeneities of the system.
Depending on the type of disorder, even weak inhomogeneities
might obscure or even destroy the DP transition.
Therefore, the most promising experiments are those where
quenched disorder is irrelevant on large scales. This is the case,
for example, in wetting experiments and systems of flowing granular matter.

Although there is certainly a lack of experimental evidence, there
is no reason to believe that DP is a purely artificial model.
To be optimistic, it is helpful to recall the history of the Ising model,
which has been introduced almost one century ago. Although the Ising
model is probably the best studied system in equilibrium statistical
mechanics, there are only few experiments in which the critical
exponents have been reproduced (for a review see Ref.~\cite{Henkel99}).
For this reason, many physicists believe that DP should have
a counterpart in reality as well, mostly because of its
simplicity and robustness. In this respect, Grassbergers message
remains valid: The experimental realization of DP is an
outstanding problem of top priority.

\newpage

%##############################################################################
         \section{Other classes of spreading transitions}
%##############################################################################

This Section discusses various other types of nonequilibrium
phase transitions into absorbing states which do not belong
to the universality class of directed percolation. In
particular we will address spreading processes with long-range
interactions and additional symmetries.

%==============================================================================
\subsection{Long-range spreading processes}
\begin{figure}
\epsfxsize=120mm
%\centerline{\epsffile{levydemo.eps}}
\centerline{\epsffile{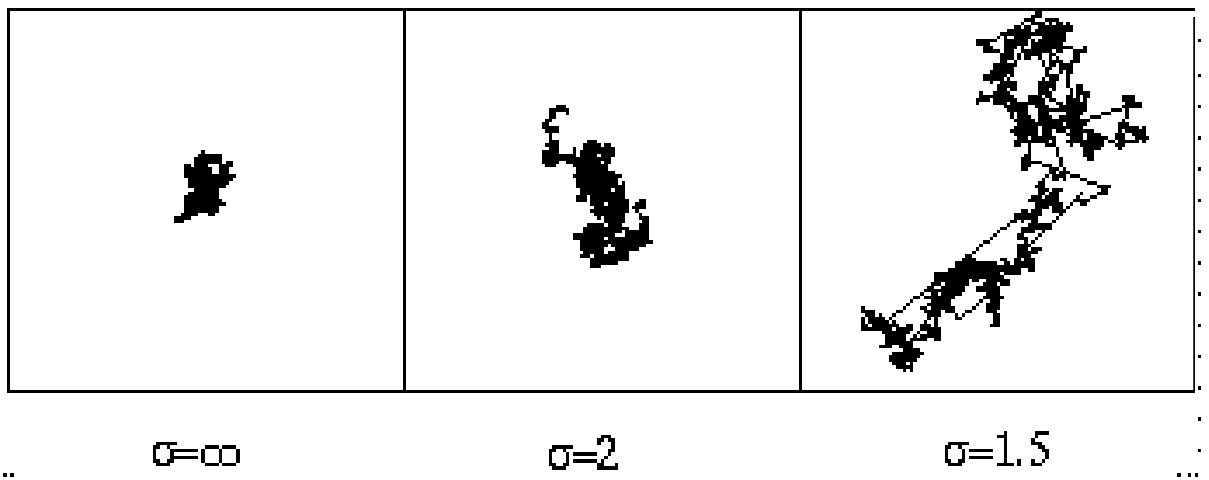}} % reduced size for condmat-version
\vspace{2mm}
\smallcaption{
\label{FIGLEVYDEMO}
Ordinary random walk ($\sigma=\infty$)
in comparison with L\'evy flights ($\sigma<\infty$).
}
\end{figure}

%==============================================================================
%
\label{ANOMALSEC}
According to the DP conjecture (see Section~\ref{ABSSCALESEC})
phase transitions in spreading models with {\it short range}
interactions generically belong to the DP universality class.
In many realistic spreading processes, however, short-range
interactions do not appropriately describe the
underlying transport mechanism. This situation emerges, for example,
if an infectious disease is transported by insects. Typically the motion
of the insects is not a random walk, one rather observes
occasional flights over long distances before the next infection
occurs. Similar phenomena are expected when the spreading
agent is subjected to a turbulent flow. Intuitively it is
clear that occasional spreading over long distances will
significantly alter the spreading properties.
On a theoretical level such a {\it super-diffusive}
transport may well be described by L\'evy flights~\cite{BouchaudGeorges90},
i.e., by uncorrelated random moves over long distances $r$ which
are algebraically distributed as
\begin{equation}
\label{ProbDis}
P(r) \sim 1/r^{d+\sigma} \ ,
\qquad (\sigma>0) \ .
\end{equation}
The exponent $\sigma$ is a free parameter that
controls the characteristic shape of the distribution.
The algebraic tale leads to occasional
long-distance flights, as shown in Fig.~\ref{FIGLEVYDEMO}.

Anomalous directed percolation, as originally
proposed by Mollison~\cite{Mollison77}
in the context of epidemic spreading,
is a generalization of DP in which the spreading agent
is transported by L\'evy flights. As in the case of ordinary
DP, we expect anomalous DP to be characterized
by certain universal critical exponents
$\beta$, $\nu_{\perp}$, and $\nu_{||}$.
The question is how these exponents depend on~$\sigma$,
whether they are independent from one
another, and how they cross over to the exponents of ordinary DP.
Based on field-theoretic considerations,
Grassberger~\cite{Grassberger86}
claimed that the critical exponents of
anomalous DP should depend continuously on the control
exponent $\sigma$. Very recently this work has been considerably
clarified and extended by Janssen~{\it et~al.}~\cite{JOWH98}, who
presented a comprehensive field-theoretic analysis of
anomalous spreading processes with and
without immunization.

%---------------------------------------------------------------------------
\headline{Anomalous directed percolation: Field-theoretic predictions}
%---------------------------------------------------------------------------
%
In order to include long-range spreading in the Langevin
equation~(\ref{DPLangevinEquation}), the Laplacian has to be
replaced by a non-local expression.
This term can be written as an integral that describes
L\'evy flights over the distance $r$
according to the probability distribution $P(r)$:
\begin{equation}
\label{IntegralLangevin}
\begin{split}
\timederivative  \phi(\xvec,t)
&=
\crit \, \phi(\xvec,t) - \lambda \, \phi^2(\xvec,t) + \noise(\xvec,t)\\
&+ D \int d^dx' \, P(|\xvec-{\bf x'}|) \Bigl[\phi({\bf x'},t)
-\phi(\xvec,t)\Bigr] \ .
\end{split}
\end{equation}
The two contributions in the integrand describe gain and
loss processes, respectively.
Keeping the most relevant terms in a small momentum
expansion~\cite{JOWH98}, this equation may be written as
\begin{equation}
\label{AnomLangevin}
\timederivative  \phi(\xvec,t) =
\Bigl(D_N \nabla^2 + D_A \nabla^\sigma_A + \crit \Bigr) \phi(\xvec,t)
- \lambda \phi^2(\xvec,t) + \noise(\xvec,t) \ ,
\end{equation}
where the noise correlations are assumed to be the same
as in Eq.~(\ref{DPNoise}). $D_N$ and $D_A$ are the rates for
normal and anomalous diffusion, respectively. The anomalous diffusion
operator $\nabla_A^\sigma$ describes moves over long distances
and is defined through its action in momentum space
\begin{equation}
\label{AnomalousDiffusion}
\nabla_A^\sigma \, e^{i {\bf k} \cdot \xvec} =
- k^\sigma \, e^{i {\bf k} \cdot \xvec} \ ,
\end{equation}
where $k=|{\bf k}|$.
The standard diffusive term $D_N \nabla^2$ takes the
short-range component of the L\'evy distribution into account.
Note that even if this term were not
initially included, it would still be generated under
renormalization of the theory. The mean-field theory of anomalous DP
is completely analogous to that of ordinary DP.
For $\sigma<2$ a scaling analysis yields the mean field results
\begin{equation}
\label{AnomalousMF}
d_c=2\sigma \ , \qquad
\beta^{MF}=1
\ , \qquad
\nu^{MF}_{\perp}=1/\sigma
\ , \qquad
\nu^{MF}_{\parallel}=1 \ .
\end{equation}
For $\sigma\geq 2$ these exponents cross over
smoothly to the ordinary DP mean-field exponents~(\ref{DPMF}).
The mean field approximation is expected to be quantitatively
accurate above the upper critical dimension $d_c$, while
for $d\leq d_c$ fluctuation effects have to be taken into account.
By using standard techniques one can derive the
effective action
\begin{equation}
\label{AnomEffectiveAction}
S[\phi,\tilde{\phi}]
=
\int d^dx ~ dt ~
\Bigl[\tilde{\phi}(\timederivative - \crit - D_N \nabla^2 - D_A
\nabla_A^\sigma)\phi
\nonumber
+  \frac{\namp}{2}(\tilde\phi\phi^2-\tilde{\phi}^2\phi) \Bigr] \ .
\end{equation}
This expression differs
from the usual action of Reggeon field
theory~(\ref{DPFreePart})-(\ref{DPInteraction})
by the addition of a term representing anomalous diffusion.
Simple power counting on this action confirms that the upper critical
dimension is $d_c=2\sigma$, below which
fluctuation effects become important.
The field-theoretic RG calculation  basically follows
the same lines as in the case of DP.
The resulting critical exponents to one-loop
order in $d=2\sigma-\epsilon$ dimensions are given by
\begin{align}
\label{AnomCriticalExponents}
\beta       &= 1-2\epsilon/7\sigma + O\left( \epsilon^{2}\right)
\nonumber \, , \\
\nu_{\perp}     &= 1/\sigma + 2\epsilon/7\sigma^2
 + O\left( \epsilon^{2}\right) \, , \\
\nu_{||}    &= 1 + \epsilon/7\sigma + O\left( \epsilon^{2}\right)
\, .   \nonumber
\end{align}
Moreover, it can be shown that the hyperscaling
relation~(\ref{HyperscalingRelation})
holds for arbitrary values of $\sigma$.
Thus, to one-loop order, $\theta$ and $\delta$ are given by
\begin{equation}
\label{AnomEtaDelta}
\theta=\epsilon/7\sigma + O\left( \epsilon^{2}\right)
\ , \qquad
\delta=1-3\epsilon/7\sigma + O\left( \epsilon^{2}\right)
\ .
\end{equation}
Finally, since $D_A$ will not be renormalized,
one can prove the additional {\em exact} scaling relation
\begin{equation}
\label{AnomalScalingRelation}
\nu_{||} - \nuperp(\sigma-d)-2\beta = 0 \ .
\end{equation}
This equation implies that anomalous DP is
described by {\em two} rather than
three independent critical exponents.
Moreover, it has another surprising
consequence. Assuming that $\beta$, $\nuperp$
and $\nu_{||}$ change continuously with $\sigma$ and
cross over smoothly, it predicts for fixed $d$
the value $\sigma_c$ where the system should cross
over to ordinary DP. In order
to compute $\sigma_c$ we simply have to insert
the numerically known values of the DP exponents into
Eq.~(\ref{AnomalScalingRelation}). Surprisingly one obtains
$\sigma_c=2.0766(2)$ in one, $\sigma_c\simeq 2.2$ in two,
and $\sigma_c=2+\epsilon/12$ in $d=4-\epsilon$ spatial
dimensions. Thus, the crossover takes place at $\sigma_c>2$ which
collides with the intuitive argument that the anomalous
diffusion operator $\nabla_A^\sigma$ should only be relevant if
$\sigma<2$. But, as pointed out in Ref.~\cite{JOWH98}, this
naive argument may be wrong in an {\it interacting} theory where the
critical behavior is determined by a nontrivial fixed point of
a RG transformation. The field-theoretic calculation rather
predicts anomalous diffusion to be relevant in the
range $2 \leq \sigma < \sigma_c(d)$ for $d<4$. The surprising
conclusion would be that
\begin{equation}
\nabla_A^2 \neq \nabla^2
\end{equation}
in certain interacting theories. Loosely speaking, the tendency
to correlate particles in local spots of activity makes a DP process
more sensitive to long-range flights. Therefore, the relevance
of L\'evy flights sets in earlier than in the case of simple diffusion.

%---------------------------------------------------------------------------
\headline{A lattice model for anomalous directed percolation}
%---------------------------------------------------------------------------
%
%
\begin{figure}
\epsfxsize=140mm
\centerline{\epsffile{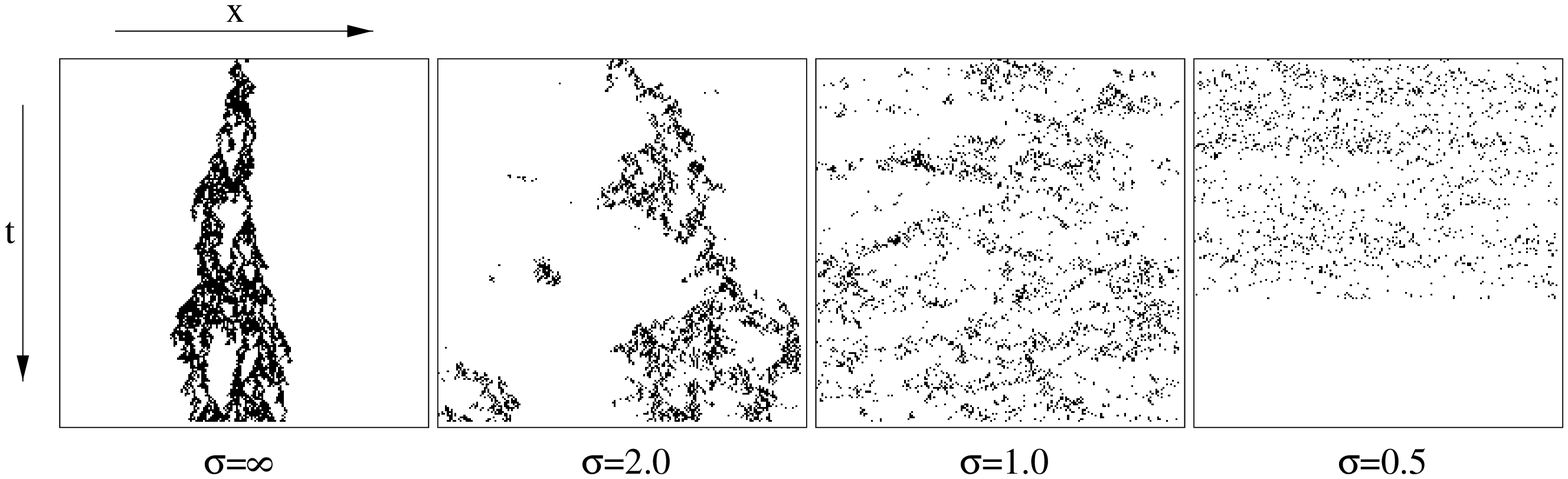}}
\vspace{2mm}
\smallcaption{
\label{FIGANOMDP}
Critical anomalous directed percolation
in 1+1 dimensions for different
values of $\sigma$. The figure shows
typical clusters starting
in the center of the lattice.
The case $\sigma=\infty$ corresponds
to ordinary DP. As $\sigma$ decreases, spatial
structures become more and more smeared out until in
the mean field limit $\sigma=1/2$ the
particles appear to be randomly
distributed over the whole system.
For small values of $\sigma$
the system quickly reaches the absorbing
state due to extremely strong
finite size effects.
}
\end{figure}
The field-theoretic predictions can be verified numerically by
studying a lattice model for anomalous DP that generalizes
directed bond percolation~\cite{HinrichsenHoward98}. The model is
defined on a tilted square lattice and evolves by parallel
updates. As usual, a binary variable $s_i(t)$ is attached to each
lattice site $i$. $s_i=1$ means that the site is active (infected)
whereas $s_i=0$ denotes an inactive (healthy) site. The dynamic
rules depend on two parameters, namely the control exponent
$\sigma>0$ and the bond probability $0\leq p \leq 1$. For a given
configuration $\{s_i(t)\}$ at time $t$, the next configuration
$\{s_i(t+1)\}$ is constructed as follows. First the new
configuration is initialized by setting $s_i(t+1):=0$. Then a loop
over all active sites $i$ in the previous configuration is
executed. In the (1+1)-dimensional case this loop consists of the
following steps:
\begin{enumerate}

\item   Generate two random numbers $z_L$ and $z_R$ from a flat
    distribution between $0$ and $1$.

\item   Define two real-valued spreading distances
    $r_L = z_L^{-1/\sigma}$  and $r_R = z_R^{-1/\sigma}$,
    for spreading to the left~(L) and to the right~(R). The
    corresponding integer spreading distances $d_L$ and $d_R$
    are defined as the largest integer numbers that are
    smaller than $r_L$ and $r_R$, respectively.

\item   Generate two further random numbers $y_L$ and $y_R$
    drawn from a flat distribution between $0$ and $1$, and assign
    $s_{i+1-2d_L}(t+1):=1$ if $y_L<p$ and
    $s_{i-1+2d_R}(t+1):=1$ if $y_R<p$.
        In finite systems the arithmetic operations in the indices
    are carried out modulo $N$ by assuming periodic boundary
    conditions, i.e. $s_i \equiv s_{i\pm N}$.

\end{enumerate}

\begin{figure}
\centerline{
\epsfxsize=80mm\epsffile{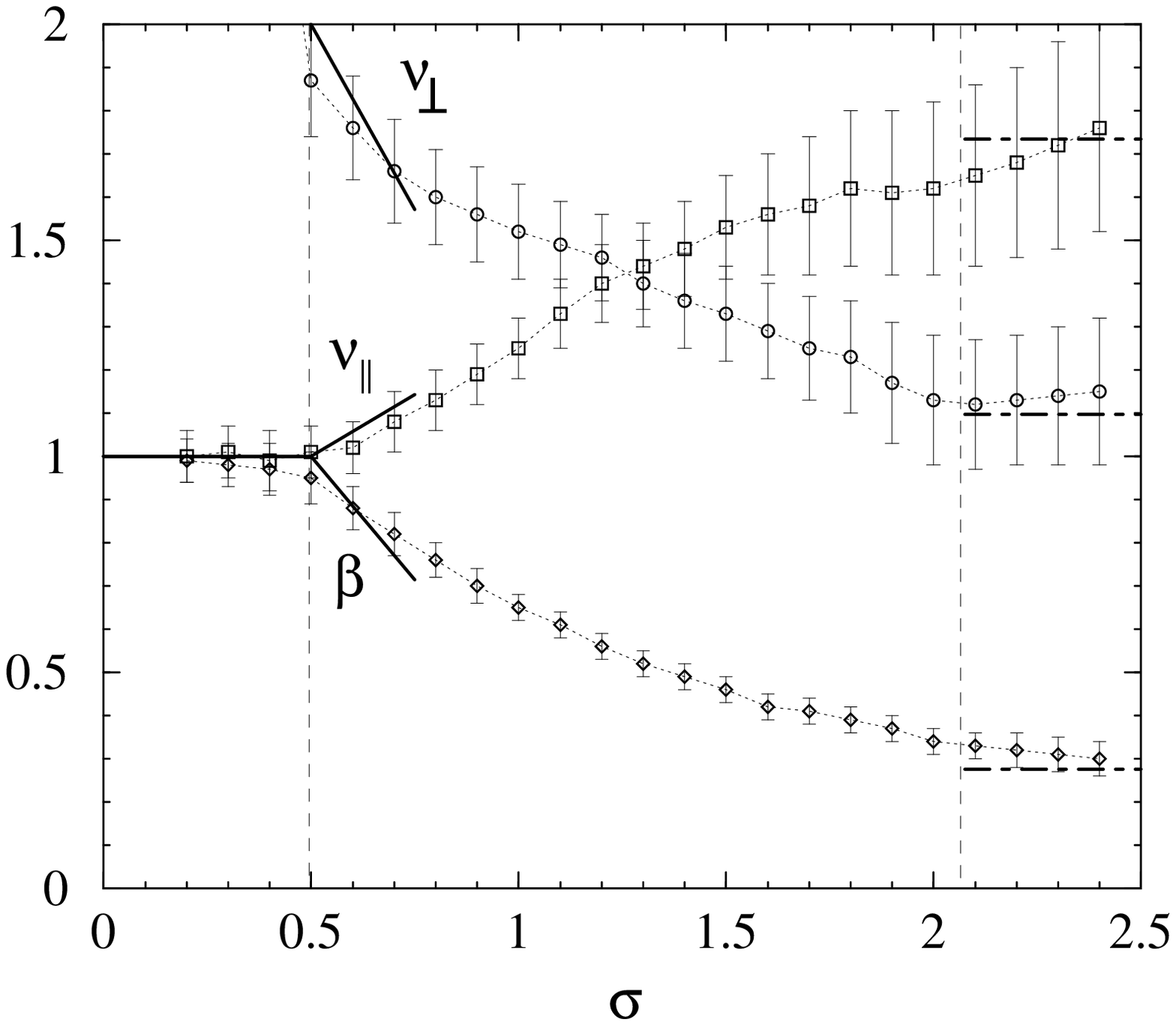}}
%\vspace{2mm}
\smallcaption{
\label{FIGANOMEXP}
Numerical estimates for the dynamic critical exponents $\beta$,
and the derived exponents $\nuperp$ and $\nu_{||}$
in comparison with the field-theoretic
predictions (solid lines) and the DP exponents
(dot-dashed lines). }
\end{figure}

\noindent 
This model includes two special cases. For $\sigma
\rightarrow \infty$ it reduces to ordinary directed bond
percolation (see Sec.~\ref{DPINTROSEC}). On the other hand, for
$\sigma \rightarrow 0$ the interaction becomes totally random. In
this case the mean-field approximation becomes exact with a
transition taking place at $p_c=1/2$. In between, the spreading
properties of the model change drastically, as demonstrated in
Fig.~\ref{FIGANOMDP}. As can be verified easily, the assignment $r
= z^{-1/\sigma}$ reproduces the normalized probability
distribution
\begin{equation}
P(r) =
\left\{
\begin{array}{cl}
\sigma/r^{1+\sigma} & \text{if \ \ } r>1 \ , \\
0 & \text{if \ \ } r \leq 1 \ .
\end{array}
\right.
\end{equation}
As usual, the distribution has a lower cutoff
at $r_{min}=1$, representing the lattice spacing.
Yet, in contrast to other models~\cite{Albano96,Cannas97},
no upper cutoff is introduced.
In order to reduce finite-size effects,
the target site is determined by assuming
periodic boundary conditions, i.e., the
particle may `revolve' several times around the system.

An interesting aspect of anomalous DP is the possibility to choose
$\sigma$ in such a way that the critical dimension $d_c=2\sigma$
approaches the actual physical dimension where the simulations are
performed. Even in one spatial dimension this allows the one-loop
results~(\ref{AnomCriticalExponents}) to be verified. For example,
if $\sigma=1/2+\mu$, the critical dimension of the system is
$d_c=1+2\mu$. Hence the exponents in a (1+1)-dimensional system
change to first order in $\mu$ as
\begin{equation}
\label{AnomTheoryPredictions}
\begin{array}{ccl}
\beta &=& 1-8 \mu / 7 + O\left( \mu^{2}\right) \ ,
\\
\nuperp &=& 2-12 \mu / 7+ O\left( \mu^{2}\right) \ ,
\\
\nu_{||} &=& 1+4 \mu / 7+ O\left( \mu^{2}\right) \ ,
\end{array}
\qquad\qquad
\begin{array}{ccl}
z &=& 1/2 + 5 \mu / 7+ O\left( \mu^{2}\right) \ ,
\\
\delta &=& 1-12 \mu / 7+ O\left( \mu^{2}\right) \ ,
\\
\theta &=& 4 \mu / 7+ O\left( \mu^{2}\right) \ .
\end{array}
\end{equation}
In Fig.~\ref{FIGANOMEXP}, the
predicted initial slopes are indicated by solid
lines. Clearly they are in fair agreement with the numerical
estimates, confirming the field-theoretic results
of Eq.~(\ref{AnomCriticalExponents}).
This is one of the rare cases where we can
directly `see' the field-theoretic results
in the simulation data.

\begin{figure}
\epsfxsize=80mm
\centerline{\epsffile{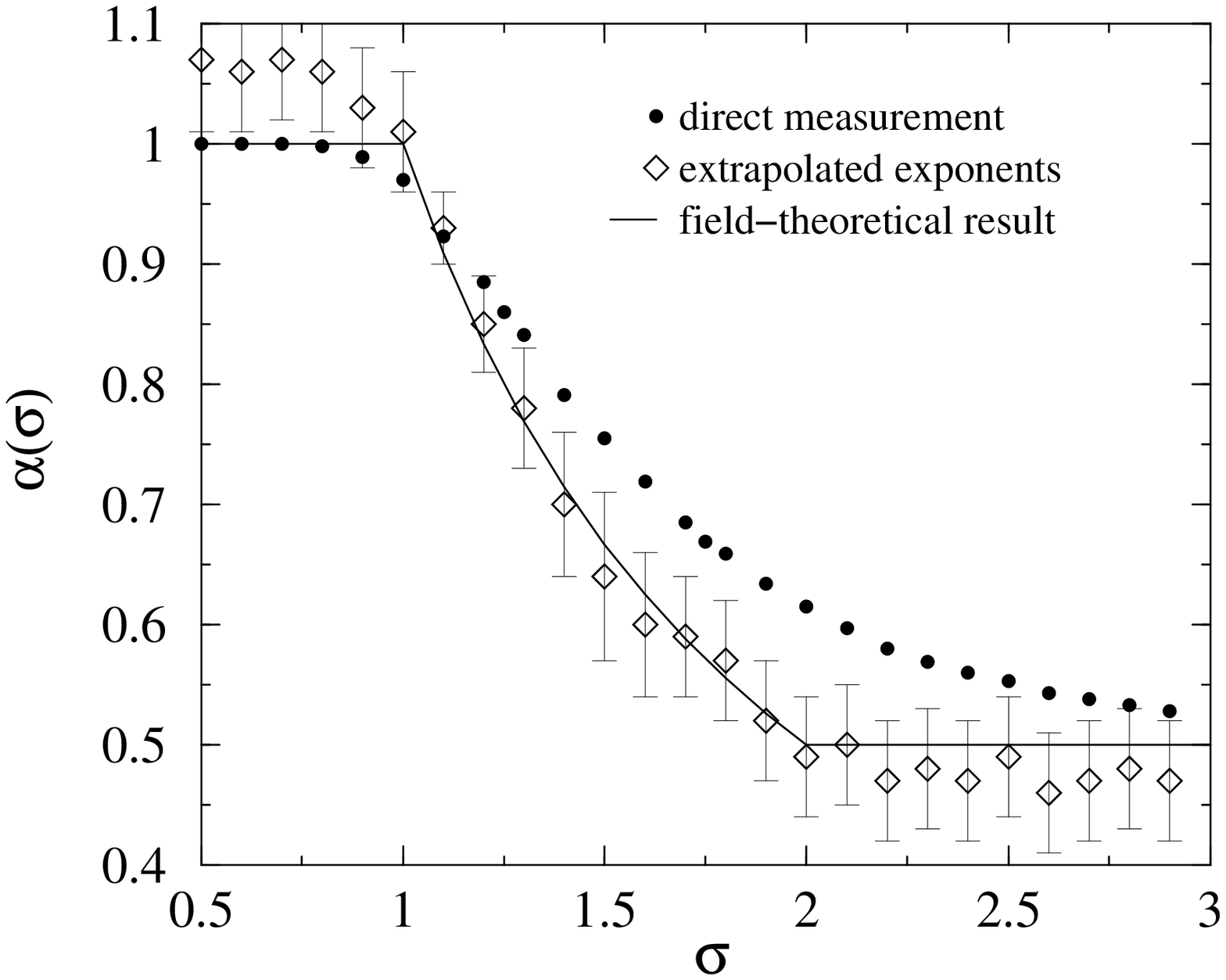}}
\smallcaption{
\label{FIGANOMANNH}
Anomalous annihilation process.
The graph shows direct and extrapolated estimates for the decay
exponent $\alpha$, as a function of $\sigma$. The solid line
represents the exact result (neglecting logarithmic corrections at
$\sigma=1$).}
\end{figure}
%
%
%---------------------------------------------------------------------------
\headline{Anomalous annihilation process}
%---------------------------------------------------------------------------
%
The more simple case of anomalous pair annihilation
$A+A \rightarrow \emptyset$ with long-range
hopping~\cite{ZumofenKlafter94} can be solved exactly by a
similar field-theoretic analysis.
In the ordinary annihilation process~\cite{Lee94}
with short-range interactions, the average particle density
is known to decay as in Eq.~(\ref{AnnhDens}).
The L\'evy-flight case may be described
theoretically by inserting an additional
operator $\nabla_A^\sigma$ into the
field-theoretic action for pair annihilation~\cite{Lee94}.
The resulting action, which can also be derived
systematically~\cite{HinrichsenHoward98}, reads
\begin{equation}
\label{AnnihilationAction}
S[\phi,\tilde{\phi}]
=
\int d^dx \, dt \,\Bigl\{
\tilde{\phi} (
\timederivative - D_N \nabla^2 - D_A \nabla_A^\sigma
)\phi
+  2 \lambda \tilde{\phi}\phi^2 + \lambda \tilde{\phi}^2\phi^2 -
\phi_0\tilde{\phi}\delta(t)
\Bigr\} \ ,
\end{equation}
where $\phi_0$ represents the initial (homogeneous) density at $t=0$.
An analysis of this action follows very closely that
of Ref.~\cite{Lee94}. For $\sigma<2$,
power counting reveals the upper critical dimension
of the model to be $d_c=\sigma<2$.
For $d>d_c$ mean-field theory is expected
to be quantitatively accurate, with an
asymptotic density decay $\sim t^{-1}$.
Below $d_c$, however, the renormalized reaction rate
flows to an order $\epsilon=\sigma-d$
fixed point. The decay of the
density can therefore
be predicted via dimensional arguments (see Fig.~\ref{FIGANOMANNH}):
\begin{equation}
\label{AnnhilationResult}
\rho(t) \sim \left\{ \begin{array}{ll}
    t^{-d/\sigma}       & {\rm for \ } d<\sigma \ , \\
    t^{-1} \ln t             & {\rm for \ } d=d_c=\sigma \ , \\
    t^{-1}              & {\rm for \ } d>\sigma \ .
    \end{array} \right.
\end{equation}
%
%
%
%==============================================================================
\subsection{Absorbing-state transitions in systems with additional symmetries}
%==============================================================================
%
\label{SEVABSSEC}
As outlined previously, directed percolation is the canonical
universality class for phase transitions into a single absorbing states.
According to the DP-conjecture, we may therefore expect non-DP behavior
to occur in systems with {\em several} absorbing states.
However, it is important to note that
the existence of several absorbing
states alone does not automatically lead to
non-DP behavior at the transition point.
As we have seen in Sec.~\ref{RELSEC}, even models with
infinitely many absorbing states may still belong
to the DP universality class. Non-DP critical behavior emerges
only if there is a {\em symmetry} among different absorbing states.

The first examples of non-DP transitions
into absorbing states were discovered by
Grassberger {\it et~al.}~\cite{GKT84,Grassberger89b}
who observed ``a new type of kinetic critical phenomenon''
in certain one-dimensional stochastic cellular automata.
The  density exponent $\beta \simeq 0.6(2)$ in 1+1 dimensions
was found to differ significantly from the usual
DP exponent $\beta \simeq 0.277$.
Partially because of the complicated dynamic rules of these models
it took almost ten years until the mechanism behind this type of
non-DP behavior was clearly identified.

Up to now two universality classes of spreading transitions with
non-DP behavior have been found,
namely the so-called {\em parity-conserving} class (PC) and
{\em $Z_2$-symmetric} directed percolation (DP2).
The PC class is represented most prominently by
branching-annihilating random walks with even number
of offspring (BAWE)~\cite{ZhongAvraham95}, where the number of particles
is preserved modulo $2$. The DP2 class, on the other hand,
which is also referred to as the directed Ising class,
introduces  {\it two symmetric} absorbing states. As we will see
below, both universality classes coincide in one spatial dimension
wherefore they are usually considered to be identical.
However, it is important to note that
they differ from each other in higher dimensions.

One may speculate whether systems with a symmetry among
several spreading agents will also be able to display novel
critical properties. However, such multi-color spreading processes
were found to belong to the DP class as well~\cite{Lipowski96}.
A field-theoretic analysis confirms this
observation~\cite{Janssen97a} and predicts
that the symmetry among the spreading agents
may be spontaneously broken.

%---------------------------------------------------------------------------
\headline{The parity-conserving universality class}
%---------------------------------------------------------------------------
%
%
\begin{figure}
\epsfxsize=110mm
\centerline{\epsffile{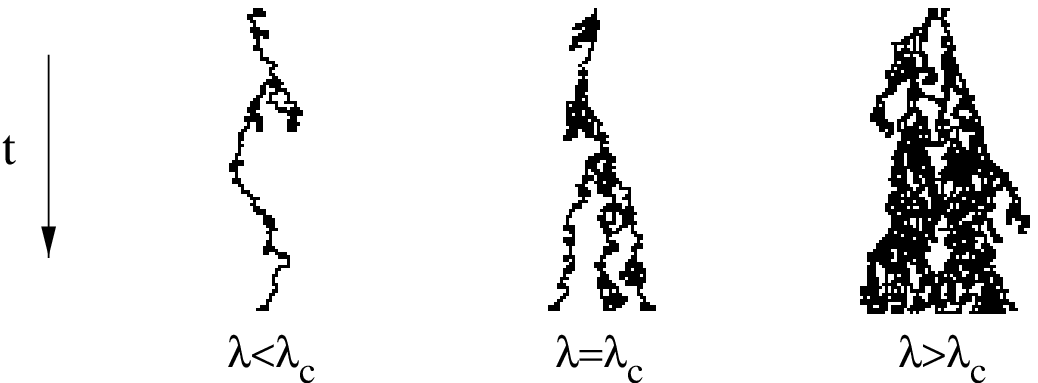}}
\smallcaption{
\label{FIPC}
The parity-conserving universality class.
Typical space-time trajectories of a branching annihilating random
walk with two offspring below, at and above the critical point.
}
\end{figure}
The parity-conserving universality class is represented most
prominently by branching annihilating walks
with an even number of offspring
\cite{ALR93,Jensen93c,Jensen94a,ZhongAvraham95,CardyTauber96,CardyTauber98}.
These non-conserved random walks are defined by the
reaction-diffusion scheme
\begin{equation}
\label{PC}
A \underset{\lambda}{\rightarrow} (n+1)A
\ , \qquad
2A \underset{\annh}{\rightarrow} \vacancy
\ ,
\end{equation}
where $n=2,4,6,\ldots$ denotes the number of offspring. As
an essential feature, this process
conserves the number of particles modulo $2$.
Early numerical studies in 1+1 dimensions
assumed instantaneous on-site
annihilation $\annh=\infty$. In this case the model
displays a continuous phase transition only for
$n \geq 4$~\cite{ALR93,Jensen93c,Jensen94a}, while there is
no such transition for $n=2$.
Later Zhong and ben-Avraham demonstrated that a phase transition also
emerges in the case of two offspring,
provided that the annihilation rate $\alpha$
is finite~\cite{ZhongAvraham95}.
Fig.~\ref{FIPC} shows a typical cluster grown from a
single seed. In contrast to DP,
a PC process starting with an odd number of particles
cannot reach the empty state since at least one particle is
left. Therefore, initial states with even and odd number
of particles are expected to lead to different cluster
morphologies.

The relaxational properties of PC models in the subcritical
phase differ significantly from the standard DP behavior.
While the particle density in DP models decays
exponentially as $\rho(t) \sim e^{-t/\xi_\parallel}$,
in PC models it decays algebraically in the
long-time limit. More precisely, the temporal
evolution of PC processes in the inactive phase is governed
by the annihilation process $2A\rightarrow \vacancy$.
Under RG transformations we therefore expect
the system to flow towards the fixed point of particle annihilation.
Consequently, in the subcritical phase the particle density
decays algebraically as in Eq.~(\ref{AnnhDens}).

A systematic field theory for PC models has been presented in
Refs.~\cite{CardyTauber96,CardyTauber98}, confirming the existence
of such an annihilation fixed point. However, the field-theoretic
treatment of the (1+1)-dimensional case poses considerable
difficulties. They stem from the presence of {\em two} critical
dimensions: $d_c=2$, above which mean-field theory applies, and
$d^\prime_c \approx 4/3$, where for $d>d^\prime_c$
($d<d^\prime_c$) the branching process is relevant (irrelevant) at
the annihilation fixed point. Therefore, the physically
interesting spatial dimension $d=1$ cannot be accessed by a
controlled $\epsilon$-expansion down from upper critical dimension
$d_c=2$. Nevertheless the usual scaling theory is still valid
below $d_c^\prime$. Currently the best numerical estimates of the
critical exponents in 1+1 dimensions are
\begin{alignat}{2}
\label{PCExponents}
\beta&=0.92(2)\,,& \qquad\qquad \delta+\theta&=0.286(2)\,, \nonumber \\
\nupar&=3.22(6)\,, & \qquad\qquad 2/z&=1.15(1)\,, \\
\nuperp&=1.83(3)\,. \nonumber
\end{alignat}
The actual values of $\delta$ and $\theta$ in dynamic simulations
depend on the initial condition. If the process starts with a single
particle, it will never stop, hence $\delta=0$.
On the other hand, if the initial seed consists of two particles,
one observers that the roles of $\delta$ and $\theta$ 
are exchanged, i.e. $\theta=0$.

It has been customary to investigate whether the numerical estimates
of the critical exponents can be fitted by simple rational numbers.
In fact, the estimates of Eq.~(\ref{PCExponents}) are in good
agreement with the rational values~\cite{Jensen94a}
$\beta=12/13$, $\nupar=42/13$, $\nuperp=24/13$,
$\delta+\theta=2/7$, and $\tilde{z}=8/7$.
In particular, $\beta/\nuperp$, the exponent for the decay
of spatial correlations at criticality, should be equal to $1/2$.
In the past, however, rational values were also proposed for
the DP exponents and later disproved by more accurate numerical
estimates~\cite{Jensen96b}.

\begin{figure}
\epsfxsize=90mm \centerline{\epsffile{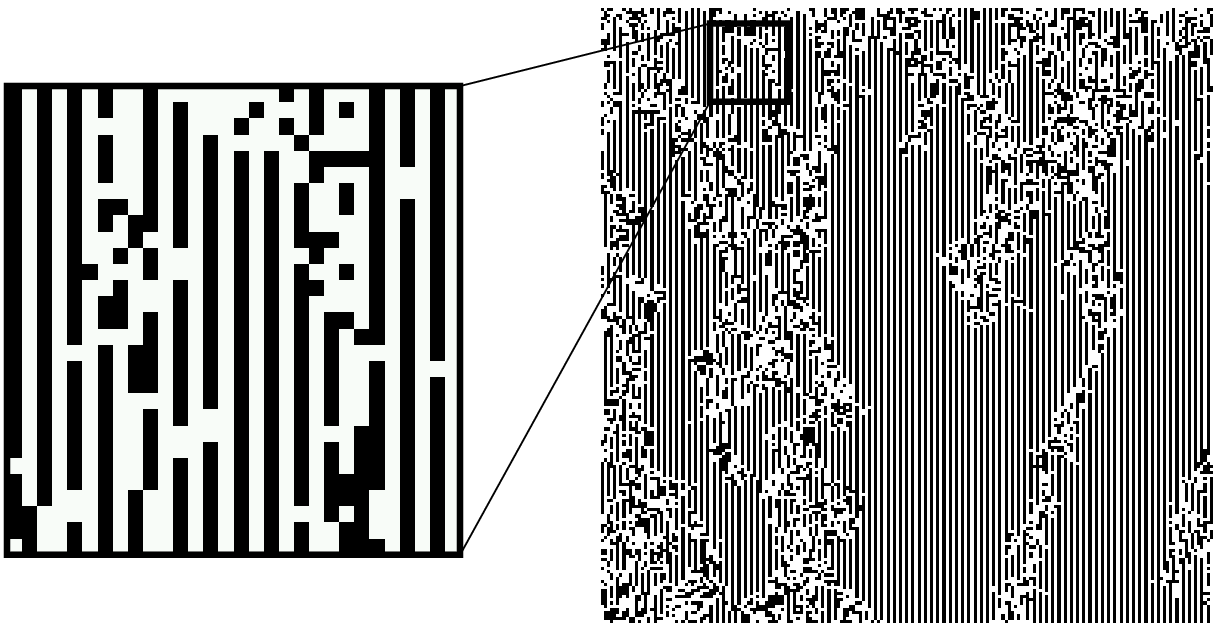}} \vspace{2mm}
\smallcaption{ \label{FIGMDM} Temporal evolution of the
(1+1)-dimensional interacting monomer-dimer model of
Ref.~\cite{KimPark94}. The two absorbing states emerge as
alternating configurations with $A$-particles at even and odd
sites, respectively. }
\end{figure}
%
%
%---------------------------------------------------------------------------
\headline{The DP2 universality class}
%---------------------------------------------------------------------------
%
The DP2 universality class describes phase transitions in
spreading models with two {\em symmetric} absorbing states. Since
the two absorbing states compete one another, the resulting
critical behavior is different from ordinary directed
percolation\footnote{Without symmetry, one of the two absorbing
states will dominate so that the critical behavior crosses over to
DP after sufficiently long time.}. In various models, for example
in certain cellular automata~\cite{GKT84,Grassberger89b} as well
as in interacting monomer-dimer
models~\cite{KimPark94,PKP95,HKPP98,Bhattacharyya98}, the two
absorbing states emerge as checkerboard-like configurations of
particles at even or odd sites, respectively (see
Fig.~\ref{FIGMDM}). Other DP2 models explicitly introduce two
symmetric inactive states, as, for example, nonequilibrium Ising
models~\cite{Menyhard94,MenyhardOdor96}, $Z_2$-symmetric
generalizations of the DK model and the 
contact process~\cite{Hinrichsen97}, monomer-monomer
surface reaction models with two absorbing states~\cite{BBB97},
and certain cellular automata~\cite{BBP98}. Even in
monomer-monomer models with three absorbing states a DP2
transition emerges at certain points in the parameter
space~\cite{BasslerBrowne96,BasslerBrowne97}.

On a phenomenological level, spreading transitions with several
absorbing states may be defined by introducing a single active
state $A$ and $n$ symmetric absorbing states $I_1,\ldots,I_n$
which can be regarded as having different colors. The system evolves
according to the following descriptive rules:

\begin{enumerate}
\item   Spreading of activity:
    Active sites turn their inactive nearest neighbors into
    the active state.
\item   Spontaneous recovery:
    Active sites turn spontaneously into an inactive
    state of a randomly chosen color.
\item   Boundaries between inactive domains of different
    colors are free to separate again, leaving active sites
    in between.
\end{enumerate}
Rules 1 and 2 resemble the usual infection and recovery
processes of DP.
Rule 3 is new and distinguishes different colors.
Roughly speaking, this rule ensures that inactive domains of
different colors cannot stick together irreversibly, rather
they will always be separated by fluctuating active `interfaces'.
The symmetry under global permutation of the colors ensures
that absorbing domains of different colors compete one another, leading
to interesting critical behavior.

Following these descriptive rules, we can introduce a generalized
version of the Domany-Kinzel cellular
automaton (see Sec.~\ref{LATTICESEC})
with $n$ absorbing states~\cite{Hinrichsen97}. It uses the same
type of lattice and is defined by the conditional probabilities
\begin{equation}
\label{GeneralizedDK}
\begin{split}
P(I_k\,|\,I_k,I_k)&=1\,, \\
P(A|A,A)=1-n\,P(I_k|A,A)&=p_2\,, \\
P(A|I_k,A)=P(A|A,I_k)&=p_1\,,  \\
P(I_k|I_k,A)=P(I_k|A,I_k)&=1-p_1\,,\\
P(A|I_k,I_l)&=1\,,
\end{split}
\end{equation}
where $k,l=1,\ldots,n$ and $k\neq l$.
Notice that rule 3 is implemented by transition $I_kI_l \rightarrow A$,
ensuring that active sites are created between
two inactive domains of different colors.
For $n=1$ the above model reduces to the original
Domany-Kinzel model. For $n=2$ it displays a DP2 phase
transition at the critical threshold $p_{1,c}=p_{2,c}=0.5673(5)$.
\begin{figure}
\epsfxsize=130mm
\centerline{\epsffile{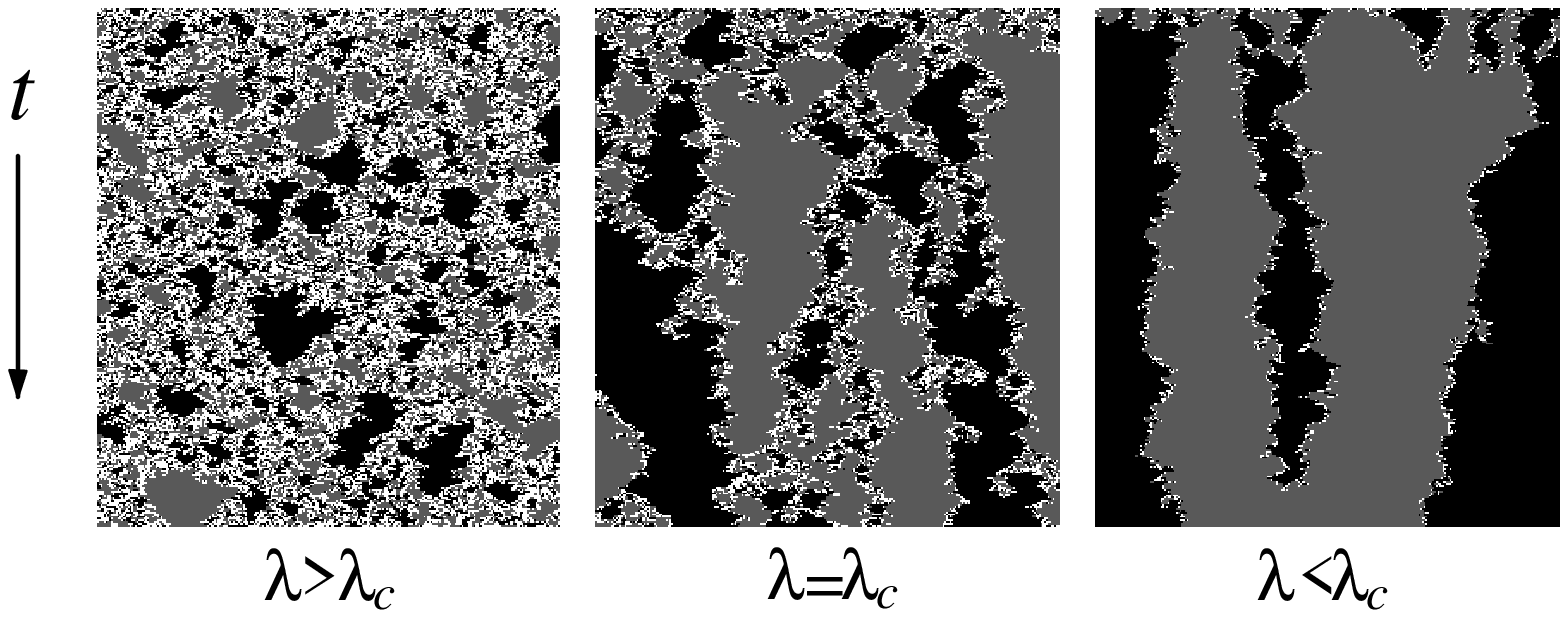}}
\smallcaption{
\label{DP2SIMUL}
Simulation of the generalized contact process with $n=2$
absorbing states starting from a random initial condition.
The two different types of inactive domains are shown in black and grey.
The active sites between the domains are represented by white pixels.
}
\end{figure}
Similarly it is possible to define a generalized contact process
by the rates
\begin{equation}
\begin{split}
w_{AA \rightarrow AI_k}=w_{AA \rightarrow I_kA} &= 1/2n\,,\\
w_{AI_k \rightarrow I_kI_k}=w_{I_kA \rightarrow I_kI_k} &= 1/2\,,\\
w_{AI_k \rightarrow AA}=w_{I_kA \rightarrow AA}&= \lambda/2\,,\\
w_{I_kI_l \rightarrow I_kA}=w_{I_kI_l \rightarrow AI_l} &= \lambda/2\,,
\end{split}
\end{equation}
Here the last equation implements rule 3.
For $n=1$ this model reduces to the usual
contact process introduced in Sec.~\ref{LATTICESEC}. For $n=2$ the
model undergoes a DP2 transition at the critical point
$\lambda_c=1.592(5)$.
A typical evolution of the generalized contact process in
1+1 dimensions is shown in Fig.~\ref{DP2SIMUL}. In the active phase
$\lambda>\lambda_c$ small inactive islands of
random color are generated which survive only for a short time.
Approaching the phase transition their size and
lifetime grows while the density of active sites decreases.
Notice that according to rule 3 a thin film of active
sites separates different inactive domains. As expected, the
numerical estimates of the critical exponents are in agreement
with the PC exponents~(\ref{PCExponents}).
Both models can easily be generalized to higher dimensions.
However, in higher dimensions the phase transition
is presumably characterized by mean field behavior. 
Similarly, increasing the number of absorbing
state does not lead to new universality classes. Simulations with
$n \geq 3$  symmetric absorbing states in 1+1 dimensions indicate that the
system is again described by mean field exponents.

If the $Z_2$ symmetry of DP2 models is broken by an external field,
the critical behavior at the transition crosses over to
ordinary DP~\cite{ParkPark95}. Roughly speaking the external field
favors one of the absorbing states so that
pairs of kinks between oppositely oriented
inactive domains form bound `dipoles' of a certain
size. Interpreting these dipoles as composite particles, they
recombine and produce a single offspring at certain rates,
resembling an ordinary DP process on large scales.

\begin{figure}
\epsfxsize=120mm
\centerline{\epsffile{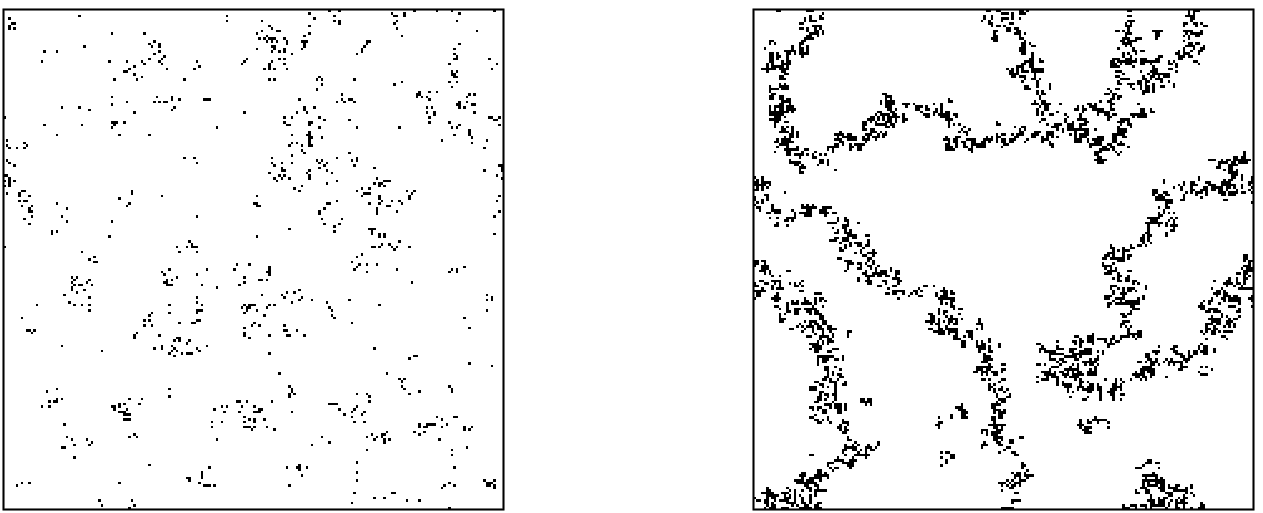}}
\smallcaption{
\label{FIGCOMPARE}
Configurations of an almost critical 
PC (left) and a DP2 process (right) in $d=2$ dimensions.
}
\end{figure}

%---------------------------------------------------------------------------
\headline{The difference between the PC and DP2 universality classes}
%---------------------------------------------------------------------------
%
In the DP2 class the `kinks' between differently colored domains may
be interpreted as branching-annihilating particles with even number of
offspring. Although the number of active sites is generally not
conserved modulo 2, we may associate with each active island
between differently colored inactive domains a particle $X$.
Obviously, these particles perform an effective branching-annihilating random
walk $X \rightarrow 3X, 2X \rightarrow \vacancy$.
Therefore, the DP2 class and the PC class coincide in
1+1 dimensions. However, it is important to note that they
are different in higher dimensions. Active sites of PC models
in $d\geq 2$ dimensions can be considered
as branching-annihilating {\em walkers},
whereas DP2 models describe the dynamics of
branching-annihilating {\em interfaces}
between oppositely oriented inactive domains
(see Fig.~\ref{FIGCOMPARE}). Therefore,
the corresponding field theories are expected to be different.
A field theory for the PC class was presented in
Ref.~\cite{CardyTauber96,CardyTauber98},
whereas the development of field theories for branching-annihilating
interfaces started only recently~\cite{Cardy98}.

In order to understand the difference between PC and DP2,
it is helpful to consider two other universality classes which
also coincide in 1+1 dimensions, namely the annihilation process
$A+A \rightarrow \vacancy$ and the voter
model~\cite{Durrett88}. The voter model is a two-state
model with spins $s_i=\pm 1$. It evolves by random-sequential updates
$+- \rightarrow ++/--$ with equal rates. Interpreting
$+-$ kinks as particles $A$, the voter model and the
annihilation process coincide in 1+1 dimensions. However,
in higher dimensions they are different. In fact, even their
Langevin equations turn out to be different. As shown in
Sec.~\ref{FLUCSEC}, the Langevin equation of the annihilation
process reads
\begin{align}
&\timederivative \rho(\xvec,t) = - \lambda \rho^2(\xvec,t) +
\diff \nabla^2 \rho(\xvec,t) + \noise(\xvec,t)\ , \\
&\langle \noise(\xvec,t) \noise(\xvec',t')  \rangle = \nonumber
\namp \, \rho^2(\xvec,t) \, \delta^d(\xvec-\xvec') \, \delta(t-t')
\ ,
\end{align}
where $\rho(\xvec,t)$ represents the coarse-grained density of
$A$-particles. On the other hand, the Langevin equation of the
voter model~\cite{DickmanTretyakov95} is given by
\begin{align}
&\timederivative \rho(\xvec,t) = \diff \nabla^2 \rho(\xvec,t) + \noise(\xvec,t) \\
&\langle \noise(\xvec,t) \noise(\xvec^\prime,t^\prime)\rangle = \nonumber
\namp \, \rho(\xvec,t)\,[1-\rho(\xvec,t)]\,\delta^d(\xvec-\xvec^\prime)
\delta(t-t^\prime)
\ ,
\end{align}
where $\rho(\xvec,t) \in [0,1]$ represents the local orientation
of the domain. Obviously, the two equations stand for
different universality classes.

%
%

%---------------------------------------------------------------------------
\headline{DP2 surface critical behavior}
%---------------------------------------------------------------------------
%
The influence of an absorbing wall in systems with a DP2
transition was studied in Ref.~\cite{LFH98}. Analyzing the
generalized Domany-Kinzel model~(\ref{GeneralizedDK})
with two absorbing states in a
semi-infinite geometry, it turned out that absorbing and
reflective boundary conditions play complementary roles. Moreover,
since $\beta$ and $\beta^\prime$ may be different in the DP2
class, one has to introduce {\em two} different surface exponents
$\beta_s$ and $\beta_s^\prime$. For absorbing boundary conditions
the numerical estimates in a (1+1)-dimensional DP2 process are
\begin{equation}
\beta_s=1.34(2) \, , \qquad \beta_s^\prime=2.06(2)  \ .
\end{equation}
For reflecting boundary conditions the two values are
simply exchanged. This property is related to a duality transformation
in parity-conserving processes~\cite{MSS98}. It is quite remarkable
that the two surface exponents seem to obey the scaling relation
\begin{equation}
\frac12 (\beta_s + \beta_s^\prime) = \nupar - 1  \, , \qquad (d=1)
\end{equation}
generalizing the conjecture $\beta_s = \nupar-1$
in the case of DP.

%==============================================================================
\subsection{Activated random walks}
%==============================================================================
%
\label{ARWSEC}
So far we considered nonequilibrium phase transitions where a
parameter (e.g. the percolation probability $p$) has to be tuned
to criticality. Other systems with conserved dynamics
can be tuned to criticality by varying the particle density
in the initial state.
An interesting example is the activated random walk
of $n$ particles~\cite{DMVZ00}. In this model each site can be
occupied with arbitrarily many particles. Sites with at least
two particles are active, i.e., their particles may move
independently to randomly selected neighbors. Sites with only
one particle are frozen. On an infinite lattice,
this model has infinitely many absorbing states. The control
parameter is the density of particles. For a low density,
the model quickly evolves into one of the absorbing states,
whereas it remains active for high particle densities.
Near the transition, the stationary density of active
sites $\rho_{stat}$ scales as
\begin{equation}
\rho_{stat} \sim (\zeta-\zeta_c)^\beta\, ,
\qquad
\zeta=n/N \, ,
\end{equation}
with the critical point
$\zeta_c \simeq 0.9486$ in 1+1 dimensions~\cite{DMVZ00}.
However, unlike other models with infinitely many absorbing state
(see Sec.~\ref{RELSEC}) the transition
does not belong to the DP universality class. For example,
in 1+1 dimensions the measured exponent $\beta=0.43(1)$ differs
significantly from the expected DP value $\beta_{DP} \simeq 0.277$.
Obviously, this deviation is due to the conservation law.
Hence, activated random walks provide a new universality class
of phase transitions into absorbing states, caused by an
additional symmetry, namely particle conservation.

%==============================================================================
\subsection{Absorbing phase transitions and self-organized criticality}
%==============================================================================
%
In contrast to ordinary transitions into absorbing states,
self-organized critical systems exhibit long-range correlations
and power laws without being tuned to a certain critical point.
In some sense the  annihilation process
$A+A\rightarrow \vacancy$ (see Sec.~\ref{REAC}) can be considered
as a simple example of self-organized criticality. Without tuning
of a parameter the annihilation process
generates long-range correlations with
power-law characteristics. Like many other coarsening processes and
growth phenomena the driving mechanism is a stationary state
which is approached but never reached.

The term `self-organized criticality' (SOC) refers to a different
type of models that are attracted to a {\it stationary} critical
state without being tuned to a critical point. The chief examples
are the sandpile model~\cite{BTW87}
and the Bak-Sneppen model~\cite{BakSneppen93}.
The concept of SOC has been used to explain the large variety of
power laws observed in nature, for example $1/f$ noise,
the distribution of earthquakes, and the dynamics of financial markets,
to name only a few~\cite{Bak96}.
For more than one decade SOC was considered as
a quite separate field of theoretical statistical physics, being more
or less unrelated to conventional phase transitions with a tuning
parameter. Recently, it was pointed
out~\cite{NarayanMiddleton94,SJD95,DVZ98}
that SOC is in fact closely related
to ordinary phase transitions into (infinitely many) absorbing
states (for a survey see~\cite{DMVZ00}).
More precisely, two classes of SOC models have to be distinguished.
The first class of SOC models, exemplified by the Bak-Sneppen
model, employs extremal dynamics. In this case the site with an
extremal value is selected for the next update, i.e., the
dynamic rules provide a mechanism of global supervision.
In the second class, which is represented by the sandpile model,
the bulk dynamics is {\it conserved}. Here a slow driving force competes
with the loss of particles at the systems boundaries and drives the
system to criticality. Hence the process of self-organization 
is characterized by a {\em separation of time scales} 
for avalanches and driving.

In the latter case, SOC is related to a conventional
absorbing-state transition as follows. As explained in
Ref.~\cite{DMVZ00}, any system with conserved local
dynamics and a continuous absorbing-states
transition can be converted into a SOC model by (1) adding 
a process for increasing the density in infinitesimal steps,
and (2) implementing a process for decreasing the density 
infinitesimally while the system is active. For example, let us 
consider the activated random walk. Adding a process for random
deposition of particles and a process for loss of particles
during avalanches at the boundaries, we obtain the so-called
Manna sandpile model~\cite{Manna91} which is known to exhibit SOC.
Using a deterministic variant of the same model, one obtains the famous
Bak-Tang-Wiesenfeld model~\cite{BTW87}. However, in order to
create a SOC counterpart for directed percolation, a slightly different
recipe has to be used, as demonstrated in 
Ref.~\cite{SJD95,SornetteDornic96}.

%==============================================================================
\subsection{The annilation/fission process}
%==============================================================================
%
\label{PCPD}
As shown in Sec.~\ref{FLUCSEC}, 
the influence of fluctuations in (1+1)-dimensional
systems can be very different. In the annihilation process, for
example, the particles become anticorrelated, i.e., they try to be
far away from each other. As shown in Ref.~\cite{HowardTauber97},
this type of fluctuations corresponds to `imaginary' noise in
the Langevin equation. In a DP process, however, the particles are
highly correlated and the noise turns out to be real. Three years
ago Howard and T{\"a}uber posed the question whether it is
possible to {\it interpolate} between real and imaginary noise. As
a prototype of such a transition, they considered the
annihilation/fission process
\begin{equation}
2A \rightarrow 3A \,, \qquad 2A \rightarrow \vacancy
\end{equation}
with diffusion of single particles. Note that this is a binary
process, i.e., at least two particles are required to meet at the
same (or at neighboring) sites in order to self-destruct or to
create offspring. Moreover, there are no exceptional symmetries
such as parity conservation on the microscopic level. The model
exhibits a nonequilibrium phase transition between an active phase
and two non-symmetric absorbing states, namely the empty lattice
and the state with only one diffusing particle. Performing a
field-theoretic Howard and T{\"a}uber argued that this transition
should belong to an independent yet unknown universality class of
phase transitions.

The annihilation/fission process may be interpreted as a pair
contact process plus diffusion of single particles. 
As a consequence, the static background shown in
Fig.~\ref{FIGPCP} begins to fluctuate. Since the decay of the
particle density in the subcritical phase is governed by the
annihilation process $2A \rightarrow \vacancy$, it is natural
to expect that the transition does not belong to the DP class. 
Similarly, the possibility of DP2 critical behavior seems 
to be unlikely since there is no parity conservation or 
$Z_2$-symmetry in the system~\cite{CHS00,Hinrichsen00b}. 
Thus, the annihilation/fission process may represent 
a new type of nonequilibrium critical phenomenon which 
has not yet been studied before.
%
%
%==============================================================================
\subsection{The Lipowski model}
%==============================================================================
%
%
\label{LIPOWSKI}
A particularly interesting model with a non-DP transition in
$d=2$ spatial dimensions has been introduced recently by
Lipowski~\cite{Lipowski99}. It has infinitely many absorbing
states and is defined by extremely simple dynamic rules.

\begin{figure}
\epsfxsize=50mm
\centerline{\epsffile{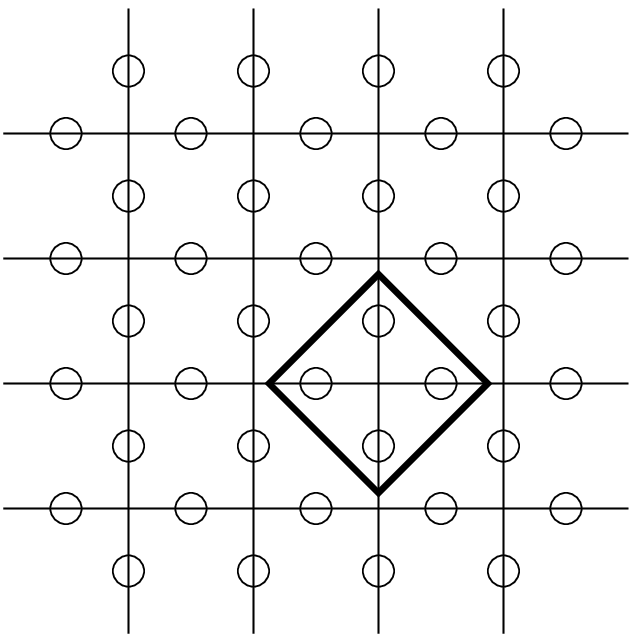}}
\vspace{0mm}
\smallcaption{
\label{FIGLIPOWSKI}
The Lipowski model. Each bond of a square lattice
carries a weight between zero and one. 
A plaquette of four bonds (hollow circles) is
randomly selected. If the sum of the four weights exceeds
a certain threshold, they are replaced by four independent
random numbers between zero and one.
}
\end{figure}

The model is defined as follows. The state of the system is specified
by real numbers $y_{ij} \in (0,1)$ which reside on the bonds of a
$d$-dimensional square lattice (see Fig.~\ref{FIGLIPOWSKI}). 
Initially all these variables are
randomly distributed between $0$ and $1$. The model evolves
by random sequential updates depending on a control parameter
$p$ which plays the role of a percolation probability.
For each update attempt a site $i$ is selected at random.
Its local `field' $h_i$ is defined by the sum of the $y$-variables
of the four connecting bonds. If $h_i>p$ the four variables
are replaced by random numbers drawn from a flat distribution
between $0$ and $1$. Otherwise the update attempt is abandoned.
As usual in models with random-sequential dynamics, each
update attempt corresponds to a time increment of
$\Delta t=1/N$, where $N$ is the total number of sites.
A site is called `active' if it is susceptible for a replacement,
i.e. $h_i>p$. Measuring the density of active sites in a
numerical simulation, the model displays a continuous phase
transition between a fluctuating active and a frozen phase.
Clearly, the model has infinitely many absorbing states.

The Lipowski model poses a puzzle: In one dimension the static
exponent $\beta$ coincides with the DP exponent $\beta\simeq 0.276$.
In two dimensions, however, the static exponent $\beta$ is found
to be much smaller than the expected value $0.58$. Surprisingly,
the measured value seems to coincide with the DP value in one
dimension. Lipowski argued that the model should
provide a mechanism for {\em dimensional reduction},
placing the two-dimensional critical phenomenon into a
one-dimensional universality class. However, the observed
coincidence may well be accidental.
Moreover, it is not obvious how such a mechanism should work.
It is therefore an interesting open question whether dimensional
reduction can be observed in stochastic lattice models.

\newpage

%##############################################################################
\section{Damage spreading}
%##############################################################################
%
%
\label{DamageSection} One of the central problems of dynamic
system theory is the dependence of the system's temporal evolution
on the initial conditions. It is well known that nonlinear systems
with deterministic dynamics may be extremely sensitive to small
perturbations of the initial state. Even in simple systems such as
in a periodically driven pendulum, a small variation of the initial
parameters can change the entire trajectory completely. If the
distance between two infinitesimally different trajectories
diverges during the temporal evolution, a dynamic system is said
to exhibit {\em chaotic} behavior.

The notion of {\em chaos} has been introduced in the context of
deterministic systems where a trajectory is uniquely determined
by the initial condition. In random processes, however, trajectories
are not uniquely determined; the time evolution of such a system
is not reproducible and therefore the usual definition of chaotic behavior
does not apply. Nevertheless, one may pose the question how the
system responds to changes in the initial condition.
It would be interesting to know, for example, how the biological
evolution on earth would have been affected if the initial
conditions were slightly different. In order to address this
question, Kauffman introduced the concept of
{\em damage spreading} (DS)~\cite{Kauffman69,Kauffman84}.
In damage spreading simulations~\cite{Herrmann90}
two copies (replicas) $S,S^\prime$ of a stochastic model
evolve simultaneously. Initially the two copies differ
only on a small number of sites. This difference
is considered as a small perturbation (damage)
in one of the two systems. Moreover, it is assumed
that the replicas evolve under identical realizations
of thermal noise, i.e., both copies use the same sequence of
random numbers in the simulation. If the number of sites
in different states, the so-called Hamming distance
\begin{equation}
\label{HammingDistance}
\Delta(t)=\sum_i \Delta_i(t) = \sum_i 1-\delta_{s_i(t),s_i^\prime(t)}
\end{equation}
does not go to zero in the long-time limit, damage is said to
spread, indicating high sensitivity with respect to the initial
condition. Otherwise, if $\Delta(t)$ vanishes,
damage is said to heal, indicating
a weak influence of the initial condition.
In order to get statistically meaningful results, the Hamming
distance has to be averaged over many
realizations of randomness.

Damage spreading first appeared in physics literature in the mid
eighties~\cite{Martin85,DerridaStauffer86,Creutz86,SSKH87,DerridaWeisbuch87}
and attracted considerable interest and attention.
The main reason behind this initial enthusiasm was the hope
that damage may spread in some regions of a system's parameter
space and disappear elsewhere, indicating
the existence of chaotic and regular phases in stochastic
systems. The initial enthusiasm abated during subsequent years,
the main reason being an apparent lack of an {\it objective
measure} whether damage does or does not
spread in a given system. More precisely, it was realized
that the location of the phase boundaries may depend on
details of the algorithmic implementation.
However, if spreading or healing of damage
indicated some intrinsic property of the system,
one would not expect the result to depend on details
of the algorithm used to generate its dynamics.
Meanwhile DS has been applied to a large variety of models (see
Table~\ref{DSAPPLTAB}). In view of the vast literature on DS simulations,
it is therefore necessary to carefully analyze the conceptual problems
of this technique.

\begin{table}
\footnotesize
\begin{center}
\begin{tabular}{|l|l|}
\hline &\\[-3mm]
{\bf Equilibrium models:} & Ref. \\[1mm]
\hline
Ising ferromagnet &
\cite{Creutz86,SSKH87,MHA90,GSS90,GlotzerJan91,JanArcangelis95} \\
Heisenberg model  & \cite{MirandaParga91}  \\
XY model          & \cite{ChiuTeitel90}  \\
Potts and Ashkin-Teller models & \cite{Mariz90,STM97,BMM97}  \\
discrete $N$-vector ferromagnets & \cite{MST93}  \\
spin glasses & \cite{ACH89,CCC89,CampbellArcangelis91,Campbell93,CampbellBernardi94,WKS96,ABC95,WVS97} \\
Ising model with microcanonical constraints & \cite{MSM96}  \\
diluted Ising model & \cite{Grassberger96,SMNC97} \\
Ising model in quenched random fields & \cite{Vojta97b}  \\
Frustrated Ising systems & \cite{JuniorNobre97}  \\
Layered Ising systems & \cite{LLC97}  \\
\hline  &\\[-3mm]
{\bf Nonequilibrium models:} & \\[1mm]
\hline
Kauffman model & \cite{Kauffman69,Kauffman84,CSJ87,Stauffer89b}  \\
The game of life & \cite{BRR91,MonettiAlbano95}  \\
ZGB model & \cite{Albano94b}  \\
Ohta-Jasnov-Kawasaki and Ginzburg-Landau model & \cite{Graham94}  \\
Irreversible reaction-diffusion processes & \cite{Albano94c,Albano95c}  \\
Models for surface reactions & \cite{Albano95d}  \\
Restricted solid on solid growth models & \cite{KLK96}  \\
Models of the PC class & \cite{OdorMenyhard98}  \\
SOC models & \cite{Bhowal97}  \\
Traveling salesman problem & \cite{MirandaParga89}  \\
Boolean random networks & \cite{LuqueSole97}  \\
\hline
\end{tabular}
\end{center}
\smallcaption{
\label{DSAPPLTAB}
Some applications of damage spreading.
}
\end{table}

In the following we discuss several aspects of DS.
First we present a simple example in order to explain
how a DS simulation depends on the algorithmic implementation. From
a more mathematical point of view, this phenomenon can also be
understood by analyzing the joint master equation of the two replicas.
Because of their algorithmic dependence, DS simulations are ambiguous
and cannot be used as a criterion for `chaotic' and regular phases.
However, to some extent the ambiguity of DS can be overcome by an
algorithm-independent definition of DS phases. In this approach
the entire family of physically legitimate algorithms
for a given dynamic system is considered as a whole.
Furthermore, we summarize what is known about the universal
properties of DS transitions. Finally we discuss several applications of
DS simulations.

%==============================================================================
\subsection{Damage spreading phases}
\begin{figure}
\epsfxsize=120mm
\centerline{\epsffile{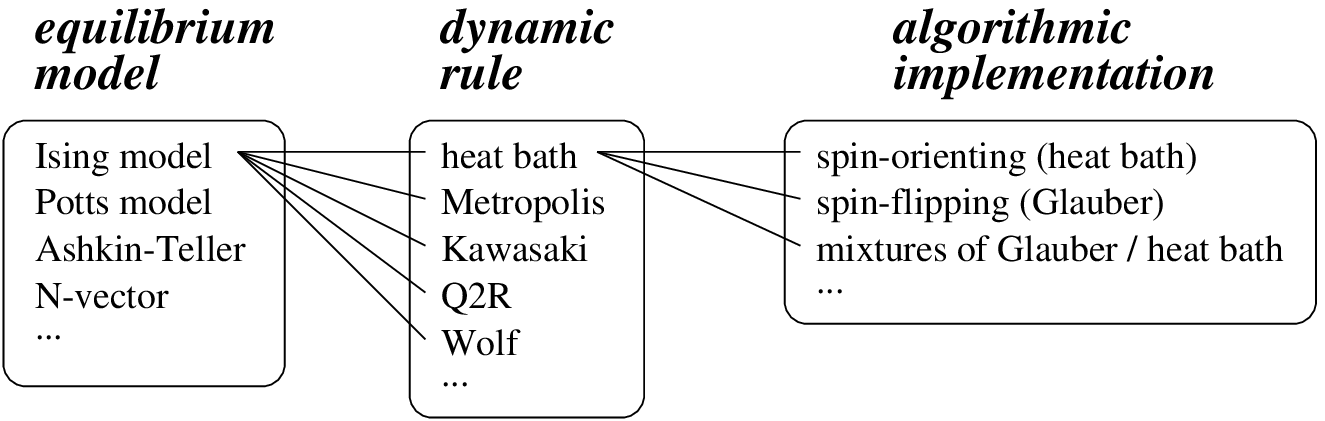}}
\vspace{0mm}
\smallcaption{
\label{FIGHIERARCHY}
The thermodynamic ensemble of an equilibrium model (left) may be generated
by different dynamic procedures (center) that are characterized
by specific transition probabilities. Each of these dynamic
procedures in turn can be realized by different algorithms
(right). These algorithms are fully equivalent and
cannot be distinguished by an observer of a
single system. However, their damage spreading properties
turn out to be different (see text).
}
\end{figure}

%==============================================================================
\vspace{-4mm}
%
%---------------------------------------------------------------------------
\headline{Algorithmic dependence of damage spreading simulations}
%---------------------------------------------------------------------------
%
The conceptual problem of DS was first discovered in the
Domany-Kinzel cellular automaton (cf. Fig.~\ref{FIGDKPHASE}).
Martins {\it et~al.}~\cite{MRTM91}
observed that in a certain region of the active phase
damage spreads and heals elsewhere.
Subsequently several other authors determined the boundary
of this region with increasing
accuracy~\cite{ZebendePenna94,MZPT94,RSS94,Grassberger95b}.
Independently, mean-field type approximations of
varying complexity confirmed the existence of this
`chaotic phase'~\cite{RSS94,KohringSchreckenberg92,Bagnoli96,Tome94}.
Its boundary, however, was shown to depend on the manner in which
the dynamic procedure of the DK model is implemented on a
computer~\cite{KohringSchreckenberg92,Bagnoli96}, while
the evolution of a single replica is completely insensitive to
the algorithmic implementation. This prompted
Grassberger~\cite{Grassberger95b} to observe that
\begin{quote}
{\it ``it is misleading to speak of different phases in the DK
automaton, ...instead these are different phases
for very specific algorithms for simulating
pairs of such automata''.}
\end{quote}
To understand the algorithmic dependence of DS simulations,
let us consider a much simpler system, namely the Ising model.
In this context it is important to note
that there are different levels of variety
in stochastic lattice models (see Fig.~\ref{FIGHIERARCHY}).
As explained in  Sec.~\ref{THEQU},
the equilibrium ensemble of the Ising model can be
generated by various different dynamic rules. For example, heat bath
and Metropolis dynamics represent two different dynamic systems
which have the same stationary state.
Initially it was hoped that DS would not depend
on the intrinsic dynamics, allowing regular and chaotic phases
to be identified as equilibrium
properties~\cite{Creutz86,SSKH87,DerridaWeisbuch87}. However,
later it was realized that different dynamic procedures
(such as Glauber versus Metropolis~\cite{MHA90,JanArcangelis95},
Q2R~\cite{GSS90}, or Kawasaki dynamics~\cite{GlotzerJan91})
exhibit different DS properties. Yet, this is not surprising since
the dynamic rules, although generating the same equilibrium ensemble,
represent different dynamic systems. Since DS is a dynamic phenomenon
it is quite natural that it depends 
on the dynamics under consideration. Similarly,
it is not surprising that DS depends on the type of
updates~\cite{NMS92,GPJ93} and the interaction range~\cite{Manna90}.

The real conceptual problem of DS occurs at the second level
in Fig.~\ref{FIGHIERARCHY}: On a computer each dynamic rule can
be realized by several algorithms. Regarding a single system,
these algorithms are fully equivalent and cannot be distinguished.
However, in DS simulations they lead to different results.
To understand this apparent paradox, let us consider
the Ising model with spin-orienting
(standard heat bath) and spin-flipping (Glauber)
dynamics. Both algorithmic prescriptions represent the {\em same}
dynamic rule which mimics the contact of the Ising model
with a thermal reservoir by means of local spin dynamics.
More precisely, an observer of a single system who analyzes the
trajectories would be unable to decide whether its temporal
evolution was generated by standard heat bath or Glauber dynamics.
This can already be seen in the example of a single-spin Ising model
at infinite temperature. Since in this case the spin $\sigma(t)=\pm 1$
changes randomly in time, the transition rates are simply given by
$w_{-1 \rightarrow +1}=w_{+1 \rightarrow -1}=1$. These transition
probabilities {\em define} our dynamic system.
However, this system can be realized by two different algorithms,
namely by spin-orienting updates (standard heat bath dynamics)

\tt
\qquad \qquad z=rnd(0,1); \\ \indent
\qquad \qquad if (z<0.5) $\sigma(t+dt)$=1; else $\sigma(t+dt)$=-1;
\rm

\noindent
and by spin-flipping updates (Glauber dynamics)

\tt
\qquad \qquad z=rnd(0,1); \\ \indent
\qquad \qquad if (z<0.5) $\sigma(t+dt)$=$-\sigma(t)$;
else $\sigma(t+dt)$=$\sigma(t)$ \ .
\rm

\noindent
It is obvious that both procedures are fully equivalent on a single
replica, i.e., an observer would be unable to decide which of the
two algorithms has been used. The difference between the two algorithms
may become evident only if we observe the evolution of two
replicas $S$ and $S'$ of the system in a DS simulation:
For spin-orienting updates an initial `damage'
$\sigma(0)=-\sigma(0)^\prime$ heals immediately while it
is preserved when spin-flipping updates are used.

A similar algorithmic dependence can be observed in the full
Ising model~(\ref{IsingModel}) at finite temperature.
Defining the transition probability
(cf. Eq.~(\ref{IsingTransitionProb}))
\begin{equation}
p_i(t)= \frac{e^{h_i(t)}}{e^{h_i(t)}+e^{-h_i(t)}}
\end{equation}
we may express the update rules of heat bath dynamics by
\begin{equation}
\sigma_i(t+1)={\rm sign}[p_i(t)-z_i(t)] \,,
\end{equation}
where $z_i(t)\in (0,1)$ are equally distributed random numbers.
The corresponding update rule for Glauber dynamics is given by
\begin{equation}
\label{Glauber}
\sigma_i(t+1)=
\left\{ \begin{array}{ll}
+{\rm sign}[p_i(t)-z] & \mbox{if $ \sigma_i(t)=+1$} \\
-{\rm sign}[1-p_i(t)-z] & \mbox{if $ \sigma_i(t)=-1$}
\end{array} \right. .
\end{equation}
It is easy to verify that for given
$\{\sigma_{i-1}(t),\sigma_i(t),\sigma_{i+1}(t)\}$,
the probability to get $\sigma_i(t+1)= +1$ is the
same in both cases. Hence, by observing the temporal evolution of
a {\it single} Ising system, one cannot tell which of
the two methods was used to generate the trajectory in
configuration space. The two algorithms can only be
distinguished when two copies are simulated in parallel,
i.e., by studying damage spreading.

Investigating the two-dimensional Ising model with Glauber
dynamics, Stanley {\it et al.}~\cite{SSKH87} and Mariz {\it
et~al.}~\cite{MHA90} found that damage spreads in the disordered
phase $T>T_c$ and heals elsewhere. Performing more precise
simulations, Grassberger~\cite{Grassberger95c} realized that the
DS transition occurs slightly below~$T_c$. This observation was
also be supported by a mean field theory~\cite{Vojta97a}.
Moreover, the critical point of the DS transition was found to
vary continuously when mixtures of Glauber and heat bath dynamics
are used~\cite{MarizHerrmann89}. Very similar properties are
observed in the three-dimensional Glauber
model~\cite{LeCaer89a,LeCaer89b,Gropengiesser94,Grassberger95c}.
Turning to the Ising model with spin orienting (standard heat
bath) dynamics, it is possible to prove that damage does not
spread at any temperature in any
dimension~\cite{MHA90,DerridaWeisbuch87}.

%
%
%---------------------------------------------------------------------------
\headline{The master equation of damage spreading simulations}
%---------------------------------------------------------------------------
%
In order to understand the algorithmic dependence
of DS simulations from a more fundamental
point of view, let us consider two copies
$S_1$ and $S_2$ of an arbitrary nonequilibrium system with
asynchronous dynamics (random-sequential updates).
Each of the two systems evolves according to a master
equation with a given Liouville operator ${\cal L}$.
The total system $S=(S_1,S_2)$ constitutes a new
nonequilibrium system that evolves according
to a joint master equation with a certain Liouville operator ${\cal M}$.
If both systems were using independent random numbers,
${\cal M}$ would be given by the
tensor product ${\cal L} \otimes {\cal L}$.
However, in DS simulations the use of the same sequence
of random numbers leads to nontrivial correlations
between $S_1$ and $S_2$.

According to the definition of damage spreading,
the Liouville operator ${\cal M}$ of the total system
$(S_1,S_2)$ is restricted by certain physical constraints.
On the one hand, each replica is required to
evolve according to its own natural dynamics. Hence, by integrating
out the degrees of freedom of one of the replicas
we obtain the Liouville operator of the other system:
\begin{align}
\label{DamageRestriction1}
&\sum_{s_1} \langle s_1,s_2 | {\cal M} | s_1^\prime,s_2^\prime \rangle
= \langle s_2 | {\cal L} | s_2^\prime \rangle
\qquad \mbox{for all } s_2,s_1^\prime,s_2^\prime \ ,\\
& \sum_{s_2} \langle s_1,s_2 | {\cal M} | s_1^\prime,s_2^\prime \rangle
= \langle s_1 | {\cal L} | s_1^\prime \rangle
\qquad \mbox{for all } s_1,s_1^\prime,s_2^\prime
\ .
\end{align}
These restrictions already imply probability conservation for
the total system. On the other hand, the trajectories
of the two replicas,  once they have reached the same
state (no damage), have to be identical:
\begin{equation}
\label{DamageRestriction2}
\langle s_1,s_2 | {\cal M} | s^\prime,s^\prime \rangle
= \langle s_1 | {\cal L} | s^\prime \rangle \,
\delta_{s_1,s_2} \ .
\end{equation}
The ambiguity of damage spreading simulations is due to the fact that
the restrictions (\ref{DamageRestriction1})-(\ref{DamageRestriction2})
do not fully determine the Liouville operator ${\cal M}$ of the
total system. This can easily be verified by counting the degrees of
freedom. For a system  with $n$ configurations the restrictions
determine less than $3n^3$ of the $n^4$ matrix elements of ${\cal M}$
so that damage spreading is ambiguous for $n \geq 3$. But even in the
case of a single-spin Ising model with
$n=2$ states one can show that only $14$ of $24$ equations are linearly
independent so that two out of $16$ matrix elements of ${\cal M}$
can be chosen freely. The remaining degrees of freedom are the
origin of the algorithmic dependence of DS simulations.
A similar ambiguity occurs in the joint transfer matrix of
models with parallel dynamics~\cite{HWD97}.

%---------------------------------------------------------------------------
\headline{An algorithm-independent definition of damage spreading}
%---------------------------------------------------------------------------
%
In order to overcome these conceptual difficulties,
let us consider the entire {\em family} of dynamic
procedures consistent with certain physically dictated
constraints~\cite{HWD97}. Then for any particular system one
of the {\it three} possibilities may hold:
\begin{enumerate}
\item
Damage spreads for any member of the family of dynamic procedures.
\item
Damage heals for any member of this family.
\item
Damage spreads for a subset of the possible dynamic procedures, and heals
for the complementing subset.
\end{enumerate}
\begin{figure}
\epsfxsize=85mm
\centerline{\epsffile{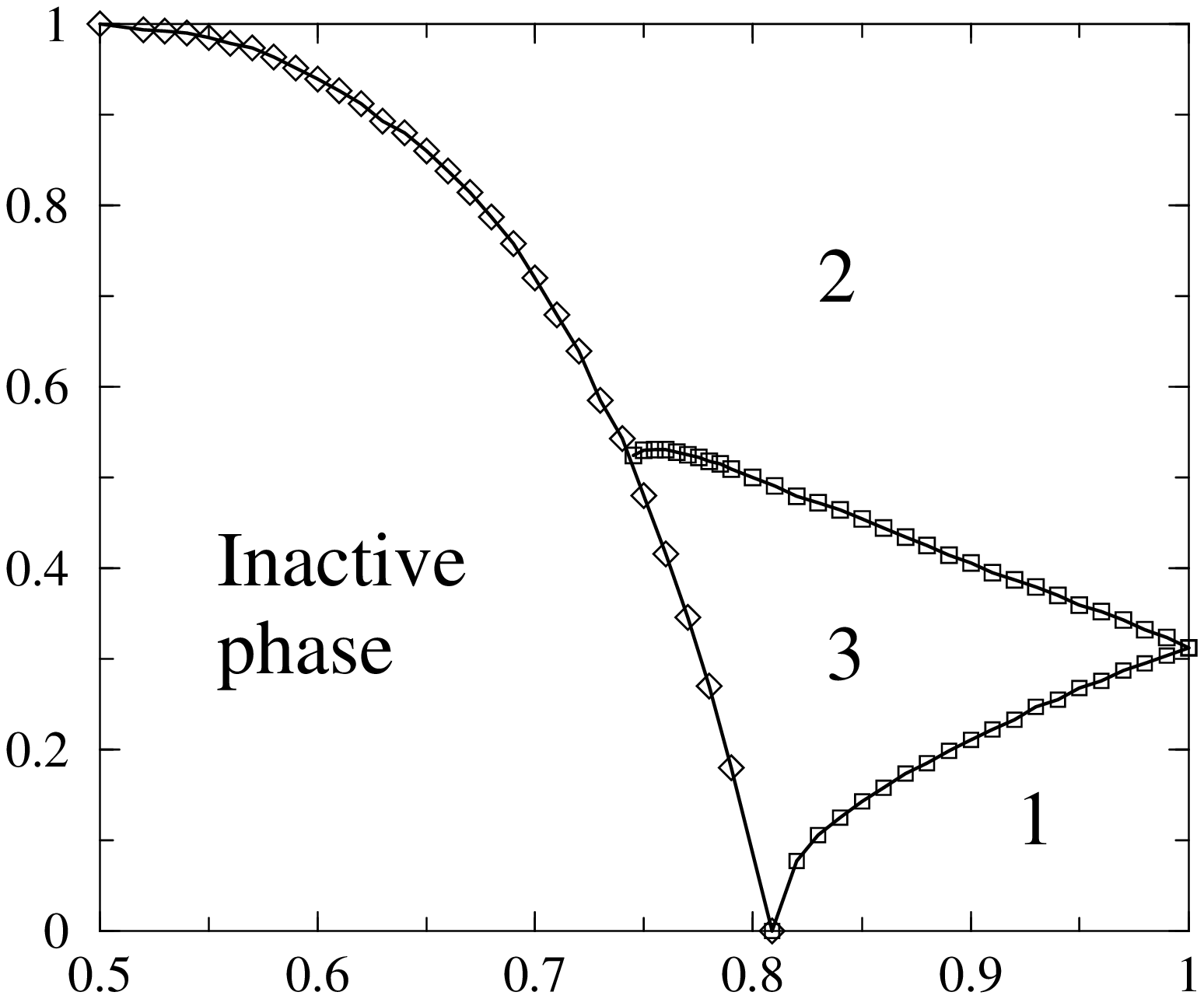}}
\smallcaption{
\label{FIGDKDAMAGE}
Algorithm-independent damage spreading phases in the
Domany-Kinzel model. 1: Damage never spreads.
2: Damage may spread, depending on the algorithm.
3: Damage always spreads.
}
\end{figure}
Obviously this allows us to classify damage spreading properties
in an algorithm-indepen\-dent manner. To this end we must,
however, consider simultaneously the {\it entire set} of
possible algorithms (dynamic procedures) which are consistent with the
physics  of the model under consideration. This set of physically
`legitimate' algorithms may be defined by certain restrictions that
are dictated by the dynamics of the single evolving system:

\newpage

\begin{itemize}
\item[a)]
Definition of DS: The dynamic rules for the evolution of the pair
of replicas are such that a single replica evolves
according to its `natural' dynamics. Once both replicas
have reached the same configuration, their temporal evolution
will be identical.
\item[b)]
Interaction range: The transition probabilities
for the combined system at site $i$ may depend only
on those sites that affect the evolution
of site $i$ under the dynamic rules of a single system.
\item[c)]
Symmetry: The rules that govern the evolution for
the pair of systems do not break any of the
symmetries of the single-replica dynamics.
\end{itemize}
The first restriction is simply a verbal formulation of
Eqs.~(\ref{DamageRestriction1})-(\ref{DamageRestriction2}).
The second condition tells us that the interaction range
in the combined system of two replicas must not exceed
the interaction range of a single system.
The third rule implies, for example, that
if there is a left-right symmetry in the evolution
of a single system, the same must hold for the pair of
replicas. It can be shown that these conditions suffice to
unambiguously determine a set of physically `legitimate'
algorithms. This was demonstrated for the one-dimensional
Domany-Kinzel model~\cite{HWD97} for which the phase diagram
in Fig.~\ref{FIGDKDAMAGE} was found.

Clearly, the subjectivity in defining DS phases,
as described before, has now been shifted to
selecting the restrictions defining
which DS procedure is `legitimate'. However, the specification
of such a family by means of physically motivated criteria
appears to be less arbitrary than choosing,
at random, one out of a continuum of
physically equivalent update procedures.
It should also be emphasized that the algorithm-independent
definition of DS phases does not mean that DS is
reflected in the dynamic behavior of a single system,
so that Grassberger's observation still holds:
DS is a property of a {\it pair} of stochastic systems.

\newpage 
%===========================================================================
\subsection{Universality of damage spreading transitions}
\begin{figure}
\epsfxsize=135mm \centerline{\epsffile{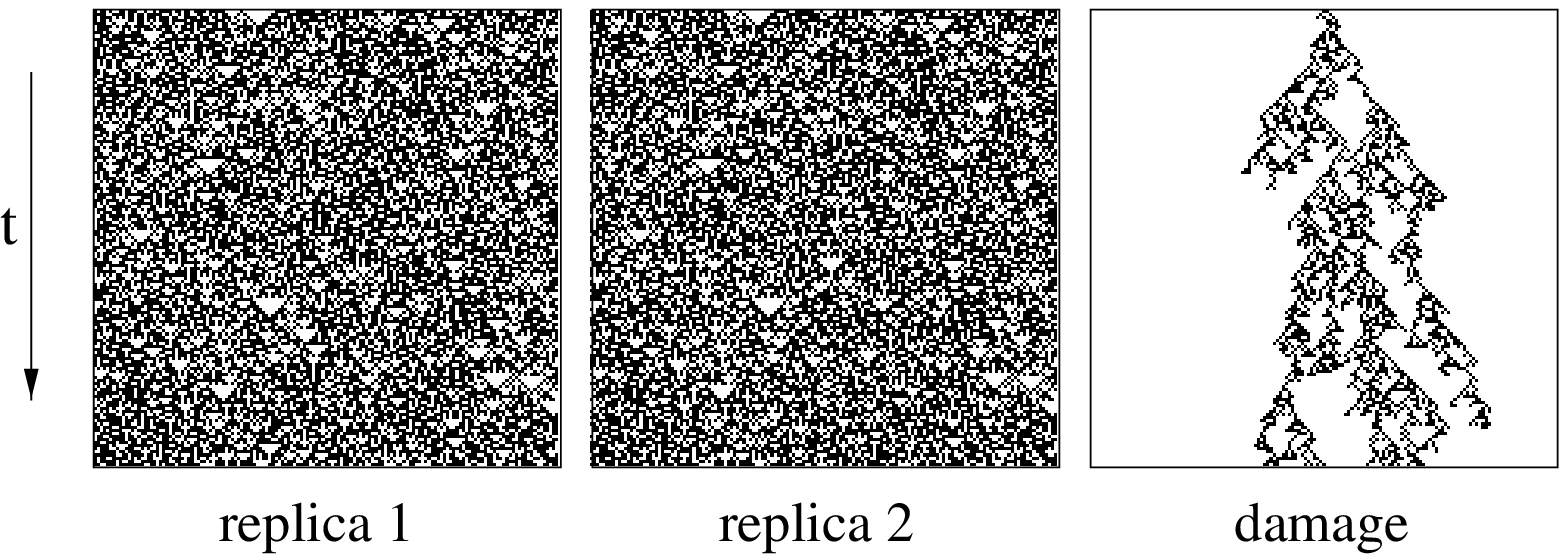}} \vspace{2mm}
\smallcaption{ \label{FIGDSUNIVDEMO} Damage spreading in the
(1+1)-dimensional Domany-Kinzel cellular automaton evolving in the
stationary active state at $p_1=0$, $p_2=0.313$. An initial damage
is introduced at the lattice site in the center. }
\end{figure}

%===========================================================================
%
\vspace{-4mm}
\label{DSUNIV}
%---------------------------------------------------------------------------
\headline{The DP conjecture for damage spreading}
%---------------------------------------------------------------------------
%
As can be seen in Fig.~\ref{FIGDSUNIVDEMO}, spreading
of damage is in many respects similar to spreading of
activity in a DP process. As in DP damage spreads to
nearest neighbors and heals spontaneously.
Once both copies have reached the same configuration
(no damage), their evolution will be identical,
i.e., they are confined to some `absorbing' subspace
from where they cannot escape.
In contrast to DP the spreading process depends
crucially on the actual evolution of the two replicas,
providing a fluctuating background in which the
spreading process takes place.
Nevertheless DS follows the same spirit
as the spreading of activity in a DP model.
This observation led Grassberger to the conjecture
that damage spreading transitions belong
generically to the directed percolation universality
class~\cite{Grassberger95b}.

So far analytical support for this conjecture came
from approximate mean-field arguments~\cite{Bagnoli96}
and an exact statement by
Kohring and Schreckenberg~\cite{KohringSchreckenberg92},
who noted that on the $p_2=0$ line the dynamics of
damage spreading in the DK automaton is precisely identical
to the evolution of the DK automaton itself, hence on this line DS
is trivially in the DP universality class. To prove this
statement, consider two replicas of $S$ and $S^\prime$
of a DK automaton evolving according to Eq.~(\ref{DKUpdateAlgorithm})
with equal random numbers $z_i(t)=z_i^\prime(t)$. The probabilities
$P_D(\Delta_i=1|s_{i-1},s_{i+1}; \ s_{i-1}^\prime,s_{i+1}^\prime)$
to generate damage at site~$i$ are
listed in Table~\ref{DAMAGETAB}. For $p_2=0$ they may be expressed as
\begin{equation}
P_D(\Delta_i=1|s_{i-1},s_{i+1}; \ s_{i-1}^\prime,s_{i+1}^\prime)
=\left\{
\begin{array}{l}
p_1 \ \ \mbox{if} \ s_{i-1}=s_{i+1} \ \mbox{and} \
          s_{i-1}^\prime \neq s_{i+1}^\prime \ , \\
p_1 \ \ \mbox{if} \ s_{i-1}\neq s_{i+1} \ \mbox{and} \
          s_{i-1}^\prime = s_{i+1}^\prime \ , \\
0 \ \ \ \ \mbox{otherwise}
\end{array}
\right.
\end{equation}
or equivalently
\begin{equation}
P_D(\Delta_i=1|s_{i-1},s_{i+1}; \ s_{i-1}^\prime,s_{i+1}^\prime)=
p_1(1-\delta_{\Delta_{i-1},\Delta_{i+1}})\,.
\end{equation}
Therefore, the damage variables  $\Delta_i(t)$ evolves
precisely according to the probabilistic rules of
a single DK automaton. Hence for $p_2=0$ the DS
transition belongs to the DP universality class.
This mapping of DS to DP was later
extended to other regions in the phase diagram of
the DK model~\cite{HWD97}. Although such an exact mapping is usually
not available for other models, various numerical simulations
show that most DS transitions are indeed characterized by DP exponents,
supporting Grassberger's conjecture. The same applies to
deterministic cellular automata with chaotic behavior
if a small noise is added~\cite{BRR92}.

Different DS properties may be expected in models with
cluster dynamics~\cite{WangSwendsen90}.
However, as pointed out in Refs.~\cite{Stauffer89a,Stauffer91},
it is difficult to extend the definition of
`using the same random numbers' to cluster algorithms
since the number of clusters may differ on both replicas.
This difficulty can be overcome by introducing a random
background field~\cite{HDS98}. Assigning a local random
number to each lattice site the new orientation
of a cluster depends on whether the sum of its random
numbers positive or negative. Although cluster algorithms
involve long-range correlations, DS transitions still belong
to the DP universality class unless they coincide with
the thermodynamic phase transition of the Ising system.

\begin{table}
\small
\begin{center}~
\begin{tabular}{|c||c|c|c|c|}
\hline
&\multicolumn{4}{c|}{$\sigma^\prime_{i-1}, \sigma^\prime_{i+1}$} \\
\hline
$\sigma_{i-1},\sigma_{i+1}$ & \bf 00 & \bf 01 & \bf 10 & \bf 11 \\
\hline \hline
\bf 00 & $\hspace{7mm} 0\hspace{7mm} $ & $p_1$ & $p_1$ & $p_2$ \\
\bf 01 & $p_1$ & $0$ & $0$ & $\max(p_1,p_2)$ \\
\bf 10 & $p_1$ & $0$ & $0$ & $\max(p_1,p_2)$ \\
\bf 11 & $p_2$ & $\max(p_1,p_2)$ & $\max(p_1,p_2)$ & $0$ \\
\hline
\end{tabular}
\end{center}
\smallcaption{
\label{DAMAGETAB}
Probabilities
$P_D(\Delta_i=1|s_{i-1},s_{i+1}; \ s_{i-1}^\prime,s_{i+1}^\prime)$
for the generation of damage in the DK model.
}
\end{table}
%

%---------------------------------------------------------------------------
\headline{DS transitions with non-DP behavior}
%---------------------------------------------------------------------------
%
The critical properties of a DS transition are expected to change
if one of the four conditions of the DP conjecture is violated.
For example, DS transitions in models with frozen randomness (such
as the Kauffman model~\cite{Kauffman69,Kauffman84}) do not belong
to the DP class. Non-DP behavior is also expected when the DS
order parameter exhibits an additional $Z_2$ symmetry (cf.
Sec.~\ref{SEVABSSEC}). In the context of damage spreading it is
important to note that such a symmetry should be a property of the
DS order parameter, i.e., the Hamming distance. For example, the
$Z_2$ symmetry of Ising systems is not sufficient -- inverting all
spins in {\it both} replicas does not change the Hamming distance
between the two configurations. Similarly, models with a non-DP
transition do not automatically exhibit non-DP damage spreading.
For example, Grassbergers cellular automaton A, which has a DP2
phase transition, displays an ordinary DS transition belonging to
DP~\cite{OdorMenyhard98}.

Non-DP behavior at the DS transition can be observed in systems with
a symmetry between two `absorbing states' of $\Delta_i(t)$,
one without damage $\Delta=0$ and the other with full damage $\Delta=1$.
The simplest example of such a symmetry is the Ising
model with Glauber dynamics. In fact, if both replicas are
in opposite states (full damage), they will always evolve through
opposite configurations. Unfortunately, there is no DS transition
in the one-dimensional Glauber model. However, by exploiting the
algorithmic freedom, it is possible to construct a modified
dynamic rule which exhibits a DP transition~\cite{HinrichsenDomany97}.
This rule depends on a parameter $\lambda$ and is defined by
\begin{align}
\label{MixDynamics}
&\sigma_i(t+1) =
\left\{
\begin{array}{ll}
+{\rm sign}(p_i(t)-z) & \mbox{if} \ y = 1 \\
-{\rm sign}(1-p_i(t)-z) & \mbox{if} \ y =-1 \\
\end{array}
\right. \,, \\
&y=\frac12 \sigma_i\Bigl[\Bigl(1+\sigma_{i-1}(t)\sigma_{i+1}(t)\Bigr)+
\Bigl(1-\sigma_{i-1}(t)\sigma_{i+1}(t)\Bigr) {\rm sign}(\lambda-\tilde{z})
\Bigr] \nonumber \ ,
\end{align}
where $z,\tilde{z}\in (0,1)$ are two independent random numbers.
On a single replica this update rule is fully
equivalent to Glauber and heat bath dynamics
for all $0 \leq \lambda \leq 1$. However, the effective rate
for spreading of damage depends on $\lambda$. For fixed
temperature $J/k_BT=0.25$ a DS transition with non-DP exponents
occurs at the critical value $\lambda^*=0.82(1)$.
This example demonstrates that additional symmetries of the
DS order parameter may lead to non-DP behavior at the DS
transition. A similar situation emerges in the
nonequilibrium Ising model introduced by
Menyh\'ard~\cite{OdorMenyhard98}.

%===========================================================================
\subsection{Applications of damage spreading}
%===========================================================================
%
\vspace{-4mm}
%---------------------------------------------------------------------------
\headline{Measurement of critical exponents in equilibrium models}
%---------------------------------------------------------------------------
%
Damage spreading simulations can also be used to determine certain static
and dynamic properties of systems at thermal equilibrium. This
application of DS was first demonstrated by Coniglio
{\it et~al.}~\cite{CAHJ89}
who showed that there exists an exact relation between
the Hamming distance $\Delta$  and
certain correlation functions. The essential idea is to
consider two copies $S,S^\prime$ of an Ising model with spins
$\sigma_i,\sigma_i^\prime=\pm 1$ and introducing a small damage
by keeping a single spin in one of the systems fixed
during the whole temporal evolution, say $\sigma_0^\prime=-1$.
Since this perturbation
breaks the symmetry between the two copies,
it is important to distinguish two different types of damage at site $i$,
namely $\sigma_i=1,\sigma_i^\prime=-1$
and $\sigma_i=-1,\sigma_i^\prime=1$.
The probabilities of finding these different types of damage
in the stationary state is given by
\begin{equation}
d^{+-}=\langle(1-s_i)s_i^\prime\rangle \ , \qquad
d^{-+}=\langle s_i(1-s_i)^\prime\rangle \ ,
\end{equation}
where $s_i=\frac12(\sigma_i-1)$ (see also Ref.~\cite{Herrmann90}).
Notice that these quantities can be understood as two-point
correlation functions between the two copies. However, their difference
\begin{equation}
\Gamma_i=d_i^{+-}-d_i^{-+}=\langle s_i \rangle - \langle s_i^\prime \rangle
\end{equation}
is a combination of one-point functions,
i.e., by taking a certain combination of the damage probabilities one
obtains quantities that describe the properties of a {\em single} system.
Therefore, such quantities do not depend on the algorithmic implementation.
In fact, using detailed balance and ergodicity one can prove that
\begin{equation}
\Gamma_i=\frac{C_{0,i}}{2(1-m)} \ ,
\end{equation}
where $C_{0,i}$ is the two-point correlation function and $m$ the
magnetization of the Ising model at thermal equilibrium. This relation
is exact and does not depend on the specific algorithmic implementation
used in the simulation.
Moreover, it can be shown by monotonicity arguments
that for standard heat bath dynamics the total damage $\Delta$ is related
to the static susceptibility $\chi$ by
\begin{equation}
\chi=2(1-m)\sum_i \Gamma_i\ .
\end{equation}
Thus, damage spreading simulations can be used to determine the
{\em static} exponents $\beta$ and $\nu$ and the critical temperature
of the Ising model (see
Refs.~\cite{SilvaHerrman88,BoissinHerrman91,BatrouniHansen92}).
Similar relations between Hamming distance
and correlation functions were found in certain lattice models
with absorbing states~\cite{DrozFrachebourg90}.

DS is also used as a tool for accurate measurements
of the dynamic exponent $z$
at the phase transition of equilibrium systems, provided
that the thermodynamic transition and the DS transition
coincide~\cite{GPJ92,WangSuzuki96}.
In this case the spreading exponent $z$ is not
given by the DP exponent $\nupar/\nuperp$,
instead it is expected to coincide with the dynamic
exponent $z$ of the model under consideration.
The knowledge of $z$ is important in order to estimate
the critical slowing down of given dynamic system.
In a series of papers the dynamic exponent
of the Ising model with heat bath dynamics has been measured in
two~\cite{MacIsaacJan92,HAMPJ93,Stauffer93,WHS95,MontaniAlbano95} and three
dimensions~\cite{PooleJan90,HAMPJ93,MHJ94,WHS95}.
After an initial controversy, Grassberger~\cite{Grassberger95d}
compared several simulation methods and found the estimates
$z=2.172(6)$ in two and $z=2.032(4)$ in three dimensions.
Later refined simulations confirmed these results~\cite{Gropengiesser95}.

\begin{figure}
\epsfxsize=120mm
\centerline{\epsffile{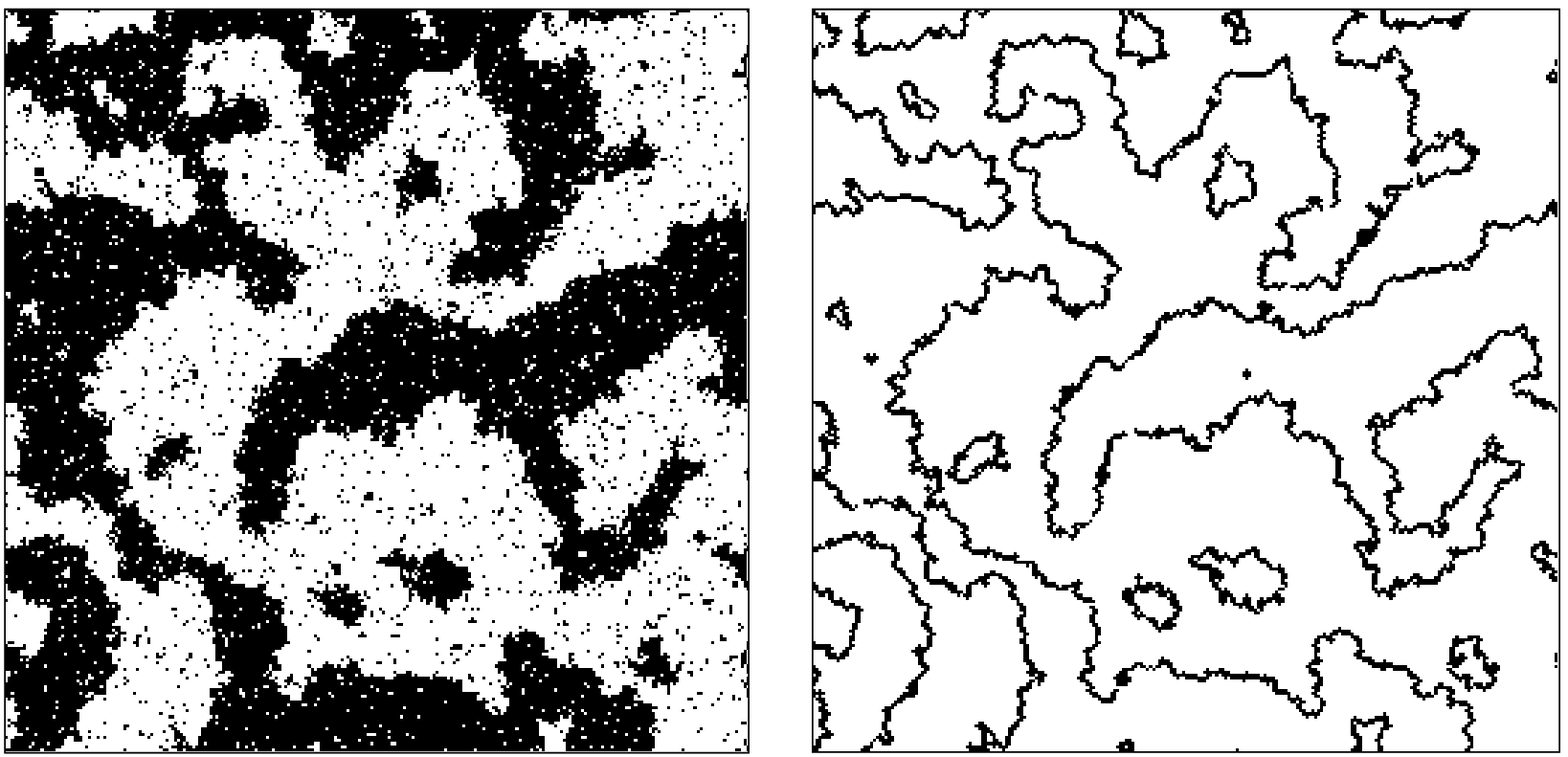}}
\vspace{2mm}
\smallcaption{
\label{FIGDOMAINWALL}
Identification of domain walls.
Left: The two-dimensional Ising model evolving by heat bath dynamics.
Right: Corresponding domain walls detected by the
observable $\Delta_i$ defined in Eq.~(\ref{DomainWallObs}).
Figure reprinted from Ref.~\cite{HinrichsenAntoni98}.
}
\end{figure}
%
%
%---------------------------------------------------------------------------
\headline{Identification of domain walls in coarsening systems}
%---------------------------------------------------------------------------
%
Damage spreading techniques can also be used to
identify domain walls in coarsening systems.
Coarsening phenomena~\cite{HohenbergHalperin77} occur
in various dynamic systems as, for example, in
the Ising model with Glauber dynamics. In the ordering phase,
starting with random initial conditions, patterns of ferromagnetic
domains are formed whose typical size grows with time as $t^{1/2}$.
For zero temperature, these domains are completely ordered and the domain
walls can be identified as bonds between oppositely
oriented spins. For nonzero temperature, however, it is difficult to
define domain walls as one has to distinguish
between `true' domains and islands of the minority phase generated
by thermal fluctuations. In order to identify coarsening domains
for $0<T<T_c$, Derrida~\cite{Derrida97} proposed to compare
two identical copies evolving under the same thermal noise.
One copy starts with random initial conditions and begins
to coarsen, whereas the other copy starts from a fully magnetized
state and remains ordered as time evolves. It is
assumed that all spin flips occurring in ordered replica can be
regarded as thermal fluctuations. Therefore, when a spin flip occurs
simultaneously in both replicas, it can be considered as a thermal
fluctuation, while it is a signature the coarsening process
otherwise. This method was used to determine the fraction of
persistent spins of the Ising model~\cite{DBG94}
as a function of time, confirming that the persistence
exponent does not change for $0 \leq T<T_c$.

Since Derrida's method allows only one type of domains to be
detected, it is impossible to identify the precise location
of domain walls. This can be overcome by comparing
three replicas instead of two~\cite{HinrichsenAntoni98}. As
before, the first copy starts with random initial conditions
and serves as the master copy in which the coarsening process
takes place. The other copies start from fully ordered initial
conditions with positive and negative magnetization, respectively.
Fluctuations in the first copy, which do not occur in the other two
replicas, indicate the presence of a domain wall. More precisely,
domain walls may be detected by the observable
\begin{equation}
\label{DomainWallObs}
\Delta_i(t)=
\biggl(
1-\prod_j\frac{1+\sigma^{(1)}_j(t)\sigma_j^{(2)}(t)}{2}
\biggr)
\biggl(
1-\prod_j\frac{1+\sigma^{(1)}_j(t)\sigma_j^{(3)}(t)}{2}
\biggr)\,,
\end{equation}
where the superscripts denote the replica. As shown in
Fig.~\ref{FIGDOMAINWALL}, this method works surprisingly well.
However, it turns out that the appearance depends on the
algorithmic implementation, i.e., Glauber and heat bath dynamics
leads to different results. This is not surprising as the method
relies on the same ideas as damage spreading.

%==============================================================================
\subsection{Damage spreading and experiments}
%==============================================================================
%
\label{DAMEXPERIMENTS}
In this Section we have seen that the concept of damage spreading
depends on the algorithmic implementation and therefore lacks
a well-defined meaning. The suggested algorithm-independent
definition of DS resolves these difficulties only  partly, since
it is based on certain (physically motivated) {\it ad hoc} assumptions.
Therefore, DS does not provide a strict definition of `chaotic'
and `regular' phases in stochastic systems. Nevertheless DS has
been useful to estimate critical exponents of certain nonequilibrium
phase transitions and to identify domain walls in coarsening system
at nonzero temperature. Concerning the question whether DS is relevant
for experiments, there is a clear answer:
DS is an artificial concept
which does {\em not} exist in nature. In particular,
there is no meaning of `using the same sequence of random
numbers' in an experimental system. DS is rather a simulation
technique for a pair of systems, taking advantage of our ability to use
deterministic pseudo random number generators. It is indeed not
surprising that such a concept, for which there is no correspondence in
nature, expresses its incompleteness by certain inconsistencies.

%---------------------------------------------------------------------------

\newpage

%##############################################################################
            \section{Interface growth}
%##############################################################################
%
\label{GrowthSection}
A further important field of statistical physics is the
study of crystal growth and transitions between different
morphologies of moving interfaces~\cite{Meakin88}.
During the last two decades there has
been an enormous progress in the understanding of growth
processes (for reviews see e.g.
Refs.~\cite{KrugSpohn91,MDVW95,BarabasiStanley95,Krug97}).
In this Section we will focus on certain classes of depinning
transitions which are closely related to nonequilibrium
phase transitions into absorbing states.

%==============================================================================
\subsection{Roughening transitions -- a brief introduction}
%==============================================================================
%
Models of growing interfaces may be realized
either on a discrete lattice
or by continuum equations. Discrete {\em solid-on-solid} (SOS)
models are usually defined on a $d$-dimensional square
lattice of $N=L^d$ sites associated with integer variables~$h_i$
marking the actual height of the interface at site $i$.
Clearly, this description does not allow for overhangs of
the interface.
The set of all heights $\{h_i\}$ determines the state of the
system which evolves according to certain stochastic rules
for adsorption and desorption. In {\em restricted} solid-on-solid
models (RSOS) the dynamic rules satisfy the additional constraint
\begin{equation}
\label{RSOS}
|h_{i}-h_{i+1}| \leq 1 \ ,
\end{equation}
i.e., adjacent sites may differ by at most one height step.
To characterize the evolution of the interface, it is useful to
introduce the mean height $\bar{h}(t)$ and the width $w(t)$:
\begin{equation}
\label{Width}
\bar{h}(t) = \frac{1}{N} \sum_i h_i(t) \ , \qquad
w(t) = \Bigl[
\frac{1}{N} \, \sum_i \,h^2_i(t)  -
\Bigl(\frac{1}{N} \, \sum_i \,h_i(t) \Bigr)^2
\Bigr]^{1/2}
\,.
\end{equation}
A surface is called `smooth' if the
heights $h_i$ are correlated over arbitrarily large distances.
Otherwise, if distant heights become uncorrelated, the surface
is said to be rough. A rough surface is typically characterized
by a diverging width when the system size is taken to infinity.

In many cases growing interfaces exhibit simple scaling laws.
In a finite system, starting from a flat interface, the
width first increases algebraically as $w\sim t^\gamma$,
where $\gamma>0$ is the {\em growth exponent} of the
system\footnote{In most textbooks the growth exponent is denoted
by $\beta$. Here we use a different symbol in order to
distinguish this exponent from the density exponent of DP.}.
In the long-time limit, when the correlation length reaches
the system size, the width saturates at some constant
value $w_{sat} \sim L^\alpha$, where $\alpha$ is the {\em roughness
exponent}. This type of scaling behavior is known as
{\em Family-Vicsek} scaling~\cite{FamilyVicsek85},
as described by the scaling form
\begin{equation}
\label{FVScaling}
w(N,t) \sim N^\alpha f(t/N^z) \ ,
\end{equation}
where $z=\alpha/\gamma$ is the dynamic exponent of the model.
$f(u)$ is a universal scaling function with the asymptotic behavior
$f(u)\sim u^\gamma$ for $u \rightarrow 0$ and $f(u)=\text{const}$
for $u \rightarrow \infty$. Notice that this scaling form is invariant under
rescaling
\begin{equation}
\label{GrowthScaling}
\xvec \rightarrow \scalefac \xvec \ , \qquad
t \rightarrow \scalefac^z t \ , \qquad
h(\xvec,t) \rightarrow \scalefac^{\alpha} h(\scalefac\xvec,\scalefac^z t) \ ,
\end{equation}
where $\scalefac$ is a dilatation parameter. Notice that in
contrast to the order parameter of a spreading process~(\ref{Rescaling}),
the fluctuations in the heights {\em increase} under rescaling.

The critical exponents $\gamma$ and $z$ are used to categorize
various {\em universality classes} of roughening interfaces.
Typically each of these universality classes is characterized by 
certain symmetry properties of the system and may be associated
with a specific stochastic differential equation. This
equation describes the growth of a continuous height
field $h(\xvec,t)$ and consists of the most relevant operators
under rescaling~(\ref{GrowthScaling})
that are consistent with the symmetries of the system.
For example, postulating the symmetries
\begin{quote}
1. translational invariance in space $\xvec \rightarrow \xvec+\Delta \xvec$, \\
2. translational invariance in time  $t \rightarrow t+\Delta t$, \\
3. translational invariance in height direction  $h \rightarrow h+\Delta h$, \\
4. reflection invariance in space $\xvec \rightarrow -\xvec$, \\
5. up/down symmetry $h \rightarrow -h$,
\end{quote}
one is led to the {\em Edwards-Wilkinson} (EW) universality
class~\cite{EdwardsWilkinson82},
as described by the equation
\begin{equation}
\label{EWEquation}
\frac{\partial h(\xvec,t)}{\partial t} = v + \sigma\nabla^2 h(\xvec,t) +
\noise(\xvec,t) \ ,
\end{equation}
where $v$ denotes the mean velocity and $\sigma$ the surface tension.
$\noise(\xvec,t)$ is a zero-average Gaussian noise field with variance
\begin{equation}
\label{GaussianNoise}
\langle\noise(\xvec,t)\noise(\xvec',t')\rangle=
2D\delta^{d}(\xvec-\xvec')\delta(t-t') \
\end{equation}
taking the stochastic nature of deposition into account.
This equation is linear and thus exactly solvable.
Scaling invariance~(\ref{GrowthScaling})
implies that the critical exponents are given by
\begin{equation}
\alpha=1-d/2 \ , \qquad \gamma=1/2-d/4 \ , \qquad z=2 \ .
\qquad (\mbox{EW})
\end{equation}
Edwards-Wilkinson growth processes are invariant under up/down
reflection of the interface 
$h\rightarrow -h$. However, if atoms are adsorbed
from a gas phase above the interface there is no particular
reason for the system to be up/down symmetric. In that case
the above equation has to be extended by the most relevant
term that breaks the up/down symmetry, leading to the
Kardar-Parisi-Zhang (KPZ) equation~\cite{KPZ86,HalpinZhang95}
\begin{equation}
\label{KPZEquation}
\frac{\partial h(\xvec,t)}{\partial t} = v + \sigma\nabla^2 h(\xvec,t)
+ \lambda (\nabla h(\xvec,t))^2 + \noise(\xvec,t) \ .
\end{equation}
For a one-dimensional interface
the critical exponents of the KPZ universality class
are given by
\begin{equation}
\alpha=1/2 \ , \qquad \gamma=1/3 \ , \qquad z=3/2 \ .
\qquad (\mbox{KPZ in 1d})
\end{equation}
whereas in $d\geq 2$ dimensions only numerical estimates
are known (see Ref.~\cite{BarabasiStanley95}).

It is particularly interesting to study {\em roughening
transitions} between a smooth and a rough phase.
A roughening transition is usually accompanied
by a diverging spatial correlation length $\xi_\perp$.
In the smooth phase this correlation length provides a typical
scale below which the interface appears to be rough. On larger
scales, however, the interface turns out to be smooth. Approaching
the roughening transition the correlation length $\xi_\perp$
diverges whereby the entire interface becomes rough.
One of the simplest models displaying a
roughening transition is
the two-dimensional discrete Gaussian SOS
model~\cite{BarabasiStanley95}. Another important
example is the KPZ equation which exhibits a roughening
transition in $d>2$ spatial dimensions.

In the following we will focus on certain growth models
with {\em depinning transitions}. In particular we will discuss
depinning transitions in random media, polynuclear growth
processes, and solid-on-solid growth processes with evaporation
at the edges of plateaus. In the pinned phase of these
models the interface is smooth and does not propagate.
Varying a control parameter the interface undergoes a
depinning transition; it starts moving and evolves into
a rough state. As we will see below, various depinning 
transitions are closely related to phase transitions 
into absorbing states.

%==============================================================================
\subsection{Depinning transitions of driven interfaces}
%==============================================================================
%
\label{DepinningSubsection}
An interesting class of depinning transitions can be observed in experiments
of driven interfaces in random media~\cite{BarabasiStanley95}.
In these experiments a liquid is pumped through a porous medium.
If the driving force $F$ is sufficiently low the liquid cannot
move through the medium since the air/liquid interface is
pinned at certain pores. Above a critical threshold,
however, the interface starts moving through
the medium with an average velocity $v$. Close to the transition,
$v$ is expected to scale as
\begin{equation}
v \sim (F-F_c)^\theta \ ,
\end{equation}
where $\theta$ is the velocity exponent.
One of the first experiments in 1+1 dimensions was performed by
Buldyrev {\it et~al.}, who studied the wetting of paper in a basin
filled with suspensions of ink or coffee~\cite{BBCHSV92}.
Here the driving force $F$ is a result of capillary forces
competing with the total weight of the absorbed suspension. Consequently,
the interface becomes pinned at a certain height where
$F \simeq F_c$. Once the interface has stopped,
the width should scale as
\begin{equation}
w(\ell,t) \sim \ell^\alpha f(t/\ell^{\tilde{z}}) \,,
\end{equation}
where $\ell$ is a box size and $\alpha$ the roughening exponent.
Measuring the interface width Buldyrev {\it et~al.} found the roughness
exponent $\alpha=0.63(4)$. In various other experiments
the values are scattered between $0.6$ and $1.25$. This
is surprising since the Kardar-Parisi-Zhang (KPZ)
class~\cite{KPZ86} predicts the much smaller value
$\alpha=1/2$.

It is believed that the large values of $\alpha$ are due
to inhomogeneities of the porous medium. Because of these
inhomogeneities, the interface does not propagate uniformly
by local fluctuations as in the KPZ equation, it rather
propagates by {\it avalanches}. In the literature two
universality classes for this type of interfacial growth have been
proposed. In case of {\it linear} growth the interface
should be described by a random field Ising model~\cite{Nattermann92},
leading to the exponents $\alpha=1$, ${\tilde{z}}=4/3$, and $\theta=1/3$
in 1+1 dimensions. In the presence of a KPZ-type nonlinearity, however,
the roughening process should exhibit a {\it depinning transition}
which is related to DP~\cite{TangLeschhorn92,MBLS98}. The underlying
DP mechanism differs significantly from an ordinary directed
percolation process in a porous medium subjected to a gravitational
field (cf. Sec.~\ref{DPEXPERIM}). In ordinary DP the spreading
agent represents active sites and percolates along the given direction.
In the present case, however, water may flow not only
in the direction of the pumping force but also
in opposite direction. Moreover, the DP process runs
{\it perpendicular} to the direction of growth,
as will be explained below.

%
% FIGURE DEPINNING MODEL
%
\begin{figure}
\epsfxsize=85mm
\centerline{\epsffile{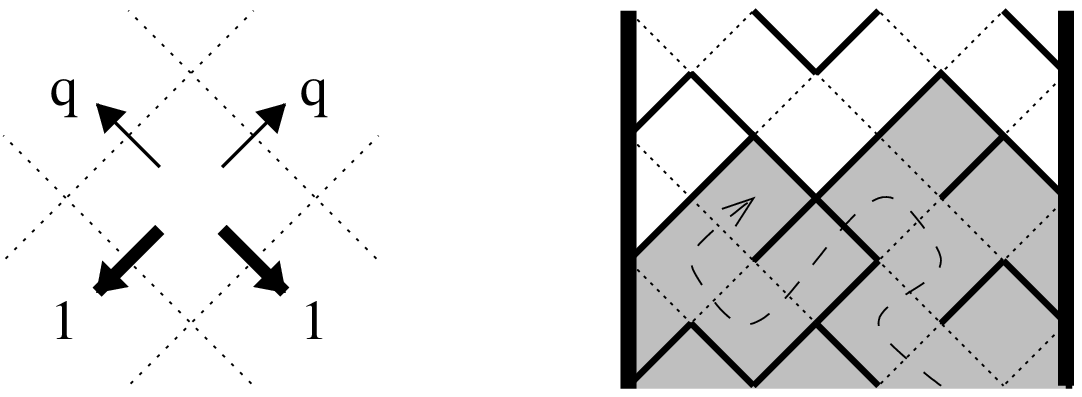}}
\smallcaption{
\label{FIGDEPINNINGMODEL}
Simple model for interface depinning in random media.
The pores of the medium are
represented by cells on a diagonal square lattice. The
permeability  across the edges of the cells depends on the
direction of flow: In downward direction all edges are
permeable, whereas in upward direction they are permeable
with probability $q$ and impermeable otherwise.
The right panel of the figure shows a particular configuration
of open (dashed) and closed (solid) edges.
Pumping in water from below, the interface becomes
pinned along a directed path of solid lines,
leading to a finite cluster of wet cells (shaded region).
The dashed arrow represents an open path in order to illustrate
a possible flow.
}
\end{figure}

A simple model exhibiting a depinning transition is shown in
Fig.~\ref{FIGDEPINNINGMODEL}. In this model the pores of the
material are represented by cells of a diagonal square lattice.
The liquid can flow to neighboring cells by crossing the edges of
a cell. Depending on the direction of flow these edges can either
be permeable or impermeable. For simplicity we assume that all
edges are permeable in downwards direction, whereas in upwards
direction they can only be crossed with a certain probability $q$.
Thus, by starting with a horizontal row of wet cells at the
bottom, we obtain a compact cluster of wet cells, as illustrated
in Fig.~\ref{FIGDEPINNINGMODEL}. The unrestricted flow downwards
ensures that the cluster has no overhangs. Clearly, the size of
the cluster (and therefore the penetration depth of the liquid)
depends on $q$. If $q$ is large enough, the cluster is infinite,
corresponding to a moving interface. If $q$ is sufficiently small,
the cluster is bound from above, i.e., the interface becomes
pinned.

\begin{figure}
\epsfxsize=85mm
\centerline{\epsffile{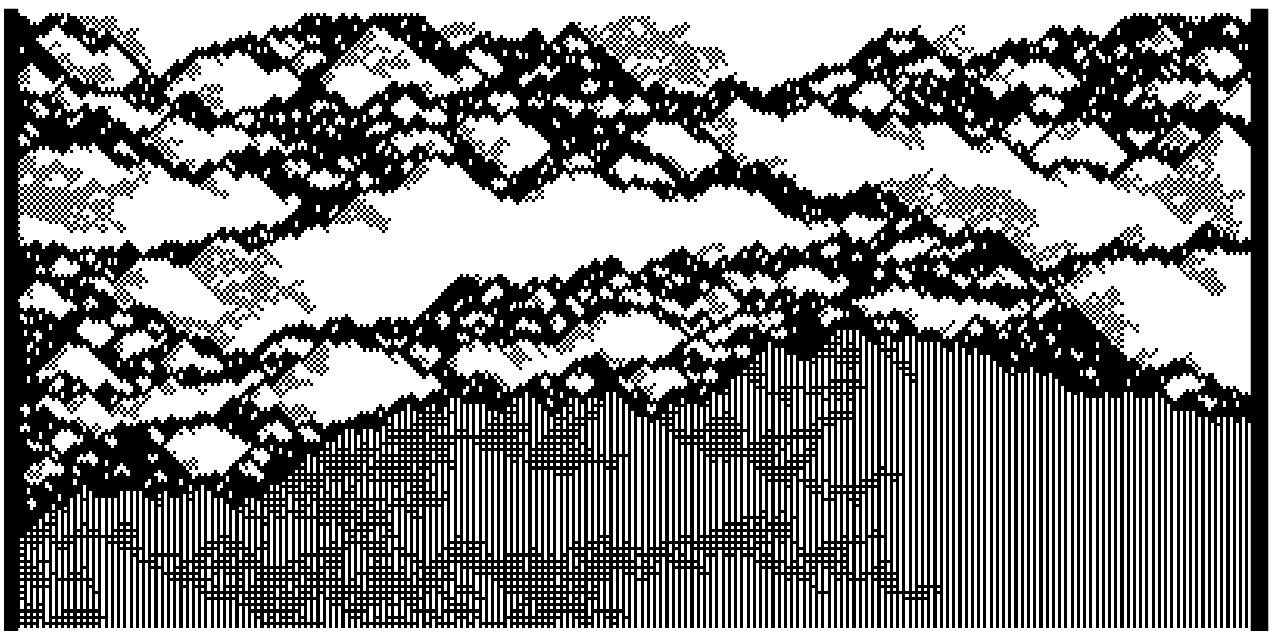}}
\smallcaption{
\label{FIGDEPINNING}
Pinned interface and the underlying DP process.
The figure is explained in the text.
}
\end{figure}

The depinning transition is related to DP as follows.
As can seen in the figure, a pinned interface
may be interpreted as a {\it directed} path along impermeable
edges running from one boundary of the system
to the other. Obviously, the interface becomes pinned only if
there exists a directed path of impermeable bonds
connecting the boundaries of the system. Hence the depinning
transition is related to an underlying directed bond
percolation process with probability $p=1-q$
running {\it perpendicular} to the direction of growth.
The pinning mechanism is illustrated in Fig.~\ref{FIGDEPINNING},
where a supercritical DP cluster propagates
from left to right. The cluster's {\em backbone},
consisting of bonds connecting the two boundaries,
is indicated by bold dots. The
shaded region denotes the resulting cluster of wet cells.
As can be seen, the interface becomes pinned at the lowest
lying branch of the DP backbone. Therefore, the roughening exponent
coincides with the meandering exponent of the backbone
\begin{equation}
\label{DepinningExponent}
\alpha = \nuperp/\nupar \, .
\end{equation}
Moreover, by analyzing the dynamics of the moving interface,
it can be shown that the dynamic critical
exponents are given by $\theta=\alpha$ and ${\tilde{z}}=1$.
Thus, depinning transitions in inhomogeneous porous media may serve
as possible experimental realizations of the DP universality class.

%
% FIGURE DEPINNING MECHANISM AND BACKBONE
%

The prediction (\ref{DepinningExponent}) matches surprisingly
well with the experimental result $\alpha=0.63(4)$ obtained by
Buldyrev {\it et~al.}~\cite{BBCHSV92}. Therefore, it is near at
hand to regard this experiment as a first quantitative
experimental evidence of DP exponents. However, only {\em one}
exponent has been verified, and it is not fully clear how accurate
and reproducible these exponents are. Moreover, pressure
differences may cause long-range correlations in the bulk,
leading to a flat interface on large scales. This means that
gravity could destroy the asymptotic critical behavior. In fact,
in subsequent experiments the estimates for the critical exponents
are scattered over a wide range. For example, Dougherty and Carle
measured the dynamical avalanche distribution of an air/water
interface moving through a porous medium made of glass
beads~\cite{DoughertyCarle98}. According to the DP hypothesis,
the distribution $P(s)$ of avalanche sizes $s$ should decrease
algebraically. In the experiment, however, a stretched exponential
behavior $P(s) \sim s^{-b} e^{-s/L}$ is observed even for small
flowing rates. The estimates for the exponent $b$ are
inconclusive; they depend on the time window of the measurement
and vary between $-0.5$ and $0.85$. Even more recently Albert {\it
et~al.} proposed to identify the universality class by measuring
the propagation velocity of locally tilted parts of the
interface~\cite{ABCD98}. Their results suggest that interfaces
propagating in glass beads are not described by a DP depinning
process, instead they seem to be related to the random-field Ising
model. Altogether the emerging picture is not yet fully
transparent and further experimental effort in this direction would
be desirable.

Depinning experiments were also carried out in 2+1 dimensions
with a spongy-like material used by florists,
as well as fine-grained paper rolls~\cite{BBHKSX92}.
In this case, however, the exponent $\alpha$ is not
related to 2+1-dimensional DP, instead it corresponds to
the dynamic exponent of percolating directed interfaces
in 2+1 dimensions. In experiments as well as in numerical
simulations a roughness exponent $\alpha=0.50(5)$ was obtained.

%==============================================================================
\subsection{Polynuclear growth}
%==============================================================================
%
\label{PNGSECTION}
A completely different type of depinning transition takes
place in models for polynuclear growth
(PNG)~\cite{KerteszWolf89,LRWK90,Toom94a,Toom94b}.
A key feature of PNG models is the use of {\em parallel updates}, leading
to a maximal propagation velocity of one monolayer per time step.
For a high adsorption rate the interface of PNG models
is smooth and propagates  at maximal velocity $v=1$.
Roughly speaking, this means that the interface is
`pinned' at the light cone of the dynamics.
Decreasing the adsorption rate below a
certain critical threshold, PNG models exhibit a
roughening transition to a rough phase with $v<1$.
In contrast to equilibrium roughening transitions, which only
exist in $d \geq 2$ dimensions, PNG models display a
roughening transition even in one spatial dimension.

%
% FIGURE PNG MODEL
%
\begin{figure}
\epsfxsize=100mm
\centerline{\epsffile{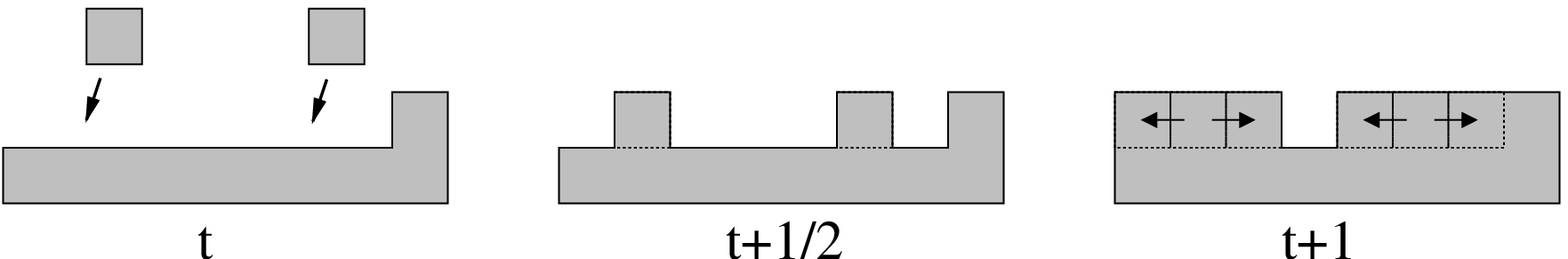}}
\vspace{2mm}
\smallcaption{
\label{FIGPNG}
Polynuclear growth model. In the first half time step atoms
are deposited with probability $p$. In the second half time step
islands grow deterministically by one step and coalesce.
}
\end{figure}

One of the simplest PNG model investigated
so far is defined by the following dynamic
rules~\cite{KerteszWolf89}.
In the first half time step atoms `nucleate' stochastically
at the surface by
\begin{equation}
\label{nucleation}
h_i(t+1/2) = \left\{
\begin{array}{ll}
h_i(t)+1 \ & \text{with prob. } p \ , \\
h_i(t)   \ & \text{with prob. } 1-p \ .
\end{array}
\right.
\end{equation}
In the second half time step the islands grow deterministically in
lateral direction by one step. This type of growth may be expressed
by the update rule
\begin{equation}
\label{coalescense}
h_i(t+1) = \max_{j \in <i>} \bigl[ h_i (t+1/2) , h_j (t+1/2) \bigr] \ ,
\end{equation}
where $j$ runs over the nearest neighbors of site $i$.

The relation to DP can be established as follows.
Starting from a flat interface $h_i(0)=0$, let us
interpret sites at maximal height $h_i(t)=t$ as active
sites of a DP process. Obviously, the adsorption
process~(\ref{nucleation}) turns active sites
into the inactive state with probability $1-p$,
while the process~(\ref{coalescense}) resembles
offspring production. Therefore, if $p$ is large
enough, the interface is smooth and propagates with maximal
velocity $v=1$. This situation corresponds to the active
phase of DP. Approaching the phase transition,
we expect the density of sites at
maximal height to scale as
\begin{equation}
\label{PNGScaling}
n_{max}=
\frac{1}{N} \sum_i \, \delta_{h_i-t} \sim (p-p_c)^\beta,
\end{equation}
where $N$ denotes the system size and $\beta \simeq 0.277$ is the 
density exponent of DP. Below a critical threshold, however,
the density of active sites at the maximum height $h_i(t)=t$
vanishes after some time, the growth velocity is smaller than~$1$,
and the interface evolves into a rough state.
Although this mapping to DP is not exact, numerical simulations
strongly support the validity of Eq.~(\ref{PNGScaling}).
As will be shown below, PNG models are actually a realization
of unidirectionally  coupled DP processes~\cite{THH98,GHHT99}.

Polynuclear growth models have also been used to describe
the growth of colonial organisms such as fungi
and bacteria~\cite{LopezJensen98}. This study was motivated
by recent  experiments with the yeast
{\em Pichia membranaefaciens} on
solidified agarose film~\cite{Sams97}.
By varying the concentration of polluting
metabolites, different front morphologies
were observed. The model proposed in~\cite{LopezJensen98}
aims to explain these morphological transitions on a
qualitative level. It is easy to verify that this
model follows the same spirit as the PNG model defined above.
They both employ parallel dynamics and exhibit a
DP-related roughening transition.

Concerning experimental realizations of PNG models, one major
problem -- apart from quenched disorder -- is the use of parallel
updates. The type of updates in these models is crucial; by using
random-sequential updates the transition is lost since in this
case there is no maximum velocity. However, in realistic
experiments atoms do not move synchronously, rather the adsorption
events are randomly distributed in time. Therefore, random
sequential updates might be more appropriate to describe such
experiments. It thus remains an open question to what extent PNG
processes can be realized in nature.

%==============================================================================
\subsection{Growth with evaporation at the edges of plateaus}
%==============================================================================
%
\label{GREDGES}
DP-related roughening transitions can also be observed in certain
solid-on-solid growth processes with random-sequential
updates~\cite{AEHM96,AEHM98}. As a key feature of these models,
atoms may desorb exclusively at the {\em edges} of existing layers,
i.e., at sites which have at least one neighbor at a lower height.
This corresponds to a physical situation where the binding energy
of atoms at the edges is much smaller than the binding energy 
of atoms in completed layers.
By varying the growth rate, such growth processes display a
roughening transition from a non-moving smooth phase to
a moving rough phase.

A simple solid-on-solid model for this
type of growth is defined by the following dynamic
rules~\cite{AEHM96}: For each update a site~$i$
is chosen at random and an atom is adsorbed
\begin{equation}
\label{adsorption}
h_i \rightarrow h_{i}+1 \; \; \text {with probability } q
\end{equation}
or desorbed at the edge of a plateau
\begin{equation}
\begin{array}{ll}
\label{desorption}
&h_i \rightarrow \text{min}(h_{i},h_{i+1}) \; \;
\text {with probability } (1-q)/2 \ , \\
&h_i \rightarrow \text{min}(h_{i},h_{i-1}) \; \;
\text {with probability } (1-q)/2 \ .
\end{array}
\end{equation}
Moreover, the growth process is assumed to be {\it restricted},
i.e., updates are only carried out if the resulting configuration obeys
the constraint~(\ref{RSOS}).

%
% FIGURE RSOS MODEL WITH EVAPORATION AT TERRACES
%
\begin{figure}
\epsfxsize=150mm
\centerline{\epsffile{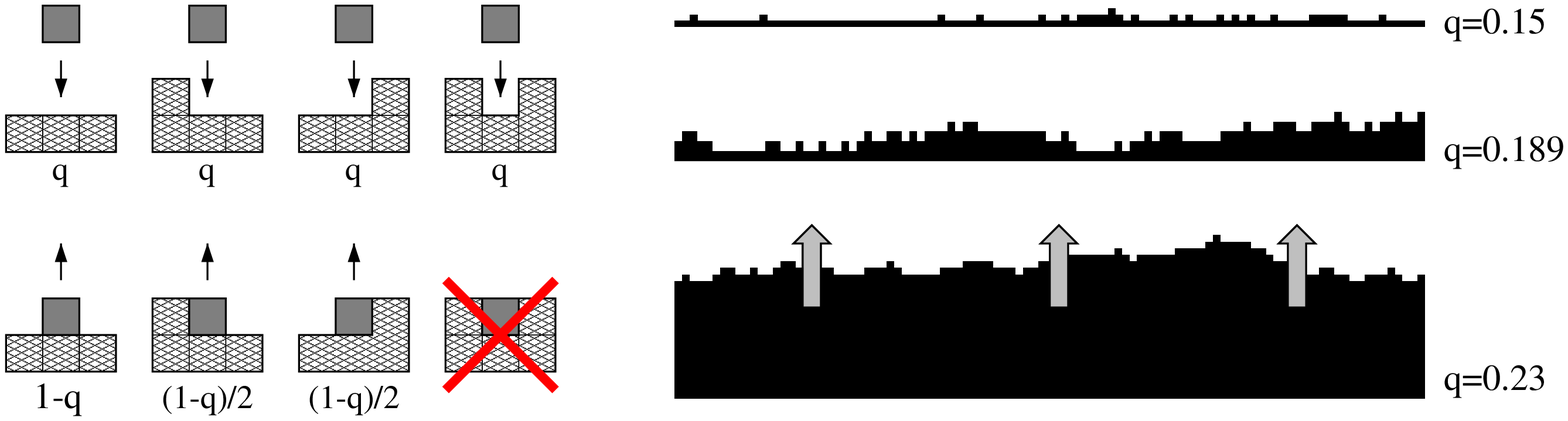}}
\vspace{1mm}
\smallcaption{
\label{FIGMONOMER}
Restricted solid-on-solid growth model exhibiting a roughening
transition from a non-moving smooth to a moving rough phase.
Monomers are randomly deposited whereas desorption takes place
only at the edges of plateaus.
}
\end{figure}
The qualitative behavior of this model
is illustrated in Fig.~\ref{FIGMONOMER}.
For small~$q$ the desorption processes (\ref{desorption}) dominate.
If all heights are initially set to the same value, this level will
remain the bottom layer of the interface. Small islands will grow on
top of the bottom layer but will be quickly eliminated
by desorption at the island edges.  Thus, the interface is effectively
anchored to its bottom layer and a smooth phase is maintained.
The growth velocity $v$ is therefore zero in the thermodynamic limit.
As $q$~is increased, more islands on top of the bottom layer
are produced until above~$q_c \simeq 0.189$, the critical value of~$q$, they
merge forming new layers at a finite rate, giving rise
to a finite growth velocity.

Interestingly, this model can be interpreted
as a driven diffusion process of two
oppositely charged types of particles. The charges
\begin{equation}
\label{ChargedRep}
c_{i,i+1} \;=\; h_{i+1}-h_i \;\in\; \{-1,0,+1\}
\end{equation}
are bond variables and represent a change
of height between two adjacent interface sites.
In this representation the dynamic rules
(\ref{adsorption})-(\ref{desorption}) can be implemented
by randomly selecting two neighboring bonds
and performing the following processes with probabilities
indicated on the arrows:
\begin{equation}
\vacancy+ \ \overset{q}{\underset{(1-q)/2}{\rightleftarrows}} \ +\vacancy \ , \qquad
\vacancy\vacancy \ \overset{q}{\underset{1-q}{\rightleftarrows}} \ +-\ , \qquad
-\vacancy \ \overset{q}{\underset{(1-q)/2}{\rightleftarrows}} \ \vacancy- \ , \qquad
-+ \ \overset{q}{\rightarrow} \ \vacancy\vacancy \ .
\end{equation}
In the smooth phase $q<q_c$,
the charges are arranged as closely bound $+-$ dipoles.
For $q>q_c$, the dipoles become unbound wherefore the fluctuations
in the total charge, measured over a distance of order~$N$,
diverge with~$N$.  Thus the transition can be described in
terms of correlations between charged particles.

%----------------------------------------------------------------------------
\headline{Relation to directed percolation}
%----------------------------------------------------------------------------
%
At the transition the dynamics of the model is related to DP as follows.
Starting with a flat interface at zero height, let us consider all sites
with $h_i=0$ as particles $A$ of a DP process. Growth according to
Eq.~(\ref{adsorption}) corresponds to spontaneous annihilation
$A\rightarrow\vacancy$. Conversely, desorption may be regarded as
a particle creation process. However, since atoms may only desorb
at the edges of plateaus, particle creation requires a neighboring
active site to be present. This process therefore
corresponds to offspring production $A \rightarrow 2A$.
Clearly, these processes resemble the dynamic rules of a DP
process. In contrast to PNG models, the DP process takes place at
the bottom layer of the interface. Moreover, the roughening
transition does not depend on the use of either parallel or
random-sequential updates.

In the model without the restriction~(\ref{RSOS}), the relation to
DP can be proven exactly. In this case the processes at the bottom
layer decouple from the evolution of the interface at higher levels.
Introducing local variables $s_i=\delta_{h_i,0}$ it is possible to map
Eqs.~(\ref{adsorption})--(\ref{desorption})
exactly onto the dynamic rules
\begin{alignat*}{2}
& \text{if} \ s_i = 1
\hspace{5mm} && s_i \rightarrow 0 \; \;\text{with prob. } q \ , \\
& \text{if} \ \{ s_{i-1}, s_i, s_{i+1}\} = \{ 0, 0, 1 \}
\hspace{5mm} && s_i \rightarrow 1 \; \; \text{with prob. } (1-q)/2 \ , \\
& \text{if} \ \{ s_{i-1}, s_i, s_{i+1}\} = \{ 1, 0, 0 \}
\hspace{5mm} && s_i \rightarrow 1 \; \; \text{with prob. } (1-q)/2 \ , \\
& \text{if} \ \{ s_{i-1}, s_i, s_{i+1}\} = \{ 1, 0, 1 \}
\hspace{5mm} && s_i \rightarrow 1  \; \; \text{with prob. } 1-q \ .
\end{alignat*}
These rules define a contact process on a square lattice
(cf. Fig.~\ref{FigureCP}) in which particles self-annihilate
at unit rate and create offspring with rate $\lambda/2 = (1-q)/2q$.
Since the percolation threshold of the contact process
$\lambda_c = 3.29785(8)$ is known, we can predict the
critical point of the unrestricted growth to be given by $q_c = 0.232675(5)$.
model. For the restricted model there is no exact mapping
to a contact process. Nevertheless numerical simulations
strongly suggest that the bottom layer still exhibits DP behavior.

The underlying DP transition is the origin of the roughening
transition. In the active phase of DP the interface
fluctuates close to the bottom layer so that the interface
is smooth. Approaching criticality the bottom layer
occupation $n_0$ vanishes as
\begin{equation}
n_0 \sim  (q_c-q)^\beta \ ,
\end{equation}
where $\beta$ is the density exponent of DP.
In the inactive phase of DP the interface detaches from
the bottom layer and evolves into a rough state.
Therefore, the front velocity $v$ for $q>q_c$ is
proportional to the characteristic survival time of a
DP process in the inactive phase
\begin{equation}
\label {eq v}
v \sim  1/\xi_\parallel \sim (q-q_c)^{1/{\nupar}} \ .
\end{equation}
With respect to the microscopic rules
the roughening transition in these models seems to be as
robust as a DP transition. Including the dynamics at higher levels
of the interface, the models turn out to be described by
unidirectionally coupled DP processes (see below).

%----------------------------------------------------------------------------
\headline{Scaling of the interface width}
%----------------------------------------------------------------------------
%
Turning to the scaling properties of the interface width~(\ref{Width})
the emerging picture is less clear and may indicate the
presence of many length scales.
With only a single length scale one would expect
the saturation width in finite systems to scale as
$w_{sat}(N) \sim N^{\alpha}$ in the growing phase and
$w_{sat}(N) \sim N^{\alpha^\prime}$ at criticality,
where $\alpha$ and $\alpha^\prime$ are generally different
critical exponents. Thus the expected scaling form would read
\begin{equation}
w_{sat}(N,q) = N^{\alpha^\prime} \ g\bigl(N^{1/\nuperp} (q-q_c)\bigr) \ ,
\end{equation}
where $g(u)$ is a universal scaling function with the asymptotic
behavior
\begin{equation}
g(u) =
\begin{cases}
|u|^{-\alpha^\prime\nuperp} & \text{for } u \rightarrow - \infty  \ ,\\
\text{const} & \text{for } u \rightarrow \pm 0 \ , \\
|u|^{(\alpha-\alpha^\prime)\nuperp} & \text{for } u \rightarrow + \infty \ .
\end{cases}
\end{equation}
\begin{figure}
\epsfxsize=130mm
\centerline{\epsffile{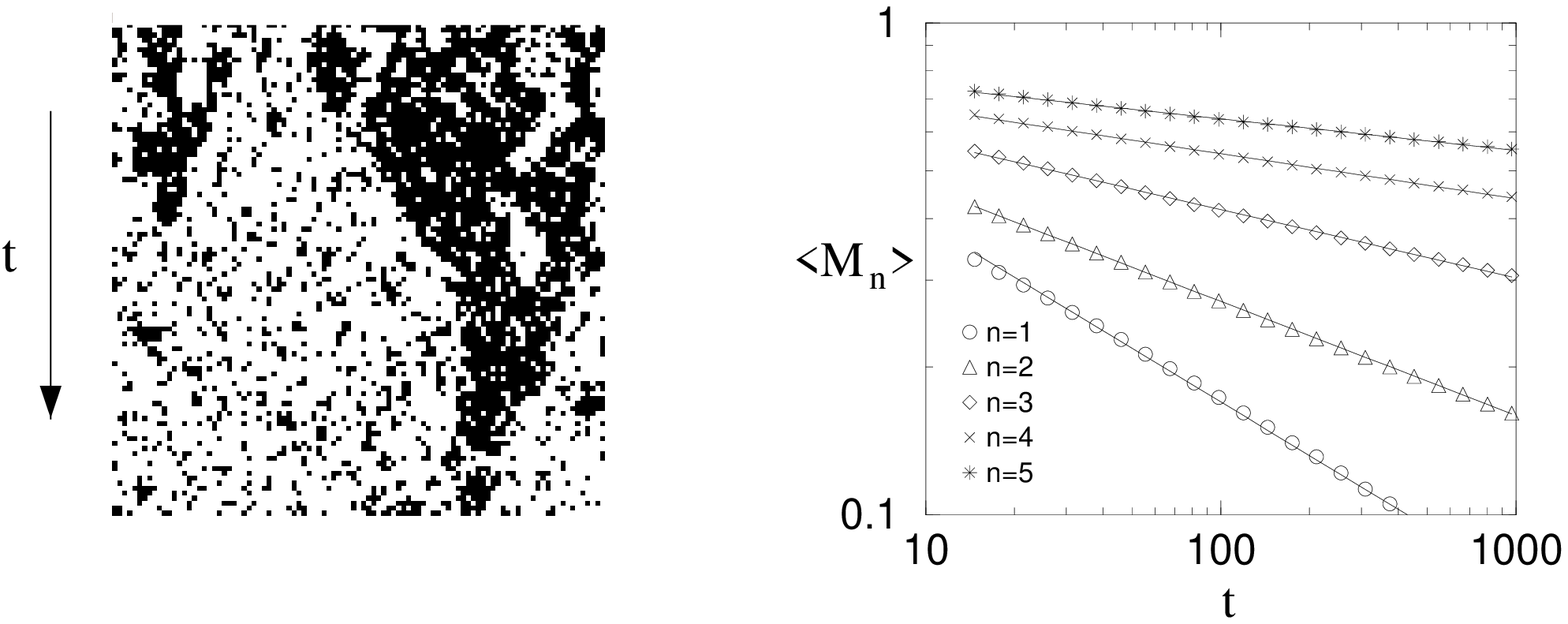}}
\vspace{2mm}
\smallcaption{
\label{FIGSSB}
Spontaneous symmetry breaking.
Left: Simulation in the smooth phase starting from
a random interface configuration. Black (white) pixels
indicate even (odd) heights. As time evolves the
systems 'coarsens', providing a robust mechanism for
the elimination of minority islands.
Right: Order parameters $M_n$ versus time at criticality
starting from a flat interface.
}
\end{figure}

This scaling form implies that the width in a finite size system
saturates at
\begin{equation}
w_{sat}(N,q) =
\begin{cases}
(q_c-q)^{-\alpha^\prime\nuperp} & \text{ if } q<q_c \ , \\
N^{\alpha^\prime} & \text{ if } q=q_c \ , \\
N^{\alpha} (q_c-q)^{(\alpha-\alpha^\prime)\nuperp} & \text{ if } q<q_c \ .
\end{cases}
\end{equation}
However, the width at criticality actually increases
as $w(t) \sim \left( \ln t \right)^\gamma$ until it
saturates at
\begin{equation}
w_{sat} \sim \left( \ln N \right)^\kappa\,,
\label{gammasat}
\end{equation}
where $\kappa \simeq 0.43$, suggesting that $\alpha^\prime=0$.
Moreover, as the growth rate is increased and the interface
starts to move, the width diverges as
$w_{sat}(q) - w_{sat}(q_c) \sim (q-q_c)^{0.95}$,
suggesting that $\alpha \simeq 0.9$. This value
differs from the expected value $\alpha=1/2$
for growing interfaces in one spatial dimension.
Therefore, the scaling behavior of the interface width
is presumably characterized by many
different length scales. This point of view is confirmed
by a numerical analysis of correlation functions at
different levels~\cite{AEHM98}.

%----------------------------------------------------------------------------
\headline{Spontaneous symmetry breaking}
%----------------------------------------------------------------------------
%
The roughening transition in growth models with evaporation
at edges of islands is accompanied by spontaneous symmetry breaking
of translational invariance in height direction.
This symmetry is associated with a family of
non-conserved  magnetization-like order parameters
\begin{equation}
M_n\;=\; \Bigl|\frac{1}{N} \, \sum_{j=1}^N \exp\Bigl(\frac{2\pi i
h_j}{n+1}\Bigr) \Bigr| \ .
\end{equation}
The order parameter $M_1=|\frac1N \sum_j (-1)^{h_j}|$, for example,
measures the difference between the densities of sites at even
and odd heights, respectively (see Fig.~\ref{FIGSSB}).
In the smooth phase the order parameters $M_n$
tend towards stationary positive values.
Approaching the phase transition these values vanish
algebraically as
\begin{equation}
\label {eq M}
\langle M_n \rangle  \sim (q_c-q)^{\theta_n} \ ,
\end{equation}
where $\theta_1 \approx 0.65$, $\theta_2\approx0.40$,
and $\theta_3 \approx 0.23$.
At criticality, starting from a flat interface, the order parameters
decay as $M_n(t) \sim t^{-\theta_n/\nupar}$,
as shown in the right hand graph of Fig.~\ref{FIGSSB}.
In the rough phase $q>q_c$, where the heights $h_i$ become
uncorrelated over large distances, $M_n=0$
in the thermodynamic limit.

\begin{figure}
\epsfxsize=65mm
\centerline{\epsffile{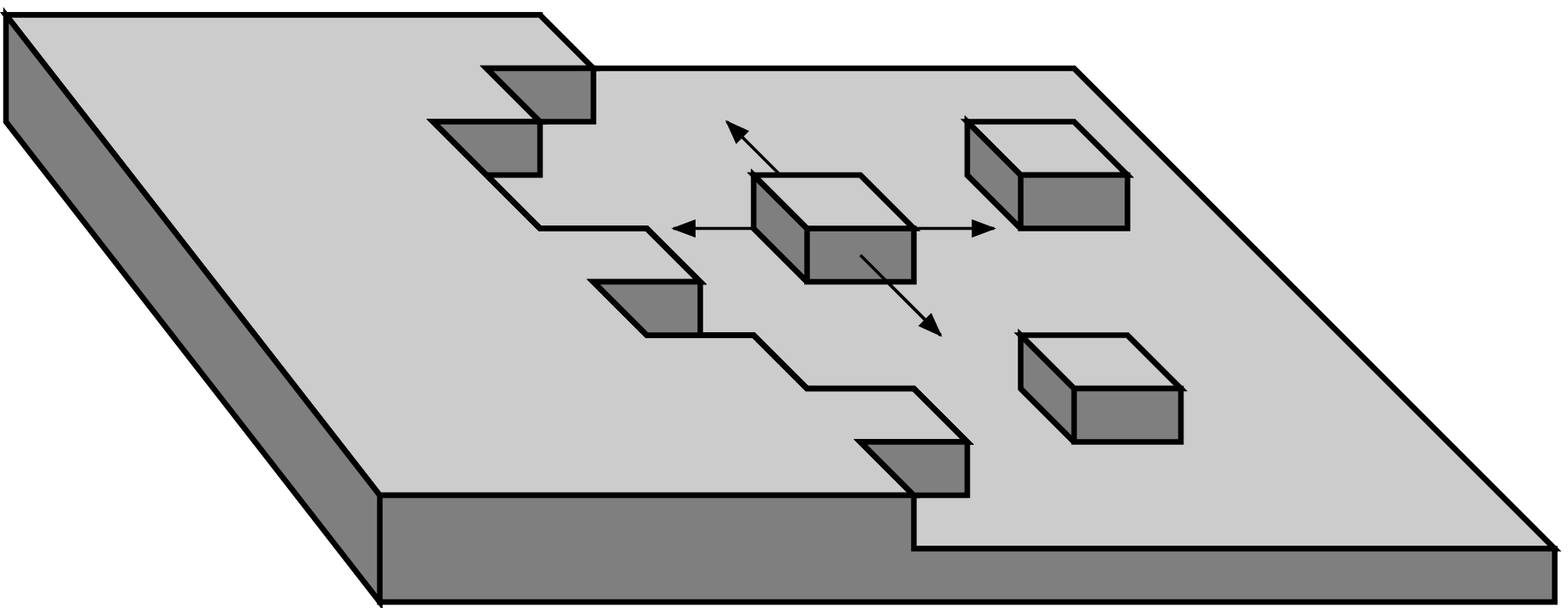}}
\vspace{2mm}
\smallcaption{
\label{FIGPERSPECTIVE}
Possible experimental realization of the growth process defined
in Eqs.~(\ref{adsorption}) and (\ref{desorption}) by lateral growth
of a monolayer on a substrate.
}
\end{figure}

%----------------------------------------------------------------------------
\headline{Experimental realizations}
%----------------------------------------------------------------------------
%
With respect to experimental realizations of the dynamic rules
(\ref{adsorption})-(\ref{desorption}) it is important to 
note that atoms are not allowed to diffuse on the surface.
This assumption is rather unnatural since in most
experiments the rate for surface diffusion is much higher
than the rate for desorption back into the gas phase.
Therefore, it will be difficult to realize this type of
homoepitaxial growth experimentally. However, in a different
setup, the above model could well be relevant~\cite{OferBiham99}.
As illustrated in Fig.~\ref{FIGPERSPECTIVE}, a laterally growing
monolayer could resemble the dynamic rules~(\ref{adsorption})
and~(\ref{desorption}) by identifying the edge of the monolayer
with the interface contour of the growth model. In this case
`surface diffusion', i.e. diffusion of atoms along
the edge of the monolayer, is highly suppressed. Moreover,
in single-step systems (such as fcc(100) surfaces)
it would also be possible to implement the restriction~(\ref{RSOS}).
%
% FIGURE LATERAL GROWTH OF A MONLAYER
%
%
%

%==============================================================================
\subsection{Unidirectionally coupled directed percolation processes}
%==============================================================================

\label{GRCOUPLEDSEC}
So far we have seen that
the scaling properties of quantities involving only the
bottom level of the interface may be adequately described by using DP
exponents. In particular it was shown that the density
of exposed sites at the bottom layer decreases as
$n_0\sim (q_c-q)^\beta$ until it vanishes at the transition.
Let us now turn to the scaling properties of the first few
layers $k=1,2,\ldots$ above the bottom layer.
Since the scaling properties at the bottom layer $k=0$
in the smooth phase are completely characterized by the
three DP exponents $\beta$, $\nuperp$, and $\nu_{||}$, it is
natural to assume that the next layers obey similar scaling laws with
analogous exponents, $\beta^{(k)}$, $\nuperp^{(k)}$, and
$\nu_{||}^{(k)}$. For example, the densities
\begin{equation}
\label{DensitiesNk}
n_k = \frac{1}{N} \sum_{j=0}^{k} \sum_i \, \delta_{h_i,j}
\ , \qquad
k=0,1,2,\ldots
\end{equation}
of sites at height $\leq k$ above the bottom layer are expected to scale as
\begin{equation}
\label{CoupledDensityScaling}
n_k \sim (q_c-q)^{\beta^{(k)}} \ . \qquad \qquad k=1,2,3,\ldots
\end{equation}
In principle all these exponents could be different and independent
from each other. However, extensive numerical simulations
and field-theoretic considerations~\cite{AEHM98} suggest that
the scaling exponents $\nuperp^{(k)}$ and $\nu_{||}^{(k)}$
are {\em identical on all levels} and equal to the DP exponents
$\nuperp$ and $\nu_{||}$. This remarkable property implies that the
growth process at criticality is characterized by a {\em single}
dynamic exponent $z=\nu_{||}/\nuperp$. The density exponents
$\beta^{(1)},\beta^{(2)},\beta^{(3)},\ldots$, however,
turn out to be different and considerably reduced compared
$\beta^{(0)}=\beta$. These exponents appear to be non-trivial in
the sense that they are not simply related to DP exponents.

\begin{figure}
\epsfxsize=150mm
\centerline{\epsffile{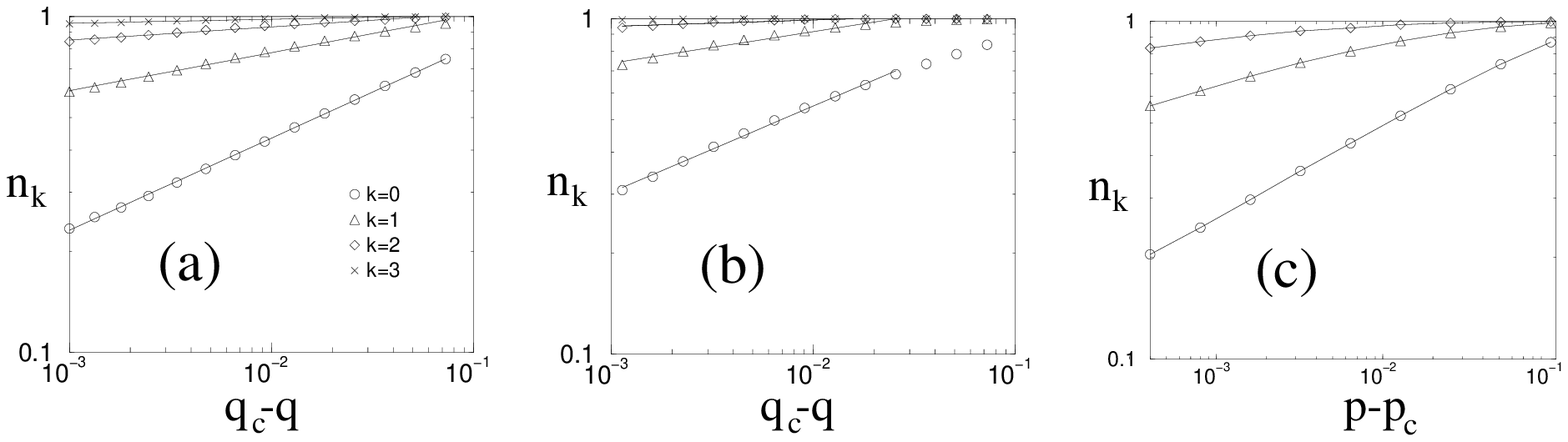}}
\smallcaption{
\label{FIGCOUPLED}
Coupled directed percolation. Stationary densities $n_k$
as functions of the distance from criticality in
(a) the unrestricted growth model, (b) the restricted
growth model, and (c) a unidirectionally coupled sequence
of directed bond percolation processes. }
\end{figure}

In order to explain the reduced values of $\beta^{(k)}$ it is
useful to introduce the following particle interpretation.
Assuming that the minimal height of the interface is zero,\
lattice sites at height $h_i\leq 0,1,2,\ldots$ may be associated
with particles $A,B,C,\ldots$, respectively. By definition,
particles of different species are allowed to occupy the same site
simultaneously. For example, the presence of an $A$-particle
induces simultaneous presence of all other particle species at the
same site. As shown before, the $A$-particles of the unrestricted
model evolve independently according to the dynamic rules of a
contact process. Similarly, in absence of $A$-particles, the
$B$-particles will evolve according to the rules of a contact
process. In presence of an $A$-particle, however, $B$-particles
are instantaneously created at the same site, giving rise to an
effective reaction $A \rightarrow A+B$ at infinite rate. As this
reaction does not modify the configuration of the $A$-particles,
it couples the two subsystems in one direction without feedback.
Similarly, the $B$-particles induce the creation of $C$-particles
by an effective reaction $B \rightarrow B+C$. Thus we obtain a
{\it unidirectionally coupled} sequence of contact processes
corresponding to the reaction-diffusion scheme
\begin{alignat*}{4}
A &\leftrightarrow 2A &\qquad B &\leftrightarrow 2B
&\qquad C &\leftrightarrow 2C  \\
A &\rightarrow A+B      &\qquad B &\rightarrow B+C
&\qquad C &\rightarrow C+D  & \qquad\text{etc.}
\end{alignat*}
Notice that this sequence can be truncated at any level without
changing the dynamics of the lower levels. In fact, numerical
simulations of the growth model and a unidirectionally coupled
sequence of three (1+1)-dimensional directed bond percolation
processes yield compatible estimates for the critical exponents
(cf. Fig.~\ref{FIGCOUPLED}):
\begin{equation}
\beta^{(0)} = 0.28(1) \ , \qquad
\beta^{(1)} = 0.13(2) \ , \qquad
\beta^{(2)} = 0.05(1) \ . \qquad
\end{equation}
Thus, `coupled DP' is a {\em universal} phenomenon and
should play a role even in more general
contexts, namely whenever DP-like processes are coupled
unidirectionally {\em without} feedback.

%----------------------------------------------------------------------------
\headline{Mean field approximation}
%----------------------------------------------------------------------------
%
The simplest set of Langevin equations for
unidirectionally coupled DP reads~\cite{AEHM98}
\begin{equation}
\label{CoupledLangevin}
\timederivative{\phi_k}(\mbox{\bf x},t) \;=\;
\crit_k \phi_k(\mbox{\bf x},t) -
\lambda \phi_k^2(\mbox{\bf x},t)  +
\diff \nabla^2 \phi_k(\mbox{\bf x},t) +
\coupling \phi_{k-1}(\mbox{\bf x},t) +
\noise_k(\mbox{\bf x},t) \ ,
\end{equation}
where $\noise_k$ are independent multiplicative noise fields
with correlations
\begin{equation}
\label{CoupledNoise}
\langle \noise_k(\mbox{\bf x},t) \rangle \;=\; 0 \,,\qquad
\langle \noise_k(\mbox{\bf x},t) \noise_l(\mbox{\bf x}',t')  \rangle \;=\;
2 \namp \, \phi_k(\mbox{\bf x},t) \, \delta_{k,l} \,
\delta^d(\mbox{\bf x}-\mbox{\bf x}') \, \delta(t-t')
\ .
\end{equation}
Assuming that $\phi_{-1} \equiv 0$, the lowest equation for $k=0$
reduces to the ordinary Langevin equation of
DP (\ref{DPLangevinEquation}). The parameters $\crit_k$ control the rates
for offspring production at level~$k$. In principle all these
parameters could be different, leading to interesting
{\em multicritical} behavior~\cite{THH98}. Here we will restrict
to the simplest case where the parameters $\crit_k=\crit$ coincide.

The reduced values of $\beta^{(1)},\beta^{(2)},\ldots$ can already
be understood on the level of a simple 
mean field approximation. Determining
the stationary solutions of the mean field equations
\begin{equation}
\label{CoupledMF}
\timederivative{\phi_k} =
\crit \phi_k - \lambda \phi_k^2 + \coupling \phi_{k-1}
\end{equation}
and expanding the result for small values of $\crit$,
the fields $\phi_k$ are found to scale asymptotically
with the mean field exponents
\begin{equation}
\beta_{MF}^{(k)} = 2^{-k} \ .
\end{equation}
In addition, dimensional analysis of
Eqs.~(\ref{CoupledLangevin})-(\ref{CoupledNoise}) reveals
that the mean field
scaling exponents $\nuperp^{(k)}$ and $\nupar^{(k)}$
coincide with the DP exponents $\nuperp^{MF}=1/2$ and
$\nupar^{MF}=1$ on all levels, i.e., they do not depend on $k$.

%----------------------------------------------------------------------------
\headline{Field-theoretic treatment}
%----------------------------------------------------------------------------
%
The above mean field approximation is expected to hold above
the critical dimension $d_c=4$. In less than four dimensions
fluctuation corrections to $\beta^{(k)}$ have to be taken into
account. These corrections can be approximated by field-theoretic
renormalization group techniques~\cite{THH98,GHHT99}.
To this end the master equation is mapped onto a second-quantized
bosonic operator representation,
leading to the effective action (cf. Sect~\ref{FTHSEC})
\begin{equation}
\label{CoupledDP}
\begin{split}
S[\phi_0,\tilde{\phi}_0,\phi_1,\tilde{\phi}_1,\ldots]
= \sum_{k=0}^{K-1} \int d^dx\,dt \,\Biggl\{
&  \tilde{\phi}_k(\xvec,t) \,
   \Bigl(\tau \partial_t - \diff \nabla^2  - \crit \Bigr)\,\phi_k(\xvec,t)
   -\coupling \tilde{\phi}_k \phi_{k-1} \\
&+ \frac{\namp}{2}\,\tilde{\phi}_k(\xvec,t)
\Bigl(\phi_k(\xvec,t) - \tilde{\phi}_k(\xvec,t) \Bigr) \phi_k(\xvec,t)
\Biggr\} \ ,
\end{split}
\end{equation}
where $K$ is the total number of levels in the hierarchy. Because
of the unidirectional structure the truncation of the hierarchy
at finite $K$ does not affect the temporal evolution of the
subsystems $k \leq K$. The field-theoretic
treatment turns out to be rather complex, even a one-loop calculation
for a two-level system involves $34$ different diagrams.
Details of these calculations are given in Ref.~\cite{GHHT99},
whose main results we summarize here. A one-loop calculation in the inactive
phase reveals that the RG flow is characterized by two fixed lines.
One of them is unstable and corresponds to a situation where the two
systems are decoupled. The other one is stable and describes
the interacting case. One can show that strongly ultraviolet-divergent
contributions cancel along the stable fixed line. To determine
the exponent $\beta^{(1)}$, similar calculations have to be performed in
the active phase. Choosing a particular point along the fixed line
it is possible to derive the result
\begin{equation}
\label{CoupledFTResult}
\beta^{(0)} = 1 - \epsilon/6 \,, \qquad
\beta^{(1)} = 1/2 - \epsilon/8 \,,
\end{equation}
where $\epsilon=4-d$. This result explains the downward correction
of the critical exponent $\beta^{(1)}$ in less than four dimensions.

The field-theoretic analysis involves
various technical and conceptual problems. On the one hand,
in the active phase several infrared-divergent diagrams are
encountered~\cite{Goldschmidt98}, without being clear to what extent
they will affect the physical properties of coupled DP. On the other
hand, the coupling constant $\coupling$ between different levels is shown
to be a {\em relevant} quantity, i.e., it grows and finally diverges
under RG transformations. This may be the cause for numerically
observed violations of scaling. In fact, the curves in
Fig.~\ref{FIGCOUPLED} for $k\geq 1$ are not perfectly straight but
slightly bent. This curvature is neither related to transients
nor to finite size effects. A careful analysis shows that the
field-theoretic prediction seems to apply in an {\em intermediate}
scaling regime, whose size depends on the coupling constant $\coupling$.
The breakdown of scaling may be an artifact of the lattice realization
which, in contrast to the field-theoretic prediction,
seems to limit the value of $\coupling$.

%==============================================================================
\subsection{Parity-conserving roughening transitions}
%==============================================================================
%
\label{GRDIMERSEC}
So far we discussed two examples of depinning transitions related
to DP, namely interface depinning in driven random media, polynuclear
growth, and growth without evaporation at the edges of terraces.
It is therefore interesting to ask whether it is possible to find
examples of roughening transitions with non-DP behavior in one
spatial dimension.

As discussed in Sec.~\ref{SEVABSSEC} non-DP phase transitions into
absorbing states can be observed in systems with additional
symmetries. For example, non-DP behavior is observed in
{\em parity-conserving} branching processes.
Therefore, the question arises whether it is possible
to replace the underlying DP transition in the previously discussed
growth models by a parity-conserving mechanism. As shown in
Ref.~\cite{HinrichsenOdor98b}, this can be done by considering
the growth of {\em dimers} with evaporation at edges of plateaus.
As dimers consist of two atoms, the number of particles at each
height level is preserved modulo~$2$. This definition of the
dynamic rules mimics a BAWE at the
bottom layer of the interface. It turns out that
some of the critical properties of the roughening transition
in this model are indeed characterized by PC exponents
(cf. Eq. (\ref{PCExponents})).

\begin{figure}
\epsfxsize=150mm
\centerline{\epsffile{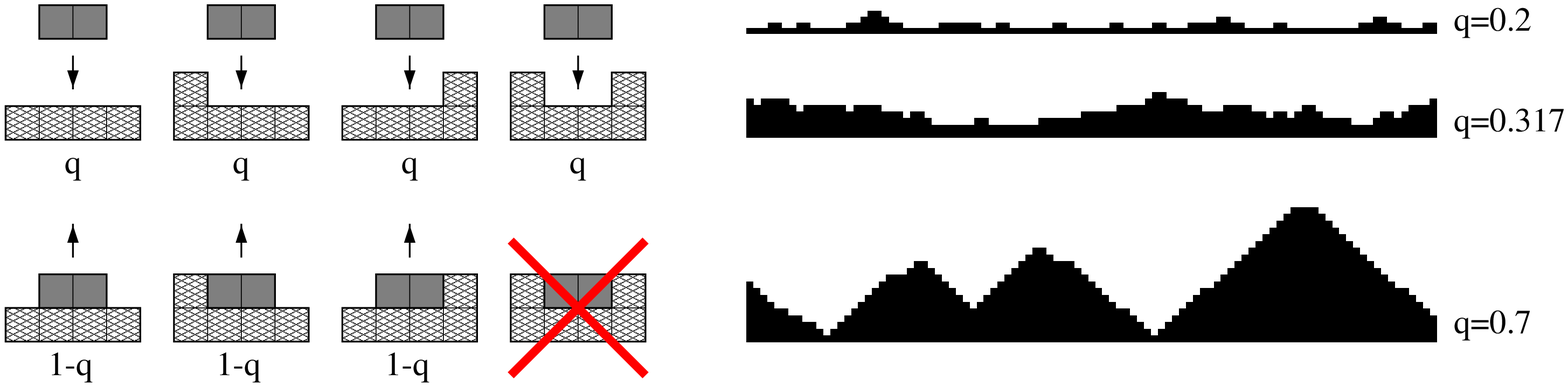}}
\vspace{2mm}
\smallcaption{
\label{FIGGRDIMER}
Roughening transition in a model for dimer adsorption
and desorption.
Left: Dynamic rules for adsorption and desorption
at the edges of terraces.
Right: Typical interface configurations below, at,
and above criticality.
}
\end{figure}

As shown in Fig.~\ref{FIGGRDIMER}, the model generalizes
the growth model introduced in Sec.~\ref{GREDGES},
replacing monomers by dimers.
In each attempted update two adjacent sites
$i$ and $i+1$ are selected randomly. If the heights
$h_{i}$ and $h_{i+1}$ are equal, one of the following
moves is carried out. Either a dimer is adsorbed with
probability $p$
\begin{equation}
\label{DimerAdsorption}
h_{i} \rightarrow h_{i}+1 \,, \qquad \ h_{i+1} \rightarrow h_{i+1}+1
\end{equation}
or desorbed with probability $1-p$:
\begin{equation}
\label{DimerDesorption}
h_{i},h_{i+1} \rightarrow \min(h_{i-1},h_{i},h_{i+1},h_{i+2}) \, .
\end{equation}
An update will be rejected if it leads to a violation
of the RSOS restriction~(\ref{RSOS}).

The transition of the model is illustrated in
Fig.~\ref{FIGGRDIMER}: If $q$ is very small,
only a few dimers are adsorbed for a short time
so that the interface is smooth and pinned at the (spontaneously selected)
bottom layer. As $q$ increases, a growing number of dimers covers
the surface and large islands of several layers stacked
on top of each other are formed. When $q$ exceeds
the critical value~$q_c \simeq 0.317(1)$, the mean size of the islands
diverges and the interface evolves into a rough state.

\begin{figure}
\epsfxsize=150mm
\centerline{\epsffile{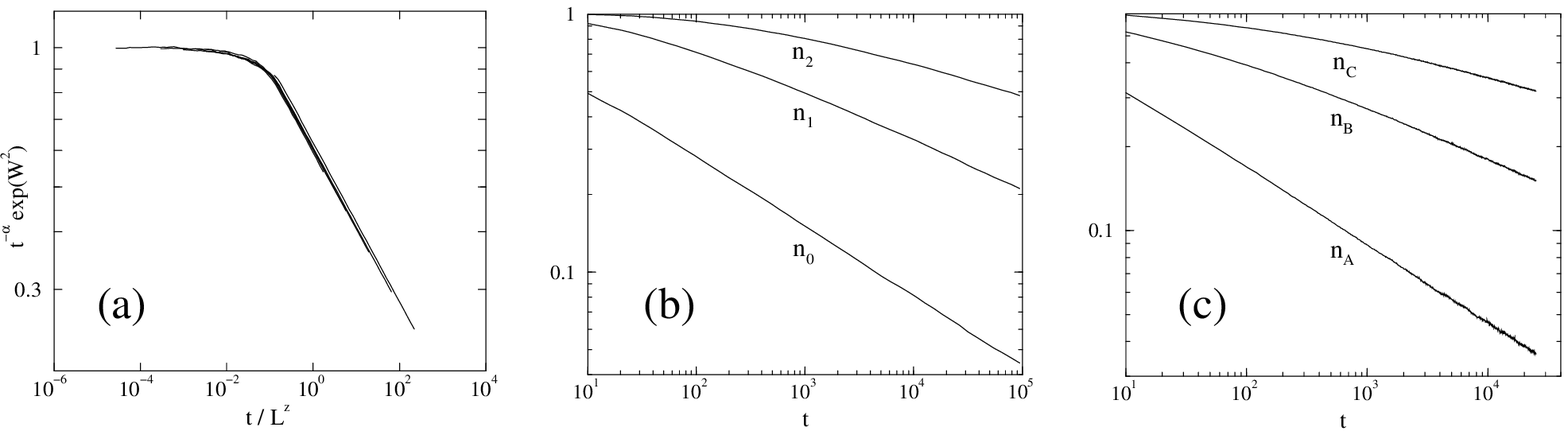}}
\smallcaption{
\label{FIGDIMERSIMUL}
Dimer model: (a) Finite size scaling of the width.
(b) Decay of the densities $n_0,n_1,n_2$ as a function
of time. (c) Decay of the corresponding densities
in unidirectionally coupled BAWE's.
}
\end{figure}

Naively one may expect the interface to detach from the bottom layer
in the rough phase, resulting in a finite propagation velocity.
In the present case it turns out, however, that the interface remains
pinned to the initial height, i.e., it does {\em not} propagate
at constant velocity. This is due to the fact that a
stochastic deposition process cannot create a
dense packing of dimers. Instead the emerging configurations
are characterized by a certain density of
defects (solitary sites at the bottom layer) where dimers cannot be
adsorbed. Because of the RSOS condition~(\ref{RSOS})
these defects act as `pinning centers'
which prevent the interface from growing, leading to 
triangular configurations as shown in Fig.~\ref{FIGGRDIMER}.
The pinning centers cannot disappear spontaneously,
they can only diffuse by interface
fluctuations and recombine in pairs so that their
number is expected to decrease very slowly.
Therefore, the interface of an infinite system in the rough phase
does not propagate at constant velocity. Instead the squared
width and the average height grow {\it logarithmically} with
time as $\bar{h}(t) \sim w(t) \sim \log t$.
At criticality, this behavior crosses over to
$w^2(t) \sim \bar{h}(t) \sim \log t$. Thus the expected scaling
form reads
\begin{equation}
\label{WidthScaling}
w^2(L,t) \simeq \log \Bigl[ t^{-\gamma}
\, F(t/L^z) \Bigr]\, ,
\end{equation}
where $F$ is a universal scaling function
and $\gamma$ plays the role of a growth exponent.
For $\gamma=0.172(10)$ and $z=1.76(5)$ we obtain
a fairly accurate data collapse
(see Fig.~\ref{FIGDIMERSIMUL}a), supporting the supposition
that $z$ is the dynamic exponent of the PC universality class.
Similarly the densities $n_k$ (see Eq.~(\ref{DensitiesNk}))
are found to decay at criticality as $n_k(t) \sim t^{-\delta_k}$ with
\begin{equation}
\label{exponents}
\delta_0=0.280(10)
\, , \
\delta_1=0.200(15)
\, , \
\delta_2=0.120(15)
\, ,
\end{equation}
where $\delta_0=\beta/\nupar$ is the usual cluster
survival exponent of the PC class.

Using the particle interpretation of Sec.~\ref{GRCOUPLEDSEC},
the temporal evolution of the $A$-particles resembles a BAWE.
Similarly, the $B$-particles perform an effective BAWE on top of
inactive islands of the $A$-system. Since $A$-particles
instantaneously create $B$-particles, the two subsystems are
coupled by the effective reaction $A \rightarrow A+B$ at infinite rate.
As this reaction does not modify the configuration of the $A$-particles,
it couples the two subsystems only in one direction without
feedback. On the other hand, the RSOS condition~(\ref{RSOS})
introduces an effective feedback so that the $A$-particles are
not completely decoupled from the $B$-particles. However,
it seems that the inhibiting influence of the $B$-particles
does not affect the critical behavior of the $A$-particles.
Similarly, the $C$-particles are coupled to the
$B$-particles by the effective reaction $B \rightarrow B+C$.
Therefore, the dimer model resembles a unidirectionally coupled
sequence of BAWE's, corresponding to the reaction scheme
\begin{alignat*}{4}
&A \rightarrow 3A &\qquad &B \rightarrow 3B
&\qquad &C \rightarrow 3C  \\
&2A \rightarrow \vacancy &\qquad &2B \rightarrow  \vacancy
&\qquad &2C \rightarrow \vacancy  \\
&A \rightarrow A+B      &\qquad &B \rightarrow B+C
&\qquad &C \rightarrow C+D  & \qquad\text{etc.}
\end{alignat*}
In fact, as shown in Fig.~\ref{FIGDIMERSIMUL} b-c, a coupled
hierarchy of three BAWE's shows the same critical behavior
as the dimer model at the first few layers.

%==============================================================================
\subsection{Nonequilibrium wetting transitions}
%==============================================================================
%
\label{GRWETTING}
Wetting phenomena are observed in various physical systems where
a bulk phase (e.g. a gas or liquid) is brought into contact with
a wall or a substrate. Because of interactions between bulk phase
and surface, a thin layer of another phase may 
be formed which is attracted
to the substrate. The thickness of the layer fluctuates
and may depend on various parameters such as temperature
or chemical potential. As some of these parameters are
varied, the thickness of the layer may diverge, leading to a
wetting transition.
Theoretical models for wetting phenomena ignore the details
of molecular interactions between surface, layer and bulk phase. Instead,
they characterize the system by an interface without overhangs
that separates the two phases.
The configuration of the interface is given in terms
of the height $h(\xvec) \geq 0$ of the interface at point $\xvec$
on the surface. Within this approach, wetting transitions 
may be viewed as the unbinding of the interface from the wall.

\begin{figure}
\epsfxsize=60mm
\centerline{\epsffile{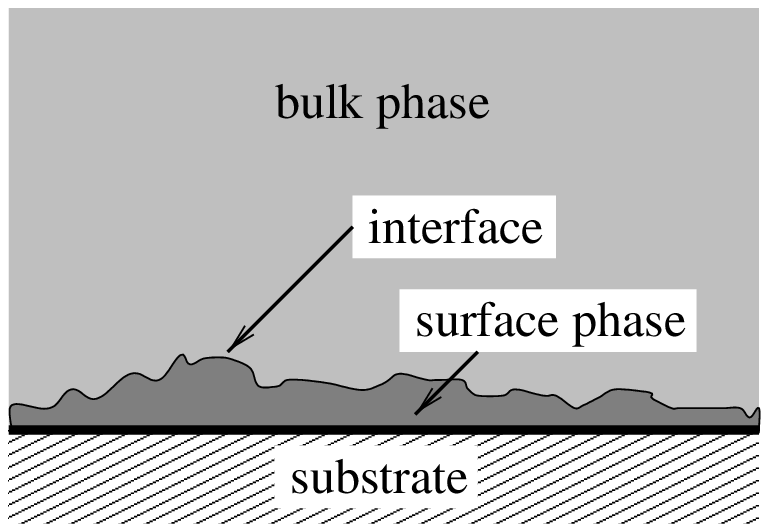}}
\smallcaption{
\label{FIGWETTING}
Wetting of a surface -- schematic illustration.
}
\end{figure}

Wetting transitions at thermal equilibrium have been
theoretically studied and experimentally observed
in a large variety of systems (for a review, see
Ref.~\cite{Dietrich88}).  Models for equilibrium wetting
are usually defined by an effective Hamiltonian of the form
\begin{equation}
\label{Hamiltonian}
{\cal H} = \int d^{d-1}x \ \bigl[ \frac{\sigma}{2}
(\nabla h)^2 + V[h(\xvec)] \bigr] \,,
\end{equation}
where $\sigma$ denotes the surface tension of the interface
and $d-1$ the dimension of the
interface~\cite{KrollLipowsky82,BHL83,FisherHuse85}.
The potential $V[h(\xvec)]$ yields the effective interaction between
wall and interface. Usually it contains an attractive component
binding the interface to the wall. However, as temperature
or other parameters are varied, the attractive component
of the potential may become so weak that the potential is longer
able to bind the interface, leading to a wetting transition.
In $d=2$ dimensions one usually distinguishes between
{\em critical} and {\em complete} wetting.
Critical wetting refers to the divergence of
the interface width when the wetting transition
is approached by moving along the coexistence curve of
the bulk and surface phases. On the other hand,
complete wetting refers to the divergence of the
interface width when the chemical potential difference
between the two phases is varied, moving towards
the coexistence curve. Critical and complete wetting transitions are
associated with generally different critical exponents.

While equilibrium wetting transitions are well understood,
the investigation of wetting transitions under {\em nonequilibrium}
conditions has started only recently. Here the surface layer is
adsorbed to the wall by a growth process whose dynamics,
unlike in equilibrium processes, does not obey detailed
balance. As expected, the resulting critical exponents differ from
those observed at thermal equilibrium.

%
%---------------------------------------------------------------------------
\headline{A lattice model for nonequilibrium wetting}
%---------------------------------------------------------------------------
%
A simple lattice model for nonequilibrium wetting can be defined as
follows~\cite{HLMP97}. As in Sec.~\ref{GREDGES}, the interface
is given by height variables~$h_i=0,1,\ldots,\infty$. For
each update a site $i$ is selected randomly and one of the
following moves is carried out:
\begin{alignat}{2}
\label{WettingRules}
& h_i \rightarrow h_i+1 & \text{ with prob. } &  q/(q+p+1) \ , \nonumber \\
& h_i \rightarrow \min(h_{i-1},h_{i},h_{i+1})
    & \text{ with prob. } & 1/(q+p+1)  \ , \\    \nonumber
& h_i \rightarrow h_i-1 \qquad  \text{if} \ \ h_{i-1}=h_{i}=h_{i+1} \qquad
    & \text{ with prob. } & p/(q+p+1) \ .
\end{alignat}
The selected move will be rejected if it would result
in a violation of the RSOS constraint $|h_i-h_{i+1}| \leq 1$.
In addition, a hard-core wall at zero height $h=0$ is introduced,
i.e., a process is only carried out if the resulting interface
heights are non-negative. Generally the processes defined above
do not satisfy detailed balance.

\begin{figure}
\epsfxsize=90mm
\centerline{\epsffile{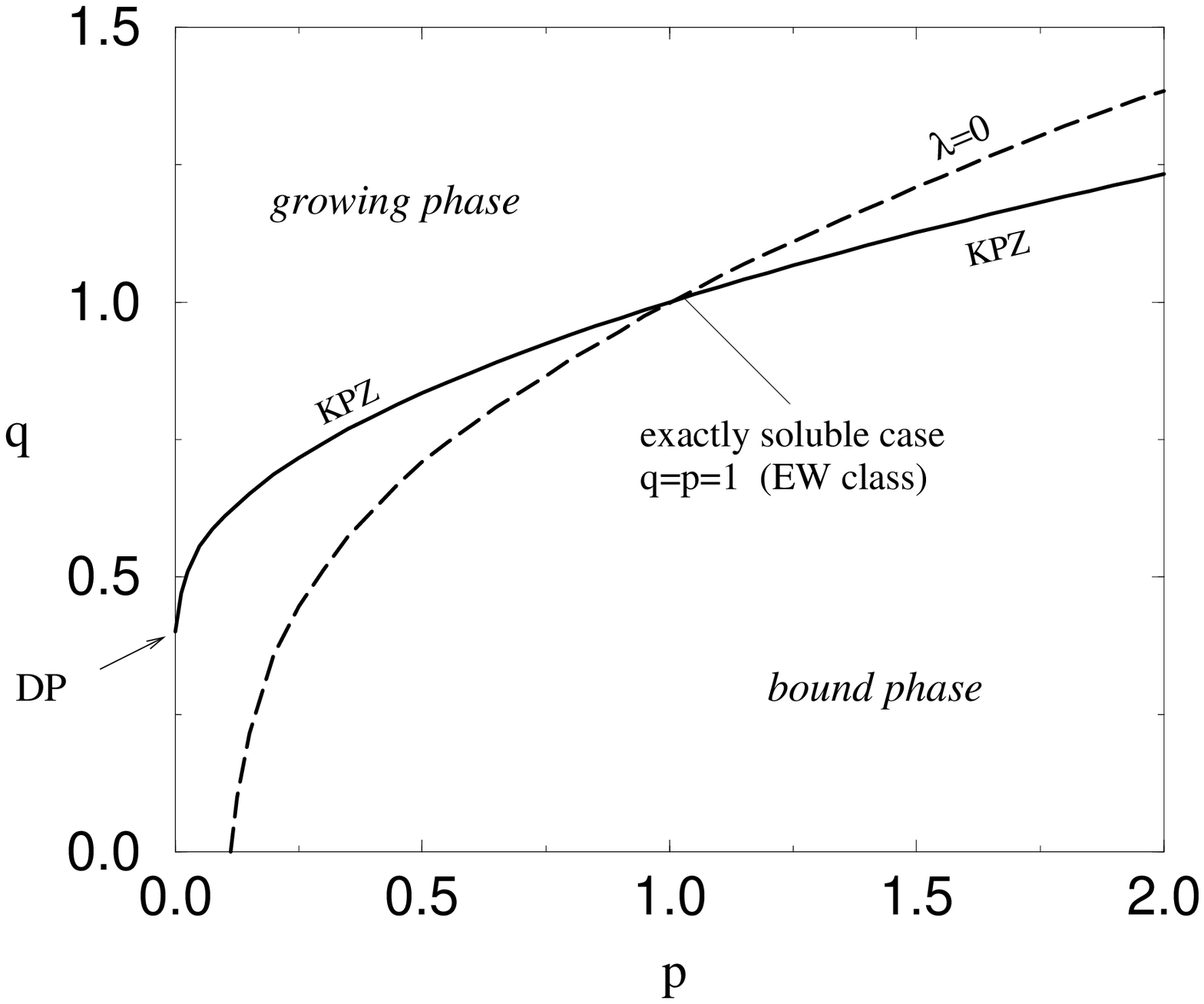}}
\smallcaption{
\label{FIGWETPH}
Phase diagram of the nonequilibrium wetting model.
The wetting transition takes place along the solid line.
Along the dashed line the growth velocity in the model without
a wall does not depend on a global tilt of the interface,
indicating that the effective KPZ nonlinearity vanishes.
Both lines intersect in the point $q=p=1$,
where the microscopic processes obey detailed balance.}
\end{figure}
By varying the relative rates of these processes, a transition
from a binding to a non-binding phase is found. This wetting
transition can be understood as follows.
Without the wall for fixed $p>0$, the parameter $q$ controls the
mean growth velocity of the interface. This velocity may be positive or
negative and vanishes at some critical value $q=q_c$.
On large time scales a lower wall will  only affect
the interface dynamics if the interface moves downwards,
i.e. $q \le q_c$, leading to a smooth interface.
In the moving phase $q>q_c$, however, the interface
does not feel the wall and evolves into a rough state
(see Fig.~\ref{FIGWETPH}). It is obvious that in the moving phase
the interface velocity scales as
$v\sim(q-q_c)^y$ with $y=1$. In the smooth phase $q<q_c$,
the expected scaling for bottom layer occupation and width is
given by
\begin{equation}
\rho_0\sim(q_c-q)^{x_0}\,, \quad\quad
w\sim(q_c-q)^{-\gamma}\,,
\end{equation}
where $x_0$ and $\gamma$ are certain critical exponents.
The above model includes two special cases, namely $p=0$,
where the interface cannot move below its actual minimum height,
and $p=1$, where the dynamic rules satisfy detailed balance.

%---------------------------------------------------------------------------
\headline{The exactly solvable case $p=1$}
%---------------------------------------------------------------------------
%
For $p=1$ the dynamic rules~(\ref{WettingRules})
can be mapped onto an exactly solvable equilibrium model
which exhibits a transition to complete wetting.
For $q<1$  the probability of finding the interface in
the configuration $\{h_1,\ldots,h_N\}$ is given by the distribution
\begin{equation}
\label{EquilibriumDist}
P(h_1,\ldots,h_N) = P(H) = Z^{-1} \,  q^{H(h_1,\ldots,h_N)}\,,
\end{equation}
where $H=H(h_1,\ldots,h_N)=\sum_{i=1}^Nh_i$ is
the sum of all heights and $Z=\sum_{h_1,\ldots,h_N} q^{H}$
denotes the partition sum running over all possible
interface configurations. Eq.~(\ref{EquilibriumDist})
can be proven by verifying the detailed-balance
condition
\begin{equation}
w(\sigma_H \rightarrow \sigma_{H+1}) /
w(\sigma_{H+1} \rightarrow \sigma_H) = q\,. \quad \quad (p=1)
\end{equation}
which is consistent with the hard wall constraint $h_i \geq 0$.
The steady state distribution~(\ref{EquilibriumDist}) does not
exist for $q > 1$ where the interface propagates at constant
velocity. The critical exponents
$x_0$ and $\gamma$ can be computed by analyzing
the transfer matrix acting in spatial
direction~\cite{LeeuwenHilhorst81,Burkhardt81}
\begin{equation}
\label{TransferMatrixDefinition}
T_{h,h'} = \left\{
\begin{array}{ccl}
q^{h} && \mbox{if} \, |h-h'| \leq 1 \\
0 && \mbox{otherwise}
\end{array} \right.\,,
\end{equation}
where $h,h'\geq 0$. Steady state properties can be derived
from the eigenvector $\phi$ that corresponds to the
largest eigenvalue $\mu$ of the transfer matrix
$\sum_{h'=0}^\infty T_{h,h'}\phi_{h'} = \mu\,\phi_h$. From
the squares of the eigenvector components various
steady state quantities can be derived. For example,
the probability $\rho_h$ of finding the interface at height $h$
is given by $\rho_h=\phi_h^2/\sum_{h'}\phi_{h'}^2$.
Close to criticality, where $\epsilon=1-q$ is small, one can carry out
the continuum limit $\phi_h \rightarrow \phi(\tilde{h})$, replacing the
discrete heights $h$ by real-valued heights~$\tilde{h}$. Then,
the above eigenvalue problem turns into a differential
equation~\cite{LeeuwenHilhorst81} which, to leading order
in $\epsilon$, is given by
\begin{equation}
\label{DifferentialEquation}
\Bigl( \frac{\partial^2}{\partial \tilde{h}^2} + (3-\mu) -
3 \epsilon \tilde{h} \Bigr)\, \phi(\tilde{h}) = 0\,.
\end{equation}
Simple dimensional analysis indicates that the height
variables scale as $h \sim \epsilon^{-1/3}$ wherefore the
width diverges as $w^2 \sim \epsilon^{-2/3}$. Similarly
one can show by elementary calculations that
$\rho_0 \sim \epsilon$. The critical exponents
for $p=1$ are thus given by
\begin{equation}
\label{EWCriticalExponents}
x_0=1\,,\quad\quad
\gamma=1/3\,.
\end{equation}
%
%
%The critical behavior associated with the transition at $p=1$
%is that of {\em complete} wetting, in contrast to critical wetting
%transitions where the interface width diverges with $\gamma = 1$.
%This observation is plausible from the physical point of view as
%$1-q$ may be interpreted as a chemical potential difference between
%the two coexisting phases.

%---------------------------------------------------------------------------
\headline{The generic case $p \neq 1$}
%---------------------------------------------------------------------------
%
In the nonequilibrium case $p \neq 1$ the critical exponents
can only be approximated numerically. For $p=0$, where
the interface cannot move below its actual minimum height,
the hard-core wall becomes irrelevant and the
model reduces to a growth process similar to the one discussed
in Sec.~\ref{GREDGES}, belonging to the universality
class of unidirectionally coupled directed percolation.
For $p=1$ the numerical estimates are consistent with
the previously derived exact results.
For $0 < p < 1$ a very slow {\it crossover} to a different behavior
with critical exponents $y=1.00(3)$, $x_0=1.5(1)$, and
$\gamma=0.41(3)$ is observed. Similar results are obtained for $p>1$
except for $x_0$ which is close to 1 in this case.

It is believed that this type of nonequilibrium wetting can be
modeled by the KPZ equation~(\ref{KPZEquation}) with an additional
term for the effective interaction between the wall and the
interface~\cite{TGM97,MunozHwa98}
\begin{equation}
\label{GeneralizedKPZEquation}
\frac{\partial h(r,t)}{\partial t} = v + \sigma\nabla^2 h(r,t)
+ \lambda (\nabla h(r,t))^2 + \noise(r,t) -
\frac{\partial V[h(r,t)]}{\partial h(r,t)}\ .
\end{equation}
Dimensional analysis suggests that the width exponent
described by this equation should be given by
$\gamma = (2-z)/(2z-2)$, where $z=3/2$ is the dynamic exponent
of the KPZ universality class, yielding $\gamma =1/2$. The numerical
estimates of $\gamma \simeq 0.41$ are smaller, presumably because
of the very slow crossover to the exactly solvable case $\gamma=1/3$.
In addition, Eq.~(\ref{GeneralizedKPZEquation}) suggests that
the bottom layer occupation $\rho_0$ should be proportional to the
inverse correlation length, hence $x_0=\nuperp = 1/(2z - 2)=1$.
However, this scaling argument seems to hold only for $p>1$, whereas for
$0<p<1$ much larger values for $x_0$ are observed. As shown in
Ref.~\cite{HLMP97}, the changing sign of the
coefficient $\lambda$ in Eq.~(\ref{GeneralizedKPZEquation}) leads
to {\em different} universality classes in both cases,
corresponding to the distinction between an `upper' and
a `lower' wall in Ref.~\cite{MunozHwa98}. In fact, comparing the
growth velocities of a flat and a tilted interface in
absence of a wall it is found that the
nonlinear term  $(\nabla h(r,t))^2 $ is indeed non-vanishing
in the $(p,q)$ plane, except for the dashed line shown in
Fig.~\ref{FIGWETPH}. Therefore, moving along the transition line,
the sign of $\lambda$ changes precisely at the integrable point,
leading to a different exponent $x_0$.

%---------------------------------------------------------------------------
\headline{Nonequilibrium wetting of an attractive substrate}
%---------------------------------------------------------------------------
%
%
%
\begin{figure}
\epsfxsize=150mm
\centerline{\epsffile{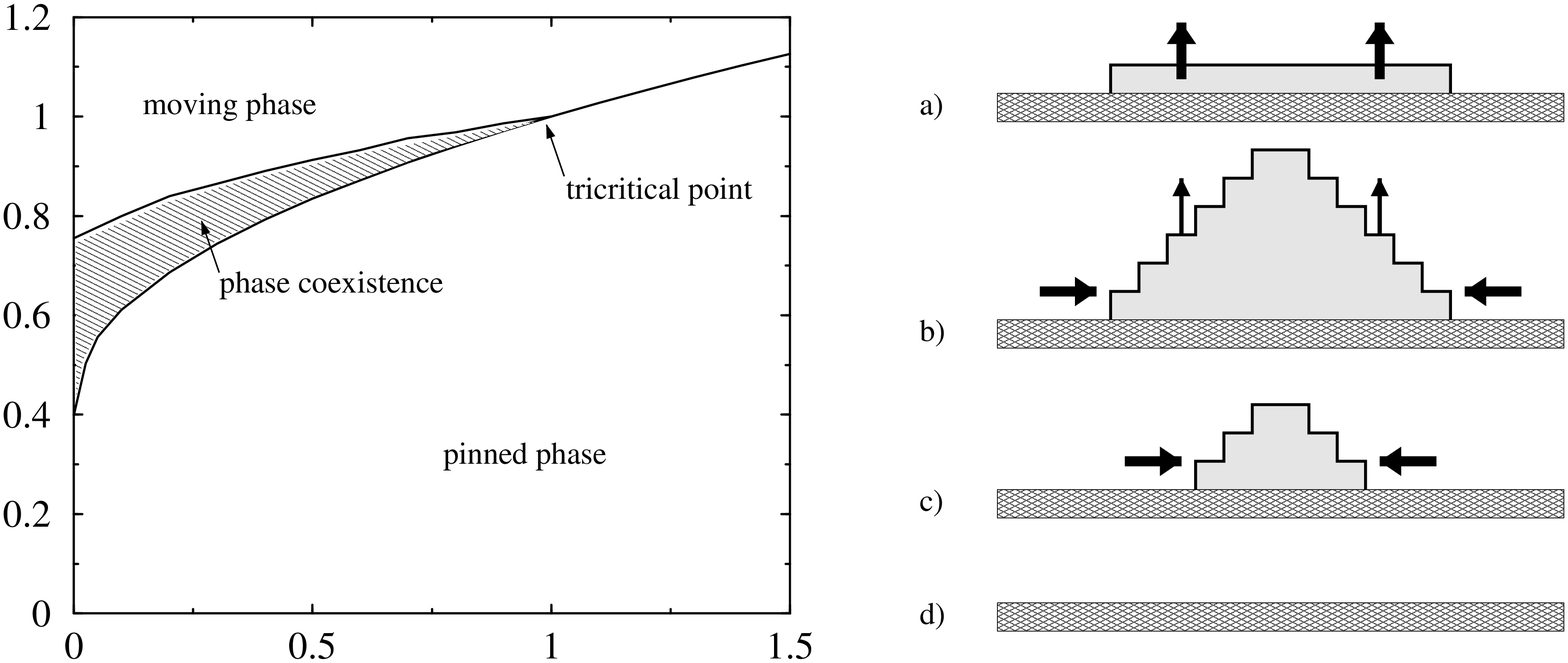}}
\smallcaption{
\label{FIGFIRSTORDER}
Phase coexistence in a nonequilibrium wetting model.
Left: Phase diagram of the wetting model with an attractive
substrate $q_0=0.2$. Right: Elimination of an island in
the coexistence phase. The flat island (a) grows quickly
until it reaches a triangular shape (b). Because of the
KPZ nonlinearity, the deposition rate decreases, leading
to a laterally shrinking island (c). Finally the island
is eliminated, ensuring phase coexistence.
}
\end{figure}
In the above wetting model the substrate is introduced as a
hard wall at zero height. Therefore, the model neglects
interactions between the substrate in the surface layer.
Loosely speaking, the free energies of the exposed
and the wetted substrate are assumed to be equal. In order to
describe more realistic situations, the model has to be generalized
by taking interactions between the substrate and the surface
layer into account. This can be done by introducing a
{\em modified growth rate} $q_0$ at zero height~\cite{HLMP00}.
Obviously the attractive short-range interaction at the bottom
layer is a surface effect. Therefore, the critical point
$q_c$ remains unchanged.
However, if the interaction is strong enough, the
transition becomes discontinuous. In the equilibrium case
$p=1$ it can be proven by using transfer matrix methods
that for $q_0<2/3$ there is a first-order
wetting transition.

In the nonequilibrium case the morphology of the phase transition
depends on the sign of the KPZ nonlinearity. For $p>1$ the
emerging picture is essentially the same as for $p=1$, although
with different critical exponents. For $p<1$, however, there is a
whole region in the phase diagram where the pinned and the moving
phase {\em coexist}. As illustrated in Fig.~\ref{FIGFIRSTORDER},
islands generated by fluctuations quickly grow until they reach an
almost triangular shape. Since the KPZ nonlinearity is negative,
adsorption processes at the inclined edges of the islands are strongly
suppressed, allowing the attractive interaction to reduce the size
of the island in lateral direction. This ensures the stability of
the pinned phase in parts of the phase diagram where a free
interface would move away from the wall. This model demonstrates
that nonequilibrium effects do not only lead to different
critical exponents but may also change the whole 
phase structure of a model.

%===========================================================================
\vspace{5mm} \noindent {\Large \bf Acknowledgments}
%===========================================================================

\vspace{4mm} \noindent This review is based on my Habilitation
thesis written at the Max-Planck-Institut f{\"u}r Physik
komplexer Systeme and submitted to the Freie Universit{\"a}t
Berlin in June 1999. I would like to thank I.~Peschel, who made
this work possible, as well as E.~Domany and D.~Mukamel, who
introduced me to the field of nonequilibrium phase transitions. I
am also grateful to V. Rittenberg, whom I owe my experience in the
field of integrable equilibrium systems. I am indebted to 
U.~Alon,
M.~Antoni, 
M.~R. Evans, 
M.~Falcke, 
Y.~Y.~Goldschmidt, 
M.~Henkel,
M.~J.~Howard, 
H.~K.~Janssen, 
A.~Jim{\'e}nez-Dalmaroni,
H.~M.~Koduvely, 
K.~Krebs, 
R.~Livi,
G.~\'Odor, 
M.~Pfannm\"uller, 
A.~Politi, 
Y.~Rozov,
S.~Ruffo, 
S.~Sandow,
G.~Schliecker, 
H.~Simon, 
D.~Stauffer, 
U.~C.~T\"auber,
B.~Wehefritz, and 
J.~S.~Weitz, 
who essentially contributed as my
collaborators to the presented results. I also express my thanks
to many other colleagues for numerous stimulating discussions.
Particularly I would like to thank P. Fulde and the
Max-Planck-Institut f\"ur Physik komplexer Systeme for generous
support.

%%%%%%%%%%%%%%%%%%%%%%%%%%%%%%%%%%%%%%%%%%%%%%%%%%%%%%%%%%%%%%%%%%%%%%%%%%%%
%               APPENDICES
%%%%%%%%%%%%%%%%%%%%%%%%%%%%%%%%%%%%%%%%%%%%%%%%%%%%%%%%%%%%%%%%%%%%%%%%%%%%

\newpage
\appendix

%===========================================================================
\section{Vector space notation and tensor products}
%===========================================================================
%
\label{APPVECSEC}
A one-dimensional lattice model, whose sites $i=1,\ldots,N$
are either occupied ($s_i=1$) or vacant ($s_i=0$), can
be in $2^N$ different states $s=\{s_1,s_2,\ldots,s_N\}$.
In order to represent the probability distribution $P_t(s)$
as a vector in a $2^N$-dimensional vector space let us define
an orthogonal set of basis vectors $|s\rangle$ corresponding
to the configurations of the system. Using the representation
\begin{equation}
\label{LocalBasisVectors}
|0\rangle = \begin{pmatrix} 1\\0 \end{pmatrix}
\ , \qquad
|1\rangle = \begin{pmatrix} 0\\1 \end{pmatrix}
\end{equation}
the basis vectors are given by
\begin{equation}
|s\rangle = |s_1\rangle \otimes |s_2\rangle
\otimes \ldots \otimes |s_N\rangle \ ,
\end{equation}
where '$\otimes$' denotes the tensor product of two vectors:
\begin{equation}
\begin{pmatrix} a_1\\a_2 \end{pmatrix} \otimes
\begin{pmatrix} b_1\\b_2 \end{pmatrix} =
\begin{pmatrix} a_1b_1\\a_1b_2\\a_2b_1\\a_2b_2 \end{pmatrix}
\ .
\end{equation}
Row vectors $\langle s|$ are the transposed vectors of $|s\rangle$.
In this vector space the probability distribution $P_t(s)$ can
be represented by the vector
\begin{equation}
|P_t\rangle = \sum_s P_t(s)\,|s\rangle \, .
\end{equation}
Defining the sum vector over all states
\begin{equation}
\sumstate = \sum_s \langle s|= (1,1)^{\otimes N} = (1,1,\ldots,1)
\end{equation}
the normalization of the probability distribution
can be simply expressed as $\sumstate P_t\rangle=1$.
Similarly the ensemble average $\langle A(t) \rangle$
of any observable $A$ can be expressed as
\begin{equation}
\langle A(t) \rangle = \sumstate  A | P_t\rangle \ .
\end{equation}
The empty lattice is represented by the vector
$|vac\rangle=|0\rangle^{\otimes N}$.
Local operators act only on a finite number of adjacent sites.
For example, a single-site operator $A_i$ can be written as
\begin{equation}
A_i = {\bf 1} \otimes {\bf 1} \otimes \ldots
\underbrace{\otimes A \otimes}_{i\mbox{-th position}}
\ldots \otimes {\bf 1} \ ,
\end{equation}
where ${\bf 1}$ and $A$ are $2\times 2$ matrices.
Here the tensor product of two matrices is defined by
\begin{equation}
\begin{pmatrix}
a_1 & a_2 \\ a_3 & a_4
\end{pmatrix} \otimes \begin{pmatrix}
b_1 & b_2 \\ b_3 & b_4
\end{pmatrix} = \begin{pmatrix}
a_1b_1 & a_1b_2 & a_2b_1 & a_2b_2 \\
a_1b_3 & a_1b_4 & a_2b_3 & a_2b_4 \\
a_3b_1 & a_3b_2 & a_4b_1 & a_4b_2 \\
a_3b_3 & a_3b_4 & a_4b_3 & a_4b_4
\end{pmatrix}
\, .
\end{equation}
Similarly
one can define two-site operators $B_{i,i+1}$, where
$B$ is a $4 \times 4$ matrix.
The above notations can be easily generalized to systems with
$n>1$ particle species by introducing local vectors
with $n$ components in Eq.~(\ref{LocalBasisVectors}).

%===========================================================================
\section{Derivation of the effective action}
%===========================================================================
%
\label{ACTION}
In order to derive the effective action of Reggeon field theory
by integration of the noise, we first introduce a
{\em response field} $\tilde{\phi}(\xvec,t)$. This allows the
$\delta$-function to be expressed as an oscillating integral
\begin{equation}
Z \sim \int D\noise\,P[\noise] \int D\phi\,D\tilde{\phi}\,I[\phi,\tilde{\phi}] \,
\exp\Bigl[ i \int d^dx\,dt\,\, \tilde{\phi} \bigl(
\timederivative\phi - \diff \nabla^2 \phi -
\crit \phi + \lambda\phi^2 - \noise \bigr) \Bigr]
\ ,
\end{equation}
where $I[\phi,\tilde{\phi}]$ denotes the Jacobian.
After a Wick rotation in the complex plane the $\phi$-dependent noise
contribution can be separated by
\begin{equation}
\begin{split}
Z \sim \int D\phi\,D\tilde{\phi}\,I[\phi,\tilde{\phi}] \,
& \exp\Bigl[ - \int d^dx\,dt\,\, \tilde{\phi} \bigl(
\timederivative\phi - \diff \nabla^2 \phi -
\crit \phi + \lambda\phi^2 \bigr) \Bigr]  \\
& \times \,
\int D\noise\, P[\noise]\,\exp\Bigl(
\int d^dx\,dt\,\, \tilde{\phi} \noise \Bigr)
\ .
\end{split}
\end{equation}
Because of the correlations~(\ref{DPNoise}) the probability
distribution $P[\noise]$ is given by
\begin{equation}
P[\noise] \;=\; f[\phi]\,\exp
\Bigl(-\int d^dx\,dt\,\,\frac{\noise^2(\xvec,t)}
{2\namp\phi(\xvec,t)}\Bigr)
\ ,
\end{equation}
where $f[\phi]$ is a (field-dependent) normalization factor. This
allows the noise to be integrated
\begin{equation}
\begin{split}
\int D\noise\, P[\noise]\,\exp\Bigl(
\int d^dx\,dt\,\, \tilde{\phi} \noise \Bigr) \;&=\;
f[\phi]\int D\noise\,\exp\Bigl[\int d^dx\,dt
\Bigl( \tilde{\phi}\noise-\frac{\noise^2}{2 \namp \phi} \Bigr) \Bigr]
\;\\&=\;
f[\phi]\int D\noise\,\exp\Bigl[\int d^dx\,dt
\Bigl( \frac{\namp}{2}\tilde{\phi}^2\phi-
\frac{\noise^2}{2 \namp \phi} \Bigr) \Bigr]
\;\\&=\;
\bar{f}[\phi] \, \exp \Bigl( \frac{\namp}{2}
\int d^dx\,dt \,\, \phi \tilde{\phi}^2 \Bigr)
\ ,
\end{split}
\end{equation}
where we used Gaussian integration of the form
\begin{equation}
\int_{-\infty}^{+\infty} d\eta \,
\frac{1}{\sqrt{2 \pi \noise \phi}} \, \exp\Bigl(\tilde{\phi} \eta -
\frac{\eta^2}{2 \noise \phi}\Bigr) \;=\;
\exp\Bigl(\frac12 \noise \phi \tilde{\phi}^2\Bigr) \ .
\end{equation}
The resulting partition function reads
\begin{equation}
\label{DPZsum}
Z \sim \int D\phi \, D\tilde{\phi}\, I'[\phi,\tilde{\phi}]\,
\exp\Bigl[-\int d^dx\,dt \,\, \Bigl(
\tilde{\phi}\,[\timederivative\phi-\diff \nabla^2 \phi - \crit \phi] \,+\,
\lambda \tilde{\phi} \phi^2 \,-\,
\frac{\namp}{2} \tilde{\phi}^2 \phi \Bigr) \Bigr]
\ .
\end{equation}
It is convenient to symmetrize the cubic terms in the
partition function. To this end we rescale the fields
by
\begin{equation}
\phi'=\sqrt{2\lambda/\namp}\phi \ , \qquad
\tilde{\phi}'=\sqrt{\namp/2\lambda}\tilde{\phi}\ , \qquad
\namp'=\sqrt{2\namp\lambda} \ .
\end{equation}
In order to keep the action symmetrized during
the RG procedure, one has to introduce an additional coefficient
$\tau$ in front of the time derivative.
Dropping the primes the effective action
$S=S_0+S_{int}$ is given by the expressions
(\ref{ActionS0}) and (\ref{ActionSint}).
Alternatively the action may directly be derived from the master equation
of a contact process by introducing bosonic creation and annihilation
operators (for this standard procedure we refer
to Refs.~\cite{Doi76,GrassbergerScheunert80,Peliti84}).

%===========================================================================
\section{Shell integration}
%===========================================================================

\label{SHELLINT}
\noindent
The one-loop integrals in Wilson's renormalization group
approach take the form
\begin{equation}
I(k) \;=\; \frac{1}{(2\pi)^d}
\int_> \,d^d k' \, \, f(k^2 ,k \scalarprod k',{k'}^2 )
\end{equation}
where '$>$' denotes integration in the momentum shell
$\cutoff(1-\smalldev) < |k'| \leq \cutoff$ .
This integral can be written as
\begin{equation}
\label{ShellInt}
I(k) \;=\; \frac{\smalldev\cutoff^d}{(2\pi)^d}\, S_{d-1}
\int_0^\pi \, d\theta \, \sin^{d-2}\theta \,\,
f(k^2,\cutoff\,|k|\,\cos\theta, \cutoff^2) \ ,
\end{equation}
where $S_d$ and $V_d$ denote surface area and volume
of a $d$-dimensional sphere:
\begin{equation}
S_d \;=\; \frac{2\,\pi^{d/2}}{\Gamma(d/2)}
\,, \qquad
V_d \;=\; \frac{S_d}{d} \ .
\end{equation}
We also use the notation $K_d=S_d/(2\pi)^d$. For easy reference
we listed some of the values for $S_d$, $K_d$, and $V_d$
in Table~\ref{TableKd}.
\begin{table}
\begin{center}
\begin{tabular}{|c||c|c|c|c|c|c|}
\hline
$d$   & $1$ & $2$ & $3$ & $4$ & $5$ & $6$
\\ \hline \hline
$S_d$ & $2$ & \ $2\pi$ \ & $4\pi$ & $2\pi^2$ & $8\pi^2/3$ & $\pi^3$
\\ \hline
$K_d$ & $\frac{1}{\pi}$ & $\frac{1}{2\pi}$ & $\frac{1}{2\pi^2}$ &
    $\frac{1}{8\pi^2}$ & $\frac{1}{12\pi^3}$ & $\frac{1}{64\pi^3}$
\\[1mm] \hline
$V_d$ & $2$ & \ $\pi$ \ & $4\pi/3$ & $\pi^2/2$ & $8\pi^2/15$ & $\pi^3/6$
\\ \hline
\end{tabular}
\smallcaption{
\label{TableKd}
Surface area $S_d$, $K_d=S_d/(2\pi)^d$,
and the volume $V_d$ of a $d$-dimensional sphere.}
\end{center}
\end{table}

To evaluate Eq.~(\ref{ShellInt}) it is often helpful to use the formulas
\begin{eqnarray}
\int_0^\pi \, d\theta \,\, \sin^{d-2} \theta
&=& \frac{S_d}{S_{d-1}} \ , \\
\int_0^\pi \, d\theta \,\, \sin^{d-2} \theta \, \cos^2\theta
&=& \frac{S_d}{d\,S_{d-1}} \ .
\end{eqnarray}
In particular, if the function $f$ does not depend on
$\theta$ the integral $I(k)$ reduces to
\begin{equation}
I(k) = \frac{1}{(2\pi)^d}
\int_> \,d^dk' \, \, f(k^2,{k'}^2) =
\smalldev K_d \cutoff^d \, f(k^2,\cutoff^2).
\end{equation}
\newpage
%===========================================================================
\section{One-loop integrals for directed percolation}
%===========================================================================
%
\label{INTEGRALS}
\noindent
The integral for propagator renormalization reads
\begin{eqnarray}
J^P &=& \int_> Dk'\omega' \,\, G_0(\frac{k}{2}+k', \frac{\omega}{2}+\omega')\,
 G_0(\frac{k}{2}-k', \frac{\omega}{2}-\omega')\\
 &=&  \int_> Dk'\omega' \,\,  \nonumber
 \frac{1}{\Bigl( \diff(\frac{k}{2}+k')^2-\crit-i\tau
 (\frac{\omega}{2}+\omega')\Bigr)\,
 \Bigl( \diff(\frac{k}{2}-k')^2-\crit-i\tau (\frac{\omega}{2}-\omega')\Bigr)\,}
 \ .
\end{eqnarray}
Denoting $A^\pm = \diff(\frac{k}{2}\pm k')^2-\crit-i\tau\frac{\omega}{2}$ and
integrating with respect to the pole $i\tau \omega'=A^+$ we obtain
\begin{eqnarray}
\label{JPIntegral}
J^P &=& \frac{1}{(2\pi)^{d+1}} \int_> d^dk' \int_{-\infty}^{\infty} d\omega'\,\,
\frac{1}{(A^+-i \tau \omega')\,(A^-+i\tau\omega')} \nonumber \\
&=& \frac{2 \pi i}{(2 \pi)^{d+1} i \tau} \int_> 
d^dk' \frac{1}{A^++A^-}\nonumber \\
&=& \frac{1}{2\,(2\pi)^{d} \,\tau} \int_> d^d k' \frac{1}
{\frac{1}{4} \diff k^2 + \diff {k'}^2 - \crit - \frac{i}{2} \tau \omega} \\
&=&
\frac{\smalldev K_d \cutoff^4}{2\tau\Bigl( \nonumber
\frac{1}{4} \diff k^2 + \cutoff^2\diff  - \crit - \frac{i}{2} \tau \omega \Bigr)}
\ .
\end{eqnarray}
The integral for vertex renormalization is given by
\begin{align}
J^V &= \int_> Dk\omega G_0^2(k,\omega)\,G_0(-k,-\omega) \nonumber \\
&= \frac{1}{(2 \pi)^{d+1}}\int_>d^dk \int_{-\infty}^\infty d\omega
\frac{1}{(\diff k^2-\crit-i\tau \omega)^2(\diff k^2-\crit+i \tau \omega)}
\ .
\end{align}
Integration with respect to the poles $i \tau \omega = \diff k^2-\crit$
yields the result
\begin{equation}
J^V \;=\; \frac{2 \pi i}{(2 \pi)^{d+1} i \tau}
\int_> d^dk \frac{1}{4(\diff
k^2-\crit)^2} \;=\;
\frac{\smalldev K_d \cutoff^d}{4 \tau (\cutoff^2\diff-\crit)^2}
\ .
\end{equation}
\newpage
%
%===========================================================================
\section{Notations}
%===========================================================================
%
%
\headline{Frequently used symbols}
\begin{tabular}{ll}
$\hspace{15mm}$     & \\

$A,B,C,\ldots$  & particle species \\
$\vacancy$	& vacant site \\
$\alpha$        & roughness exponent \\
$\beta$         & density exponent \\
$\beta^{(k)}$   & density exponents at different hierarchy levels \\
$\delta$        & exponent for temporal decay \\
$\diff$         & diffusion constant \\
$\namp$         & noise amplitude \\
$d$             & spatial dimension of the system \\
$D,E,\bar{D},\bar{E}$   & matrices for matrix product states \\
$\Delta$        & distance from criticality \\
$\Delta(t)$	& Hamming distance (damage) \\
$\gamma$        & growth exponent \\
$\namp$		& noise amplitude \\
$h_i(t)$        & interface height at site $i$ \\
${\cal H}$      & Hamiltonian, internal energy \\
$<i>$           & set of nearest neighbors of site $i$ \\
$\theta$        & critical initial slip exponent \\
$\theta_n$      & critical exponents of the order parameters $M_n$ \\
$I_\ell$        & probability for empty interval of length $\ell$ \\
$J$         & coupling constant in equilibrium systems \\
$\crit$         & rate for offspring production in Langevin equations \\
$L$         & length of the system  \\
$\lambda$       & coefficient for nonlinear term in Langevin equations \\
$\rho(\xvec,t)$     & particle density \\
$\scalefac$     & dilatation parameter for scaling transformations \\
$\coupling$     & coupling constant between different levels \\
${\cal L}$      & Liouville operator \\
$M_n$           & order parameters for spontaneous symmetry breaking \\
$n_k(t)$        & density of sites at height $k$ above the bottom layer \\
$N$             & system size (total number of sites $N=L^d$) \\
$\nupar$        & temporal scaling exponent \\
$\nuperp$       & spatial scaling exponent \\
$p$             & percolation probability \\
$P_t(s)$        & probability to find the system
                  in state $s$ at time $t$ \\
$P_l(t)$        & local persistence probability \\
$P_g(t)$        & global persistence probability \\
$q$             & probability for interface growth \\
$s$             & state of the model \\
$s_i$           & local state $s_i=0,1,\ldots$ of lattice site $i$ \\
\end{tabular}

\newpage
\begin{tabular}{ll}
$\hspace{15mm}$     & \\
${\cal S}$      & field-theoretic action \\
$\sigma$        & control exponent for anomalous DP \\
$\sigma_i$      & local Ising spin $\sigma_i=\pm 1$ at site $i$ \\
$T$             & temperature \\
$v$             & interface velocity \\
$w_{s \rightarrow {s'}}$& transition rate from state $s$ to state $s'$ \\
$w(N,t)$        & interface width \\
$\cutoff$	& cutoff in momentum space \\
$\noise(\xvec,t)$   & noise field in Langevin equations \\
$\xi_\perp$     & spatial correlation length \\
$\xi_\parallel$ & temporal correlation length \\
$z$             & dynamic critical exponent \\
$Z$		& partition sum \\

\end{tabular}

\vspace{7mm}\noindent{\bf Abbreviations}\\*[3mm]
\begin{tabular}{ll}
BAWE    & branching-annihilating random walk with even number of offspring \\
CDP     & compact directed percolation \\
DK  & Domany-Kinzel (model) \\
DMRG  & density matrix renormalization group \\
DP  & directed percolation \\
DP2 & directed percolation with two absorbing states \\
DS  & damage spreading \\
EW  & Edwards-Wilkinson \\
IMF     & improved mean field \\
IPDF    & interparticle distribution function \\
KPZ & Kardar-Parisi-Zhang \\
MC  & Monte Carlo \\
MF  & mean field \\
MPS & matrix product state \\
PNG & polynuclear growth \\
RG  & renormalization group \\
RSOS    & restricted solid on solid \\
SOS & solid on solid
\end{tabular}
\newpage

%%%%%%%%%%%%%%%%%%%%%%%%%%%%%%%%%%%%%%%%%%%%%%%%%%%%%%%%%%%%%%%%%%%%%%%%%%%%%%
%               REFERENCES
%%%%%%%%%%%%%%%%%%%%%%%%%%%%%%%%%%%%%%%%%%%%%%%%%%%%%%%%%%%%%%%%%%%%%%%%%%%%%%

\renewcommand{\sAppendix}[1]{%
{\flushleft\large\bfseries\flushleft#1}}
\addcontentsline{toc}{section}{References}{}
\bibliographystyle{review}
\footnotesize
\baselineskip 0mm
\bibliography{review}
\end{document}